\documentclass[12pt]{article}
\usepackage{amssymb}
\usepackage{graphics}
\usepackage{psfrag}

\DeclareFontFamily{U}{rsf}{}
\DeclareFontShape{U}{rsf}{m}{n}{
  <5> <6> rsfs5 <7> <8> <9> rsfs7 <10-> rsfs10}{}
\DeclareMathAlphabet\Scr{U}{rsf}{m}{n}

\catcode`\@=11
\def\@citex[#1]#2{%
\if@filesw \immediate \write \@auxout {\string \citation {#2}}\fi
\@tempcntb\m@ne \let\@h@ld\relax \def\@citea{}%
\@cite{%
  \@for \@citeb:=#2\do {%
    \@ifundefined {b@\@citeb}%
      {\@h@ld\@citea\@tempcntb\m@ne{\bf ?}%
      \@warning {Citation `\@citeb ' on page \thepage \space undefined}}%
      {\@tempcnta\@tempcntb \advance\@tempcnta\@ne%
      \@tempcntb\number\csname b@\@citeb \endcsname \relax%
      \ifnum\@tempcnta=\@tempcntb %Number follows previous--hold on to it
        \ifx\@h@ld\relax%
          \edef \@h@ld{\@citea\csname b@\@citeb\endcsname}%
        \else%
          \edef\@h@ld{\ifmmode{-}\else--\fi\csname b@\@citeb\endcsname}%
        \fi%
      \else%   %  non-successor--dump what's held and do this one
        \@h@ld\@citea\csname b@\@citeb \endcsname%
        \let\@h@ld\relax%
      \fi}%
    \def\@citea{,\penalty\@highpenalty\,}%
  }\@h@ld
}{#1}}

\def\@citeb#1#2{{[#1]\if@tempswa , #2\fi}}
\def\@citeu#1#2{{$^{#1}$\if@tempswa , #2\fi }}
\def\@citep#1#2{{#1\if@tempswa , #2\fi}}

\def\bcites{         % cite with []'s
        \catcode`\@=11
        \let\@cite=\@citeb
        \catcode`\@=12
}

\def\upcites{         % cite with exponents
        \catcode`\@=11
        \let\@cite=\@citeu
        \catcode`\@=12
}

\def\plaincites{      % cite without brackets
        \catcode`\@=11
        \let\@cite=\@citep
        \catcode`\@=12
}

\newcount\hour
\newcount\minute
\newtoks\amorpm
\hour=\time\divide\hour by 60
\minute=\time{\multiply\hour by 60 \global\advance\minute by-\hour}
\edef\standardtime{{\ifnum\hour<12 \global\amorpm={am}%
        \else\global\amorpm={pm}\advance\hour by-12 \fi
        \ifnum\hour=0 \hour=12 \fi
        \number\hour:\ifnum\minute<10 0\fi\number\minute\the\amorpm}}
\edef\militarytime{\number\hour:\ifnum\minute<10 0\fi\number\minute}

\def\draftlabel#1{{\@bsphack\if@filesw {\let\thepage\relax
   \xdef\@gtempa{\write\@auxout{\string
      \newlabel{#1}{{\@currentlabel}{\thepage}}}}}\@gtempa
   \if@nobreak \ifvmode\nobreak\fi\fi\fi\@esphack}
        \gdef\@eqnlabel{#1}}
\def\@eqnlabel{}
\def\@vacuum{}
\def\marginnote#1{}
\def\draftmarginnote#1{\marginpar{\raggedright\scriptsize\tt#1}}
\def\draft{
        \pagestyle{plain}
        \overfullrule=2pt
        \oddsidemargin -.5truein
        \def\@oddhead{\sl \phantom{\today\quad\militarytime} \hfil
        \smash{\Large\sl DRAFT} \hfil \today\quad\militarytime}
        \let\@evenhead\@oddhead
        \let\label=\draftlabel
        \let\marginnote=\draftmarginnote
        \def\ps@empty{\let\@mkboth\@gobbletwo
        \def\@oddfoot{\hfil \smash{\Large\sl DRAFT} \hfil}
        \let\@evenfoot\@oddhead}
        \def\@eqnnum{(\theequation)\rlap{\kern\marginparsep\tt\@eqnlabel}%
        \global\let\@eqnlabel\@vacuum}  }

\def\section{\@startsection {section}{1}{\z@}{3.ex plus 1ex minus
 .2ex}{2.ex plus .2ex}{\large\bf}}
\def\subsection{\@startsection{subsection}{2}{\z@}{2.75ex plus 1ex minus
 .2ex}{1.5ex plus .2ex}{\bf}}

\def\appendix{{\newpage\section*{Appendix}}\let\appendix\section%
        {\setcounter{section}{0}
        \gdef\thesection{\Alph{section}}}\section}

\def\abstract{\if@twocolumn
\section*{Abstract}
\else %\small
\begin{center}
{\bf Abstract\vspace{-.5em}\vspace{0pt}}
\end{center}
\quotation
\fi}

\catcode`\@=12

\newcommand{\beq}{\begin{equation}}
\newcommand{\eeq}{\end{equation}}
\newcommand{\beqa}{\begin{eqnarray}}
\newcommand{\eeqa}{\end{eqnarray}}
\newcommand{\dd}{{\rm d}}

\newcommand{\Z}{{\bf Z}}

\newcommand{\R}{{\bf R}}

\newcommand{\C}{{\bf C}}

\newcommand{\e}{\,{\rm e}}
\newcommand{\CP}{{\bf CP}}

\newcommand{\be}{\begin{equation}}
\newcommand{\ee}{\end{equation}}
\newcommand{\bea}{\begin{eqnarray}}
\newcommand{\eea}{\end{eqnarray}}

\def\to{\rightarrow}

\def\longlongrightarrow{\relbar\joinrel\relbar\joinrel\rightarrow}

\def\lae{\mathrel{\mathop{\smash{\lower .5 ex \hbox{$\stackrel<\sim$}}}}}
\def\lae{\mathrel{\mathop{\smash{\lower .5 ex \hbox{$\stackrel>\sim$}}}}}

\def\l:{\mathopen{:}\,}
\def\r:{\,\mathclose{:}}

\catcode`\@=11
\def\theequation{\arabic{equation}}
\catcode`\@=12

\bcites

\catcode`\@=11
\def\theequation{\thesection.\arabic{equation}}
\@addtoreset{equation}{section}
\@addtoreset{footnote}{section}
\@addtoreset{footnote}{subsection}
\catcode`\@=12

\newcommand{\btheta}{\overline{\theta}}

\newcommand{\bepsilon}{\overline{\epsilon}}
\newcommand{\bPhi}{\overline{\Phi}}

\newcommand{\bphi}{\overline{\phi}}

\newcommand{\bpsi}{\overline{\psi}}

\newcommand{\bi}{\overline{\imath}}
\newcommand{\bj}{\overline{\jmath}}
\newcommand{\bl}{\overline{l}}
\newcommand{\bz}{\overline{z}}
\newcommand{\bu}{\overline{u}}

\newcommand{\blambda}{\overline{\lambda}}
\newcommand{\bsigma}{\overline{\sigma}}
\newcommand{\bSigma}{\overline{\Sigma}}
\newcommand{\bareta}{\overline{\eta}}
\newcommand{\nn}{\nonumber}

\newcommand{\frakg}{\mathfrak{g}}
\newcommand{\frakz}{\mathfrak{z}}
\newcommand{\ttt}{\mathfrak{t}}
\newcommand{\bXi}{\overline{\Xi}}
\newcommand{\vs}{\sigma}
\newcommand{\vssm}{\sigma}
\newcommand{\bfR}{\mbox{\boldmath $J$}}
\newcommand{\bfsmR}{\mbox{\scriptsize\boldmath $J$}}
\newcommand{\bartial}{\overline{\partial}}
\newcommand{\mc}{{\rm g}}
\newcommand{\MMM}{\mathfrak{M}}
\newcommand{\dbar}{\overline{\partial}}
\newcommand{\Deps}{\Delta_{\varepsilon}}
\newcommand{\bxi}{\overline{\xi}}
\newcommand{\rmq}{{\rm q}}
\newcommand{\rmQv}{{\rm Q}^{\vee}}

\newcommand{\by}{\mathbf{y}}
\newcommand{\vps}{\varepsilon}

\newcommand{\rmm}{{\rm m}}
\newcommand{\half}{{1\over 2}}

\begin{document}

\begin{titlepage}

\begin{center}

\today\hfill
{\tt\normalsize KIAS-P14039}
\\

\vskip 2.5 cm
{\large \bf Witten Index and Wall Crossing}
\vskip 1 cm

Kentaro Hori\footnote{\tt kentaro.hori@ipmu.jp}$^\diamond$,
Heeyeon Kim\footnote{\tt hykim@phya.snu.ac.kr}$^\dagger$,
and Piljin Yi\footnote{\tt piljin@kias.re.kr}$^\ddagger$

\vskip 5mm
$^\diamond${\it Kavli Institute for the Physics and Mathematics of the Universe
(WPI),\\
University of Tokyo, Kashiwa, Chiba 277-8583, Japan}
\vskip 2mm $^\dagger${\it Department of Physics and Astronomy, Seoul
National University, \\Seoul 151-147, Korea}
\vskip 2mm $^\ddagger${\it School of Physics, Korea Institute
for Advanced Study, \\Seoul 130-722, Korea}

\end{center}

\vskip 0.5 cm
\begin{abstract}
We compute the Witten index of one-dimensional gauged linear sigma models
with at least ${\mathcal N}=2$ supersymmetry.
In the phase where
the gauge group is broken to a finite group, the index is expressed as
a certain residue integral.
It is subject to a change
as the Fayet-Iliopoulos parameter is varied through the phase boundaries.
The wall crossing formula is expressed as an integral at infinity of
the Coulomb branch.
The result is applied to many examples, including
quiver quantum mechanics that is relevant for
 BPS states in $d=4$ ${\mathcal N}=2$ theories.
%, as well as $2d$ ${\mathcal N}=(2,2)$ theories.

\end{abstract}

\end{titlepage}

\pagestyle{empty}
\tableofcontents

\newpage

\pagestyle{plain}
\setcounter{page}{1}
\section{Introduction}

The present paper is about Witten index \cite{Windex}
of one dimensional gauge theories with at least
${\mathcal N}=2$ supersymmetry.
We do not attempt to consider the most general gauge theories but
restrict our target to a class of theories called
``gauged linear sigma models''.
A theory in this class has gauge group with at least one $U(1)$ factor
and has the Fayet-Iliopoulos (FI) D-term
\beq
-\zeta(D)
\eeq
in its Lagrangian. The space of FI parameter $\zeta$ is decomposed into
chambers called ``phases'' by the pattern of classical gauge symmetry breaking.
Typically, the gauge group is broken to its finite subgroup
 when $\zeta$ is inside a phase, and a continuous unbroken subgroup
shows up when $\zeta$ reaches its boundary.
With ${\mathcal N}=2$ or more supersymmetry, the vector multiplet
has at least one scalar component, and hence the appearance of continuous
unbroken gauge symmetry means emergence of non-compact flat direction,
called the Coulomb branch.
In particular, we expect a phase transition as $\zeta$ moves from one phase
to another through the mutual boundary. This is unlike in two dimensional
models \cite{Wphases} where the actual transition is avoided by turning on
the theta angle which can lift the Coulomb branch.

Witten index enjoys the usual good properties such as integrality
and deformation invariance, under the condition that
there is no flat direction. Suppose the only possible
source of non-compactness is unbroken gauge symmetry.
Then, the index enjoys the good property and
does not depend on $\zeta$ as long as it is inside a phase
in which the gauge symmetry is broken to a finite subgroup.
However, as $\zeta$ approaches a phase boundary, the index ceases to have
that property. If $\zeta$ enters another phase,
it may regain that property but there is no reason for it
to have the same value as in the original phase.
Namely, Witten index may jump as $\zeta$ moves from one phase to another.
This behaviour, called the ``wall crossing'',
is a distinguished feature of one dimensional models.
In two dimensions, wall crossing will not happen
since the flat direction can be avoided by turning on the theta angle.
Absence of wall crossing is also observed in
three dimensional models \cite{IntSei}.

The goal of the present paper is
to compute the Witten index in one dimensional gauged linear sigma models,
to see how it depends on the phase, and
to understand the physics of the wall crossing.

Apart from its own interest, the present work has a
strong motivation in string theory.
An effective theory of D-particles in Type II superstring theory is
provided by the ${\cal N}=16$ maximally supersymmetric $SU(n)$ matrix
quantum mechanics \cite{Witten:1995im}, or by
a class of gauged linear sigma models called quiver quantum mechanics
\cite{DouglasMoore,DouglasADHM,DGM}, depending on the background.
BPS bound states of D-particles correspond to supersymmetric ground states
of the effective quantum mechanics, and the index computation in
the latter is an extremely important problem.
For example, the index for ${\mathcal N}=16$ theories
\cite{Yi:1997eg,Sethi:1997pa} is relevant for the M-theory conjecture.
The spectrum of BPS states can change discontinuously as
the theory is deformed. Such behavior, also called ``wall-crossing,"
was initially discovered for BPS solitons in supersymmetric field theory
($d=2$ BPS kinks \cite{Cecotti:1992qh}, $d=4$ BPS solitons
in strongly coupled 
\cite{Seiberg:1994rs,Seiberg:1994aj,Ferrari:1996sv}
and weakly coupled
\cite{Bergman:1997yw,Lee:1998nv,Bak:1999da,Gauntlett:1999vc,Stern:2000ie}
regimes)
and then for BPS black holes in supergravity \cite{Denef:2000nb}.

When the effective theory is given by a gauged linear sigma model,
it is natural to expect that the two notions of wall crossing are related.
After the pioneering works on D-brane stability and the
r\^ole of FI parameter \cite{DouglasStab,KachruMcgreevy},
the relationship was first discussed in detail by F. Denef \cite{Denef}
when the effective theory is a simple class
of ${\mathcal N}=4$ quiver quantum mechanics.
The decay of BPS bound
states, which had been known to be due to infinite separation
of constituents \cite{Lee:1998nv,Bak:1999da,Gauntlett:1999vc,Denef:2000nb},
was rephrased by Denef as the run away of the ground state
wavefunctions to infinity in the Coulomb branch of the quiver theory
\cite{Denef,Denef:2007vg}. Around the same time, Reineke gave
a master formula for counting the index in the geometric Higgs
description \cite{Reineke}, although this was limited to
tree-like quiver theories.

Recently, there have been renewed efforts for systematic and explicit
counting of index for all ${\mathcal N}=4$ quiver quantum mechanics
\cite{deBoer:2008zn,Manschot:2010qz,Lee:2011ph,Kim:2011sc,Manschot:2011xc,Lee:2012sc,Lee:2012naa,Manschot:2012rx,Lee:2013yka},
but much of these efforts rely on approximate schemes, either Coulomb
or Higgs. The Coulomb approach in particular is physically
attractive, as it explains the origin of wall-crossing behavior
intuitively, and, for theories without superpotential, has been
successfully demonstrated to be equivalent \cite{Sen:2011aa} to
predictions via the Kontsevich-Soibelman algebra \cite{KS,GMN1,GMN2}.
For theories with superpotentials, such an approximate scheme often
fails dramatically and only a limited subset of such theories have
been explored so far
\cite{Denef:2007vg,Lee:2012sc,Lee:2012naa,Manschot:2012rx,Bena:2012hf}.
Nevertheless these new studies lead to a glimpse of very rich and surprising
vacuum structure, whereby emerged the concept of quiver invariants and their
proposed role as microstates of BPS black holes in $d=4$ ${\cal N}=2$
theories. In view of these new developments, the need for a more efficient and
faithful method for computing the index of gauged quantum mechanics is
all the more pressing.

The present paper offers a sweeping new approach to
the index computation, which
does not resort to any truncation or approximation to one sector,
and thus is capable of computing full quantum mechanics index
exactly, with wall-crossing taken into account.
For quiver quantum mechanics, in particular, this opens
up a new line of attack where both wall-crossing behavior and
quiver invariants can be computed simultaneously.

To compute the index,
we employ the method of supersymmetric localization,
extending the recent computation of
elliptic genus of two-dimensional $(0,2)$ gauge theories
\cite{BEHT1,BEHT2} to quantum mechanics.
In one dimension, the classical Coulomb branch cannot be
lifted by quantum corrections, unlike in two dimensions where it can
\cite{HoTo,Hduality}.
This is one reason to limit our target to gauged linear sigma models
with phases in which the gauge symmetry is classically broken to its
finite subgroup. We may also turn on twist by a global symmetry
that commutes
with an ${\mathcal N}=2$ supersymmetry. This is necessary especially when
the theory has a non-compact Higgs branch.
If the global symmetry has a compact set of fixed points,
the twist or the corresponding mass term lifts the flat direction,
and the twisted index regains the good property.
The index to compute is
\beq
I(\e^{2\pi i {\bf G}^F(z)})\,\,=\,\,
{\rm Tr}^{}_{\mathcal H}(-1)^F \e^{2\pi i{\bf G}^F(z)}\e^{-\beta H},
\label{IndI}
\eeq
where $\e^{2\pi i {\bf G}^F(z)}$ is the twist. Outline of computation
is as follows.

\begin{figure}[htb]
%\psfrag{ka}{$k_A$}
\centerline{\includegraphics{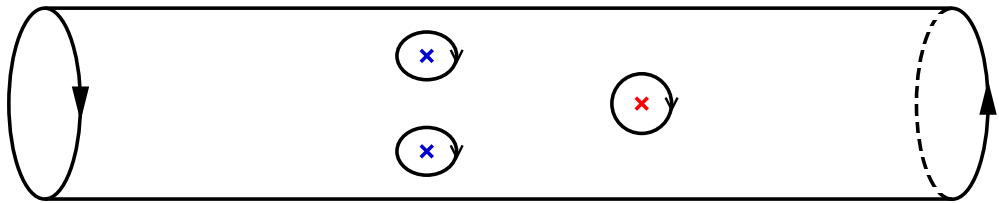}}
\caption{The moduli space $\MMM$ for a $U(1)$ gauge theory}
\label{fig:MMM}
\end{figure}

The index is realized as the path integral on the circle,
and the main idea of localization is to take the limit of vanishing
gauge coupling constant $e\to 0$.
The path integral first localizes on the supersymmetric configurations for the
${\mathcal N}=2$ vector multiplet, which consists of a pair of
the holonomy on the circle and the constant scalar that 
commute with each other. The moduli space is $\MMM=(T\times \ttt\,)/W$,
where $T$ is a maximal torus of the gauge group,
$\ttt$ is its Lie algebra and $W$ is the Weyl group.
To be precise, we need to excise from $T\times \ttt=:\tilde\MMM$
a neighborhood $\Deps$ of the union of {\it singular hyperplanes}
in which the scalar components of the ${\mathcal N}=2$ chiral multiplets
have zero modes.
It turns out that the integrand is a total derivative, and by
Stokes theorem we have an integral on the boundary of
$\tilde\MMM\setminus\Deps$.
The boundary consists of $-\partial \Deps$ as well as the boundary at
infinity $\partial \tilde\MMM=T\times \partial\,\ttt$,
 which exists because $\ttt$ is non-compact.
See Fig.~\ref{fig:MMM}.
Presence of the latter boundary is the main new feature
in the one-dimensional theories compared to the two-dimensional theories
in which the moduli space
was a compact space $(T\times T)/W$. And this presence of infinity
is the main reason for the wall crossing.
This is in accord with the picture of wall crossing that is associated
with the emergence of the Coulomb branch, and this non-compact $\ttt$
is indeed the Coulomb branch itself!

At each boundary component, we have an integral of the Cartan zero mode
of the auxiliary D-field, which is the ${\mathcal N}=2$ superpartner of
the coordinates of $\MMM$. The integral is of the form
\beq
\int {\dd Q(D)\over Q(D)} \;
\exp\left(-\frac{1}{2e^2}D^2-i\zeta(D)\right),
\label{DintI}
\eeq
where $Q$ is an element of $i\ttt^*$ that depends on the component
of $\partial(\tilde\MMM\setminus\Deps)$ under consideration.
At the boundary $-\partial \Deps$, the integral can be processed in exactly
the same way as in \cite{BEHT1,BEHT2}. There, we have chosen
an arbitrary generic element $\eta\in i\ttt^*$ that specifies the way to
deform the $D$ integration contour.
That procedure resulted
in having a sum over the isolated intersections of singular hyperplanes,
where the summand is the so called {\it JK residue} \cite{JK,BV,SV}
of a meromorphic form on $\tilde\MMM$ that is obtained by the one-loop
integral.
There, the integral (\ref{DintI}) is either zero or picks up the simple
pole at $Q(D)=0$.
In either way, the $\zeta$ dependence is washed away.
The residues depends only on $\eta$.
The main question is what to do with the component at infinity
$\partial\tilde\MMM$
in which we put $Q=Q_{\infty}$. There,
we need to evaluate (\ref{DintI}) directly,
along a contour specified by $\eta$.
The simplest way to proceed is to set
\beq
\eta=\zeta,\qquad Q_{\infty}=\zeta.
\label{choicezeta}
\eeq
Then, it is the integral over $\zeta(D)$ along the contour $\R-i\epsilon$.

At this stage we recall that we are taking the limit $e\to 0$, which
is potentially singular. The limit is regular, however,
if $\zeta$ is inside a phase where the gauge group is broken to a
finite subgroup, and if the Higgs mass scale $M_H=e\sqrt{|\zeta|}$ is held
non-zero. That is, at the same time as $e\to 0$, we scale up $\zeta$ so that
 $\zeta'=e^2\zeta$ is held at a fixed value inside the phase.
In this scaling limit, which may be called the Higgs scaling,
the $\zeta(D)$ integral vanishes. To see this,
we scale the $\zeta$ component of $D$ as $e^2D'$.
Then, (\ref{DintI}) is proportional to
\beq
\int_{\R-i\epsilon}{\dd \zeta'(D')\over \zeta'(D')}
\exp\left(-{e^2\over 2}D^{\prime 2}-i\zeta'(D')\right)
\stackrel{e\to 0}{\longlongrightarrow}
\int_{\R-i\epsilon}{\dd \zeta'(D')\over \zeta'(D')}
\exp\left(-i\zeta'(D')\right)=0.\label{intInt}
\eeq

To summarize, with the choice (\ref{choicezeta}) and in the Higgs scaling,
all the integrals at infinity vanish.
We are therefore left with the sum over isolated intersections
of singular hyperplanes,
\beq
I(\e^{2\pi i{\bf G}^F(z)})\,\,=\,\,
{1\over |W|}\sum_{p}\mathop{\mbox{JK-Res}}_p(Q(p),\zeta)
\Bigl[\,g(u,z)\dd^{\ell}u\Bigr],
\label{resultI}
\eeq
where
$Q(p)$ is the set of charges of the chiral multiplets
that define the singular hyperplanes that meet at $p$,
$g(u,z)$ is the result of one-loop
integral under the supersymmetric background,
and $\dd^{\ell}u$ is a correctly normalized holomorphic volume form
on $\tilde\MMM$ ($\ell$ is the rank
of the gauge group).

What happens if $\zeta$ moves from one phase to another?
If we keep $\eta$ and $Q_{\infty}$ at the initial value of $\zeta$
during the process, nothing happens to the sum
over isolated intersections of the singular hyperplanes.
On the other hand, the integrals at infinity may become non-zero,
since ${\rm Im}\zeta(D)$ may go from negative to positive:
The integral (\ref{intInt}) does not vanish if the contour $\R-i\epsilon$
is replaced by $\R+i\epsilon$.
Therefore, the change of the index can be expressed
as a certain sum over integrals at infinity.
This matches with the picture of wall crossing due to
the emergence of the Coulomb branch.
In fact, one can sum up the integrals at infinity
and make the relation to the Coulomb branch more precise, at least for a
``simple wall crossing''.
A phase boundary is called simple when the unbroken gauge group there
is isomorphic to $U(1)$. Any phase boundary is simple for a general theory
with Abelian gauge groups, and many are simple also in non-Abelian
theories.
At such a simple phase boundary, the states that are responsible
for the wall crossing can be analyzed reliably. They have wavefunctions
supported on
the mixed Coulomb-Higgs branch, and the analysis yields a wall crossing
formula. And that formula matches precisely
with the wall crossing formula obtained by summing up the
integrals at infinity!

The rest of the paper is organized as follows.

In Section~\ref{sec:models}, we introduce the one-dimensional
 gauged linear sigma models with at least ${\mathcal N}=2$ supersymmetry
and describe their symmetries. We define the indices we are going to
study in this paper. To be precise, (\ref{IndI}) is only a mnemonic,
and the actual definition is only in (\ref{indtwo}) in general. We show that
the index depends holomorphically on the twist parameter, thus motivating
that mnemonic. We also prove that, in an effectively compact
${\mathcal N}=4$ theory, the ground states have charge zero under
${\mathcal N}=4$ flavor symmetries. We introduce the notion of phases,
Coulomb, Higgs, mixed branches. We end with presenting
three classes of examples illustrating the phase structures.

In Section~\ref{sec:Coulomb}, we describe the effective theory on the
Coulomb branch near the phase boundary of $U(1)$ theories
as well as the effective theory on the
mixed branch near simple phase boundaries of a general theory.
This analysis is new for ${\mathcal N}=2$ theories but is of course a review
of \cite{Denef} for ${\mathcal N}=4$ theories. A complete detail is
presented for the ${\mathcal N}=2$ $\CP^{N-1}$ model. We present a
wall crossing formula for the index, for the case of simple wall crossing,
as a result of the Coulomb and mixed branch analysis.

Section~\ref{sec:Index} is the main section where we compute
the Witten index, as outlined above.

In Section~\ref{sec:WC}, we derive the wall crossing formula
at simple phase boundary, first in $U(1)$ theories, next in a general Abelian
theory and then in a general non-Abelian theory. Complete agreement with
Coulomb or mixed branch formula is observed.

The remaining sections are devoted to examples that illustrate
our findings in the earlier sections.
In Section~\ref{sec:4Ex}, we compute the index for ${\mathcal N}=4$
theories, including the linear sigma models for Grassmannians and
hypersurfaces in projective spaces, models with compact and non-compact
Higgs branches with flavor symmetries, and
 ``the two parameter model''. We see that
Coulomb and mixed branch analysis can tell us also about Hodge diamond,
which carries finer information than the index.
We illustrate flavor decoupling in compact theories, and
flavor non-decoupling in non-compact theories. In the non-compact
theories, we find that the index without R-symmetry twist does
not depend on the detail of the flavor twist even though its presence
is certainly needed even to define the index.
We also study the full Coulomb branch in ``the two parameter model''
and examine the meaning of the result.

Section~\ref{sec:4Quiver} is devoted to ${\mathcal N}=4$ quiver theories
which are important for 4d ${\mathcal N}=4$ BPS states counting and
wall crossing phenomena.
We start by illustrating the computation of the index for cyclic Abelian
quivers and triangle non-Abelian quivers.
The computation reproduces the known results on Abelian quivers, while
it produces new results for non-Abelian quivers.
We then move on to examine the idea of quiver invariants
\cite{Lee:2012sc,Lee:2012naa}
and an existing proposal \cite{Manschot:2013sya} on how these enter
the full indices of quiver theories as building blocks,
with a brief introduction to that subject.
Our new results provide a non-trivial test of this over-determined system.

In the final Section~\ref{sec:2Ex}, we compute the index for
${\mathcal N}=2$ theories, including the linear
sigma models for $\CP^{N-1}$ and Grassmannians, Distler-Kachru model, and
triangle Abelian quivers. In $\CP^{N-1}$ and Grassmannian $G(k,N)$,
we see that the result agrees with the Weyl character formula for
representations of $SU(N)$, in agreement with Borel-Weil-Bott theorem.
In Distler-Kachru model, we observe an interesting process of wall crossing,
where a pair of bosonic and fermionic states go down together to have
zero energy before entering the wall crossing regime.
We also illustrate the mixed branch wall crossing formula
$\Delta I=I^{({\rm H})}\ast \Delta_CI^{({\rm C})}$.

In Appendix we describe 1d ${\mathcal N}=2$ superfield formalism.

\vskip 5mm

After this work was completed, presented \cite{Talks}, and at the stage
of being edited, two papers with partial overlaps \cite{Cordova,Hwang}
appeared in the ArXiv.

\subsection*{Notational convention:}

{\bf 1.}~In the rest of this paper, we denote a compact Lie group
by an upper case
roman character and its Lie algebra by the lower case Gothic character,
$G$ and $\frakg$ for example. We denote their complexifications
by putting subscript $\C$, as $G_{\C}$ and $\frakg_{\C}$.
We call elements of $\frakg\subset \frakg_{\C}$ {\it pure imaginary},
while elements of $i\frakg\subset \frakg_{\C}$ are called {\it real}.
This is motivated by the fact that the Lie algebra of $U(1)$
is naturally the vector space $i\R$ of pure imaginary numbers.
For a given compact Lie group, say $G$,
we shall write $T$ and $\ttt$ for a maximal torus of $G$ and
its Lie algebra,  $N_T$ for the normalizer of $T$ in $G$,
$W$ for the associated Weyl group $N_T/T$,
$Z_G$ and $\mathfrak{z}$ for the center and its Lie algebra.

{\bf 2.} ~For a meromorphic function $f(z)$ of one
variable, ${\rm res}_{z=a}f(z)$ is defined with a factor of
extra $2\pi i$  compared to the standard,
\beq
\mathop{\rm res}_{z=a}f(z)=\oint_a\dd z\,f(z).
\eeq
For example, ${\rm res}_{z=0}(1/z)=2\pi i$. The same applies to
JK-Res as in (\ref{resultI}).
 It is $(2\pi i)^{\ell}$ times the standard one.

\section{Gauged Linear Sigma Models in One Dimension}
\label{sec:models}

\subsection{The Models}\label{subsec:models}

We consider an ${\mathcal N}=2$ supersymmetric quantum mechanics with
a compact gauge group $G$ and a certain set of matter multiplets.
We first describe the variables, supersymmetry and Lagrangians.
Much of the construction is obtained from the dimensional
reduction of $2d$ ${\mathcal N}=(0,2)$ gauged linear sigma models
\cite{Wphases}, while a part is borrowed from the boundary interactions
of $2d$ ${\mathcal N}=(2,2)$ theories \cite{HHP}.
We shall use the component expressions.
See Appendix~\ref{app:SUSY} for the description in superspace.

\subsection*{\rm \underline{${\mathcal N}=2$ Supersymmetry}}

An ${\mathcal N}=2$ supersymmetric quantum mechanics has
the supercharges ${\bf Q},\overline{\bf Q}$
and the Hamiltonian $H$ obeying
\beq
{\bf Q}^2=\overline{\bf Q}^2=0,\quad
\{{\bf Q},\overline{\bf Q}\}=H.
\label{N2SUSY}
\eeq
We may also have the R-charge $\bfR$ which obeys
\beq
[\bfR,{\bf Q}]=-{\bf Q},\quad
[\bfR,\overline{\bf Q}]=\overline{\bf Q},\quad
[\bfR,H]=0.
\eeq
${\bf Q}$ and $\overline{\bf Q}$ are the adjoints of each other
while $H$ and $\bfR$ are self adjoint.
We shall write the supersymmetry transformation of field variables by
\beq
\delta{\mathcal O}
=[-i\epsilon{\bf Q}+i\bepsilon\overline{\bf Q},{\mathcal O}],
\eeq
where $\epsilon$ and $\bepsilon$ are
complex conjugate pair of fermionic variational parameters
which have R-charges $+1$ and $-1$ respectively.

\subsection*{\rm \underline{Vector Multiplet}}

The vector multiplet consists of a gauge field
$v_t$, a scalar field $\vs$, a complex conjugate pair
of fermions $\lambda, \blambda$, and an auxiliary field $D$, which are
all valued in the adjoint representation of $G$.
($v_t$, $\vs$ and $D$ are ``real'' in the convention stated
in the introduction.)
They transform under the ${\mathcal N}=2$ supersymmetry as
\beqa
\delta v_t&=&-\delta\vs~=~{i\over 2}\epsilon\blambda
+{i\over 2}\bepsilon\lambda,\nn\\
\delta \lambda&=&\epsilon\Bigl(D_t\vs+iD\Bigr),\nn\\
\delta D&=&{1\over 2}\epsilon D_t^{(+)}\blambda
-{1\over 2}\bepsilon D_t^{(+)}\lambda.
\label{Vector}
\eeqa
Here and in what follows, for a field ${\mathcal O}$ in a representation of $G$,
we write $D_t{\mathcal O}:=\dot{\mathcal O}+iv_t{\mathcal O}$
for the covariant derivative
and put
\beq
D^{(\pm)}_t{\mathcal O}\,\,:=\,\,D_t{\mathcal O}\pm i\vs {\mathcal O}.
\eeq
Note that $\lambda$ is chiral,
$\{\overline{\bf Q},\lambda\}=0$.
%The transformation of $\blambda=\lambda^{\dag}$
%is obtained by the usual rule $(AB)^{\dag}=B^{\dag}A^{\dag}$.
The supersymmetric kinetic term is
\beq
L_{\rm gauge}~=~
{1\over 2e^2}{\rm Tr}\Bigl[\, (D_t\vs)^{2}+i\blambda\,D_t^{(+)}\lambda+D^2
\,\Bigr].
\label{Lgauge}
\eeq
For an adjoint invariant linear form
$\zeta:i\mathfrak{g}\to \R$, we have the Fayet-Iliopoulos term
\beq
L_{\rm FI}~=~-\zeta(D).
\label{FI}
\eeq
Note that $\zeta$, which belongs to $i(\mathfrak{g}^*)^G\cong i(\ttt^*)^W$,
can be regarded as an element of $i\mathfrak{z}^*$ since
$(\ttt^*)^W\hookrightarrow \ttt^*\to \mathfrak{z}^*$ is an isomorphism.

\subsection*{\rm \underline{Chiral Multiplet}}

A chiral multiplet consists of
a scalar field $\phi$ and a fermion $\psi$ which are valued in a unitary
representation $V_{\rm chiral}$ of $G$.
The supersymmetry transformations are
\beqa
\delta\phi&=&-\epsilon\psi,\nn\\
\delta\psi&=&i\bepsilon D_t^{(+)}\phi.
\label{Chiral}
\eeqa
Note that $\phi$ is chiral, $[\overline{\bf Q},\phi]=0$.
The supersymmetric kinetic term is
\beq
L_{\rm chiral}~=~
D_t\bphi D_t\phi
+i\bpsi\,D_t^{(-)}\psi
+\bphi \left\{D-\vs^{\,2}\right\}\phi
-i\bphi\lambda\psi+i\bpsi\,\blambda\phi.
\label{Lchiral}
\eeq

\subsection*{\rm \underline{Fermi Multiplet}}

A fermi multiplet consists of a fermion $\eta$ and an auxiliary field
$F$ which are valued in a unitary
representation $V_{\rm fermi}$ of $G$.
The supersymmetry transformations are
\beqa
\delta\eta&=&\epsilon F+\bepsilon E(\phi),\nn\\
\delta F&=&\bepsilon \left(-iD_t^{(+)}\eta+\psi^i\partial_iE(\phi)\right),
\label{Fermi}
\eeqa
where $(\phi,\psi)$ is a chiral multiplet in a representation
$V_{\rm chiral}$ of $G$
 and $E:V_{\rm chiral}\to V_{\rm fermi}$ is a $G$-equivariant holomorphic
map, i.e., a $V_{\rm fermi}$-valued holomorphic function $E(\phi)$ of $\phi$
satisfying $E(g\phi)=gE(\phi)$.
Note that
$\{\overline{\bf Q},\eta\}=-iE(\phi)$.
Thus, $\eta$ is chiral only when $E(\phi)$ is trivial,
in which case the multiplet may be called
a chiral fermi multiplet.
The supersymmetric kinetic term is
\beq
L_{\rm fermi}~=~
i\bareta\,D_t^{(+)}\eta+\overline{F} F
-\overline{E(\phi)}E(\phi)
-\bareta\partial_iE(\phi)\psi^i
-\bpsi^{\bi}\partial_{\bi}\overline{E(\phi)}\eta.
\label{Lfermi}
\eeq

\subsection*{\rm \underline{Superpotential}}

For a chiral multiplet $(\phi,\psi)$ in a representation $V_{\rm chiral}$
and a fermi multiplet $(\eta,F)$ in another representation $V_{\rm fermi}$
as above,
let $J:V_{\rm chiral}\to V_{\rm fermi}^*$ be a $G$-equivariant holomorphic map,
such that $J(\phi)E(\phi)=0$.
Then, $J(\phi)\eta$ is chiral, $\{\overline{\bf Q},J(\phi)\eta\}=0$.
In such a case, we have a supersymmetric interaction terms,
\beq
L_J=
\psi^i\partial_iJ(\phi)\eta-J(\phi)F
\,\,\,+\,\,{\rm c.c.}.
\label{2superpotential}
\eeq
We shall call this the F-term associated to the superpotential
$\mathfrak{W}=J(\phi)\eta$.

\subsection*{\rm \underline{Global Anomaly}}

As always in gauge theory, we need to impose the Gauss law constraint.
Here we would like to point out that a non-trivial condition
is required for that to be possible.

Let us consider a (not necessarily supersymmetric)
gauge theory with a gauge group $G$, a bosonic variable in some representation
of $G$ and a fermion pair, $\eta$ and $\bareta$, which take values
in a unitary representation $V_F$ of $G$ and its dual $V_F^*$ respectively.
We assume Lagrangian of the form
\beq
L=i\bareta D_t\eta+\cdots
=i\sum_{\alpha,\beta=1}^{d_F}\bareta_{\alpha}\left(\delta^{\alpha}_{\,\beta}
{\dd \over \dd t}+i(v_t)^{\alpha}_{\,\beta}\right)\eta^{\beta}+\cdots,
\eeq
where $d_F=\dim V_F$ and
$+\cdots$ are terms that do not involve
the time derivatives of $\eta$ and $\bareta$.
Quantization of the fermionic variables is standard.
The canonical anticommutation relation is
$\{\eta^{\alpha},\bareta_{\beta}\}=\delta^{\alpha}_{\,\beta}$,
$\{\eta^{\alpha},\eta^{\beta}\}=\{\bareta_{\alpha},\bareta_{\beta}\}=0$,
and the hermiticity is $\eta^{\alpha\dag}=\bareta_{\alpha}$.
The space of states ${\mathcal H}_F$
is built on a ``vacuum'' $|0\rangle$ annihilated
by all $\eta^{\alpha}$'s,
\beq
|0\rangle,\quad\bareta_{\alpha}|0\rangle,\quad
\bareta_{\alpha}\bareta_{\beta}|0\rangle,\quad\ldots,\quad
\bareta_1\cdots\bareta_{d_F}|0\rangle.
\eeq
Gauss law requires physical states to be $G$-invariant.
For this, we need to specify how $G$ acts on the states.
Throughout the paper, we take the standard quantization rule where
$\eta$ and $\bareta$ play symmetric r\^oles. 
At the infinitesimal level, this yields the following expression for
the gauge charge corresponding to $\xi\in\frakg$
\beq
{\bf G}(\xi)={1\over 2}[\bareta,\xi\eta]+\cdots
={1\over 2}\xi^\alpha_{\,\beta}[\bareta_\alpha,\eta^\beta]+\cdots \ ,
\label{standardQ}
\eeq
where $+\cdots$ is the bosonic contribution which is unambiguous.
For example, ${\bf G}(\xi)$ acts on the ``vacuum'' $|0\rangle$
by multiplication of $-{1\over 2}{\rm tr}^{}_{V_F}(\xi)$.
The question is whether this $\frakg$-action lifts to a $G$-action.
The condition is that
$g\in G\mapsto  \det^{{1\over 2}}_{V_F\!}g\in U(1)$ is a well-defined
group homomorphism. That is, the one-dimensional representation
$\det V_F=\wedge^{d_F}V_F$ must have a square root.
Absence of a square root can be regarded as a global anomaly.
Suppose there is one, $\det^{1\over 2}V_F$.
(If $G$ is connected, it is unique when it exists.
If there is an ambiguity, we make a choice.)
Then, as the space of (fermionic) states we can take
\beq
{\mathcal H}_F\cong {\det}^{-{1\over 2}} V_F\otimes\wedge V_F.
\eeq
%Note that it is invariant under swap of the r\^oles of
%$\eta$ and $\bareta$, $\wedge V_F\cong \det V_F\otimes \wedge V_F^*$.

This global anomaly is in fact the same as the more familiar form
of global anomaly,
which is usually discussed in the path-integral quantization.
For illustration, let us take the system of a single
$\eta,\bareta$ pair where $\eta$ has charge one under $G=U(1)$,
with the simple Lagrangian $L=i\bareta D_t\eta$, and consider
the partition function on the Euclidean circle of
circumference $\beta$ with anti-periodic
boundary condition.
Assuming that $v_{\tau}$ is a constant $-2\pi a/\beta$, the Dirac operator
is diagonalized by the Fourier modes $\e^{-2\pi i\rmm \tau/\beta}$,
and the partition function is the product of the eigenvalues
\beq
Z_{S^1}(a)\,=\,
C\prod_{\rmm\in\Z+\half}\left({2\pi m\over \beta}+{2\pi a\over\beta}\right).
\label{partFF}
\eeq
where $C$ is a normalization constant.
$a$ and its integer shifts are all gauge equivalent since they are
related by large gauge transformations. Indeed the set of eigenvalues
is invariant under the shifts.
However, the partition function is not:
As we continuously deform $a$ to $a+1$,
(\ref{partFF}) changes by a sign, since
a single eigenvalue of the Dirac operator goes from negative to positive.
This is the standard statement of global anomaly.
To relate this to the above discussion, we note that $Z_{S^1}(a)$ is
identified as
\beq
{\rm Tr}^{}_{{\mathcal H}_F}\e^{2\pi i{\bf G}(a)},
\eeq
and also that (\ref{partFF}) can be computed as
\beq
C'\prod_{\rmm=\half}^{\infty}\left(1-\left({a\over\rmm}\right)^2\right)
={C'\over 2}(\e^{\pi i a}+\e^{-\pi i a}),
\label{computZ}
\eeq
for another normalization constant $C'$.
We see that the symmetric quantization (\ref{standardQ})
agrees with this result, and that
the globaly anomaly in the canonical quantization
is really the same things as the globaly anomaly in the path-integral.

Let us apply this to our ${\mathcal N}=2$ supersymmetric system with
gauge group $G$, a chiral multiplet and a fermi multiplet
with values in $V_{\rm chiral}$ and $V_{\rm fermi}$.
The fermion in this theory takes values in
$V_F=\frakg_{\C}\oplus V_{\rm chiral}\oplus V_{\rm fermi}$.
Note that $\det\frakg_{\C}$ has a square root. Thus, the anomaly free
condition is
\beq
\det (V_{\rm chiral}\oplus V_{\rm fermi})\quad\mbox{has
a square root}.
\label{anomalyfree}
\eeq

\subsection*{\rm \underline{Wilson Line}}

An important class of supersymmetric interactions is provided
by a supersymmetric version of the Wilson line.
Let $M$ be a $\Z_2$ graded vector space with a hermitian
inner product, on which the gauge group elements act as unitary
operators, $\rho:G\to U(M)$. We suppose that there is a
$G$-equivariant holomorphic map
$Q:V_{\rm chiral}\to \,{\rm End}^{\rm od}(M)$,
whose values square to zero,
\beqa
&Q(g\phi)=\rho(g)Q(\phi)\rho(g)^{-1},\nn\\
&\partial_{\bi}Q(\phi)=0,\nn\\
&Q(\phi)^2=0.
\eeqa
Let us put
\beq
{\mathcal A}_t=\rho(v_t+\vs)-\psi^i\partial_iQ(\phi)
+\bpsi^{\bi}\partial_{\bi} Q(\phi)^{\dag}
+\{Q(\phi),Q(\phi)^{\dag}\}.
\label{calA}
\eeq
Then, under the supersymmetry transformations (\ref{Vector}) and
(\ref{Chiral}),
letting the variational parameters $\epsilon$, $\bepsilon$ depend on
$t$,
 it transforms as
\beq
\delta {\mathcal A}_t
=\Scr{D}_t(-i\bepsilon Q-i\epsilon Q^{\dag})
+i\dot{\bepsilon}Q+i\dot{\epsilon}Q^{\dag}
\label{stA}
\eeq
with $\Scr{D}_tY:={\dd\over\dd t}Y
+i[{\mathcal A}_t,Y]$.
This means that matrix factor
\beq
{\rm Pexp}\left(-i\int {\mathcal A}_t\dd t\,\right)
\label{CP}
\eeq
has ${\mathcal N}=2$ supersymmetry and can be placed as a
path-integral weight. Moreover, the $\dot{\epsilon}$ terms in (\ref{stA})
shows that the supercharges act on the space $M$ as
\beq
\overline{\bf Q}|_M=-i Q(\phi),\qquad
{\bf Q}|_M=iQ(\phi)^{\dag}.
\label{QCP}
\eeq

When the anomaly free condition (\ref{anomalyfree}) is met,
$\rho:G\to U(M)$ must be a genuine representation.
One advantage of the Wilson line is that,
even if the condition (\ref{anomalyfree}) is violated,
the anomaly may be cancelled by choosing $\rho$ which fails to be
a genuine representation of $G$ ---
we only need
${\mathcal H}_F\otimes (M,\rho)$ to be a well-defined representation.

A class of Wilson lines may be induced from a fermi
multiplet.
If we quantize the fermi multiplet, we obtain
the space of states
\beq
M\,\,=\,\,{\det}^{-{1\over 2}}V_{\rm fermi}\otimes\wedge V_{\rm fermi}.
\label{Mfer}
\eeq
This factor itself may or may not be a genuine representation of $G$,
but that does not matter as long as the total system is anomaly free.
The interaction terms in (\ref{Lfermi}) and (\ref{2superpotential})
 are nothing but $-{\mathcal A}_t$
with
\beq
Q(\phi)\,=\,\bareta E(\phi)+J(\phi)\eta\,=\,E(\phi)\wedge+J(\phi)\lrcorner.
\label{Qfer}
\eeq
Note that $Q(\phi)^2=0$ is equivalent to the condition $J(\phi)E(\phi)=0$.
Note also that this expression for
$Q(\phi)$ is consistent with the supersymmetry transformation
--- compare (\ref{Fermi}) and (\ref{QCP}).
This construction, however, is special in that the dimension of $M$
is a power of $2$.

\subsection*{\rm \underline{${\mathcal N}=4$ Supersymmetric Systems}}

\newcommand{\bfepsilon}{\mbox{\boldmath $\epsilon$}}
\newcommand{\bfbepsilon}{\overline{\mbox{\boldmath $\epsilon$}}}
\newcommand{\bfx}{\mbox{\large\bf x}}
\newcommand{\bfpsi}{\mbox{\boldmath $\psi$}}
\newcommand{\bfbpsi}{\overline{\mbox{\boldmath $\psi$}}}
\newcommand{\bfsigma}{\mbox{\boldmath $\sigma$}}
\newcommand{\bfbsigma}{\overline{\mbox{\boldmath $\sigma$}}}
\newcommand{\bflambda}{\mbox{\boldmath $\lambda$}}
\newcommand{\bfblambda}{\overline{\mbox{\boldmath $\lambda$}}}
\newcommand{\bfL}{\mbox{\boldmath $L$}}
\newcommand{\bfD}{\mbox{\boldmath $D$}}
\newcommand{\bfA}{\mbox{\boldmath $A$}}
\newcommand{\bfsmA}{\mbox{\scriptsize\boldmath $A$}}
\newcommand{\bfC}{\mbox{\boldmath $C$}}
\newcommand{\bfV}{\mbox{\boldmath $V$}}
\newcommand{\bfdelta}{\mbox{\boldmath $\delta$}}
\newcommand{\vsf}{\mbox{\large\sl x}_{}}
\newcommand{\vscf}{\mbox{\large\sl x}}
\newcommand{\vssmf}{\mbox{\sl x}}

An important class of systems have ${\mathcal N}=4$ supersymmetry ---
in addition to ${\bf Q}={\bf Q}_+$ and
$\overline{\bf Q}=\overline{\bf Q}_+$
there is another set of supercharges
${\bf Q}_-$ and $\overline{\bf Q}_-$ which obey the same algebra
as (\ref{N2SUSY}) and anticommute with ${\bf Q}_+$ and $\overline{\bf Q}_+$.

The ${\mathcal N}=4$ vector multiplet consists of
the ${\mathcal N}=2$ vector multiplet $(v_t,\vscf_3,\lambda_-,D)$
and an ${\mathcal N}=2$ chiral multiplet $(\sigma,i\blambda_+)$
in the complexified adjoint representation.\footnote{We shall use
a special notation for ${\mathcal N}=4$ theories:
$(\vs,\lambda)$ of the ${\mathcal N}=2$ vector part of
the ${\mathcal N}=4$ vector multiplet is denoted by $(\vscf_3,\lambda_-)$
and  instead , ``$\sigma$'' is used
for the scalar component of the ${\mathcal N}=2$
chiral part of the ${\mathcal N}=4$ vector multiplet.}
An ${\mathcal N}=4$ chiral multiplet consists of
an ${\mathcal N}=2$ chiral multiplet $(\phi,\psi_+)$ and
an ${\mathcal N}=2$ fermi multiplet $(\psi_-,F)$ in the same representation,
say $V$, with $E(\sigma,\phi)=\sigma\phi$.
The ${\mathcal N}=4$ invariant kinetic terms are just the sum of
the ${\mathcal N}=2$ invariant kinetic terms:
\beqa
\bfL_{\rm gauge}\!\!&=&\!\!L_{\rm gauge}(v_t,\vscf_3,\lambda_-,D)
+{1\over 2e^2}L_{\rm chiral}(\sigma,i\blambda_+,\ast),
\label{N4Lvec}\\
\bfL_{\rm chiral}\!\!&=&\!\!L_{\rm chiral}(\phi,\psi_+,\ast)
+L_{\rm fermi}(\psi_-,F,\ast),
\label{N4Lchi}
\eeqa
where ``$\ast$'' stands for the
${\mathcal N}=2$ vector multiplet fields
or $E=\sigma\phi$ with its superpartner.
For a $G$-invariant holomorphic function $W(\phi)$ of $\phi\in V$,
an ${\mathcal N}=4$ invariant superpotential term is obtained from
the ${\mathcal N}=2$ F-term with the superpotential
$\mathfrak{W}=-\partial_iW(\phi)\psi^i_-$
(i.e. set $J(\phi)=-\dd W(\phi)$).
The FI-term (\ref{FI}) is ${\mathcal N}=4$ supersymmetric by itself.
Note that $JE=0$ follows from the $G$-invariance of $W$ and
the anomaly free condition (\ref{anomalyfree})
is automatically satisfied.

${\mathcal N}=4$ invariant systems are obtained
from $4d$ ${\mathcal N}=1$ or $2d$ ${\mathcal N}=(2,2)$ systems
 by dimensional reduction.
To write the Lagrangians in a more familiar form, let us introduce
a triplet scalar $\bfx=(\vscf_1,\vscf_2,\vscf_3)$ for
$\sigma=\vscf_1-i\vscf_2$,
and doublet fermions
\beq
\bfblambda=(\blambda_-,\blambda_+),\quad
\bflambda=\left(\begin{array}{c}
\lambda_-\\
\lambda_+
\end{array}\right),\,\quad
\bfbpsi=(\bpsi_-,\bpsi_+),\quad
\bfpsi=\left(\begin{array}{c}
\psi_-\\
\psi_+
\end{array}\right).
\eeq
We also write
\beq
D+{1\over 2}[\sigma,\bsigma]=\bfD.
\eeq
Then, the ${\mathcal N}=4$ invariant Lagrangians can be written as
\beqa
\bfL_{\rm gauge}
\!\!&=&\!{1\over 2e^2}{\rm Tr}\left[\,(D_t\bfx)^2+i\bfblambda D_t\bflambda
+\bfD^2+\sum_{i<j}[\vscf_i,\vscf_j]^2
-\bfblambda [\not\!\bfx,\bflambda]\,\right],\label{LgN4}\\
\bfL_{\rm chiral}
\!\!&=&\!D_t\bphi D_t\phi+i\bfbpsi D_t\bfpsi +\overline{F}F
+\bphi\left\{\bfD-\bfx^2\right\}\phi
-\bfbpsi\! \not\!\bfx\bfpsi
+i\bphi\bflambda\bfpsi-i\bfbpsi\bfblambda\phi.~~~\label{LchN4}\\
\bfL_{W}
&=&F^i\partial_iW(\phi)
-{1\over 2}\bfpsi^i\bfpsi^j\partial_i\partial_jW(\phi)
\,\,+\,\,{\rm c.c.},\label{LWN4}\\
\bfL_{\rm FI}
&=&-\zeta(\bfD).
\eeqa
Here we used
$\not\!\!\bfx:=\sum_j\sigma_j\vscf_j$ where
$\sigma_1,\sigma_2,\sigma_3$ are the Pauli
matrices.
We also used the contractions
$\bflambda\bfpsi:=\lambda_+\psi_--\lambda_-\psi_+$
and $\bfbpsi\bfblambda:=-\bpsi_+\blambda_-+\bpsi_-\blambda_+$, etc,
as in \cite{WessBagger}.
The ${\mathcal N}=4$ supersymmetry transformation
\beq
\bfdelta{\mathcal O}
=[i\epsilon_+{\bf Q}_--i\epsilon_-{\bf Q}_+
-i\bepsilon_+\overline{\bf Q}_-+i\bepsilon_-\overline{\bf Q}_+,
{\mathcal O}] \
\eeq
is given by
\beqa
\bfdelta v_m&=&{i\over 2}\bfbepsilon\bsigma_m\bflambda
-{i\over 2}\bfblambda\bsigma_m\bfepsilon,\nn\\
\bfdelta\bflambda&=&\left(i\bfD+\sigma^{mn}v_{mn}\right)\bfepsilon,\nn\\
\bfdelta\bfD&=&{1\over 2}\bfbepsilon\bsigma^mD_m\bflambda
+{1\over 2}D_m\bfblambda \bsigma^m\bfepsilon,
\eeqa
\beqa
\bfdelta\phi&=&\bfepsilon\bfpsi,\nn\\
\bfdelta\bfpsi&=&i\sigma^m\bfbepsilon D_m\phi+\bfepsilon F,\nn\\
\bfdelta F&=&i\bfbepsilon \bsigma^mD_m\bfpsi-i\bfbepsilon\bfblambda \phi.
\eeqa
Here we follow the Wess-Bagger convention \cite{WessBagger}
for the contraction of spinors and the sigma matrices.
$m,n$ run over $0,1,2,3$, and we write
$v_0=v_t$, $v_j=\vscf_j$, $v_{0j}=D_t\vscf_j$, $v_{jk}=i[\vscf_j,\vscf_k]$,
$D_0{\mathcal O}=D_t{\mathcal O}$,
$D_j{\mathcal O}=i\vscf_j{\mathcal O}$.
The ${\mathcal N}=2$ supersymmetry is obtained by setting
$\epsilon_+=0$, $\bepsilon_+=0$,
$\epsilon_-=\epsilon$, and $\bepsilon_-=\bepsilon$.

There is a no go theorem for Wilson line (\ref{calA}) in an
 ${\mathcal N}=4$ supersymmetric theory.
In the present case, a Wilson line would be of the from
\beq
{\mathcal A}_t=\rho(v_t+\vscf_3)-\psi_+^i\partial_iQ
+i\blambda^a_+\partial_{\sigma^a}Q+\bpsi_+^{\bi}\partial_{\bi}Q^{\dag}
-i\lambda_+^a\partial_{\bsigma^a}Q^{\dag}
+\{Q,Q^{\dag}\}.
\eeq
The fields involved %in ${\mathcal A}_t$
transform under the second ${\mathcal N}=2$ supersymmetry as
\beqa
&\delta''(v_t+\vscf_3)=i(\epsilon_+\blambda_++\bepsilon_+\lambda_+),\quad
\delta''\phi=\epsilon_+\psi_-,\quad
\delta''\sigma=-i\bepsilon_+\lambda_-,\nn\\
&\delta''\psi_+=-\bepsilon_+\bsigma\phi+\epsilon_+F,\quad
\delta''\blambda_+
=-i\bepsilon_+\{(\bfD-iD_t\vscf_3)+{1\over 2}[\sigma,\bsigma]\}.\nn
\eeqa
It is clear that
$\delta''{\mathcal A}_t$ cannot be written
as ${\mathcal D}_tY=\partial_tY+i[{\mathcal A}_t,Y]$ for any $Y$.
Therefore, the Wilson line does not preserve the ${\mathcal N}=4$
supersymmetry.
Of course as remarked earlier, we do have a Wilson line if we quantize
the ${\mathcal N}=2$ fermi multiplet,
which is $(\psi_-,F)$ in the present case.
If we do so, $-{\mathcal A}_t$ constitutes
a part of the sum of (\ref{LchN4})
and (\ref{LWN4}), and is ${\mathcal N}=4$ supersymmetric only with
 the remaining part of the sum.
The theorem states that there is no
Wilson line which is ${\mathcal N}=4$ supersymmetric by itself.

\subsection*{\rm \underline{Global Symmetry and Real Mass}}

We recall that the R-symmetry group of ${\mathcal N}=2$ supersymmetry
is $U(1)$ generated by $\bfR$ under which the variational parameters
$\epsilon$ and $\bepsilon$ have charges $1$ and $-1$.
The R-charges of the component fields in the vector, chiral and
fermi multiplets
are as follows (the real variables $v_t$, $\vs$ and $D$
cannot have any R-charge):
\beq
\begin{array}{c||c|cc|cc}
&\lambda&\phi&\psi&\eta&F\\
\hline
\bfR&1&r_{\rm c}&r_{\rm c}-1&r_{\rm f}&r_{\rm f}-1
\end{array}
\label{N2Rcharges}
\eeq
$r_{\rm c}:V_{\rm chiral}\to V_{\rm chiral}$ and
$r_{\rm f}:V_{\rm fermi}\to V_{\rm fermi}$
are the R-charges of the chiral and the fermi
multiplets.
The system
has $U(1)$ R-symmetry when we can find $r_{\rm c}$, $r_{\rm f}$
and/or $r_M:M\to M$
so that $E(\phi)$, $J(\phi)$ and/or $Q(\phi)$ have R-charges $r_{\rm f}+1$,
$1-r_{\rm f}^T$, and/or $1$ respectively in the sense that,
\beqa
&E(\omega^{r_{\rm c}}\phi)\,=\,\omega^{r_{\rm f}+1}E(\phi),\\
&J(\omega^{r_{\rm c}}\phi)\,=\,J(\phi)\omega^{1-r_{\rm f}},\\
&\omega^{r_M}Q(\omega^{r_{\rm c}}\phi)\omega^{-r_M}\,=\,\omega Q(\phi).
\eeqa
The system has a discrete R-symmetry when these hold for $\omega$ in
a subset of $U(1)$.

Suppose there is a set $h$ of unitary
transformations $h_{\rm c}$, $h_{\rm f}$ and $h_M$ on $V_{\rm chiral}$,
 $V_{\rm fermi}$ and $M$ that commute with the gauge group action,
obeying
\beqa
&E(h_{\rm c}\phi)\,=\,h_{\rm f}E(\phi),\\
&J(h_{\rm c}\phi)\,=\,J(\phi)h_{\rm f}^{-1},\\
&Q(h_{\rm c}\phi)\,=\,h_MQ(\phi)h_M^{-1}.
\eeqa
Then, we have a global symmetry
that acts on the chiral and fermi multiplets
as $(\phi,\psi)\to (h_{\rm c}\phi,h_{\rm c}\psi)$
and $(\eta,F)\to (h_{\rm f}\eta,h_{\rm f}F)$ while keeping
the vector multiplet $(v_t,\vs,\lambda,D)$ intact.
It commutes with the supercharges ${\bf Q}$ and $\overline{\bf Q}$.
We shall call it a {\it flavor symmetry}.

When there is a continuous group $G_F$ of flavor symmetries,
the theory can be deformed by the following procedure:
Promote $G_F$ to a gauge group,
turn on a supersymmetric configuration
$(v_t^F,\vs^F,D^F,\lambda^F)$ of its vector multiplet, and then demote it
back to a flavor group by turning off the gauge coupling.
In view of (\ref{Vector}), the condition of supersymmetry is
$D^F_t\vs^F=0$, $D^F=0$ and $\lambda^F=0$.
For example, we may turn on constant $v_t^F$ and
$\vs^F=m$ that commute with each other.
This $m$ is an analog of real mass in 3d
${\mathcal N}=2$ theories and twisted mass in $2d$ $(2,2)$ theories,
and shall be called {\it real mass} again. $v_t^F$ may be called
{\it flavor Wilson line}.
This deformation preserves the supersymmetry but deforms
its algebra as
\beq
\{{\bf Q},\overline{\bf Q}\}\,\,=\,\,H-{\bf G}^F(v^F_t+m),
\label{defSA}
\eeq
where ${\bf G}^F(\xi)$ is the Noether charge of $\xi\in i\mathfrak{g}_F$.
The parameters $v_t^F$ and $m$ enter into $L_{\rm chiral}$,
$L_{\rm fermi}$ and ${\mathcal A}_t^F$. The latter two depends only on
the combination $v_t^F+m$.
$L_{\rm chiral}$ depends also on $v_t^F-m$ but
its variation is Q-exact: for $\xi\in i\mathfrak{g}_F$ we have
\beqa
(\delta_{\xi}^{v^F_t}-\delta_{\xi}^{m})L_{\rm chiral}
&=&-i\bphi \xi D_t^{(+)}\phi+iD_t^{(+)}\bphi \xi\phi-2\bpsi \xi\psi\nn\\
&=&i[{\bf Q},\bpsi \xi\phi]-i[\overline{\bf Q},\bphi \xi\psi].
\eeqa
This means that supersymmetric correlators depend on $v_t^F$ and
$m$ only through the combination $v_t^F+m$.

\newcommand{\wtm}{\widetilde{m}}

The R-symmetry group of ${\mathcal N}=4$ supersymmetry is $O(4)$
which rotates the four real components of the supercharges ${\bf Q}_{\pm}$
and $\overline{\bf Q}_{\pm}$.
The R-charge $\bfR$ of the ${\mathcal N}=2$ subalgebra is $\bfR_{\!+}$
that rotates the phases of ${\bf Q}={\bf Q}_+$ and
$\overline{\bf Q}=\overline{\bf Q}_+$ only.
The subgroup of elements
that commutes with ${\bf Q}_+$ and $\overline{\bf Q}_+$
is generated by the charge $\bfR_{\!-}$ which rotates the phases of
${\bf Q}_-$ and $\overline{\bf Q}_-$ only.
In fact, $\bfR_{\!\pm}$ generate the maximal torus of $U(2)\subset O(4)$.
Its $SU(2)$ subgroup is always a symmetry in the systems introduced in
the previous section, under which $\bfx$ and $\bflambda$ form a triplet
and a doublet.
In this paper, we shall consider theories which are invariant also
under the central $U(1)$ subgroup of $U(2)$.
This requires the superpotential to be homogeneous,
that is, there is an endomorphism $R:V\to V$ under which
\beq
W(\omega^{R}\phi)=\omega^2W(\phi).
\eeq
Then, the system is invariant under $SU(2)\times U(1)$ with
the following charge assignment,\footnote{The $J_3$ component of $SU(2)$
and the $U(1)$ generator {\boldmath $R$} is related to $\bfR_{\!\pm}$ as
$2J_3=-\bfR_{\!-}+\bfR_{\!+}$, $\mbox{\boldmath $R$}=\bfR_{\!-}+\bfR_{\!+}$.
$2J_3$ and \mbox{\boldmath $R$} are
respectively the axial and vector R-charges in 2d (2,2)
supersymmetry.}\label{foot:Rcharges}
\beq
\begin{array}{r||c|cc||cc|cc}
&\lambda_-&\sigma&\blambda_+&\phi&\psi_+&\psi_-&F\\
\hline
\bfR=\bfR_{\!+}
&1&1&0&{R\over 2}&{R\over 2}-1&{R\over 2}&
{R\over 2}-1\\
\bfR_{\!-}
&0&-1&-1&{R\over 2}&{R\over 2}&{R\over 2}-1&
{R\over 2}-1
\end{array}
\label{N4Rcharges}
\eeq
Since $\bfR_{\!-}$ commutes with the ${\mathcal N}=2$ supercharges
${\bf Q}_+$ and $\overline{\bf Q}_+$, it can be considered as a flavor
symmetry of the ${\mathcal N}=2$ theory.
Indeed, the $\bfR_{\!-}$ charges in the table (\ref{N4Rcharges})
are the same for the fields in each ${\mathcal N}=2$ multiplet, and
zero for $\lambda_-$ in the ${\mathcal N}=2$ vector multiplet.
If there is a unitary transformation $h:V\to V$
that leaves $W(\phi)$ invariant, then we have a flavor symmetry that
commutes with the ${\mathcal N}=4$ supersymmetry generators.
For a continuous group
of such flavor symmetries, we have a deformation of the theory
by a supersymmetric background $v_t^F$, $\bfx^F=(m_1,m_2,m_3)$ of
the flavor group.
Note that the four elements $v_t^F$, $m_1$, $m_2$, $m_3$
must commute with each other. Extending the terminology in 2d
$(2,2)$ theories, we may refer to $\wtm=m_1+im_2$ as the
{\it twisted mass}.
This deformation preserves the supersymmetry and the supersymmetry algebra
is deformed as
\beqa
&\{{\bf Q}_{\pm},\overline{\bf Q}_{\pm}\}\,=\,
H-{\bf G}^F(v^F_t\pm m_3),\label{defSAN41}\\
&\{{\bf Q}_-,\overline{\bf Q}_+\}\,=\,{\bf G}^F(\widetilde{m}),\quad
\{{\bf Q}_+,\overline{\bf Q}_-\}\,=\,{\bf G}^F(\overline{\widetilde{m}}).
\label{defSAN42}
\eeqa
When there is such a continuous flavor symmetry,
there is an ambiguity in the choice of $R$.
Unlike in theories in higher dimensions where we would like to find
the ``correct'' R-charge that corresponds to the R-symmetry of the
superconformal field theory in the infra-red limit,
there is no distinguished choice in quantum mechanics.

\subsection{The Index}\label{subsec:DefInd}

The main purpose of the present paper is to compute the Witten index
\cite{Windex} of ${\mathcal N}=2$ supersymmetric gauge theories,
possibly with a twist by a flavor symmetry,
\beq
I(h)\,\,=\,\,{\rm Tr}^{}_{\mathcal H}\left((-1)^F\,\widehat{h}
\e^{-\beta H}\right).
\label{indh}
\eeq
This can be identified as the path-integral
on the Euclidean circle of circumference $\beta$
with the boundary condition ${\mathcal O}(\tau)=h{\mathcal O}(\tau+\beta)$
on all fields.
This is equivalent to the circle with periodic boundary condition,
but under the influence of the background flavor gauge field
with holonomy $h$.
Suppose there is a continuous group $G_F$ of flavor symmetries,
and let us consider the index with a twist by an element of the form
$h=\e^{i\xi}$ for $\xi\in i\mathfrak{g}_F$.
Then, the flavor gauge field can be taken to be of the form
$v^F_\tau=-\xi/\beta$.
At the same time, we may also turn on the real mass $m\in i\mathfrak{g}_F$
that commutes with $\xi$.
Then, we have a two parameter family of Witten index
\beq
I(\xi,m)\,\,=\,\,{\rm Tr}^{}_{{\mathcal H}_m}\left((-1)^F
\e^{i{\bf G}^F_{m}(\xi)}\e^{-\beta H_m}\right),
\label{indtwo}
\eeq
where $({\mathcal H}_m,H_m,{\bf G}^F_{m})$ is for the theory deformed by
the real mass $m$.
Due to the remark given in the previous subsection, it depends on
 $-i\xi/\beta+m$ but not on $-i\xi/\beta-m$.
(Note that Wick rotation $t\to -i\tau$ does $v^F_t\to i v^F_{\tau}$.)
That is, it is a holomorphic function of $\xi+i \beta m$.

The index enjoys the well known nice properties \cite{Windex}
when the theory is effectively compact,
that is, when the spectrum is discrete and each level
consists of finite number of square normalizable states.
In particular, it receives contribution only from the finite number of
supersymmetric
ground states, $I(h)={\rm Str}_{{\mathcal H}_{\rm SUSY}}(\widehat{h})$.
For the twist by an element
$\e^{i\xi}$ of a continuous flavor group $G_F$, the function
$\e^{i\xi}\mapsto {\rm Str}_{{\mathcal H}_{\rm SUSY}}(\e^{i{\bf G}^F(\xi)})$, being
a (graded) character of a finite dimensional representation,
 has a holomorphic extension
$\e^{i(\xi+i\beta m)}\mapsto
{\rm Str}_{{\mathcal H}_{\rm SUSY}}(\e^{i{\bf G}^F(\xi+i\beta m)})$.
By the uniqueness of holomorphic extension,
it must agree with (\ref{indtwo}). That is,
\beq
I(\xi,m)\,\,=\,\,
{\rm Tr}^{}_{{\mathcal H}}\left((-1)^F
\e^{i{\bf G}^F(\xi+i\beta m)}\e^{-\beta H}\right).
\label{indidentity}
\eeq
This can also be shown directly: Using the deformed supersymmetry algebra
(\ref{defSA}), we have $\e^{i{\bf G}^F_{m}(\xi)}\e^{-\beta H_m}
=\e^{i{\bf G}_m^F(\xi+i\beta m)}\e^{-\beta\{{\bf Q},\overline{\bf Q}\}}$,
while the charge of a normalizable state is quantized and
is constant under deformation.
In either way, we have seen that the real mass plays
a minor r\^ole for the index --- it simply complexifies the twist parameter.
In other words, the index with a complex twist parameter can be
identified as the index with a unitary twist,
of the theory deformed by a real mass.

Things are subtle when the theory is not effectively compact,
which happens when the scalar potential has a non-compact flat direction.
In such a theory, the index may not even be defined ---
the path-integral may diverge by the non-compact integral without
exponential suppression; also, it is a delicate problem to define
the trace of an operator on a non-separable Hilbert space.
Here the flavor twist and/or the real mass $m$ play an essential r\^ole.
When a maximal torus of the flavor group
acts on the flat direction with a
compact set of fixed points, then the twisted path-integral will converge,
and that would provide a definition of the twisted index.
On the other hand, with only such a twist, it is not clear if
the operator definition as in the right hand side of (\ref{indh})
makes sense. If the real mass is turned on, however,
the potential term $\bphi m^2\phi$ is introduced and the
theory becomes effectively compact.
Then, the index can certainly be defined in the operator formalism by
(\ref{indtwo}).
In what follows, to simplify the notation, we shall
write the twisted index as (\ref{indidentity}) and refer to it
as the index with {\it complex flavor twist},
even if the actual definition may only be (\ref{indtwo}).

One of our main interests in ${\mathcal N}=4$ theories
is the index twisted by the R-symmetry $\bfR_-$ that commutes with
$\bfR=\bfR_{\!+}$,
\beq
I(\,\by^{2\bfsmR_{\!-}})\,\,=\,\,
{\rm Tr}^{}_{\mathcal H}\left((-1)^F\by^{2\bfsmR_{\!-}}\e^{-\beta H}\right),
\label{indN4}
\eeq
for $\by\in \C^{\times}$.
Of course, if there is a flavor symmetry, i.e.,
$h:V\to V$ that leaves $W(\phi)$
invariant, we can also twist the index by that.
For a continuous group of flavor symmetries, not only the twist but also
the triplet mass deformation $(m_1,m_2,m_3)$ may be considered. However,
the twisted mass deformation $\widetilde{m}=m_1+im_2$
is forbidden in the presence of the R-twist
$\by^{2\bfsmR_{\!-}}$. This is because $\bfR_-$ rotates the complex scalar
$\sigma$ of the ${\mathcal N}=4$ vector multiplet,
and hence would not be a symmetry
if the twisted mass is turned on.
Only the real mass $m_3$ is allowed and we have
the complex flavor twist (along with $\by^{2\bfsmR_-}$),
as in the general ${\mathcal N}=2$ theories.

When the theory is effectively compact,
the supersymmetric ground states have vanishing flavor charge.
This can be shown as follows.
Let us consider the twisted mass deformation. (This is just for the proof
of the claim.)
The deformed theory has a deformed supersymmetry algebra (\ref{defSAN42}).
Taking the $\wtm$ derivative
and then setting $\wtm=0$, we find
\beq
\{{\bf Q}_-,\delta_{\wtm}\overline{\bf Q}_+\}+
\{\delta_{\wtm}{\bf Q}_-,\overline{\bf Q}_+\}
={\bf G}^F(\delta\widetilde{m}).
\label{Opid}
\eeq
Suppose $|\Psi\rangle$ is a supersymmetric ground state.
Applying the above operator identity to $|\Psi\rangle$,
we see that ${\bf G}^F(\delta\wtm)|\Psi\rangle$ is of the form
${\bf Q}_-|\alpha\rangle+\overline{\bf Q}_+|\beta\rangle$.
Since ${\bf G}^F$ is a symmetry, ${\bf G}^F(\delta\wtm)|\Psi\rangle$ is also a
supersymmetric ground state.
If $\Pi_0$ denotes the projection to the space of supersymmetric ground states,
we find ${\bf G}^F(\delta\wtm)|\Psi\rangle=\Pi_0{\bf G}^F(\delta\wtm)
|\Psi\rangle
=\Pi_0({\bf Q}_-|\alpha\rangle+\overline{\bf Q}_+|\beta\rangle)=0$.
Since this holds for any $\delta\wtm$,
this means that the ground state $|\Psi\rangle$ has vanishing flavor charge.
Note that we needed the effective compactness of the theory so that
the spectrum changes smoothly under the twisted mass deformation.

This means that the index does not depend on the twist by a continuous
flavor symmetry. In particular, possible ambiguity in the R-charge
assignment does not affect the index.
Therefore,
``the index with just the R-twist (\ref{indN4})'' has an unambiguous meaning
and that is our primary interest in
effectively compact ${\mathcal N}=4$ theories.

When the theory is not effectively compact, the spectrum changes
discontinuously when the twisted mass deformation is turned on,
and the above argument does not apply. As remarked above in
a general ${\mathcal N}=2$ theory, for a non-compact theory,
flavor twist is very important even for having a well-defined index,
and the result of course depends on the twist parameters.

The above proof does not apply to a theory with only
${\mathcal N}=2$ supersymmetry, even if it is compact.
One might try to repeat the argument using
the deformed algebra (\ref{defSA}).
But, as the real mass $m$ is deformed, the Hamiltonian also changes
and one cannot conclude that the ground states have vanishing flavor charge.
In fact, our primary interest in ${\mathcal N}=2$ theories is
the dependence on the flavor twist. Also, presence of flavor twist
is sometimes necessary for the localization computation.

%(so that)
%the variation of operators on a fixed space of states is valid and also to
%ensure the existence of a well defined projection operator $\Pi_0$
%satisfying $\Pi_0\circ {\bf Q}_-=\Pi_0\circ \overline{\bf Q}_+=0$.

\subsection{Phases}\label{subsec:phases}

Let us consider a system of ${\mathcal N}=2$ supersymmetric
quantum mechanics with gauge group $G$,
a chiral multiplet in $V_{\rm chiral}$ and a fermi multiplet
in $V_{\rm fermi}$.
As the Lagrangian, we take the sum of
$L_{\rm gauge}$ from (\ref{Lgauge}),
$L_{\rm chiral}$ from (\ref{Lchiral}),
$L_{\rm fermi}$ from (\ref{Lfermi}),
$L_{\rm FI}$ from (\ref{FI}) and
$L_J$ from (\ref{2superpotential}), and we may also include
the Wilson line $-{\mathcal A}_t$ from (\ref{calA}).
(Not all of them has to be non-trivial.)

Our main interest is the space of supersymmetric ground states.
Classically, supersymmetric vacua correspond to
configurations of the scalar
variables, $\vs$ and $\phi$, at which the potential vanishes.
The scalar potential of the system is
\beqa
U(\vs,\phi)&=&|\vs \phi|^2+{e^2\over 2}\Bigl(\,\phi\bphi-\zeta\,\Bigr)^2
+|E(\phi)|^2+|J(\phi)|^2\nn\\
&&+\rho(\vs)+\{Q(\phi),Q(\phi)^{\dag}\}.
\label{UN2}
\eeqa
The second and the fourth terms are obtained after integrating out the
auxiliary fields $D$ and $F$ and called the $D$-term potential
and the $F$-term potential respectively.
${e^2\over 2}(\,\cdot\,)^2$ is the quadratic form on $i\frakg^*$
defined as the dual of ${1\over 2e^2}{\rm Tr}(\,\cdot\,)^2$.
$\phi\bphi$ inside that parenthesis is the moment map of
the $G$ action on $V_{\rm chiral}$ defined by
$\langle \phi\bphi,\xi\rangle:=\bphi\,\xi\phi$.
When $M$ has rank $2$ or higher, the potential is valued in
hermitian matrices.
The term $\rho(\vs)$ is not positive definite, which may appear
strange in the presence of supersymmetry.
However, as we will see, positivity does exist if
the Gauss law is taken into account.
Since that is a quantum effect, in the classical analysis which we present
now, we shall simply forget about the term $\rho(\vs)$.
Then the vacuum equations are
\beqa
&\vs\phi=0,\label{vaceq1}\\
&\phi\bphi=\zeta,\label{vaceq2}\\
&E(\phi)=0,\quad
J(\phi)=0,\quad
\{Q(\phi),Q(\phi)^{\dag}\}=0.
\label{vaceq3}
\eeqa

Notice that the FI parameter $\zeta$ enters into the second equation
(\ref{vaceq2}), called the D-term equation.
For a generic $\zeta$, this forces $\phi$ to have a non-zero value, which
breaks the gauge group $G$ to its subgroup $G_1$
and forces $\vs$ to lie in its Lie algebra.
The space of solutions to the vacuum equations
is in general a union of components labeled by the unbroken subgroup
$G_1$. 
A component is called a {\it Higgs branch} if $G_1$ is a finite
subgroup so that $\phi\ne 0$ and $\vs=0$,
a {\it Coulomb branch} if $G_1$ is $G$ itself so that $\phi=0$ and $\vs\ne 0$,
and a {\it mixed branch} if $G_1$ is a continuous proper subgroup
of $G$ so that $\phi$ and $\vs$ are both non-zero while obeying $\vs\phi=0$.
At each branch the gauge group is further broken to a subgroup of
$G_1$ by the values of $\vs$, generically to its maximal torus
--- thus the name ``Coulomb''.
The pattern of gauge symmetry breaking depends very much on the value
of $\zeta$. The space $i\frakz^*$ of FI parameters is stratified by cones
labelled by the symmetry breaking pattern. The cones of the maximal
dimension shall be called the {\it phases}, following the terminology
used in two dimensional systems. Cones of codimension one
are called {\it walls} or {\it phase boundaries}.
When the gauge group $G$ is Abelian, it is broken to a finite subgroup
in each phase and to a rank one subgroup at each wall.
When $G$ is non-Abelian, the situation can be different.
Some models contain phases with continuous unbroken subgroups.
Examples are the 1d versions of the 2d models
in \cite{HoTo,Hduality,HoKn1}.

Inside a phase where the gauge group is broken to a finite subgroup
and the vacuum manifold is a Higgs branch,
we can often find an effective description of the system that
correctly captures the supersymmetric ground states.
Since the continuous part of the gauge group is completely broken,
the vector multiplet together with the gauge orbit direction of
the chiral multiplet acquire a non-zero mass by
the Higgs mechanism.
(In quantum mechanics, ``mass'' should
better be rephrased as ``frequency'', but we shall keep
the terminology used in higher dimensions.)
By looking at the term $|\vs\phi|^2$ in the potential
and the D-term equation $\phi\bphi=\zeta$, we find that
the mass is of the order of
\beq
M_H=e\sqrt{|\zeta|},
\label{HiggsEnergy}
\eeq
where $|\zeta|$ is the typical size of the FI parameter that is responsible
for the gauge symmetry breaking.\footnote{In one dimension,
the various fields and parameters have the following dimensions:
$[D]={\rm energy}^2$, $[\zeta D]=[D^2/e^2]={\rm energy}$, and hence
$[\zeta]={\rm energy}^{-1}$ and
$[e^2]={\rm energy}^3$.}\label{foot:dimension}
Therefore, the vector multiplet
and the gauge orbit modes decouple from the physics
at the energy scale $E$ below $M_H$.
%One may say ``Deep inside the phase, we can forget about these Higgsed modes
%at low energy''.

When the FI parameter $\zeta$ is on a phase boundary,
the space of classical vacua has mixed or Coulomb branch components.
Quantum analysis is required to obtain
the effective theory at or near the phase boundary.
Simple cases will be treated in the next section.

Let us make a brief comment on the ${\mathcal N}=4$ systems.
The vector multiplet includes an ${\mathcal N}=2$ chiral multiplet
$(\sigma,i\blambda_+)$ in the adjoint,
and the chiral multiplet includes an ${\mathcal N}=2$ fermi multiplet
$(\psi_-,F)$ with $E=\sigma\phi$.
Therefore, the scalar potential reads
\beqa
U(\bfx,\phi)&=&{1\over 2e^2}|[\vscf_3,\sigma]|^2
+|\vscf_3 \phi|^2+{e^2\over 2}\Bigl(\,\phi\bphi
+{1\over 2e^2}[\sigma,\bsigma]-\zeta\,\Bigr)^2
+|\sigma\phi|^2+|\dd W(\phi)|^2\nn\\
&=&{1\over 2e^2}\sum_{i<j}|[\vscf_i,\vscf_j]|^2
+|\bfx \phi|^2+{e^2\over 2}\Bigl(\,\phi\bphi-\zeta\,\Bigr)^2
+|\dd W(\phi)|^2,
\label{UN4}
\eeqa
where we paid attention to the normalization ${1\over 2e^2}$
of the kinetic term of $(\sigma,i\blambda_+)$ in (\ref{N4Lvec}). We also
 used $\zeta([\sigma,\bsigma])=0$
and $|\vscf_3\phi|^2+{1\over 2}\bphi[\sigma,\bsigma]\phi
+|\sigma\phi|^2=|\bfx\phi|^2$ to go to the second line.
Deep inside a phase where the gauge group is broken to a finite subgroup,
the entire ${\mathcal N}=4$ vector multiplet together with
the ${\mathcal N}=4$
chiral multiplet in the gauge orbit direction decouple from
the low energy physics.
In particular, we can set all three components of $\bfx$ to zero.
This is the ${\mathcal N}=4$ Higgs branch.
At the phase boundary, the vacuum manifold has mixed or Coulomb
branch components. In each of them, the vanishing of the term
$|[\vscf_i,\vscf_j]|^2/2e^2$ requires all three components of $\bfx$ to
lie in the Lie algebra of a common maximal torus of
the unbroken gauge group $G_1$.

Now, let us describe two typical phases --- the geometric phase
and the Landau-Ginzburg phase. There can also be hybrid of the
geometric and Landau-Ginzburg phase.

\subsection*{\underline{\rm Geometric Phase}}

Suppose $G$ is completely broken,
the set of solutions to the equations of (\ref{vaceq1})-(\ref{vaceq3}) is a
compact smooth manifold ${\mathcal C}_{\rm vac}$, and the
modes transverse to the solution space have non-zero masses.
Then, as the effective description, we may
take a non-linear sigma model whose target space is the
quotient space $X={\mathcal C}_{\rm vac}/G$, which is a K\"ahler manifold.
The sigma model may have fermions valued in a holomorphic vector bundle
and/or a Wilson line determined by a complex of vector bundles.

The non-linear sigma model with a target K\"ahler manifold $(X,g)$
is described by a chiral multiplet $(\phi,\psi)$ where
$\phi$ takes values in $X$ and $\psi$ takes values in the pull back
$\phi^*T_X$ ($T_X$ is the holomorphic tangent bundle of $X$).
Its Lagrangian is
\beq
L_X=g(\dot{\bphi},\dot{\phi})+ig(\bpsi,D_t\psi) \ ,
\eeq
where $D_t\psi$ is the covariant derivative with respect to the
pulled back Levi-Civita connection.
The model may have a fermi multiplet $(\eta,F)$ with values in
a holomorphic vector bundle with a hermitian metric $({\mathcal E},h)$
on $X$. Its Lagrangian is
\beq
L_{\mathcal E}=ih(\bareta,D_t\eta)+h(\bareta,F_A(\psi,\bpsi)\eta)
+h(\overline{\tilde{F}},\tilde{F}),
\eeq
where $D_t\eta$ is the covariant derivative with respect to
the pull back of the hermitian connection $A$ of $({\mathcal E},h)$,
$F_A(\psi,\bpsi):=F_{i\bj}\psi^i\bpsi^{\bj}$
with the curvature $F_{i\bj}$ of $A$, and
$\tilde{F}:=F-\psi^jA_j\eta$ is the covariantized auxiliary field.
See Appendix~\ref{app:NLSM} for a more detailed description.
The model may also have a Wilson line associated to a
complex of vector bundles
\beq
\cdots\stackrel{Q^{i-2}}{\longrightarrow}
{\mathcal F}^{i-1}\stackrel{Q^{i-1}}{\longrightarrow}
{\mathcal F}^i\stackrel{Q^i}{\longrightarrow}
{\mathcal F}^{i+1}\stackrel{Q^{i+1}}{\longrightarrow}
\cdots.
\eeq
Each ${\mathcal F}^i$ is
equipped with a hermitian metric $h^i$ which determines a hermitian
connection $A^i$. The Wilson line is valued in
${\mathcal End}({\mathcal F},{\mathcal F})$ with
${\mathcal F}=\oplus_i{\mathcal F}^i$ and is given by
\beq
{\mathcal A}_t=-i(\phi^*A)_t-{1\over 2}F_A(\psi,\bpsi)
-\psi^i\partial_i Q(\phi)+\bpsi^{\bi}\partial_{\bi}Q(\phi)^{\dag}
+\{Q(\phi),Q(\phi)^{\dag}\}.
\eeq
The construction is the same
as in $2d$ $(2,2)$ theories with boundary. See \cite{HHP} for detail.

The anomaly free condition is reflected in the effective
description as the condition that the fermionic Hilbert space
tensored with ${\mathcal F}$,
\beq
\sqrt{K}_X\otimes\wedge T_X\otimes
{\det}^{-{1\over 2}} {\mathcal E}\otimes \wedge {\mathcal E}
\otimes {\mathcal F}
\label{HFNLSM}
\eeq
forms a well-defined vector bundle on $X$. Note that
${\mathcal F}$ should not be a well-defined vector bundle
if $\sqrt{K}_X\otimes \det^{-{1\over 2}}{\mathcal E}$ is not.
The bundle (\ref{HFNLSM}) is isomorphic to
$\Omega_X^{0,\bullet}\otimes \mathfrak{F}$
with $\mathfrak{F}:=\sqrt{K}_X\otimes
{\det}^{-{1\over 2}} {\mathcal E}\otimes \wedge {\mathcal E}
\otimes {\mathcal F}$ by the isomorphism
$T_X\cong \overline{T}_X^*$.

The states of the system are sections of the vector bundle
(\ref{HFNLSM}), or equivalently, antiholomorphic forms with values in
$\mathfrak{F}$.
As always in a theory with the supersymmetry (\ref{N2SUSY}),
the space of supersymmetric ground states is isomorphic to
the cohomology group of one of the supercharges, say $\overline{\bf Q}$.
This operator acts on the states as
\beq
i\overline{\bf Q}\,s\,\,=\,\,\bartial\,s\,+\,Q\, s.
\label{Qbaraction}
\eeq
Thus, the space of supersymmetric ground states is
\beq
{\mathcal H}^{}_{\rm SUSY}\,\cong\,
H^{0,\bullet}_{\bartial+Q}(X,\mathfrak{F}).
\eeq
Since $X$ is compact, the Witten index is given by the Riemann-Roch formula
\beq
I({\rm id})=\int_X{\rm td}_X
\e^{-{1\over 2}c_1(X)-{1\over 2}c_1({\mathcal E})}{\rm ch}(\wedge{\mathcal E})
{\rm ch}({\mathcal F}) \ ,
\label{RR}
\eeq
where ${\rm ch}(\wedge{\mathcal E})
=\sum_{i=0}^{{\rm rank}{\mathcal E}}(-1)^i{\rm ch}(\wedge^i{\mathcal E})$
and ${\rm ch}({\mathcal F})=\sum_i(-1)^i{\rm ch}({\mathcal F}^i)$.

Suppose $X$ admits a spin structure, ${\mathcal E}=0$, and
${\mathcal F}$ is just a vector bundle
(${\mathcal F}^i=0$ for $i\ne 0$).
Then, $\sqrt{K}_X\otimes \wedge T_X$ makes sense as a vector bundle
--- it is the spinor bundle of $X$.
Therefore, the states are spinors with values in ${\mathcal F}$.
One can also show that the supercharge $\overline{\bf Q}+{\bf Q}$
is the Dirac operator.
Indeed,
(\ref{RR}) is the Atiyah-Singer formula for the
Dirac index since
${\rm td}_X\e^{-{1\over 2}c_1(X)}=\widehat{\rm A}_X$.

The ${\mathcal N}=4$ supersymmetric non-linear sigma model
with the target $(X,g)$
is obtained by setting ${\mathcal E}=T_X$ and ${\mathcal F}=0$.
In this case, we have $\mathfrak{F}=K_X\otimes\wedge T_X
\cong \wedge T^*_X$ with a shift of the $\Z_2$ grading by $n:=\dim X$.
Hence the states are differential forms on $X$, and
the space of supersymmetric ground states
is the Dolbeault cohomology, which is isomorphic to
the de Rham cohomology for the K\"ahler manifold,
\beq
{\mathcal H}_{\rm SUSY}^{}\,\cong\,
\bigoplus_{p,q=0}^nH^{0,q}_{\bartial}(X,\wedge^p T^*_X[n])
\cong H^{\ast}_{\rm de\, Rham}(X,\C)[n].
\eeq
Vanishing of the flavor charge of the supersymmetric ground states
is obvious in this representation, thanks to the identity
${\mathcal L}_v=d\circ i_v+i_v\circ d$ for a vector field $v$ on $X$.
In fact, this is the origin of the general statement and its proof
based on the identity (\ref{Opid}), presented in
Section~\ref{subsec:models}.
The ground state for a $(p,q)$ class carries
 $2J_3=p+q-n$ and $\mbox{\boldmath $R$}=-p+q$.
If we take $(-1)^F=(-1)^{2J_3}$,
the canonically twisted index is
\beq
I(\by^{2\bfsmR_{\!-}})=\sum_{p,q=0}^n(-1)^{p+q-n}\by^{-2p+n}h^{p,q}(X)
=\sum_{p=0}^n(-1)^{p-n}\by^{-2p+n}\chi(X,\wedge^pT^*_X).
\label{chiy}
\eeq
This is the $\chi_y$ genus of $X$ for $y=\by^{-2}$,
up the prefactor $(-\by)^n$.
The untwisted index $I({\rm id})$ is the Euler number $\chi(X)$
times $(-1)^n$.
Indeed,
(\ref{RR}) is the Gauss-Bonnet formula since
${\rm td}_X\e^{-c_1(X)}{\rm ch}(\wedge T_X)=(-1)^{\dim X}e(T_X)$
which follows from ${x\over 1-e^{-x}}\e^{-x}(1-\e^{x})=-x$.

We should note that one of the first applications of supersymmetry
after \cite{Windex} is the derivation of these index formulae
via supersymmetric localization of the non-linear sigma models
\cite{Alv,FriWin}.

\subsection*{\underline{\rm Landau-Ginzburg Phase}}

Suppose the vacuum equations (\ref{vaceq1})-(\ref{vaceq3})
have a unique (up to gauge) solution
at which the gauge group $G$ is broken to a finite group $\Gamma$.
Then, as the effective theory, we can take the Landau-Ginzburg model
of the massless variables modulo the gauge group $\Gamma$,
that is, the Landau-Ginzburg orbifold.

A Landau-Ginzburg orbifold is obtained by the gauge theory described
in the previous subsection by taking the gauge group to be the finite
group $\Gamma$. In particular, there is no vector multiplet. The only
r\^ole of the gauge group is the Gauss law, i.e.,
$\Gamma$ invariance of physical states, which makes sense when we can find
$\det^{-{1\over 2}}(V_{\rm chiral}\oplus V_{\rm fermi})\otimes M$
as a genuine representation of $\Gamma$.
As remarked earlier, a fermi multiplet can be traded into a part of
the Wilson line. Thus, we may consider only the chiral multiplet
valued in $V_{\rm chiral}$ with a Wilson line determined by
$(M,\rho,Q)$. The anomaly free condition is then
$\mathfrak{F}=\det^{-{1\over 2}}V_{\rm chiral}\otimes M$
is a genuine representation of $\Gamma$.

The states are $\Gamma$-invariant
antiholomorphic differential forms on $V_{\rm chiral}$
with values in $\mathfrak{F}$,
and the supercharge $\overline{\bf Q}$ acts as (\ref{Qbaraction}).
By the standard argument (see for example \cite{HHP} Section~2.2.3),
the $\bartial+Q$ cohomology is isomorphic to the $Q$ cohomology acting
on the space ${\mathcal Hol}(V_{\rm chiral},\mathfrak{F})$ of
holomorphic functions on $V_{\rm chiral}$ with values in $\mathfrak{F}$. Thus,
the space of supersymmetric ground states is
\beq
{\mathcal H}_{\rm SUSY}^{}\,\cong\,
H^{\bullet}_Q({\mathcal Hol}(V_{\rm chiral},\mathfrak{F}))^{\Gamma},
\eeq
where the superscript $\Gamma$ stands for taking the $\Gamma$ invariants.
When $(M,\rho,Q)$ comes from a fermi multiplet,
as in (\ref{Mfer}) and (\ref{Qfer}),
this is the untwisted and left-handed sector of the $2d$
version of the system
\cite{KacWit,DK}.

The ${\mathcal N}=4$ Landau-Ginzburg orbifold with a
$\Gamma$-invariant superpotential $W:V\to\C$
is obtained by taking  $M=\det^{-{1\over 2}}V\otimes\wedge V$
and $Q=-\dd W\lrcorner$. In this case,
we have an isomorphism
$\mathfrak{F}=\det^{-1}V\otimes \wedge V\cong \wedge V^*$
(with a shift of $\Z_2$ grading by $n:=\dim V$)
of $\Gamma$ representations,
and $Q$ acts on ${\mathcal Hol}(V,\wedge V^*)$ by $-\dd W\wedge$.
The cohomology concentrates on the top degree forms
and we have
\beq
{\mathcal H}^{}_{\rm SUSY}\,\cong\,
\left({{\mathcal Hol}(V,\wedge^nV^*)\over \dd W\wedge
{\mathcal Hol}(V,\wedge^{n-1}V^*)}\right)^{\Gamma}[n].
\eeq
This is nothing but the untwisted RR ground states
in the $2d$ version of the system \cite{IntVaf}.

\subsection{Examples}
\label{subsec:LSMex}

\subsection*{Example 1: Projective Space}

Let us first consider the ${\mathcal N}=2$
system with $U(1)$ gauge group and
$N$ chiral multiplets $(\phi_1,\psi_1),\ldots,(\phi_N,\psi_N)$
of charge $1$. We turn on the Wilson line ${\mathcal A}_t=\rmq(v_t+\vs)$.
The anomaly free condition is
\beq
{N\over 2}+\rmq\in\Z.
\label{afProj}
\eeq
The FI parameter $\zeta$ is naturally a real number and the D-term
equation reads $|| \phi||^2=\zeta$.
When $\zeta$ is positive, the solution breaks the gauge group completely.
The $U(1)$ quotient of the space of solutions
 is the complex projective space $\CP^{N-1}$.
The modes transverse to the solution space are all massive and we are
left with the non-linear sigma model whose target space is $\CP^{N-1}$.
Thus $\zeta>0$ is a geometric phase.
The equation of motion sets $v_t$ to be the pull back of the
gauge connection of the line bundle ${\mathcal O}(1)$ and therefore the
Wilson line induces the ``line bundle'' ${\mathcal F}={\mathcal O}(\rmq)$.
This yields $\mathfrak{F}={\mathcal O}(\rmq-{N\over 2})$, which
is indeed a genuine line bundle over $\CP^{N-1}$ by the anomaly free condition
(\ref{afProj}). The space of ground states of the effective
sigma model is isomorphic to the Dolbeault cohomology
$\oplus_iH^{0,i}(\CP^{N-1},{\mathcal O}(\rmq-{N\over 2}))$.
The dimension of each component is known to be
\beqa
\rmq\geq {N\over 2}:&&
h^{0,i}(\CP^{N-1},{\mathcal O}({\textstyle \rmq-{N\over 2}}))=
\left\{\begin{array}{ll}
0&i\ne 0\\
{\rmq+{N\over 2}-1\choose N-1}&i=0,
\end{array}\right.
\label{hCPN1}\\
-{N\over 2}<\rmq<{N\over 2}:
&&h^{0,i}(\CP^{N-1},{\mathcal O}({\textstyle \rmq-{N\over 2}}))=0
\quad\forall i,
\label{hCPN2}\\
\rmq\leq -{N\over 2}:&&
h^{0,i}(\CP^{N-1},{\mathcal O}({\textstyle \rmq-{N\over 2}}))=
\left\{\begin{array}{ll}
0&i\ne N-1\\
{{N\over 2}-\rmq-1\choose N-1}&i=N-1.
\label{hCPN3}
\end{array}\right.
\eeqa
When $\zeta$ is negative, there is no solution to the D-term equation,
and we expect that there is
no supersymmetric ground state.

Let us also consider the ${\mathcal N}=4$ system with $U(1)$ gauge group
and $N$ chiral multiplets $(\phi_1,\psi_{1\pm},F_1),\ldots,
(\phi_N,\psi_{N\pm},F_N)$ of charge $1$. No superpotential is allowed
for this matter content, and there is no room for Wilson line in 
an ${\mathcal N}=4$ system as shown in Section~\ref{subsec:models}.
 The D-term equation is
still $||\phi||^2=\zeta$, see (\ref{UN4}).
When $\zeta$ is positive, the solution breaks the gauge group completely.
We are in the geometric phase where we have
the ${\mathcal N}=4$ supersymmetric non-linear sigma model
with the target space $\CP^{N-1}$ as the effective description.
The space of supersymmetric ground states is
$\oplus_{p,q}H^{p,q}(\CP^{N-1})$ whose dimension is
\beq
h^{p,q}(\CP^{N-1})=\left\{\begin{array}{ll}
1&p=q=0,\ldots, N-1,\\
0&\mbox{otherwise}.
\end{array}\right.
\eeq
When $\zeta$ is negative, there is no solution to the D-term equation,
and we expect that the supersymmetry is broken.

\subsection*{Example 2:  Hypersurface in Projective Space}

Let us consider the model which has a geometric phase
and Landau-Ginzburg phase.

\noindent
\underline{${\mathcal N}=4$ Model}

We start with a much familiar system with ${\mathcal N}=4$ supersymmetry,
with gauge group $U(1)$, $N$ chiral multiplets
$(X_1,\psi_{1\pm},F_1),\ldots,
(X_N,\psi_{N\pm},F_N)$ of charge $1$,
one chiral multiplet $(P,\psi_{P\pm},F_P)$ of charge $-d$, and
superpotential $W=Pf(X)$, where $f(X)$ is a polynomial
of degree $d$ which is generic in the sense that
the common zero of $f(X)$ and its first derivatives is $X=0$ only.
The D-term equation is
$||X||^2-d|P|^2=\zeta$, while the F-term equations are
$f(X)=0$ and $P\partial_1f(X)=\cdots=P\partial_Nf(X)=0$.

When $\zeta$ is positive, the solution to the D-term equation
has non-zero $X$ which breaks the gauge group completely.
By the genericity of $f(X)$, the F-term equations require
 $P=f(X)=0$. Massless modes are modes of $X$ multiplets
tangent to the vacuum manifold, which is the hypersurface
$X_f=(f=0)$ of the projective space $\CP^{N-1}$.
It is a geometric phase described effectively
by the ${\mathcal N}=4$ sigma model whose target is $X_f$.
The space of supersymmetric ground states is isomorphic to the
cohomology group $\oplus_{p,q}H^{p,q}(X_f)$.

When $\zeta$ is negative, the solution to the D-term equation has
non-zero $P$ which breaks the gauge groups to $\Z_d\subset U(1)$.
By the genericity of $f(X)$, the solution to the vacuum equation
is unique up to gauge: $P\ne 0, X=0$.
Only the $X$ multiplets are massless, and
we are left with the Landau-Ginzburg orbifold with superpotential
$W(X)$ and orbifold group $\Z_d$.

\noindent
\underline{${\mathcal N}=2$ Model}

We next consider a system with ${\mathcal N}=2$ supersymmetry.
This is taken from the two-dimensional version \cite{DK} and shall be
called the Distler-Kachru model.
It is a $U(1)$ gauge theory with $N$ chiral multiplets
$(X_1,\psi_1),\ldots,(X_N,\psi_N)$
of charge $1$, one chiral multiplet $(P,\psi_P)$ of charge
$-\ell$, one fermi multiplet $(\xi,F_\xi)$ of charge $-d$,
$M$ chiral fermi multiplets $(\eta^1,F^1),\ldots, (\eta^M,F^M)$
of charge $q_1,\ldots, q_M$, and the superpotential
$\mathfrak{W}=f(X)\xi+\sum_{\alpha}Pg_\alpha(X)\eta^{\alpha}$,
where $f(X), g_1(X),\ldots, g_M(X)$ are polynomials
 of degree $d, \ell-q_1,\ldots, \ell-q_M$ which are generic in the sense that
their common zero is $X=0$ only. (This requires
$M+1\geq N$.) We do not turn on
the Wilson line. The anomaly free condition is
\beq
N-\ell-d+\sum_{\alpha=1}^Mq_{\alpha}\in 2\Z.
\label{afex2}
\eeq
The model has a global $U(1)_F$ symmetry under which $X$, $P$,
$\xi$, $\eta$ have charges $0$ $1$, $0$, $-1$ respectively.
The D-term equations are $||X||^2-\ell|P|^2=\zeta$ and the F-term equations
are $f(X)=Pg_1(X)=\cdots =Pg_M(X)=0$.

When $\zeta$ is positive, the solution to the D-term equations
have non-zero $X$ which breaks the gauge group completely.
The F-term equations require $P=f(X)=0$.
Massless modes are modes of the $X$ multiplets tangent to the vacuum
manifold $X_f$ as well as the modes of $\eta$ satisfying
the equation $\sum_{\alpha}g_{\alpha}(X)\eta^{\alpha}=0$.
It is a geometric phase described by the sigma model with the target $X_f$
and with the fermi multiplet valued in the kernel ${\mathcal E}$ of
\beq
\bigoplus_{\alpha=1}^M{\mathcal O}(q_{\alpha})
\stackrel{(g_{\alpha})_{\alpha=1}^M}{\longlongrightarrow}{\mathcal O}(\ell).
\eeq
The space of states in this effective theory is the 
space of anti-holomorphic forms
with values in $\mathfrak{F}=
\sqrt{K}_{X_f}\otimes\det^{-{1\over 2}}{\mathcal E}\otimes \wedge{\mathcal E}
={\mathcal O}(-{1\over 2}(N-d+\sum_{\alpha=1}^Mq_{\alpha}-\ell))
\otimes \wedge{\mathcal E}$ which is indeed a well-defined graded
vector bundle over $X_f$ in view of (\ref{afex2}).

When $\zeta$ is negative, the solution to the D-term equation has
non-zero $P$ which breaks the gauge groups to $\Z_\ell\subset U(1)$.
The solution to the vacuum equation
is unique up to gauge: $P\ne 0, X=0$.
Massless modes are the chiral multiplet of $X$
and the chiral fermi multiplets of $\xi,\eta^1,\ldots,\eta^M$.
The $U(1)_F$ charges are
$1/\ell$ for $X$, $-d/\ell$ for $\xi$ and
$-1+q_{\alpha}/\ell$ for $\eta^{\alpha}$ --- these are obtained
by dressing the original assignment
with a gauge symmetry so that $P$ has charge zero.
We are left with the Landau-Ginzburg orbifold of these variables
with the superpotential
$\mathfrak{W}=f(X)\xi+\sum_{\alpha}g_\alpha(X)\eta^\alpha$
and the orbifold group $\Z_\ell$.

\subsection*{Example 3: Quiver Quantum Mechanics}

\begin{figure}[htb]
\psfrag{ka}{$k_A$}
\psfrag{kb}{$k_B$}
\psfrag{kc}{$k_C$}
\psfrag{X}{$X_{1,\ldots,a}$}
\psfrag{Y}{$Y_{1,\ldots,b}$}
\psfrag{Z}{$Z^{1,\ldots,c}\!\!$}
\centerline{\includegraphics{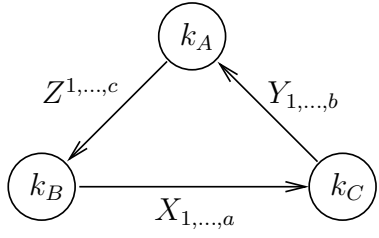}}
\caption{triangle quiver}
\label{fig:quiverT}
\end{figure}

An interesting class of examples with ${\mathcal N}=4$ supersymmetry is
quiver quantum mechanics. Let us describe the phases for the
triangle quiver as in Fig.~\ref{fig:quiverT}.
It has a gauge group
$G=(U(k_A)\times U(k_B)\times U(k_C))/U(1)_{\rm diag}$,
bifundamental matters as in Fig.~\ref{fig:quiverT},
and a generic superpotential of the form
\beq
W=\sum_{i,j,h}A^{ij}_h{\rm tr}(Z^hY_jX_i).
\eeq
We parameterize the FI parameter as\footnote{It is
an invariant linear form on
$\mathfrak{g}=\mathfrak{u}(k_A)\oplus \mathfrak{u}(K_B)\oplus
\mathfrak{u}(k_C)/\mathfrak{u}(1)_{\rm diag}$, i.e., it must be invariant
under conjugation and the shift of $(D_A,D_B,D_C)$ by
$({\bf 1}_{k_A},{\bf 1}_{k_B},{\bf 1}_{k_C})$. A general solution is
(\ref{FItriangle}).}
\beq
\zeta(D)={\zeta^2\over k_A}{\rm tr}(D_A)-{\zeta^1\over k_B}{\rm tr}(D_B)
+{\zeta^1-\zeta^2\over k_C}{\rm tr}(D_C).\label{FItriangle}
\eeq
The D-term and F-term equations read
\beq
\begin{array}{ll}
\displaystyle
YY^{\dag}-Z^{\dag}Z={\zeta^2\over k_A}{\bf 1}_{k_A},&
\displaystyle
\sum_{j,h}A^{ij}_hZ^hY_j=0,\quad\forall i,
\\
\displaystyle
ZZ^{\dag}-X^{\dag}X=-{\zeta^1\over k_B}{\bf 1}_{k_B},&
\displaystyle
\sum_{i,h}A^{ij}_hX_iZ^h=0,\quad\forall j,
\\
\displaystyle
XX^{\dag}-Y^{\dag}Y={\zeta^1-\zeta^2\over k_C}{\bf 1}_{k_C},~~~~~
&
\displaystyle
\sum_{i,j}A^{ij}_hY_jX_i=0,\quad\forall h.
\end{array}
\label{VEtriangle}
\eeq
where $XX^{\dag}:=\sum_iX_iX_i^{\dag}$ etc.

In order to find the phase structure, it is best to start looking for
possible solutions to (\ref{VEtriangle}) with unbroken gauge symmetry.
A typical unbroken subgroup is associated with
a partition of the rank vector $(k_A,k_B,k_C)$ into two,
$(k_A',k_B',k_C')+(k_A'',k_B'',k_C'')$, and includes the $U(1)$
subgroup of elements of the form
\beq
\left[
\left(\begin{array}{cc}
z'{\bf 1}_{k_A'}&0\\
0&z''{\bf 1}_{k_A''}
\end{array}\right),
\left(\begin{array}{cc}
z'{\bf 1}_{k_B'}&0\\
0&z''{\bf 1}_{k_B''}
\end{array}\right),
\left(\begin{array}{cc}
z'{\bf 1}_{k_C'}&0\\
0&z''{\bf 1}_{k_C''}
\end{array}\right)
\right].
\label{unbrokenTRQ}
\eeq
A solution invariant under this subgroup exists only when the FI parameter
vanishes on its Lie algebra,
\beq
\left({k_C'\over k_C}-{k_B'\over k_B}\right)\zeta^1+
\left({k_A'\over k_A}-{k_C'\over k_C}\right)\zeta^2
=0.
\label{lineTRQ}
\eeq
This is an empty condition when $k_A'/k_A=k_B'/k_B=k_C'/k_C$,
that is, when $k_A$, $k_B$, $k_C$ have a common divisor.
In such a case, for any FI parameter, there is a solution with
unbroken gauge group. In what follows, we assume that $k_A$, $k_B$
and $k_C$ are relatively prime. Then, the equation (\ref{lineTRQ})
defines a line in the FI-parameter space.

The line (\ref{lineTRQ}) is a candidate of the phase boundary.
In many cases, however, only half of the line is the actual phase boundary,
since the equations to be solved takes the form like $|\phi|^2=\zeta$
and has solution only for one sign of $\zeta$.
It is also possible that a solution does not exists along both halves of
the line.
Obvious examples are $k_A'$ and $k_A''$ both non-zero but $k_B''=k_C'=0$.
In such a case, $X,Y,Z$ must be of the form
\beq
X=0,\quad
Y=\left(\begin{array}{c}
0\\
Y''
\end{array}\right),\quad
Z=\left(\begin{array}{cc}
Z'&0
\end{array}\right),
\eeq
and the D-term equations include
$Y'Y^{\prime\dag}={\zeta^2\over k_A}{\bf 1}_{k_A'}$
and $-Z^{\prime\prime\dag}Z''={\zeta^2\over k_A}{\bf 1}_{k_A''}$,
which have no solution if $\zeta\ne 0$.
There are non-obvious examples where a solutions does not exist
for some range of $(a,b,c)$ because the number of equations is too large
compared to the number of
variables.

In what follows, we present the phase structure of the model for
some range of rank vectors. We assume that $a,b,c$ are all positive.

\subsection*{Rank (1,1,1)}

\begin{figure}[htb]
\psfrag{z1}{$\zeta^1$}
\psfrag{z2}{$\zeta^2$}
\psfrag{I}{I}
\psfrag{II}{II}
\psfrag{III}{III}
\psfrag{i}{}%{\footnotesize i}
\psfrag{ii}{}%{\footnotesize ii}
\psfrag{iii}{}%{\footnotesize iii}
\centerline{\includegraphics{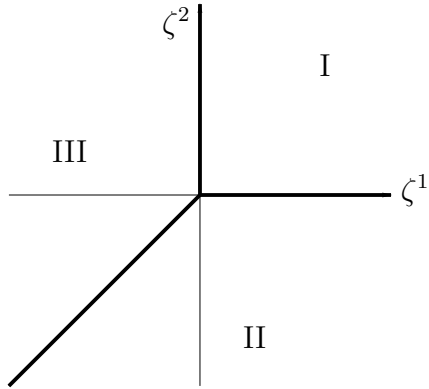}}
\caption{The phases of the model (1,1,1)}
\label{fig:q111}
\end{figure}

The model has three phases as in
Fig.~\ref{fig:q111}.

In Phase I, supersymmetry is unbroken if and only if  $a+b-2\geq c$,
in which case we have the effective theory on the Higgs branch
given by the non-linear sigma model on the complete intersection of
the $c$ hypersurfaces $\sum_{i,j}A^{ij}_hX_iY_j=0$
in $\CP^{a-1}\times \CP^{b-1}$. $Z^h=0$ there.
The other phases are similar: to go from I to II and to III,
we do the cyclic rotation $c\to b\to a$ and $Z\to Y\to X$.
The three phases are separated by three phase boundaries.
In the table below, we describe the unbroken
gauge group $G_1$, the type of vacuum configuration, and
the Higgs part of the mixed branch, at each of them:
\beq
\begin{array}{c|c|c|c}
&G_1&\mbox{type}&\mbox{Higgs part}\\
\hline
\mbox{I-II}&\{[z,1,1]\}\cong U(1)&Y=Z=0&\CP^{a-1}\\
\mbox{I-III}&\{[1,z,1]\}\cong U(1)&X=Z=0&\CP^{b-1}\\
\mbox{II-III}&\{[1,1,z]\}\cong U(1)&X=Y=0&\CP^{c-1}
\end{array}
\eeq

\subsection*{Rank (k,1,1)}

\begin{figure}[htb]
\psfrag{z1}{$\zeta^1$}
\psfrag{z2}{$\zeta^2$}
\psfrag{I}{I}
\psfrag{II}{II}
\psfrag{III}{III}
\psfrag{IV}{IV}
\psfrag{i}{}%{\footnotesize i}
\psfrag{ii}{}%{\footnotesize ii}
\psfrag{iii}{}%{\footnotesize iii}
\psfrag{i'}{}%{\footnotesize i'${}_l$}
\centerline{\includegraphics{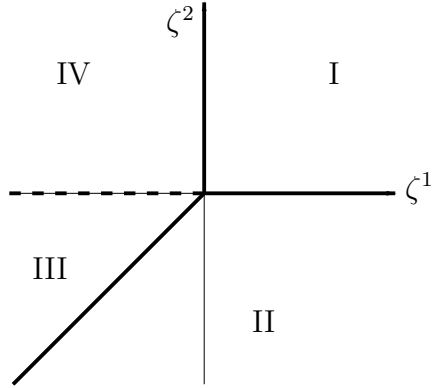}}
\caption{(Possible) phases of the model (k,1,1)}
\label{fig:qk11}
\end{figure}

The model has at most four phases as shown in Fig.~\ref{fig:qk11}.
When $b,c\geq k$, Cambers I and II are phases by themselves.
Chambers III and IV combine into one phase when
\beq
b\geq k,\quad c\geq k,\quad (b+c)(k-1)\leq a+(k-1)^2,
\label{condrank}
\eeq
and are different phases otherwise.

In chamber I, supersymmetry is unbroken when
$a\geq 1$, $b\geq k$ and $k(b-k)+a-1\geq ck$, in which case
we have the effective theory on the Higgs branch
--- the zero of the $c$ sections $\sum_{i,j}A^{ij}_hY_jX_i$
of the rank $k$ vector bundle $S_{k,b}^*\boxtimes {\mathcal O}(1)$
over $G(k,b)\times \CP^{a-1}$.
(Here $G(k,b)$ is the Grassmannian of $k$-planes in $\C^b$ and
$S_{k,b}$ is its tautological vector bundle.)
The description of chamber II is obtained from this
by the exchange $Y\leftrightarrow Z$, $b\leftrightarrow c$.
In chamber IV, supersymmetry is unbroken when
$b\geq k$, $c\geq 1$ and $k(b-k)+kc-1\geq a$, in which case
we have the effective theory on the Higgs branch ---
the complete intersection of $a$ hypersurfaces
$\sum_{j,h}A^{ij}_hZ^hY_j=0$
in ${\bf P}(S_{k,b}^{\oplus c})$.
Description in chamber III is obtained from this
by the exchange $Y\leftrightarrow Z$, $b\leftrightarrow c$.

The Higgs branches in chambers IV and III are the same as complex
manifolds if and only if the condition (\ref{condrank})
is satisfied. To see that, let us regard $Y$ and $Z$ as $k\times b$
and $c\times k$ matrices. Then the defining F-term equations are
 $\sum_{j,h}A^{ij}_h(ZY)^h_{\,\,\,j}=0$ in both chambers.
A possible difference is the condition on the rank of $Y$ and $Z$.
In chamber IV we have ${\rm rank}(Y)=k$ and ${\rm rank}(Z)\leq k$
while in chamber III we have ${\rm rank}(Y)\leq k$ and ${\rm rank}(Z)=k$.
The two manifolds are the same if
${\rm rank}(Z)$ is always $k$ in chamber IV and
${\rm rank}(Y)$ is always $k$ in chamber III.
This obviously fails when $b$ or $c$ is less than $k$.
But even when they are both bigger or equal to $k$, it fails when,
say in chamber IV, there is a solution with
${\rm rank}(Z)\leq k-1$. Since $Y$ has rank $k$ there, this is
equivalent to that the $c\times b$ matrix $ZY$ has rank $(k-1)$ or less.
This imposes $(c-(k-1))(b-(k-1))$ conditions on $ZY$.
On the other hand, the F-term equation $\sum_{j,h}A^{ij}_h(ZY)^h_{\,\,j}=0$
imposes $a$ conditions on $ZY$. Thus, there is such a solution when
$bc-1\geq (c-(k-1))(b-(k-1))+a$, which is the opposite of the last
inequality in (\ref{condrank}).

Accordingly, there are at most four phase boundaries.
In the table below, we describe the data of the (possible) mixed branches:
\beq
\begin{array}{c|c|c|c}
&G_1&\mbox{type}&\mbox{Higgs part}\\
\hline
\mbox{I-II}&\{[g,1,1]\}\cong U(k)&Y=Z=0&\CP^{a-1}\\
\mbox{I-IV}&\{[{\bf 1}_k,z,1]\}\cong U(1)&X=Z=0&G(k,b)\\
\mbox{II-III}&\{[{\bf 1}_k,1,z]\}\cong U(1)&X=Y=0&G(k,c)\\
\mbox{III-IV}&\displaystyle
\left\{\left[{\,g~~~~\choose ~~~{\bf 1}_l},1,1\right]\right\}
\cong U(k-l)
&
Y\!\!=\!\!\left(\begin{array}{c}
0\\\!\!Y''\!\!\end{array}\right),\,\,Z\!\!=\!\!\left(\begin{array}{cc}
\!\!0\!&\!\!Z''\!\!\end{array}\right)&{\mathcal H}_l
\end{array}
\eeq
The first three mixed branches exist when $a\geq 1$,
$b\geq k$ and $c\geq k$ respectively.
On the III-IV wall, $l$ ranges over $1,\ldots, k-1$.
${\mathcal H}_l$ is a variety of dimension $l(b+c)-l^2-1-a$ and is
non-empty when this number is non-negative. For $l\geq 2$,
it is singular if there is a locus where ${\rm rank}(Y'')={\rm rank}(Z'')$
drops from $l$, and the singular locus is isomorphic to ${\mathcal H}_{l-1}$.
(${\mathcal H}_l$ is smooth when ${\mathcal H}_{l-1}$ is empty.)
${\mathcal H}_1$ is always smooth and is given by the complete intersection
of $a$ hypersurfaces $\sum_{j,h}A^{ij}_hz^hy_j=0$ in $\CP^{b-1}\times \CP^{c-1}$.
The space of classical vacua is a union of the mixed branches,
\beq
{\mathcal M}_{{\rm III}\mbox{-}{\rm IV}}
=\bigcup_{l=1}^{k-1}\,\,\R^{3(k-l)}\times {\mathcal H}_l.
\label{mixedQEX}
\eeq
Note that the $l$-th and the $(l-1)$-th mixed branches touch each other
at the singular locus of ${\mathcal H}_l$ of the former
and the origin of the $\R^3$ factor of $\R^{3(k-l+1)}$ of the latter.
When $b,c\geq k$, ${\mathcal H}_{k-1}$ has the largest dimension
among ${\mathcal H}_l$'s. Therefore,
the space (\ref{mixedQEX}) is non-empty if ${\mathcal H}_{k-1}$ is non-empty,
and that is so when the last inequality of (\ref{condrank}) is violated.
Thus, III-IV wall is indeed a genuine phase boundary when the mixed branch
exists there.

Special cases of this class of triangle quivers are studied in
two dimensions in \cite{JKLMR}: the cases are $(a,b,c)=(kn,n,n)$
where the Higgs branches are Calabi-Yau manifolds ---
in that paper, $k$ here is denoted as $(n-k)$.
The above structure of mixed branches on the III-IV wall (2d version)
was worked out by KH and was conveyed to the authors of
\cite{JKLMR} who initially missed the possibility that
the III-IV wall may not be a phase boundary.
Among models with Calabi-Yau threefolds as the Higgs branches,
an example
where III-IV wall is not a phase boundary is $k=2$ and $(a,b,c)=(8,4,4)$.

\subsection*{Higher  Ranks}

\begin{figure}[htb]
\psfrag{z1}{$\zeta^1$}
\psfrag{z2}{$\zeta^2$}
\centerline{\includegraphics{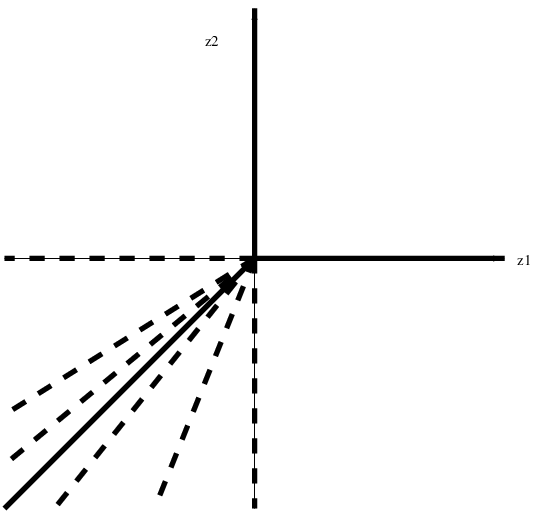}~~~~~~~~~\includegraphics{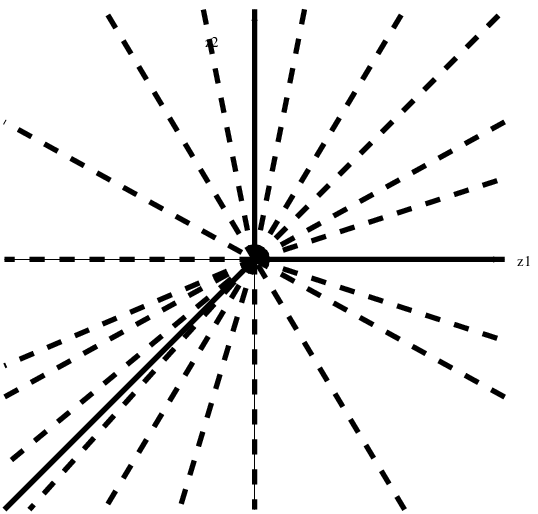}}
\caption{(Possible) phases of the models with $\vec{k}=(5,2,1)$
(Left) and $\vec{k}=(5,3,2)$ (Right)}
\label{fig:qcomp}
\end{figure}
As the rank vector increases, the number of phases proliferates,
as shown in Fig.~\ref{fig:qcomp}.
This is because the lines  (\ref{lineTRQ}) corresponding to
different partitions of the rank vector are generically different.
The case of rank vector $(k,1,1)$ was special in that
the partitions $(k,1,1)=(k-l,0,0)+(l,1,1)$ all define the same
line, which includes the III-IV wall.
For example, for the case of rank vector $(k,2,1)$,
the partitions $(k-l,1,0)+(l,1,1)$ define different lines
$\zeta^1=(2(k-l)/k)\zeta^2$, which appear, for example in $k=5$,
as the four dashed segments in the
south-west directions in Fig.~\ref{fig:qcomp}-Left.
Also, for higher ranks, it is more difficult to explicitly describe
the Higgs branch in each phase as well as the mixed branches
at each phase boundary.
The index formula which we will derive in this paper will be particularly
useful in such a circumstance, since we do not need to
solve the D-term and F-term equations nor think about computing
the cohomology on the space of solutions.

\section{Coulomb Branch}\label{sec:Coulomb}

As we have seen,
when the FI parameter $\zeta$ is deep inside a phase where the gauge group
is broken to a finite subgroup, the Higgsed modes, consisting of
the vector multiplet and the chiral multiplet in the gauge orbit direction,
decouple from the physics below the energy scale $M_H=e\sqrt{|\zeta|}$.
In this section, we would like to analyze what happens when
the FI parameter $\zeta$ comes close to the walls.

For ${\mathcal N}=4$ systems,
this was studied extensively, first by Denef \cite{Denef}
and later by many people including two of the authors of
the present paper. As far as we are aware of,
the analysis of ${\mathcal N}=2$ systems has not been done.
As we shall see, they have much richer structures than ${\mathcal N}=4$
systems.

\subsection{The Outline}\label{subsec:outline}

When the FI parameter $\zeta$ comes close to a wall,
the mass of the Higgsed modes approaches zero.
There, it is natural to look at the Coulomb branch ---
the region of the field space where
the scalar component of the vector multiplet has large values.
The chiral multiplet has a large mass in such a region and hence
can be ignored, or to be more precise, should be integrated out.
This leaves us with an effective theory of the vector multiplet.
The main task is to determine this effective theory and then
study its ground states.

We shall present the analysis for the theory with $U(1)$ gauge group.
We believe that this captures the essence of the general case, since typically
only a single $U(1)$ subgroup of the gauge group is unHiggsed at the wall.
We shall examine later if that is indeed the case.
Also, we shall start with the ${\mathcal N}=2$ systems, where the
scalar $\vs$ takes values in the real line.
Note that there are two disconnected regions with ``large $|\vs|$''
--- large positive $\vs$ and large negative $\vs$.
The analysis should be done separately.

``Integrating out the matter chiral multiplet'' is usually done
by path-integral, where we can take the one-loop approximation
for large values of $|\vs|$. However, the operator formalism is more
convenient in order to impose the Gauss law constraint.
In that formulation, we shall find the ground state of the
matter system in the background $(\vs,\lambda,D)$
of the vector multiplet. The gauge field $v_t$ is used up already
at this stage for the Gauss law.
The matter theory has ${\mathcal N}=2$ supersymmetry
in a supersymmetric background, which satisfies (see (\ref{Vector}))
\beq
\dot{\vs\,}\,=\,D\,=\,0,\quad\,\lambda\,=\,0.
\label{susybg}
\eeq
The matter system may or may not have a supersymmetric ground state.
Correspondingly, the effective theory of the vector multiplet
may or may not have a supersymmetry.
It may happen that the matter system has a multiple of supersymmetric
ground states. In that case, the effective theory decomposes into
a multiple of sectors of supersymmetric theories.
In general, the spectrum of matter ground states depends on whether $\vs$ is
large positive or large negative, and correspondingly
the character of the effective theory of the vector multiplet
depends on the regions of $\vs$.

When the matter system preserves the supersymmetry, the effective
Lagrangian must be supersymmetric.
A supersymmetric Lagrangian of $(\vs,\lambda,D)$ is of the form
\beq
{1\over 2e(\vs)^2}\left(\dot{\vs\,}^2
+{i\over 2}(\blambda\dot{\lambda}-\dot{\blambda}\lambda)
+D^2\right)-\left({1\over 4e(\vs)^2}\right)^{\!\prime}\!\!D\blambda\lambda
+h'(\vs)D-{1\over 2}h''(\vs)\blambda\lambda,
\eeq
to the second order in
${\mathcal O}(\vs)=0$, ${\mathcal O}({\dd\over \dd t})=1$,
${\mathcal O}(D)=1$, ${\mathcal O}(\lambda)={1\over 2}$
and ${\mathcal O}(\epsilon)=-{1\over 2}$. (This order assignment is
taken from \cite{Denef}.)
The prime $'$ stands for the $\vs$ derivative.
Note that the classical Lagrangian has
$e(\vs)=e$ and $h'(\vs)=-\zeta$.
We know that the effective gauge coupling constant behaves at large $|\vs|$
as $1/e(\vs)^2=1/e^2+\#/|\vs|^3+\cdots$, for some numerical constant $\#$.
At large enough values, $|\vs|^3\gg e^2$,
we may ignore the correction and
assume the following form of the effective Lagrangian,
\beqa
L_{\,\rm eff}&=&
{1\over 2e^2}\left(\dot{\vs\,}^2+i\blambda\dot{\lambda}
+D^2\right)+h'(\vs)D-{1\over 2}h''(\vs)\blambda\lambda\nn\\
&\simeq&{1\over 2e^2}\left(\dot{\vs\,}^2+i\blambda\dot{\lambda}\right)
-{e^2\over 2}\Bigl(h'(\vs)\Bigr)^2
-{1\over 2}h''(\vs)\blambda\lambda.
\label{LeffC}
\eeqa

When the matter system breaks the supersymmetry spontaneously, then
the effective Lagrangian must have a supersymmetry breaking term.
For example, the ground state energy of the matter sector,
combined with the contribution from the Wilson line,
is of the form $2N_B|\vs|$ for some positive integer $N_B$, so that
the effective potential is
%At the level of the effective potential, it is to add
%a linear potential $2N_B|\vs|$ for some positive integer $N_B$, as
\beq
U_{\rm eff}(\vs)\,=\,
{e^2\over 2}\Bigl(h'(\vs)\Bigr)^2
\,+\,2N_B|\vs|.
\eeq
The potential is positive
and we do not expect to have a zero energy state.
This only means that there is no supersymmetric ground state supported
in the region of $\vs$ under question.
It is still possible that there are supersymmetric
ground states supported on the other side of the $\vs$ line or
at $\vs=0$.

\newcommand{\neff}{N_{\rm eff}}

\begin{figure}[htb]
\psfrag{x}{$\vs$}
\psfrag{min}{${\neff\over 2\zeta}$}
\psfrag{0}{$0$}
\psfrag{Ueff}{$U_{\rm eff}(\vs)$}
\psfrag{Psi}{$\Psi_{\uparrow}$}
\psfrag{ez}{${e^2\zeta^2\over 2}$}
\centerline{\includegraphics{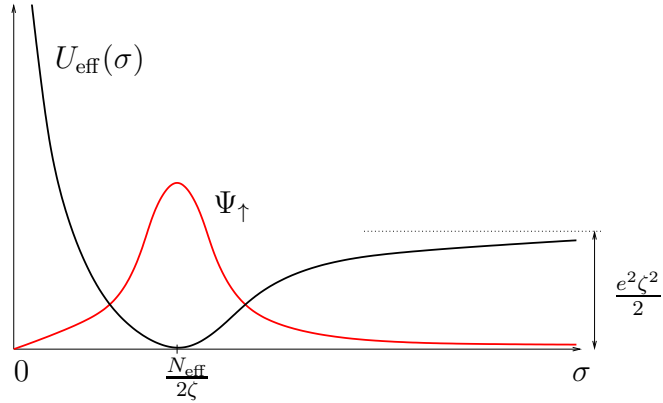}}
\caption{The effective potential and the wavefunction for the case $\zeta>0$
and $\neff>0$.}
\label{fig:graphU}
\end{figure}

Let us come back the case where the matter system preserves the
supersymmetry. It turns out that the function $h'(\vs)$ is of the form
\beq
h'(\vs)\,=\,{\neff\over 2|\vs|}-\zeta,
\label{hprime}
\eeq
for an integer $\neff$ that depends on the choice of the
matter ground state. In particular, the effective potential is given by
\beq
U_{\rm eff}(\vs)\,=\,
{e^2\over 2}\left(\,{\neff\over 2|\vs|}-\zeta\,\right)^2 \ .
\label{UeffC}
\eeq
Note that (\ref{LeffC}) is the famous system introduced in
\cite{Windex}. It is found that
the exact supersymmetric ground state, if it exists,
must be either of the following two,
\beq
\Psi_{\downarrow}=\e^{-h(\vssm)}|0\rangle_{\rm eff}\quad\mbox{or}\quad
\Psi_{\uparrow}=\e^{h(\vssm)}\blambda|0\rangle_{\rm eff},
\eeq
where $|0\rangle_{\rm eff}$ is the state annihilated by $\lambda$.
Which (or none) to be taken is decided by the requirement that
the wavefunction is square normalizable.
Integrating (\ref{hprime}) we find
\beq
\e^{h(\vssm)}=|\vs|^{{\neff\over 2}{\rm sgn}(\vssm)}\e^{-\zeta\vssm}.
\eeq
Suppose we have the supersymmetric effective theory
(\ref{LeffC}) with (\ref{hprime}) on the $\vs>0$ Coulomb branch.
If $\zeta>0$, we should take $\Psi_{\uparrow}$ for square normalizability
at $\vs\to\infty$. In addition, if $\neff>0$, then the wavefunction
$\Psi_{\uparrow}$ vanishes as $\vs\to 0$. It is square normalizable
over $\vs>0$ and is supported around $\vs=\neff/2\zeta$.
(See Fig.~\ref{fig:graphU})
Thus, it qualifies as the supersymmetric ground state on the
$\vs>0$ Coulomb branch.
If $\neff<0$ on the other hand, the wavefunction $\Psi_{\uparrow}$
diverges as $\vs\to 0$ and is not square normalizable over $\vs>0$.
That it is not square normalizable at $\vs\to 0$
may not be a serious problem
since we are interested in large values of $|\vs|$ anyway.
However, the fact that its support is toward $\vs\to 0$
disqualifies it to be a Coulomb branch vacuum.
Subtle is the case $\neff=0$: the state $\Psi_{\uparrow}$
does not diverge as $\vs\to 0$
and is perfectly square normalizable over $\vs>0$, but
the peak of the support is toward $\vs\to 0$.
Therefore it is not clear at this point if we should discard it
or not.
If $\zeta<0$, we should take $\Psi_{\downarrow}$, and it qualifies as
a supersymmetric ground state on the $\vs>0$ Coulomb branch
only when $\neff<0$.
Similarly for the case where we have the supersymmetric effective theory
(\ref{LeffC}) with (\ref{hprime}) on the $\vs<0$ Coulomb branch.
If $\zeta>0$ ({\it resp}. $\zeta<0$), we should take $\Psi_{\downarrow}$
({\it resp}. $\Psi_{\uparrow}$) and it qualifies
as a supersymmetric ground state on the $x<0$ Coulomb branch
if $\neff>0$ ({\it resp}. $\neff<0$).
Subtle is the case $\neff=0$ where it is normalizable but has a profile
which is peaked toward the origin $\vs\to 0$.

Let us see what happens to the ground states as the FI parameter
$\zeta$ is varied across zero, say from positive to negative.
For concreteness, we consider the $\vs>0$ Coulomb branch.
First suppose $\neff>0$. As we have seen,
for $\zeta>0$,  there is a supersymmetric ground state $\Psi_{\uparrow}$
which is supported around $\vs=\neff/2\zeta$.
As $\zeta$ approaches $0$ from above, the location of the support
runs away to infinity.
For $\zeta<0$, there is no supersymmetric ground state on the
$\vs>0$ Coulomb branch.
{\sl This is the wall crossing phenomenon ---
a normalizable supersymmetric ground state disappears by
exiting the field space at infinity}.
When $\neff<0$, the opposite happens:
There is no supersymmetric ground state for $\zeta>0$.
As $\zeta$ crosses zero and goes negative, a supersymmetric
ground state $\Psi_{\downarrow}$ enters from infinity.
Subtle is the case $\neff=0$: there is a square normalizable
supersymmetric ground state before and after $\zeta$ crosses zero
--- $\Psi_{\uparrow}$ for $\zeta>0$ and $\Psi_{\downarrow}$ for $\zeta<0$.
As $\zeta\to 0$, its profile does not run away to infinity
but instead spreads out.
As we will see in examples, it does play a particular
r\^ole in the wall crossing phenomenon.

The effective theory on the Coulomb branch
has a characteristic energy scale
\beq
M_C=e^2\zeta^2.
\label{CoulombEnergy}
\eeq
To show this, let us rescale the fields as $\vs\to e\vs$
and $\lambda\to e\lambda$ so that they have the canonically normalized
kinetic terms. Then, the effective Lagrangian
depends on $e$ and $\zeta$ only through
the combination $e\zeta$ which has the dimension of $\sqrt{\rm energy}$.
It depends also on $\neff$ and possibly also $N_B$
but they are integer valued discrete parameters.
All relevant energy scale is $M_C$ times a numerical constant.
Indeed, the hight of the potential $U_{\rm eff}$ at $|\vs|\to\infty$
is $e^2\zeta^2/2$ (in the supersymmetric region).
Also, the perturbative spectrum around the supersymmetric
ground state $\Psi_{\uparrow}$ discussed above has level spacing
$\omega=2e^2\zeta^2/\neff$ since the potential behaves near the bottom
$\vs={\neff\over 2\zeta}+e\xi$ as
$U_{\rm eff}\simeq {\omega^2\over 2}\xi^2$.

Let us now describe the Coulomb branch of the ${\mathcal N}=4$ systems
with a $U(1)$ gauge group. Everything is essentially obtained in \cite{Denef}.
We would like to find the effective theory of the variables
$(\bfx, \bflambda,\bfD)$ at large values of $|\bfx|$.
The matter sector always has a unique supersymmetric
ground state in the supersymmetric backgrounds,
and yields a supersymmetric Lagrangian for $(\bfx, \bflambda,\bfD)$.
It must be of the form
\beq
\bfL_{\rm eff}={1\over 2e^2}
\left(\dot{\bfx}^2+i\overline{\bflambda}\dot{\bflambda}+\bfD^2\right)
-V(\bfx)\bfD-\bfA(\bfx)\cdot\dot{\bfx}
+\bfC(\bfx)\cdot \overline{\bflambda}\bfsigma\bflambda,
\eeq
with $\bfC={1\over 2}\nabla V=-{1\over 2}\nabla\times \bfA$,
to the second order in ${\mathcal O}$, if we neglect terms that vanishes
in the regime $e^2/|\bfx|^3\ll 1$.
An explicit computation, say in path-integral, yields
\beq
V=\zeta-{\neff^{(4)}\over 2|\bfx|},\quad
\bfA=\neff^{(4)}\bfA_{\rm mono},\quad
\bfC={\neff^{(4)}\over 4}{\bfx\over |\bfx|^3},
\label{VAC}
\eeq
where
\beq
\neff^{(4)}:=\#(\mbox{positively charged fields})
-\#(\mbox{negatively charged fields}),
\label{neff4def}
\eeq
and $\bfA_{\rm mono}$ is the Dirac monopole of unit magnetic charge.
In particular, $\bfA$ has first Chen class $\neff^{(4)}$
on the sphere surrounding the origin.
The four supercharges are proportional to\footnote{There is a sign error
in \cite{Denef}}
\beq
\overline{\bflambda}\left(\bfsigma\cdot \bfD_{\! \bfsmA}-V\right),
\quad
\left(\bfsigma\cdot\bfD_{\! \bfsmA}+V\right)\bflambda,
\eeq
where $\bfD_{\! \bfsmA}$ is the covariant derivative,
$(D_{\bfsmA})_j=\partial_j+iA_j$,
for $\bfA$ in (\ref{VAC}). Note that the $SU(2)\times U(1)$
R-symmetry is present in this effective theory, since the background
($V$, $\bfA$, $\bfC$) is rotationally symmetric. $U(1)_R$ is just a phase
rotation of $\bflambda$ and $\overline{\bflambda}$.

The supersymmetric ground states are of the form
\beq
\Psi^{\alpha}(\bfx)\blambda_{\alpha}|0\rangle_{{}_{\rm osc}},
\label{4gsU1}
\eeq
for $\Psi(\bfx)$ satisfying
\beq
\left(\bfsigma\cdot \bfD_{\! \bfsmA}+V\right)\Psi(\bfx)=0.
\label{VeqN4}
\eeq
The doublet wavefunction $\Psi(\bfx)$ is a spinor on $\R^3\setminus 0$
with values in a line bundle of magnetic charge $\neff^{(4)}$.
We may regard it as a spinor on $S^2$ with values in
${\mathcal O}(\neff^{(4)})$ which depends on $r=|\bfx|$.
With respect to the chiral basis on $S^2$, the vacuum equation (\ref{VeqN4})
takes the form
\beq
\left\{\left(\begin{array}{cc}
\partial_r&0\\
0&-\partial_r
\end{array}\right)+{1\over r}\left[\left(\begin{array}{cc}
1&0\\
0&-1
\end{array}\right)+\bfD\!\!\!\!/{}_{\,S^2\!,\bfsmA\,}
\right]+V\right\}\left(\begin{array}{c}
\Psi_-\\\Psi_+
\end{array}\right)=0.
\label{VeqN42}
\eeq
$\Psi_{\pm}$ is the left/right component of $\Psi$, i.e.,
an $r$-dependent section of the
line bundle ${\mathcal O}(\neff^{(4)}\pm 1)$.
$\bfD\!\!\!\!/{}_{\,S^2\!,\bfsmA\,}$ is the Dirac operator on $S^2$
which swaps the left and the right components.
We may decompose $\Psi$ into eigenmodes of
$\bfD\!\!\!\!/{}_{\,S^2\!,\bfsmA\,}$.
A non-zero mode has both the left and the right components
and hence cannot solve (\ref{VeqN42}).
A zero mode has only the left or the right component
and can solve (\ref{VeqN42}).
Recall the index theorem
\beq
{\rm ind}\,\bfD\!\!\!\!/{}_{\,S^2\!,\bfsmA\,}=\neff^{(4)}.
\eeq
In fact the index is the actual number of zero modes.
When $\neff^{(4)}$ is zero, there is no zero mode. In particular,
there is no supersymmetric ground state for any value of $\zeta$.
When $\neff^{(4)}$ is positive, $\Psi_-$ has $\neff^{(4)}$ zero modes
which may be represented by the holomorphic sections of
${\mathcal O}(\neff^{(4)}-1)$. They form the
spin $j={\neff^{(4)}-1\over 2}$ representation of $SU(2)$ (Borel-Weil
theorem). The equation (\ref{VeqN42}) reads
\beq
\left[{\partial\over \partial r}+{1\over r}+\zeta-{\neff^{(4)}\over 2r}
\right]\Psi_-=0.
\eeq
For $\zeta>0$, it has a unique solution that vanishes at infinity,
$\Psi_-\propto r^{{\neff^{(4)}\over 2}-1}\e^{-\zeta r}$.
Note that
\beq
r^2|\Psi_-|^2\propto \left(r^{{\neff^{(4)}\over 2}}\e^{-\zeta r}\right)^2,
\eeq
which shows that it is square normalizable.
This also shows that the probability to be in the shell of radius $r$
is largest for $r=\neff^{(4)}/(2\zeta)$.
As $\zeta\searrow 0$, the peak runs away to infinity,
and for $\zeta<0$ there is no normalizable solution.
Therefore, when $\neff^{(4)}$ is positive,
$\neff^{(4)}$ supersymmetric ground states forming the
spin $j={\neff^{(4)}-1\over 2}$ representation of $SU(2)$
run away to infinity in $|\bfx|$
as $\zeta$ is sent from positive to negative.
When $\neff^{(4)}$ is negative, $\Psi_+$ has $|\neff^{(4)}|$ zero modes
represented by anti-holomorphic sections of ${\mathcal O}(\neff^{(4)}+1)$.
The equation  (\ref{VeqN42}) reads
\beq
\left[-{\partial\over \partial r}-{1\over r}+\zeta-{\neff^{(4)}\over 2r}
\right]\Psi_+=0.
\eeq
For $\zeta>0$, there is no normalizable solution.
For $\zeta<0$, there is a unique normalizable solution such that
\beq
r^2|\Psi_+|^2\propto \left(r^{-{\neff^{(4)}\over 2}}\e^{\zeta r}\right)^2.
\eeq
Note that the peak $r=\neff^{(4)}/(2\zeta)$ runs off to infinity as
$\zeta\nearrow 0$. Therefore, when $\neff^{(4)}$ is negative,
$|\neff^{(4)}|$ supersymmetric ground states forming the
spin $j={|\neff^{(4)}|-1\over 2}$ representation of $SU(2)$
come in from infinity in $|\bfx|$
as $\zeta$ is sent from positive to negative.

The above ground states,
whose wavefunctions are of the form (\ref{4gsU1}),
have vanishing $U(1)$ R-charge. This is because only $\bflambda$ and
$\overline{\bflambda}$ have nonzero R-charge, and the conjugate invariant
quantization yields R-charge $1,0,-1$ on the states
$|0\rangle_{{}_{\rm osc}}$, $\blambda_{\pm}|0\rangle_{{}_{\rm osc}}$,
$\blambda_+\blambda_-|0\rangle_{{}_{\rm osc}}$, respectively.
In view of the relation between $\bfR_{\!\pm}$ and the generators of
$SU(2)\times U(1)_R$ (see the footnote in page~\pageref{foot:Rcharges}),
we see that these ground states have $\bfR_{\!\pm}=\pm J_3$.

In what follows, we illustrate these results in some of the examples
introduced in Section~\ref{subsec:LSMex}.

\subsection{The $\CP^{N-1}$ Model}\label{subsec:CPN}

\newcommand{\barn}{\bar{n}}

Let us look into the detail of the Coulomb branch of the
$\CP^{N-1}$ model (Example 1). First, the ${\mathcal N}=2$ supersymmetric
system: $U(1)$ gauge group, $N$ chiral multiplets of charge $1$,
and Wilson line of charge $q$, obeying the anomaly free condition
(\ref{afProj}), $N/2+\rmq\in \Z$.

The Lagrangian of the system is
\beqa
L&=&{1\over 2e^2}\left(\dot{\vs\,}^2+i\blambda\dot{\lambda}+D^2\right)
-\zeta D-\rmq(v_t+\vs).
\nn\\
&&+|D_t\phi|^2+i\bpsi D_t\psi
+\bphi\left(\,{}^{}_{}D-\vs^2\,\right)\phi
+\bpsi\vs\psi
-i\bphi\lambda\psi
+\bpsi\,\blambda\phi.
\eeqa
We first quantize the matter system, with the Lagrangian on the second line,
in the supersymmetric background (\ref{susybg}), i.e.,
$D=\lambda=0$ and $\vs$ constant. Later, fluctuations away from
the background shall be taken into account as perturbation.
This matter system is a free theory and we know the exact spectrum.
Let us denote by
 $a_{\phi}$ and $a_{\bphi}$ ({\it resp}. $a_{\phi}^{\dag}$ and
$a_{\bphi}^{\dag}$) the annihilation ({\it resp}. creation) operators
of $\phi$ and $\bphi$.
The Hamiltonian and the gauge charge are
\beqa
H_{\rm matter}
&=&\omega\left(a_{\phi}^{\dag}a_{\phi}+a_{\bphi}^{\dag}a_{\bphi}+N\right)
-\vs\,{[\bpsi,\psi]\over 2},\\
{\bf G}_{\rm matter}
&=&a_{\phi}^{\dag}a_{\phi}-a_{\bphi}^{\dag}a_{\bphi}
+{[\bpsi,\psi]\over 2}.
\eeqa
The oscillator frequency is $\omega=|\vs|$ for now --- it will be replaced by
$\omega=\sqrt{\vs^2-D}$ when the fluctuation is introduced.
A general state is of the form
\beq
|n,\barn,m\rangle=\left(\prod_{i=1}^N{a_{\phi_i}^{\dag n_i}\over
\sqrt{n_i!}}{a_{\bphi_i}^{\dag \barn_i}\over
\sqrt{\barn_i!}}\bpsi_i^{m_i}\right)|0\rangle^{}_{\rm osc} \ ,
\label{genstateCPN}
\eeq
which has energy and charge
\beqa
E_{\rm matter}&=&\omega\left(|n|+|\barn|+N\right)-
\vs\left(|m|-{N\over 2}\right),\label{EspCPN}\\
Q_{\rm matter}&=&|n|-|\barn|+|m|-{N\over 2}.\label{QspCPN}
\eeqa
In the above expressions, $|0\rangle^{}_{\rm osc}$ is the state annihilated by
$a_{\phi_i}$, $a_{\bphi_i}$ and $\psi_i$, and we define
 $|n|:=\sum_{i=1}^Nn_i$, etc.
Note that $n_i$ and $\barn_i$ run over all non-negative integers
while $m_i$ runs over $0$ and $1$. In particular,
$|m|$ can range only over $0,1,\ldots, N$.
The Gauss law is $\left({\bf G}_{\rm matter}+\rmq\right)|{\rm phys}\rangle=0$,
which requires the state (\ref{genstateCPN}) to satisfy
\beq
|n|-|\barn|+|m|-{N\over 2}+\rmq=0.
\eeq
Including the linear potential $\rmq\vs$ from the Wilson line,
the state has energy
\beqa
E&=&\omega\left(|n|+|\barn|+N\right)-
\vs\left(|m|-{N\over 2}-\rmq\right)
\nn\\
&=&\left\{\begin{array}{ll}
\displaystyle
2|\vs|\left(|n|+{N\over 2}+\rmq\right)
=2|\vs|\left(|\barn|+(N-|m|)\right)&\mbox{for $\vs>0$},\\[0.4cm]
\displaystyle
2|\vs|\left(|\barn|+{N\over 2}-\rmq\right)
=2|\vs|\left(|n|+|m|\right)&\mbox{for $\vs<0$.}
\end{array}\right.
\label{ENB}
\eeqa
The energy is indeed non-negative if the Gauss
law is taken into account.
It is zero (i.e., the state is supersymmetric) when:
\beqa
\vs>0:&& |\barn|=0,\quad |m|=N,\quad
|n|=-\rmq-{N\over 2},\label{sgpos}\\
\vs<0:&& |n|=0,\quad |m|=0,\quad
|\barn|=\rmq-{N\over 2}.\label{sgneg}
\eeqa
Since $|n|$ and $|\barn|$ are non-negative, supersymmetric ground states
exist at $\vs>0$ ({\it resp}. $\vs<0$) if and only if
$\rmq\leq -{N\over 2}$ ({\it resp}. $\rmq\geq {N\over 2}$).
The degeneracy is the number of $n$'s  with $|n|=-\rmq-{N\over 2}$
({\it resp}. $\barn$'s with $|\barn|=\rmq-{N\over 2}$), which is
\beq
{-\rmq-{N\over 2}+N-1\choose N-1}\quad\left(\mbox{\it resp}.\quad
{\rmq-{N\over 2}+N-1\choose N-1}\right).
\eeq
When this condition is violated, i.e.
on $\vs>0$ for ${N\over 2}+\rmq>0$ ({\it resp}.
on $\vs<0$ for ${N\over 2}-\rmq>0$), supersymmetry is spontaneously broken.
The ground states are those with $|n|=0$
({\it resp}. $|\barn|=0$) having energy
\beq
E_0=2|\vs|\left({N\over 2}+\rmq\right)\quad
\left(\mbox{\it resp}.\quad
2|\vs|\left({N\over 2}-\rmq\right)\right).
\label{E0NB}
\eeq
There is a degeneracy with multiple possibilities for
$|\barn|=|m|-N+\left({N\over 2}+\rmq\right)$
({\it resp}. $|n|=-|m|+\left({N\over 2}-\rmq\right)$).

Let us now take into account the fluctuation from the supersymmetric
background (\ref{susybg}) as a perturbation.
We describe in detail
the region of $\vs$ where the unperturbed system has supersymmetric
ground states.
The effect of non-zero $D$ simply occurs in
$\omega=\sqrt{\vs^2-D}=|\vs|-{D\over 2|\vs|}+\cdots$. To the first order
in $D$, it shifts the ground state energy as
\beq
\Delta_D E_0=-\left(|n|+|\barn|+N\right){D\over 2|\vs|}\nn\\
=-\left({N\over 2}+|\rmq|\right){D\over 2|\vs|},
\label{effD}
\eeq
where we used (\ref{sgpos})/(\ref{sgneg}).
The effect of non-zero $\lambda$ is the perturbation
$\Delta_{\lambda} H=i\bphi\lambda\psi-i\bpsi\,\blambda\phi
=-{1\over\sqrt{2\omega}}(a_{\bphi}-a_{\phi}^{\dag})\lambda\psi
+{1\over \sqrt{2\omega}}\bpsi\,\blambda(a_{\phi}-a_{\bphi}^{\dag})$,
which can be treated in the standard way.
Since all the ground states have the same $|m|$
and $\Delta_{\lambda}H$ changes the $\psi$ number,
the energy shift vanishes to the
first order. The second order shifts are the eigenvalues
of the matrix
\beq
A_{ij}=\sum_{m\ne 0}{(\Delta_{\lambda}H)_{0_im}(\Delta_{\lambda}H)_{m0_j}\over
0-E_m},
\eeq
where $\{0_i\}$ are the basis of the ground states
and $m$ runs over the excited states, in the unperturbed system.
In fact it is diagonal with all the same eigenvalue,
\beq
\Delta_{\lambda}E_0={1\over 0-2|\vs|}
\left\{\begin{array}{ll}
{1\over 2\omega}\blambda\lambda\sum_i(n_i+1)&(\vs>0)\\
{1\over 2\omega}\lambda\blambda\sum_i(\barn_i+1)&(\vs<0)
\end{array}\right.
=-\left({N\over 2}+|\rmq|\right){{\rm sgn}(\vs)\over 4\vs^2}\blambda\lambda,
\eeq
where (\ref{sgpos})/(\ref{sgneg}) is used again.
To summarize, to the first non-trivial order,
the effect of perturbation is the shift
of the ground state energy by
\beq
\Delta E=-\left({N\over 2}+|\rmq|\right){\rm sgn}(\vs)
\left[\,{D\over 2\vs}+{1\over 4\vs^2}\blambda\lambda\,\right].
\eeq
When the unperturbed system breaks the supersymmetry,
the effect of non-zero $D$ is similar to (\ref{effD}) though the coefficient
$|n|+|\barn|+N$ depends on the choice of the degenerate vacua.
The effect of non-zero $\lambda$ is far more complicated to explain here.

Integrating out the auxiliary field $D$, we find that
the effective potential is of the form
\beq
U_{\rm eff}(\vs)={e^2\over 2}\left(\,{\neff\over 2|\vs|}-\zeta\,\right)^2
+2N_B|\vs|.
\eeq
$N_B$ is read from (\ref{ENB}) and (\ref{E0NB}).
In the region $\vs>0$ for $\rmq\leq -{N\over 2}$ and
$\vs<0$ for $\rmq\geq {N\over 2}$ where the matter system preserves the
supersymmetry, $N_B=0$ and
\beq
\neff={N\over 2}+|\rmq|.
\label{neffCP}
\eeq
The effective Lagrangian is
of the supersymmetric form (\ref{LeffC}) with (\ref{hprime}),
with $\neff$ given by (\ref{neffCP}).
In other regions where the supersymmetry is spontaneously broken in the
matter system,
$N_B$ and $\neff$ are positive integers which
depend on the choice of matter vacuum.
\begin{figure}[htb]
\psfrag{x}{\small $\vs$}
\psfrag{Ueff}{\small $U_{\rm eff}$}
\psfrag{zpos}{\small $\zeta>0$}
\psfrag{zzero}{\small $\zeta=0$}
\psfrag{zneg}{\small $\zeta<0$}
\centerline{\includegraphics{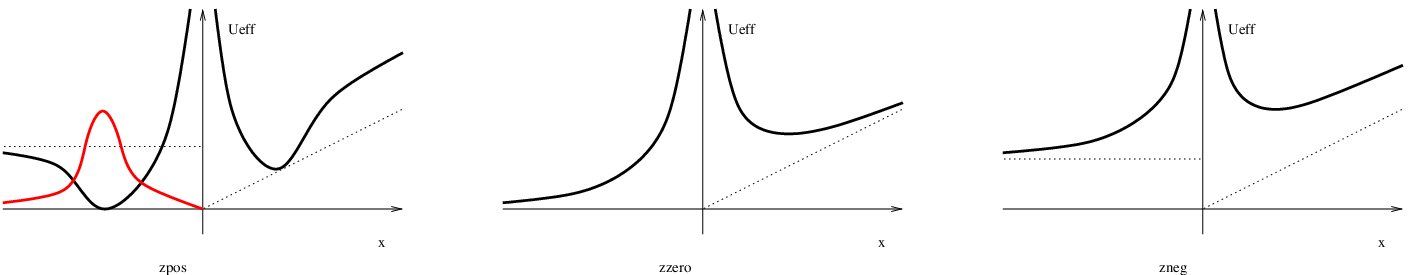}}
\caption{The effective potential for the case $\rmq\geq {N\over 2}$.}
\label{fig:CPpos}
\end{figure}
\begin{figure}[htb]
\psfrag{x}{\small $\vs$}
\psfrag{Ueff}{\small $U_{\rm eff}$}
\psfrag{zpos}{\small $\zeta>0$}
\psfrag{zzero}{\small $\zeta=0$}
\psfrag{zneg}{\small $\zeta<0$}
\centerline{\includegraphics{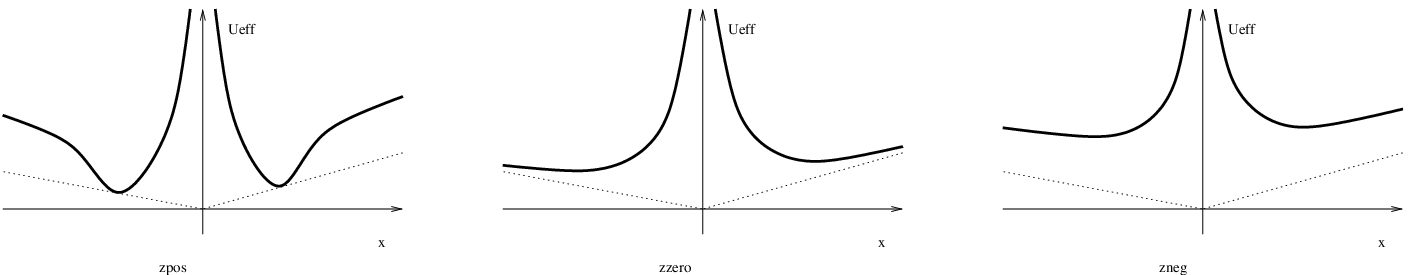}}
\caption{The effective potential for the case $-{N\over 2}<\rmq<{N\over 2}$.}
\label{fig:CPmid}
\end{figure}
\begin{figure}[htb]
\psfrag{x}{\small $\vs$}
\psfrag{Ueff}{\small $U_{\rm eff}$}
\psfrag{zpos}{\small $\zeta>0$}
\psfrag{zzero}{\small $\zeta=0$}
\psfrag{zneg}{\small $\zeta<0$}
\centerline{\includegraphics{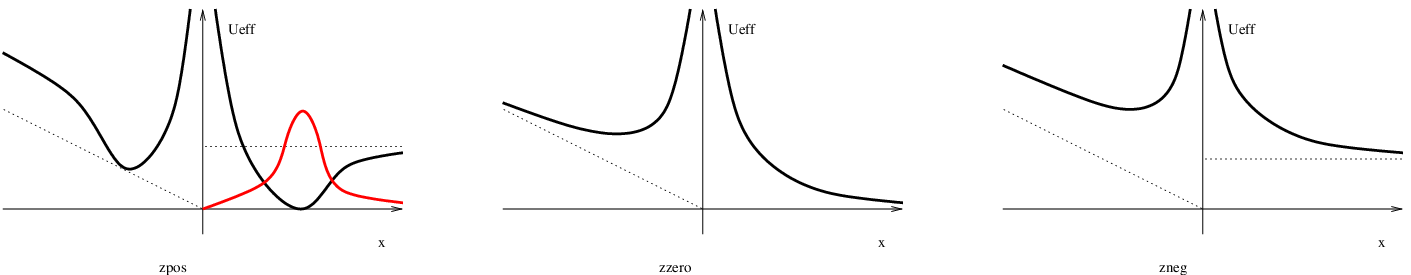}}
\caption{The effective potential for the case $\rmq\leq -{N\over 2}$.}
\label{fig:CPneg}
\end{figure}
We depict the graph of the potential
in Fig.~\ref{fig:CPpos}, \ref{fig:CPmid} and \ref{fig:CPneg}.

The wavefunctions of the supersymmetric ground states,
which exist for $\zeta>0$ and $|\rmq|\ge {N\over 2}$,
are
\beqa
\rmq\geq {N\over 2}:&&
\theta(-\vs)
|\vs|^{\neff\over 2}\e^{\zeta\vs}\left(\prod_{i=1}^N{a_{\bphi_i}^{\dag \barn_i}
\over\sqrt{\barn_i!}}\right)|0\rangle^{}_{\rm osc},\quad
|\barn|=\rmq-{N\over 2}\label{wfCPN1}\\
\rmq\leq -{N\over 2}:&&
\theta(\vs)
|\vs|^{\neff\over 2}\e^{-\zeta\vs}\left(\prod_{i=1}^N{a_{\phi_i}^{\dag n_i}
\over\sqrt{n_i!}}\right)\bpsi_1\cdots\bpsi_N\blambda|0\rangle^{}_{\rm osc},
\quad |n|=-\rmq-{N\over 2}~~~~~~~~
\label{wfCPN2}
\eeqa
where $\theta(x)$ is the step function: $1$ for $x>0$ and $0$ for $x<0$.
Note that the R-charge of the latter is bigger than the one of
the former by $(N-1)$: see (\ref{N2Rcharges}).
If we assign $r_{\rm c}=0$ to $\phi$
and take the conjugation symmetric charge assignment to
$|0\rangle^{}_{\rm osc}$, the former and the latter have R-charges
$-{N-1\over 2}$ and ${N-1\over 2}$ respectively.
This is the same as the spectrum of the supersymmetric ground states
of the theory on the Higgs branch, written in
(\ref{hCPN1}), (\ref{hCPN2}), (\ref{hCPN3}).
That is, all the supersymmetric ground states deep in the geometric phase
$\zeta\gg 0$
become the Coulomb branch ground states for small positive $\zeta$.
They disappear to infinity in the limit $\zeta\searrow 0$,
and there is no zero energy state in the Coulomb branch for small
negative $\zeta$. In particular, no supersymmetric ground state is
left for $\zeta<0$. This is consistent with the fact that there is no
solution to the D-term equation for $\zeta\ll 0$.

For completeness, we describe the Coulomb branch of
the ${\mathcal N}=4$ supersymmetric $\CP^{N-1}$ model:
$U(1)$ gauge group and $N$ chiral multiplets of charge 1.
We see that $\neff^{(4)}=N$.
Therefore, for $\zeta>0$ there are $N$ Coulomb branch vacua
constituting the spin $j={N-1\over 2}$ representation of the
$SU(2)$ subgroup of the $U(2)\subset SO(4)$ R-symmetry.
They run away to infinity as $\zeta\searrow 0$.
For $\zeta <0$, there is no supersymmetric vacua.

\subsection{General $U(1)$ Theory}\label{subsec:CgenU1}

\newcommand{\rmI}{{\rm I}}
\newcommand{\rmJ}{{\rm J}}
\newcommand{\rmK}{{\rm K}}

It is straightforward to extend the analysis to a general theory with
$U(1)$ gauge group.
Consider the $U(1)$ gauge theory with
chiral multiplets $(\phi_i,\psi_i)$ of charge $Q_i$ ($i\in \rmI$),
fermi multiplets $(\eta_j,F_j)$ of charge $q_j$ ($j\in\rmJ$),
and Wilson line of charge $\rmq$, possibly with $E(\phi)$
and/or $J(\phi)$ which are assumed to be quadratic or higher in $\phi$.
The anomaly free condition is $\rmq+\half\sum_iQ_i+\half\sum_jq_j\in \Z$.
We also assume that there is an Abelian flavor symmetry group $T_F$
which acts on the fields with the charges $Q_i^f$, $q_j^f$
and have Wilson line of charge $\rmq^f$, where
$f\in {\rm F}$ is the label of the generators of $T_F$.

The matter Hamiltonian and the gauge $U(1)$ charge are
\beqa
H_{\rm matter}&=&\sum_{i}\left\{\omega_i
\left(a_i^{\dag}a_i+a_{\bi}^{\dag}a_{\bi}+1\right)
-Q_i\vs\,{[\bpsi_i,\psi_i]\over 2}\right\}
+\sum_{j}q_{j}\vs\,
{[\bareta_{j},\eta^{j}]\over 2},\\
{\bf G}_{\rm matter}
&=&\sum_{i}Q_i\left\{a_i^{\dag}a_i-a_{\bi}^{\dag}a_{\bi}
+{[\bpsi_i,\psi_i]\over 2}\right\}
+\sum_{j}q_{j}{[\bareta_{j},\eta^{j}]\over 2},
\label{exprG}
\eeqa
to the quadratic order.
Here $\omega_i=\sqrt{(Q_i\vs)^2-Q_iD}$.
The expression for the flavor charge
${\bf G}^f$ is obtained from (\ref{exprG}) by the replacement
$Q_i\to Q_i^f$ and $q_j\to q_j^f$. Higher order terms exists
in $H$ if there is an $E(\phi)$ and/or $J(\phi)$.
In the present analysis at large $|\vs|$, however,
they have effects suppressed by inverse powers of $|\vs|$ and will be
ignored.
We label the states as in (\ref{genstateCPN}), where we write
$n_i,\bar n_i, m_i$ and $m_{j}$ for the powers of
the creation operators.
The energy and the charge of the state are
\beqa
E_{\rm matter}&=&\sum_{i}\left\{
\omega_i\left(n_i+\bar n_i+1\right)-Q_i\vs\left(
m_i-{1\over 2}\right)\right\}
+\sum_{j}q_{j}\vs
\left(m_{j}-{1\over 2}\right),\\
Q_{\rm matter}&=&\sum_{i}Q_i\left\{
n_i-\bar n_i+m_i-{1\over 2}\right\}
+\sum_{j}q_{j}\left(
m_{j}-{1\over 2}\right).
\eeqa
The expression for $Q^f$ is similar.
Under the Gauss law
$Q_{\rm matter}+\rmq=0$ and in the supersymmetric background $D=0$
where $\omega_i=|Q_i\vs|$, the energy $E=E_{\rm matter}+q\vs$
is indeed non-negative,
\beqa
E&=&\sum_i\left\{|Q_i\vs|(n_i+\bar n_i+1)-Q_i\vs(n_i-\bar n_i)
-Q_i\vs (2m_i-1)\right\}\nn\\
&=&
2|\vs|\times\left\{\begin{array}{ll}
\displaystyle
\sum_{Q_i>0}Q_i(n_i+m_i)+\sum_{Q_i<0}|Q_i|(\bar n_i+(1-m_i))
&\vs<0\\[0.6cm]
\displaystyle
\sum_{Q_i>0}Q_i(\bar n_i+(1-m_i))+\sum_{Q_i<0}|Q_i|(n_i+m_i)
&\vs>0.
\end{array}\right.
\eeqa
It vanishes under the following conditions:
\beqa
\vs<0:&&Q_i>0\Rightarrow n_i=m_i=0;\quad
Q_i<0\Rightarrow \bar n_i=1-m_i=0,\nn\\
&&\!\!\!\!\!\!\!\!\!\!\!\!\!
-\sum_{Q_i>0}Q_i\left(\bar n_i+\half\right)
+\sum_{Q_i<0}Q_i\left(n_i+\half\right)
+\sum_{j}q_{j}\left(m_{j}-{1\over 2}\right)+\rmq=0,~~~~~
\label{gveqneg}\\[0.2cm]
\vs>0:&&Q_i>0\Rightarrow \bar n_i=1-m_i=0;\quad
Q_i<0\Rightarrow n_i=m_i=0,\nn\\
&&\!\!\!\!\!\!\!
\sum_{Q_i>0}Q_i\left(n_i+\half\right)
-\sum_{Q_i<0}Q_i\left(\bar n_i+\half\right)
+\sum_{j}q_{j}\left(m_{j}-{1\over 2}\right)+\rmq=0.
\label{gveqpos}
\eeqa
For each solution to this equation,
we have a supersymmetric effective theory on the Coulomb branch
with
\beq
\neff=
\sum_{Q_i>0}(n_i+\bar n_i+1)-\sum_{Q_i<0}(n_i+\bar n_i+1).
\eeq
Suppose $\neff>0$. Then, the effective theory
has a supersymmetric ground state for $\zeta>0$,
\beqa
\vs<0:&
|\vs|^{\neff\over 2}\e^{\zeta\vs}
|n_{\phi},\bar n_{\phi},m_{\phi},m_{\eta}\rangle,&(-1)^F=(-1)^{|m|}
\label{profilespZns} \ ,\\
\vs>0:&
|\vs|^{\neff\over 2}\e^{-\zeta\vs}\blambda
|n_{\phi},\bar n_{\phi},m_{\phi},m_{\eta}\rangle,&(-1)^F=(-1)^{|m|+1} \ ,
\label{profilespZps}
\eeqa
which runs away to infinity as $\zeta\searrow 0$.
For $\zeta<0$, it has no supersymmetric ground state.
Suppose $\neff<0$. Then, the effective theory does not have a supersymmetric
ground state for $\zeta>0$.
For $\zeta<0$, it has a supersymmetric ground state
\beqa
\vs<0:&
|\vs|^{|\neff|\over 2}\e^{-\zeta\vs}\blambda
|n_{\phi},\bar n_{\phi},m_{\phi},m_{\eta}\rangle,&(-1)^F=(-1)^{|m|+1}
\label{profilesnZns}\\
\vs>0:&
|\vs|^{|\neff|\over 2}\e^{\zeta\vs}
|n_{\phi},\bar n_{\phi},m_{\phi},m_{\eta}\rangle,&(-1)^F=(-1)^{|m|},
\label{profilesnZps}
\eeqa
which runs away to infinity as $\zeta\nearrow 0$. Finally,
suppose $\neff=0$. Then, the effective theory has a zero energy state both
for $\zeta>0$ and $\zeta<0$, with the profile as in
(\ref{profilespZns})-(\ref{profilespZps}) and
(\ref{profilesnZns})-(\ref{profilesnZps}).
It is square normalizable but it is not clear if
it qualifies as the Coulomb branch vacua since
its profile is peaked near the origin.
However, the wavefunction spreads out to infinity
as $\zeta\to 0$. Hence there is a chance that they
play a r\^ole in the wall crossing phenomenon.

Let us compute the change of the index $I(y^{{\bf G}^F})$,
where $y^{{\bf G}^F}=\prod_fy_f^{{\bf G}^f}$,
as $\zeta$ goes from positive to negative.
There is no subtlety concerning the contribution from
the sector with $\neff>0$ or $\neff<0$:
the states (\ref{profilespZns}) and (\ref{profilespZps})
go away to infinity as $\zeta\searrow 0$
or the states (\ref{profilesnZns}) and (\ref{profilesnZps})
come in from infinity as $\zeta<0$ is turned on.
The contributions to the change in the index in the two cases
take the same form, $(-1)^{|m|+1}\prod_fy_f^{Q^f}$ from the $\vs<0$ branch
and $(-1)^{|m|}\prod_fy_f^{Q^f}$ from the $\vs>0$ branch, where
$Q^f$ is
\beqa
\vs<0:~~~~&&\!\!\!\!\!\!\!\!\!\!\!\!\!
-\sum_{Q_i>0}Q^f_i\left(\bar n_i+{1\over 2}\right)
+\sum_{Q_i<0}Q^f_i\left(n_i+{1\over 2}\right)
+\sum_{j}q^f_{j}\left(m_{j}-{1\over 2}\right)+\rmq^f,
\label{gQneg}\\
\vs>0:~~~~&&\!\!\!\!\!\!\!\!\!\!
\sum_{Q_i>0}Q^f_i\left(n_i+{1\over 2}\right)
-\sum_{Q_i<0}Q^f_i\left(\bar n_i+{1\over 2}\right)
+\sum_{j}q^f_{j}\left(m_{j}-{1\over 2}\right)+\rmq^f.
\label{gQpos}
\eeqa
Subtle is what to do for the sector with $\neff=0$ where the states
do not run away but simply spread out as $\zeta\to 0$.
The contribution to the change of the index takes the same form
as the $\neff\ne 0$ cases if we take the average
--- a ``half'' of
(\ref{profilespZns}) and (\ref{profilespZps})
 go away to infinity and a ``half'' of
(\ref{profilesnZns}) and (\ref{profilesnZps})
come in from infinity. If we decide to do so, then,
the change of the index takes a concise formula:
\beq
\Delta_C I(y^{{\bf G}^F})
=\sum_{(\ref{gveqneg})}(-1)^{|m|+1}
\prod_fy_f^{Q^f}
+
\sum_{(\ref{gveqpos})}(-1)^{|m|}
\prod_fy_f^{Q^f}.
\label{DeltaCI}
\eeq
We will see that this is in perfect agreement with the wall crossing
formula that will be derived in Section~\ref{sec:WC}.
This invites us to propose that, for the $\neff=0$ sector,
a ``half'' of the ground state for $\zeta>0$ plus a ``half'' of
the ground state for $\zeta<0$ contribute to the wall crossing
of the index as if they were a single state.
Recall, however, that the $\neff=0$ states have main support toward the
origin $\vs\to 0$ of the Coulomb branch,
and hence we need more information there
to say whether the state really exists.
The index result is understandable if
the state exists only in the $\zeta>0$ side
or only in the $\zeta<0$ side.

In the above computation of $\Delta_C I(y^{{\bf G}^F})$,
we assumed that the twist parameters are unitary $y_f=\e^{i\xi_f}$,
so that we do not need to turn on the real mass $m_f$.
Turning on real masses would change the form of the Hamiltonian
and we would need to start the analysis over.
However, the wall crossing part is sensitive only to the behaviour
of the wavefunction at large values of $|\vs|$ to which the finite real masses
have only a minor effect. The only effect we need to have in mind is the
the deformed supersymmetry algebra (\ref{defSA}) which dictates how
$m_f$ enters into the result: that is, we can simply
set $y_f=\e^{i(\xi_f+i\beta m_f)}$ in the result (\ref{DeltaCI}).
This can also be seen buy looking at the result
(\ref{DeltaCI}) itself. It is a Laurent polynomial in $y_f$'s and the
holomorphic extension from $y_f=\e^{i\xi_f}$ to
$y_f=\e^{i(\xi_f+i\beta m_f)}$ is straightforward.
This is so even when the theory is not
effectively compact. (Of course, in an effectively compact theory
the replacement $\e^{i\xi_f}\to \e^{i(\xi_f+i\beta m_f)}$ is valid everywhere.)
Thus, the wall crossing formula (\ref{DeltaCI}) is valid for arbitrary
complex twist parameters $y_f\in\C^{\times}$, whether or not the theory
is effectively compact.

Let us also write down the wall crossing formula
in ${\mathcal N}=4$ systems with $U(1)$ gauge group.
We recall that the Coulomb branch vacua exists only when $\zeta$ has the
same sign as $\neff^{(4)}$, defined in (\ref{neff4def}).
They are invariant under $U(1)$ R-symmetry (so that $\bfR_{\!-}=- J_3$)
and form a representation of
the $SU(2)$ R-symmetry group of spin $j={|\neff^{(4)}|-1\over 2}$,
which has character
$\chi(\by^{-2J_3})=\by^{-2j}+\by^{-2j+2}+\cdots +\by^{2j}$.
As a convention, we assign $(-1)^F=(-1)^{2J_3}$.
Therefore, as $\zeta$ goes from positive to negative, the change of the
index is
\beq
\Delta_C I(\by^{2\bfsmR_-}y^{{\bf G}^F})
=(-1)^{\neff^{(4)}}
{\rm sign}(\neff^{(4)})\left(\by^{-(|\neff^{(4)}|-1)}+\cdots
+\by^{|\neff^{(4)}|-1}\right).
\label{DeltaIC4}
\eeq
This is true even when the theory is effectively
non-compact and when there is a flavor twist, as included in the notation.
This is because the wall crossing states belong
to a discrete part of the spectrum in which the effect of the real and
twisted masses is continuous. In particular, the operator identity
(\ref{Opid}) makes sense on those states
and we can prove that all the Coulomb branch
vacua that contribute to the wall crossing have zero flavor charge.

\subsection{Simple Wall Crossing}\label{subsec:simpleWC}

Let us next consider theories with higher rank gauge groups.
On the wall between two phases, there are classical vacua with
continuous unbroken gauge symmetries.
In general, the symmetry breaking pattern is complex
and it is not clear how to analyze the wall crossing. For example,
in the III-IV wall in the triangle quiver with rank vector
$(k,1,1)$ with $k\geq 3$ (see Fig.~\ref{fig:qk11}),
the space of classical vacua is stratified by mixed branches
of different unbroken gauge groups.
However, in some cases, we have a simple symmetry breaking where
the unbroken gauge group is isomorphic to $U(1)$.
This is always the case in the interior of a wall
in a theory with an Abelian gauge group,
but there can be such walls even in non-Abelian gauge theories.
Let us analyze the wall crossing across such a wall, which we shall
refer to as {\it simple}.

Thus,
suppose we have a simple wall between two phases, with
 the unbroken gauge group
$G_1\subset G$ isomorphic to $U(1)$.
When $\zeta$ is close to such a wall region, we must look at
the effective theory on the mixed Coulomb-Higgs branch where
the $G_1$ component of the vector multiplet scalar field is very large
while the complementary gauge group $C(G_1)/G_1$ is Higgsed.
Here $C(G_1)\subset G$ is the subgroup of commutants of $G_1$.
We decompose the fields into two groups:
(C) those which are charged under $G_1$ and
(H) those which are invariant under $G_1$.
When $G$ is non-Abelian, the components of the vector multiplet
which do not commute with $G_1$ should be included in (C) ---
as fermi multiplets in ${\mathcal N}=2$ theories and as chiral multiplets
in ${\mathcal N}=4$ theories.
Accordingly we have two theories:
\begin{description}
\item[Theory (C)] gauge group $G_1$,
matters from class (C), and the FI parameter close to zero,
\item[Theory (H)] gauge group $C(G_1)/G_1$,
matters from class (H) and the FI parameter in the wall region.
\end{description}
The effective theory on the mixed branch is the ``semi-direct product''
of the Coulomb branch of Theory (C) and the
Higgs branch of Theory (H).
We say ``semi-direct'' because each vacuum of Theory (C)
may carry its own charge under $C(G_1)/G_1$,
which provides a background charge, i.e., the Wilson line,
in Theory (H).
Note that $C(G_1)/G_1$ is regarded as a flavor symmetry in Theory (C).

In ${\mathcal N}=4$ theories, however,
Coulomb branch vacua of (C) that contribute to the wall crossing
has charge zero under the flavor symmetry,
as shown right above.
Therefore, the Coulomb and the Higgs dynamics decouples and
the mixed branch is the direct product of the Coulomb and the Higgs
branches.
In particular, supersymmetric ground states are the
tensor products of the ground states on the two branches,
and the contribution to the change of the index is the simple product:
\beq
\Delta I(\by^{2\bfsmR_{\!-}}y^{{\bf G}^F})
=I_{\rm wall}^{({\rm H})}(\by^{2\bfsmR_{\!-}}y^{{\bf G}^F})\times
\Delta_CI^{({\rm C})}
(\by^{2\bfsmR_{\!-}}).
\label{4factorization}
\eeq
Let us illustrate this in the wall crossing in the Abelian triangle quiver
(Example 3, $\vec{k}=(1,1,1)$).
We look at the wall between Phase I and Phase II (see Fig.~\ref{fig:q111}).
The unbroken gauge group is $G_1=\{[z,1,1]\}\cong U(1)$
and the fields are decomposed as
\beqa
({\rm C})&&y_{1,\ldots, b}\, (1),~ z_{1,\ldots, c}\,(-1),\nn\\
({\rm H})&&x_{1,\ldots, a}\, (1).\nn
\eeqa
To the right of the fields, we put their charges with respect to
$G_1$ for (C) and $G/G_1$ for (H).
The Coulomb branch of (C) has $\neff^{(4)}=b-c$ so that we have
$\Delta_CI^{({\rm C})}(\by^{2\bfsmR_{\!-}})=(-1)^{b-c-1}
{\rm sign}(c-b)(\by^{-(|b-c|-1)}+\cdots+\by^{|b-c|-1})$.
The Higgs branch of (H) is $\CP^{a-1}$ which has
$I^{({\rm H})}(\by^{2\bfsmR_{\!-}})=(-1)^{a-1}(\by^{-(a-1)}+\cdots+\by^{a-1})$.
Thus, as $\zeta$ moves from Phase I to Phase II, the index changes as
\beq
\Delta I(\by^{2\bfsmR_{\!-}})=(-1)^{a+b+c}
(\by^{-(a-1)}\!\!+\cdots+\by^{a-1})
\cdot {\rm sign}(c-b)(\by^{-(|b-c|-1)}\!\!+\cdots+\by^{|b-c|-1}).
\label{wcAbTr}
\eeq
We will see more examples in later sections.

In a theory without ${\mathcal N}=4$ supersymmetry,
each vacuum may have a non-trivial
flavor charge and hence Theory (H) may have different background charges
for different (C) vacua. Therefore, the change in the
index is not in general the simple product, but is a ``semi-direct'' product,
which may be denoted as
\beq
\Delta I(y^{{\bf G}^F})
=(I_{\rm wall}^{({\rm H})}\ast
\Delta_CI^{({\rm C})})
(y^{{\bf G}^F}).
\label{sWC1}
\eeq
We will present illustrative examples in later sections.

\section{The Index}\label{sec:Index}

In this section, we compute the Witten index
\beq
I(y^{{\bf G}^F})
\,\,=\,\,{\rm Tr}_{\mathcal H}^{}\left((-1)^Fy^{{\bf G}^F}\e^{-\beta H}\right).
\label{Ih}
\eeq
As we have seen, the spectrum of supersymmetric ground states
depends very much on the region where the FI parameter $\zeta$ belongs.
When it is deep inside a phase
where the gauge group is broken to a finite group,
the ground state spectrum is that of the effective theory on the Higgs branch,
and in particular the index should agree with the Higgs branch index.
``Deep inside'' means that $1/\beta$ is comparable to or smaller than
the Higgs mass $M_H=e\sqrt{|\Delta\zeta|}$ (see (\ref{HiggsEnergy})),
\beq
e^2|\Delta \zeta|\,\sim\, {1\over \beta^2},
\label{HCindex}
\eeq
where $|\Delta \zeta|$ is the distance to the walls of the phase.
As $\zeta$ approaches a wall, some of the ground states
start to have supports on the Coulomb branch and eventually
disappear by running away to infinity.
As a consequence, the index must undergo a transition
as $\zeta$ goes across the wall.
The effective theory on the Coulomb branch has a characteristic energy
scale $M_C=e^2|\Delta \zeta|^2$ (see (\ref{CoulombEnergy})).
Thus, the wall crossing transition is expected within
a range where it is vanishingly small compared to $1/\beta$,
\beq
e^2|\Delta \zeta|^2\,\ll \, {1\over \beta}.
\label{CCindex}
\eeq

We shall employ the method of supersymmetric localization
to compute the index, which includes taking the limit $e^2\to 0$.
This limit is potentially singular because it turns off
the D-term potential and the Higgs mass.
The singularity is especially severe in the one-dimensional system
because of the non-compact flat direction in the Coulomb branch.
As we will see, in order to obtain a sensible answer,
we need to take particular scaling limits in which $\Delta\zeta$ is sent to
infinity. In addition, there is a subtlety in the definition of the
index associated to the non-compactness,
and some of the above expectation needs to be reexamined. 
All these are unlike the similar computation of the elliptic genus of
two-dimensional $(2,2)$ supersymmetric theories \cite{BEHT1,BEHT2}
where the Coulomb branch is lifted
by the R-symmetry twist.

\subsection{Setting Up The Computation}\label{subsec:setting}

\newcommand{\cMMM}{\tilde{\MMM}}

We consider a general ${\mathcal N}=2$ supersymmetric gauge theory
introduced in Section~\ref{subsec:models},
with the gauge group $G$, the matter contents specified by
$V_{\rm chiral}$, $V_{\rm fermi}$ and $M$, and the interaction
given by $\zeta$, $E(\phi)$, $J(\phi)$ and
$Q(\phi)$. We suppose that there is a group $G_F$ of flavor symmetries.

The index (\ref{Ih}) is equal to the path integral on the Euclidean
circle of circumference $\beta$ with
periodic boundary condition, with the flavor Wilson line $v_{\tau}^F$
and real mass $m$ turned on, for $y=\e^{-\beta(iv^F_{\tau}+m)}$,
\beq
I=\int{\mathcal D}\mu\,\exp\left(-\int_0^{\beta}L_E\,\dd\tau\right)
{\rm Str}^{}_M{\rm Pexp}
\left(-\int_0^{\beta}i{\mathcal A}_{\tau}\dd\tau\right).
\eeq
The Euclidean Lagrangian is obtained by
the Wick rotation $t\to -i\tau$, which includes
the rotation of the auxiliary fields,
$D\to iD_E$, where $D_E$ is real valued (i.e. $D_E\in i\mathfrak{g}$),
and $F\to iF_E$, $\overline{F}\to iF_E^{\dag}$:
\beqa
L_E&=&
{1\over 2e^2}{\rm Tr}\Bigl[\, (D_\tau\vs)^{2}
+\blambda\,D_{\tau}^{(+)}\lambda+D_E^2
\,\Bigr]\,\,+\,\,i\zeta(D_E)\nn\\
&&+{1\over\mc^2}\left[
\tilde{D}_\tau\bphi \tilde{D}_\tau\phi
+\bpsi\,\tilde{D}_t^{(-)}\psi
+\bphi \left\{-iD_E+\tilde{\vs}^{\,2}\right\}\phi
+i\bphi\lambda\psi-i\bpsi\,\blambda\phi\right]\nn\\
&&+{1\over\mc^2}\left[
\bareta\,\tilde{D}_{\tau}^{(+)}\eta+F^{\dag}_E F_E
+\overline{E(\phi)}E(\phi)
+\bareta\partial_iE(\phi)\psi^i
+\bpsi^{\bi}\partial_{\bi}\overline{E(\phi)}\eta\right]\nn\\
&&
-\psi^i\partial_iJ(\phi)\eta+iJ(\phi)F_F
-\bareta\partial_{\bi}J(\phi)^{\dag}\bpsi^{\bi}
+iF_E^{\dag}J(\phi)^{\dag},\\[0.3cm]
i{\mathcal A}_{\tau}&=&\rho(i\tilde{v}_{\tau}+\tilde{\vs})
-\psi^i\partial_iQ(\phi)
+\bpsi^{\bi}\partial_{\bi} Q(\phi)^{\dag}
+\{Q(\phi),Q(\phi)^{\dag}\}.
\eeqa
Here we put $\tilde{v}_{\tau}:=v_{\tau}+v^F_{\tau}$ and $\tilde{\vs}:=\vs+m$,
and in particular,
we have $\tilde{D}_{\tau}^{(\pm)}={\dd\over \dd \tau}+i\tilde{v}_{\tau}\pm
\tilde{\vs}$
We assume that there is a region of parameters $\zeta$, $v_{\tau}^F$, $m$,
$J(\phi)$, etc,
where the theory is effectively compact.
Then, the index is well-defined and is independent of parameters as long as
they are in that region (except the dependence on
$iv^F_{\tau}+m$ due to (\ref{defSA})).
In particular, it is independent of $e$ and $\mc$, and
we evaluate the path-integral in the regime where $e$ and $\mc$ are very small.
The basic strategy to evaluate the path-integral is mostly the same as
in the computation \cite{BEHT1,BEHT2} of 2d index,
but we shall see important differences.

We will eventually take the limit where the gauge coupling constant
$e$ is sent to zero.
In that limit, the path-integral is dominated by the
configurations satisfying
\beq
D_{\tau}\vs=0,\quad D_E=0.
\label{SUSYvec}
\eeq
The only observable of a gauge field on the circle is its holonomy,
$h={\rm Pexp}\left(-i\int_0^{\beta}v_{\tau}\dd\tau\right)$,
and a parallel section $\vs$ is determined by its value
at $\tau=0$ that commutes with $h$.
Therefore, the moduli space $\MMM$ of solutions to (\ref{SUSYvec})
is the space of commuting pairs in $G\times i\mathfrak{g}$
modulo the adjoint action, which is isomorphic to
\beq
\MMM\,\,=\,\,\left(T\times i\ttt\right)/W,
\eeq
where $T$ is a maximal torus of $G$ and $W$ is the Weyl group of $(G,T)$.
A solution is realized by constant profiles with values in
$i\ttt$,
$v_{\tau}\equiv v^{\ttt}_{\tau 0}$
and $\vs\equiv {\vs}_0^{\ttt}$. Since the holonomy is
$h=\e^{-i\beta v_{\tau 0}^{\ttt}}$ we see that the shift of $\beta v^{\ttt}_{\tau 0}$
by an element of $2\pi \rmQv$ is a gauge transformation,
where $\rmQv\subset i\ttt$ is the coroot lattice --- the kernel of
the map $\xi\in i\ttt\mapsto \e^{2\pi i\xi}\in T$.
It is convenient to take the supersymmetric combination
\beq
u:={\beta\over 2\pi}(-v^{\ttt}_{\tau 0}+i{\vs}_0^{\ttt})\in \ttt_{\C},
\label{defu}
\eeq
which has periodicity $\rmQv$. It defines a system of
complex flat coordinates of the cover
$\cMMM=\ttt_{\C}/\rmQv$
of the moduli space, $\MMM\cong \ttt_{\C}/(\rmQv\rtimes W)=\cMMM/W$.
In contrast to the 2d index computation,
the moduli space $\MMM$ is non-compact,
and that is responsible for the wall crossing phenomenon which is
the distinguished feature in 1d.

Note that (\ref{SUSYvec}) is the condition of supersymmetry,
$\delta\lambda=\delta\blambda=0$. Therefore, $\MMM$ is also
the moduli space of
supersymmetric background for the vector multiplet.
In fact, the above discussion guarantees that
our discussion in Section~\ref{subsec:DefInd}
on the flavor twist and real mass was general enough.
Just like (\ref{defu}), we write
\beq
z:={\beta\over 2\pi}(-v^F_{\tau}+im)\in\ttt_{F\C},
\label{defz}
\eeq
for the complex parameter of the flavor twist, $y=\e^{2\pi i z}\in T_{F\C}$.

The index can be written as the integration on the moduli space,
\beq
I=\int_{\mathfrak{M}}\dd^{2\ell}u\, F_{e,\mc}(u)
={1\over |W|}\int_{\cMMM}\dd^{2\ell}u\, F_{e,\mc}(u),
\label{intex}
\eeq
where $F_{e,\mc}(u)$ is the result of path integral over all fields
except the commuting zero modes of $v_{\tau}$ and $\vs$.
The integrand $F_{e,\mc}(u)$ depends on $e$ and $\mc$, even though the
the integral (\ref{intex}) does not.
The matter integral for a given gauge multiplet is simplified if we
can take the limit $\mc\to 0$.
This is certainly valid when the scalar component $\phi$
of the chiral multiplet does not have a zero mode. Along the supersymmetric
background specified by $u\in \ttt_{\C}$, this is the case except when
$\e^{2\pi i(u+z)}:V_{\rm chiral}\to V_{\rm chiral}$ has eigenvalue $1$.
Let
\beq
V_{\rm chiral}=\bigoplus_{i\in\rmI} \C(Q_i,Q^F_i),
\label{wtChiral}
\eeq
be the weight decomposition. Then, the $i$-th component $\phi_i$
of $\phi$ has a zero mode when
\beq
Q_i(u)+Q^F_i(z)\equiv 0\quad \mbox{modulo $\Z$}.
\label{Hidef}
\eeq
We denote the locus (\ref{Hidef}) by $H_i\subset \cMMM$
and call it a {\it singular hyperplane} for the field $\phi_i$.
If $u$ is away from any of such hyperplanes,
then, the limit $\mc\to 0$ can obviously be taken.
 We further make the following assumption that allows $\mc\to 0$
(for non-zero $e$)
even along singular hyperplanes: At each point $u_*\in \cMMM$,
let $(\phi_i)_{i\in \rmI_{u_*}}$ be the set of fields
which has zero modes at $u_*$. Then, we assume that the space of
$(\phi_i)_{i\in \rmI_{u_*}}$ solving the D-term equations is compact.
For example, for $G=U(1)$, this means that the fields having zero modes at
each point of $\cMMM=\MMM$ have either positive charges or negative charges,
but not both.
Under this assumption, the limit $\mc\to 0$ can be taken everywhere
on $\cMMM$, provided $e$ is non-zero.
Moreover, this allows us to ignore the contribution from
an infinitesimal neighborhood of the singular hyperplanes
and take a limit $e\to 0$ at the same time. To state it precisely,
let us put
\beq
\Deps:=\bigcup_{i\in \rmI}\Deps(H_i) \ ,
\label{Deps}
\eeq
where $\Deps(H_i)$ is the $\varepsilon$-neighborhood of $H_i$ in $\cMMM$.
Then, there is a double scaling limit $e\to 0$ and $\varepsilon\to 0$
under which the contribution from $\Deps$ at
$\mc=0$ vanishes, so that the integral (\ref{intex}) can be replaced by
\beq
I={1\over |W|}\lim_{e\to 0\atop\varepsilon\to 0}
\int_{\tilde{\MMM}\setminus \Delta_{\varepsilon}}\dd^{2\ell}u\,F_{e,0}(u).
\eeq

Thus, we would like to compute $F_{e,\mc}(u)$ for $u$ away from the singular
hyperplanes, in the limit $e\to 0$ and $\mc\to 0$.
Note that the moduli variable $u$, i.e.,
$v_{\tau 0}^{\ttt}$ and ${\vs}_0^{\ttt}$
are not closed under the supersymmetry (\ref{Vector}) but is
a part of the supermultiplet consisting of the constant modes of the
 Cartan part of the vector multiplet
$(v_{\tau 0}^{\ttt},{\vs}_0^{\ttt},\lambda_0^{\ttt},\blambda_0^{\ttt},
D^{\ttt}_{E0})$.
 To obtain $F_{e,\mc}(u)$, we first
do the path-integral over all the supermultiplets
which are orthogonal to this Cartan zero modes, and then
do the integration over $\lambda_0^{\ttt}$, $\blambda_0^{\ttt}$
and $D^{\ttt}_{E 0}$. In the limit $e\to 0$ and $\mc\to 0$, the first part can
be done exactly by the one-loop integral.

\subsection{One Loop Integral}\label{subsec:oneloop}

\subsubsection*{\rm\underline{Vector Multiplet}}

We write the vector multiplet modes which are orthogonal to the
Cartan zero modes as $(v_{\tau}',\vs',\lambda',\blambda{}',D'_E)$.
The gauge fixing amounts to dropping the integral over
$(v_{\tau}',\vs')$. The simplest way to see this is to note
that the Gauss law is simply $D_{\tau}\vs=0$ in the limit $e\to 0$,
which sets $v_{\tau}'=\vs'=0$ and leaves us with the integral over
the moduli space $\MMM$. Thus, the one-loop integral is
just over $(\lambda',\blambda{}',D_E')$, which yields
\beq
Z'_{\rm vector}\,\,=\,\,{\rm det}'D_{\tau 0}^{\ttt (+)} \ .
\eeq
The operator $D_{\tau 0}^{\ttt (+)}={\dd\over \dd\tau}
+{\rm ad}(iv_{\tau 0}^{\ttt}+{\vs}_0^{\ttt})
={\dd\over \dd\tau}-{2\pi i\over \beta}{\rm ad}(u)$ is diagonalized via
the Fourier modes $\e^{-2\pi i\rmm\tau/\beta}$ and the root
decomposition of $\mathfrak{g}_{\C}$, and we find
(we omit writing $(\pm i){2\pi\over \beta}$ that multiplies each eigenvalue)
\beqa
Z'_{\rm vector}&=&
\left(\prod_{\rmm\ne 0}\rmm\right)^{\ell}\cdot
\prod_{\alpha}\prod_{\rmm\in\Z}\left(\rmm+\alpha(u)\right)\nn\\
&\propto&\prod_{\alpha}\left(\e^{-\pi i\alpha(u)}-\e^{\pi i\alpha(u)}
\right).
\label{417}
\eeqa
%where $C$ is a constant that results from the infinite product.
Each root factor in (\ref{417}) can also be understood from the
operator formalism: The index of the
$\lambda_{\alpha},\blambda{}_{\alpha}$ system may be written as
$Z_{\alpha}={\rm Tr}(-1)^F\e^{2\pi i {\bf G}(u)}
\e^{-\beta H}$,
with $H=0$ and ${\bf G}(u)
={\alpha(u)\over 2}[\blambda_{\alpha},\lambda_{\alpha}]$.
If we assign $(-1)^F=1$ to the Fock vacuum annihilated by
$\lambda_{\alpha}$, then we find
$Z_{\alpha}=\e^{-\pi i\alpha(u)}-\e^{\pi i\alpha(u)}$.
It may be more natural to put $v_{\tau 0}^{\ttt}$ to ${\bf G}$
and ${\vs}_0^{\ttt}$ to $H$, but the result is the same.

In the remaining part of the one-loop integral, the fluctuation modes
$(v_{\tau}',\vs',\lambda',\blambda{}',D'_E)$ will never appear.
To simplify the notation, we denote the Cartan zero modes simply as
$(v_{\tau},\vs,\lambda,\blambda,D_E)$, dropping the super/subscripts
$\ttt/0$.
We also write
\beq
D:={\beta^2\over (2\pi )^2}D_E,
\label{DefrmD}
\eeq
for a dimensionless variable.

\subsubsection*{\rm\underline{Chiral Multiplet}}

Since we shall eventually perform the integral over the $2\ell$
gluino zero modes,
we may drop the terms $i\bphi\lambda\psi-i\bpsi\,\blambda \phi$ from the
exponential, and look only at
\beqa
Z_{\rm chiral}&=&
\int{\mathcal D}\phi{\mathcal D}\psi
\e^{-\int_0^{\beta}\dd\tau\left[
\bphi\left(-\tilde{D}_{\tau}^2+\tilde{\vssm}^2-iD_E\right)\phi
+\bpsi \tilde{D}_{\tau}^{(-)}\psi\right]}
{1\over (\ell !)^2}\left(\int_0^{\beta}\bphi\lambda\psi\dd\tau
\int_0^{\beta}\bpsi\,\blambda\phi\dd\tau'\right)^{\ell}
\nn\\
&=&
{\displaystyle {\rm det}\,\tilde{D}_{\tau}^{(-)}
\over
\displaystyle {\rm det}\left(-\tilde{D}_{\tau}^2+\tilde{\vssm}^2-iD_E\right)}
\left\langle
{1\over (\ell !)^2}\left(\int_0^{\beta}\bphi\lambda\psi\dd\tau
\int_0^{\beta}\bpsi\,\blambda\phi\dd\tau'\right)^{\ell}
\right\rangle \ .
\eeqa
In view of the weight decomposition (\ref{wtChiral}), this is equal to
\beq
Z_{\rm chiral}\,\,=\,\,
g_{\rm chiral}(u,D)\cdot  \det h(u,D)
\prod_{a=1}^{\ell}\lambda_a\blambda_a,
\eeq
where
\beq
g_{\rm chiral}(u,D)
\,\,:=\,\,{\displaystyle \prod_{i\in\rmI}\prod_{\rmm\in \Z}
\left(\rmm
+Q_i(\bu)+Q_i^F(\bz)\right)\over \displaystyle
\prod_{i\in\rmI}\prod_{\rmm\in \Z}
\left(\left|\rmm
+Q_i(u)+Q_i^F(z)\right|^2-iQ_i(D)\right)},
\eeq
and $\det h(u,D)$ is the determinant of the $\ell\times \ell$ matrix
\beq
h^{ab}(u,D)\,\,:=\,\,\sum_{i\in\rmI}\sum_{\rmm}{Q^a_iQ^b_i\over
\left(\left|\rmm
+Q_i(u)+Q_i^F(z)\right|^2-iQ_i(D)\right)
\left(\rmm
+Q_i(\bu)+Q_i^F(\bz)\right)}.
\label{hab}
\eeq
If we set $D=0$, the function $g_{\rm chiral}$ simplifies
\beqa
g_{\rm chiral}(u,0)&=&{1\over  \displaystyle
\prod_{i\in\rmI}\prod_{\rmm\in \Z}
\left(\rmm
+Q_i(u)+Q_i^F(z)\right)}\nn\\
&\propto&
{1\over  \displaystyle
\prod_{i\in\rmI}
\left(\e^{\pi i(Q_i(u)+Q_i^F(z))}
-\e^{-\pi i(Q_i(u)+Q_i^F(z))}
\right)}.\label{gchiralexp}
\eeqa
It is the index of a chiral multiplet in the supersymmetric background,
which explains the holomorphic dependence on $u$ and $z$,
see Section~\ref{subsec:DefInd}.
The precise form (\ref{gchiralexp})
can also be understood from the operator formalism.
For this purpose, it is enough to consider the $U(1)$ gauge theory
with a single chiral multiplet of charge $1$.
Quantization of the chiral multiplet in the constant background
$\vs$ has been done in Section~\ref{subsec:CPN}, as its special case $N=1$.
The energy and the charge spectrum is given in (\ref{EspCPN})
and (\ref{QspCPN}). If we assign $(-1)^F=1$ to the oscillator vacuum
$|0\rangle{}_{{}_{\rm OSC}}$, we see that the twisted index is
\beqa
{\rm Tr}(-1)^F\e^{-i\beta {\bf G}(v_{\tau})}\e^{-\beta H}
&=&\sum_{n,\bar n\in \Z}\sum_{m=0,1}
(-1)^m\e^{-i\beta v_{\tau}\left(n-\bar n+m-\half\right)}
\e^{-\beta|\vssm|(n+\bar n+1)+\beta\vssm \left(m-\half\right)}\nn\\
&=&\e^{-\beta |\vssm|}
{\displaystyle
\e^{-i\beta v_{\tau}\left(-\half\right)}\e^{\beta\vssm\left(-\half\right)}
-\e^{-i\beta v_{\tau}\left(\half\right)}\e^{\beta\vssm\left(\half\right)}
\over\displaystyle
(1-\e^{-i\beta v_{\tau}}\e^{-\beta|\vssm|})
(1-\e^{i\beta v_{\tau}}\e^{-\beta|\vssm|})}\nn\\
&=&{1\over \e^{\pi i u}-\e^{-\pi i u}}\qquad\mbox{(for both signs of
$\vs$)}\ ,
\eeqa
which is indeed of the form (\ref{gchiralexp}).

\subsubsection*{\rm\underline{Fermi Multiplet}}

Let
\beq
V_{\rm fermi}=\bigoplus_{j\in\rmJ}\C(q_j,q_j^F)
\eeq
be the weight decomposition.
Then, the one loop integral is given by
\beqa
Z_{\rm fermi}&=&{\rm det}\,D_{\tau}^{(+)}
=\prod_{j\in \rmJ}\prod_{\rmm\in\Z}(\rmm+q_j(u)+q_j^F(z))\nn\\
&\propto&
\prod_{j\in\rmJ}\left(\e^{-\pi i(q_j(u)+q_j^F(z))}
-\e^{\pi i(q_j(u)+q_j^F(z))}\right).
\label{Zfermi}
\eeqa
The final expression can also be understood from
the operator formalism.

\subsubsection*{\rm\underline{Wilson Line}}

Let
\beq
M=\bigoplus_{k\in \rmK}\C(\rmq_k,\rmq_k^F)[r_k]
\eeq
be the weight decomposition, where $r_k=0$ for the even part
and $r_k=1$ for the odd part.
Then, the Wilson line gives the following factor
\beq
Z_{\rm Wilson}\,\,=\,\,\sum_{k\in\rmK}
(-1)^{r_k}\e^{2\pi i(\rmq_k(u)+\rmq_k^F(z))}.
\label{ZWilson}
\eeq

\subsubsection*{\rm\underline{Summary}}

After performing the gluino integral, we have
\beq
I\,=\,{N_{\ell}\over |W|}\lim_{e\to 0\atop\varepsilon\to 0}
\int_{\cMMM\setminus \Deps}\dd^{2\ell\!}u
\int_{i\ttt}\dd^{\ell} D\,\,
g(u,D)\det h(u,D)
\exp\left(-{(2\pi)^4\over 2e^2\beta^3}D^2
-i{(2\pi)^2\over\beta}\zeta(D)\right),
\label{int}
\eeq
where $N_{\ell}$ is a normalization constant to be determined, and
$g(u,D)$ is the product of $Z_{\rm vector}'(u)$, $g_{\rm chiral}(u,D)$,
$Z_{\rm fermi}(u)$ and $Z_{\rm Wilson}(u)$,
which is normalized so that $g(u,D=0)$ is the product of the right hand
sides of
(\ref{417}), (\ref{gchiralexp}), (\ref{Zfermi}) and (\ref{ZWilson}),
\beq
g(u,0)=\prod_{\alpha}2i\sin(-\pi\alpha(u))
{\displaystyle
\prod_{j\in\rmJ}2i\sin(-\pi(q_i(u)+q_i^F(z)))
\over  \displaystyle
\prod_{i\in\rmI}
2i\sin(\pi(Q_i(u)+Q_i^F(z)))}
\sum_{k\in\rmK}(-1)^{r_k}\e^{2\pi i(\rmq_k(u)+\rmq_k^F(z))}.
\eeq
It remains to perform the $u$ and $D$ integrals and then take the limit.
For this purpose, there is a very useful identity
\beq
{\partial\over \partial \bu_a}g(u,D)=-i h^{ab}(u,D)D_b\,
g(u,D).\label{useful}
\eeq
As in \cite{BEHT1,BEHT2},
this can be used to turn the integral into the residue integral
around intersections of the singular hyperplanes.
We first consider a theory with $U(1)$ gauge group.

\subsection{$U(1)$ Theories}\label{subsec:U1}

If the gauge group is $U(1)$, the moduli space is one dimensional,
$\cMMM=\MMM=\C/\Z\cong\C^{\times}$, and singular hyperplanes are
points of $\MMM$.
We denote by $\MMM_{\rm sing}^{(+)}$ and $\MMM_{\rm sing}^{(-)}$
the sets of singular points corresponding to positively
charges fields and negatively charged fields respectively.
By the assumption made above, they have no overlap,
$\MMM_{\rm sing}^{(+)}\cap \MMM_{\rm sing}^{(-)}=\emptyset$.
We denote by $\Deps^{(\pm)}$ the epsilon neighborhood of
$\MMM_{\rm sing}^{(\pm)}$, so that $\Deps=\Deps^{(+)}\sqcup\Deps^{(-)}$.
We set $\beta=2\pi$ to simplify the notation.

For a $U(1)$ theory, the integral (\ref{int}) is
\beq
I=N_1\lim_{e\to 0\atop\varepsilon\to 0}
\int_{\mathfrak{M}\setminus \Deps}\dd^{2}u
\int_{\R}\dd D\,\,
g(u,D) h(u,D)
\exp\left(-{\pi\over e^2}D^2-2\pi i\zeta D\right),
\label{int1}
\eeq
and the identity (\ref{useful}) reads
\beq
g(u,D)h(u,D)={i\over D}
{\partial\over\partial\overline{u}}g(u,D).
\label{useful1}
\eeq
The plan is to insert (\ref{useful1}) into (\ref{int1}),
apply the Stokes theorem in the $u$-integral, and then
evaluate the integral at each boundary component.
It turns out to be useful to deform the domain of $D$ integral
in advance, from the
real line $\R$ to a contour in the complex $D$-plane that goes around
$D=0$. This is valid as long as the deformation does not hit
the poles of the integrand of (\ref{int1}).
Denoting the new contour by $\Gamma$
and applying the Stokes theorem,
we have
\beq
I={N_1\over 2}\lim_{e\to 0\atop\varepsilon\to 0}
\int_{\Gamma}\dd D
\oint_{\partial (\mathfrak{M}\setminus\Delta_{\varepsilon})}
\dd u\,{1\over D}
g(u,D)\exp\left(-{\pi\over e^2}D^2-2\pi i\zeta D\right).
\eeq
There is a simple pole at $D=0$ in the integrand, even though
it must go away after the $u$-integration, as $D=0$ was
perfectly regular in (\ref{int1}).
It goes away if the sum over all components of
$\partial(\mathfrak{M}\setminus \Deps)$ is taken, but survives
in the individual component.
That is why we deformed the $D$ contour in advance.
We denote by $\Gamma_-$ ({\it resp.} $\Gamma_+$)
the contour that goes below ({\it resp}. above) $D=0$.
To be specific, let us take $\Gamma=\Gamma_-$.
We emphasize that the contour deformation should be small enough so
as not to hit the poles of (\ref{int1}), i.e., those of $g(u,D)$.

Note that $\partial(\mathfrak{M}\setminus \Deps)$
has three groups of components,
\beq
\partial(\mathfrak{M}\setminus \Delta_{\varepsilon})
=\partial \mathfrak{M}-\partial\Delta_{\varepsilon}^{(+)}
-\partial\Delta_{\varepsilon}^{(-)},
\eeq
where $\partial \mathfrak{M}$ is the boundary at infinity (two circles
at $u\to \pm i \infty$).
Let us look at a component of $\partial\Delta_{\varepsilon}^{(\pm)}$
that encircles a singular point $u_*\in \mathfrak{M}_{\rm sing}^{(\pm)}$.
In addition to $D=0$,
there is a pole at $D=-iQ_i\varepsilon^2$ for each $Q_i$ satisfying
(\ref{Hidef}) for $u=u_*$. In the limit $\varepsilon\to 0$,
it collides with the pole at $D=0$.
The other poles stay away from the real axis, even after the limit.
Recall that the sign of such $Q_i$ is fixed:
it is positive if $u_*\in \mathfrak{M}_{\rm sing}^{(+)}$ and
negative if $u_*\in \mathfrak{M}_{\rm sing}^{(-)}$.
On the other hand, at $u\in \partial \mathfrak{M}$,
there is one pole at $D=0$
and other poles are on the imaginary axis and are
infinitely far away.

\begin{figure}[htb]
\psfrag{Gminus}{$\Gamma_-$}
\psfrag{minus}{$u\in\partial\Delta^{(-)}_{\varepsilon}$}
\psfrag{plus}{$u\in\partial\Delta^{(+)}_{\varepsilon}$}
\psfrag{infty}{$u\in \partial\mathfrak{M}$}
\centerline{\includegraphics{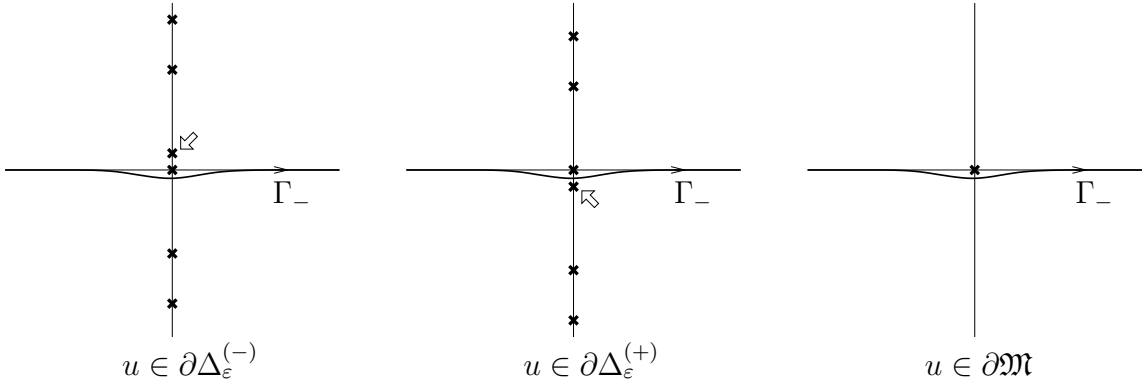}}
\caption{Poles and the contour $\Gamma_-$ in the complex $D$ plane.}
\label{fig:poles}
\end{figure}

Let us first compute the contribution from the components of
$\partial\Delta^{(-)}_{\varepsilon}$. (See Figure~\ref{fig:poles} left.)
No pole approaches $D=0$ from the lower half $D$ plane in the limit.
Thus, the contour $\Gamma_-$ can be deformed further away from $D=0$.
In particular, we can take the $\varepsilon\to 0$ limit
holding the contour $\Gamma_-$ fixed.
As a function of $u$, the integrand is continuous and its absolute value
is bounded from above for any $\varepsilon$.
Hence the integral vanishes in the $\varepsilon\to 0$ limit since the
contour simply shrinks. Thus, we find
\beq
\lim_{e\to 0}\cdot\lim_{\varepsilon\to 0}
\int_{\Gamma_-}\dd D
\oint_{\partial \Delta^{(-)}_{\varepsilon}}\dd u\,{1\over D}
g(u,D)\exp\left(-{\pi\over e^2}D^2-2\pi i\zeta  D\right)=0.
\label{vanish-}
\eeq

Let us next consider the components of $\partial\Delta^{(+)}_{\varepsilon}$.
(See Figure~\ref{fig:poles} middle.)
No pole approaches $D=0$ from the upper half $D$ plane,
but one pole $D=-iQ_i\varepsilon^2$ for each $Q_i$ satisfying
(\ref{Hidef}) approaches $D=0$ in the limit.
And the contour $\Gamma_-$ goes between $D=0$ and these poles.
Therefore, we cannot take that limit holding the contour fixed.
To avoid the complication, we decompose the contour as
\beq
\Gamma_-=\Gamma_++C_0 \ ,
\eeq
where $C_0$ is a circle of radius smaller than $\varepsilon^2$
that goes around $D=0$ counter-clockwise.
The integral along $\Gamma_+$ vanishes in the limit
for the same reason as in (\ref{vanish-}),
\beq
\lim_{e\to 0\atop \varepsilon\to 0}
\int_{\Gamma_+}\dd D
\oint_{\partial \Delta^{(+)}_{\varepsilon}}\dd u\,{1\over D}
g(u,D)\exp\left(-{\pi\over e^2}D^2-2\pi i\zeta  D\right)=0.
\label{vanish+}
\eeq
The integral along $C_0$, on the other hand, remains,
\beqa
\lefteqn{
\lim_{e\to 0}\cdot\lim_{\varepsilon\to 0}
\int_{C_0}\dd D
\oint_{\partial \Delta^{(+)}_{\varepsilon}}\dd u\,{1\over D}
g(u,D)\exp\left(-{\pi\over e^2}D^2-2\pi i\zeta D\right)}\nn\\
&=&\lim_{e\to 0}\cdot\lim_{\varepsilon\to 0}
2\pi i \oint_{\partial \Delta^{(+)}_{\varepsilon}}\dd u\,
g(u,0)\exp\left(-{\pi\over e^2}0^2-2\pi i\zeta  0\right)\nn\\
&=&
2\pi i
\oint_{\partial \Delta^{(+)}_{\varepsilon}}\dd u\, g(u,0).
\eeqa

Finally, let us study the contribution of $\partial\mathfrak{M}$.
This is the main new point compared to the 2d index computation
\cite{BEHT1} where $\MMM$ was closed and had no infinity.
In 1d, the index should undergo a transition as $\zeta$ crosses a wall,
and it is the integral on $\partial \MMM$ that is responsible for that
transition.
As we have discussed at the beginning of this section,
the result should depend on the distance $|\Delta \zeta|$
to the wall in relation to $e$ and $\beta$.
Since we are taking the limit $e\to 0$, we should also scale
$\zeta$ accordingly.
Here we present one particular scaling limit, which we shall
call the {\it Higgs scaling}. (A more detailed discussion will be presented
in the next subsection.)
It is to scale up $|\zeta|$ as the limit $e\to 0$ is taken so that
\beq
\zeta'=e^2\zeta
\eeq
is held fixed.
Then, the $D$ integral can be processed as follows:
\beqa
\lefteqn{\int_{\Gamma_-}{\dd D\over D}g(u,D)\exp\left(
-{\pi\over e^2}D^2-2\pi i\zeta D\right)}\nn\\
&=&
\int_{\Gamma_-}{\dd D'\over D'}g(u,e^2D')\exp\left(
-{\pi e^2}D^{\prime 2}-2\pi i\zeta' D'\right)\nn\\
&\stackrel{e\to 0}{\longrightarrow}&
\int_{\Gamma_-}{\dd D'\over D'}g(u,0)\exp\Bigl(
-2\pi i\zeta' D'\Bigr)
\nn\\
&=&\left\{\begin{array}{ll}
0&\zeta'>0\\
2\pi i g(u,0)&\zeta'<0
\end{array}\right.
=2\pi i \Theta(-\zeta') g(u,0).
\label{inftybdry}
\eeqa

Summing up the contributions, we find
\beqa
I&=&{N_1\over 2}\left[\,\,0
-2\pi i\oint_{\partial\Delta_{\varepsilon}^{(+)}}\dd u\,g(u,0)
+2\pi i\Theta(-\zeta')
\oint_{\partial\mathfrak{M}}\dd u\,g(u,0)
\right]\nn\\
&=&\mp N_1\pi i\oint_{\partial\Delta_{\varepsilon}^{(\pm)}}
\dd u\,g(u,0)\qquad\mbox{for $\pm\zeta' >0$}.
\label{above}
\eeqa

If we choose $\Gamma=\Gamma_+$, then, we have non-zero contribution from
$\partial\Delta_{\varepsilon}^{(-)}$, zero from
$\partial\Delta_{\varepsilon}^{(+)}$, and similar
(but opposite) contribution as (\ref{inftybdry}) from
$\partial\mathfrak{M}$.
The sum is
\beqa
I&=&{N_1\over 2}\left[
2\pi i\oint_{\partial\Delta_{\varepsilon}^{(-)}}\dd u\,g(u,0)
+0
-2\pi i\Theta(\zeta')
\oint_{\partial\mathfrak{M}}\dd u\,g(u,0)
\right]\nn\\
&=&\mp N_1\pi i\oint_{\partial\Delta_{\varepsilon}^{(\pm)}}
\dd u\,g(u,0)\qquad\mbox{for $\pm\zeta' >0$},
\eeqa
which is indeed the same as (\ref{above}).

\subsection{Scaling Limits}\label{subsec:scaling}

In the evaluation of the
integral along $\partial\MMM$, we have taken the scaling limit where
$e\to 0$, $|\zeta|\to \infty$ holding $e^2\zeta$ fixed at a non-zero value.
To see the physical meaning of the limit, let us restore the
circumference $\beta$ of the circle and also use the physical $D_E$
variable.
The important factor of the integrand is
\beq
\exp\left(-{\beta\over 2e^2}D_E^2-i\beta \zeta D_E\right).
\label{Dfactor}
\eeq
Note that the exponent is indeed dimensionless, as
the parameters and variables have the following
dimensions (see the footnote in page~\pageref{foot:dimension}):
\beq
[D_E]={\rm energy}^2,\quad
[e^2]={\rm energy}^3,\quad
[\zeta]={\rm energy}^{-1},\quad
[\beta]={\rm energy}^{-1}.
\eeq
In any computation of path integral, it is natural to neutralize the dimension
of integration variables. Since there are more than one
dimensionful parameters, there are more than one ways to do so.

The scaling of the variable $D_E$ in the above computation is, when $\beta$
is restored,
\beq
D_E=\beta e^2 D_E'.
\eeq
This $D_E'$ is indeed dimensionless and the factor (\ref{Dfactor})
is written as
\beq
\exp\left(-{\beta^3e^2\over 2}(D_E')^2-i\beta^2e^2\zeta D_E'\right).
\label{DfactorH}
\eeq
The limit we have taken is $e^2\beta^3\to 0$ holding
$\zeta'=\beta^2e^2\zeta$ fixed at a non-zero value.
Even though $e$ vanishes,
the D-term potential remains because $|\zeta|$ is sent to infinity
at the same time.
In particular, the Higgs mass $e\sqrt{|\zeta|}$ is
kept finite and non-zero, in comparison to $1/\beta$.
Moreover, the potential barrier on the Coulomb branch is infinitely high,
$e^2\zeta^2/2=[(\zeta')^2/(2e^2\beta^3)](1/\beta)$.
Therefore, this is a regular limit where the theory
at the energies below $e^2|\zeta|=\zeta'/\beta$
is well approximated by the theory on the Higgs branch.
In particular, the obtained answer must agree with the index of
the Higgs branch theory. That is why we called it the Higgs scaling.

Let us consider another scaling limit which attempts to probe the
Coulomb branch dynamics, having characteristic energy scale 
$e^2\zeta^2$.
That is to hold $\zeta''=\sqrt{\beta} e\zeta$ fixed
when taking the limit $e^2\beta^3\to 0$.
The natural rescaling of the variable $D_E$ is
\beq
D_E={e\over\sqrt{\beta}}D_E'',
\eeq
so that (\ref{Dfactor}) is written as
\beq
\exp\left(-{1\over 2}(D_E'')^2-i\sqrt{\beta} e\zeta D_E''\right).
\label{DfactorC}
\eeq
Then, the $D_E$ integration at infinity behaves as
\beqa
\lefteqn{\int_{\Gamma}{\dd D_E\over D_E}
g(u,{\textstyle {\beta^2\over (2\pi)^2}}D_E)\exp\left(
-{\beta\over 2e^2}D_E^2-i\beta\zeta D_E\right)}\nn\\
&=&
\int_{\Gamma}{\dd D''_E\over D''_E}g(u,{\textstyle {\beta^2 e\over
(2\pi)^2\sqrt{\beta}}}D''_E)\exp\left(
-{1\over 2}(D_E'')^2-i\zeta'' D''_E\right)\nn\\
&\stackrel{e\to 0}{\longrightarrow}&
\int_{\Gamma}{\dd D''_E\over D''_E}g(u,0)\exp\Bigl(
-{1\over 2}(D_E'')^2
-i\zeta'' D''_E\Bigr).
\label{CoulombSc}
\eeqa
As long as $\Gamma$ avoids the pole at $D_E=0$ and asymptotes to the real
line, this is absolutely convergent and depends continuously on
$\zeta''$. If we choose $\Gamma=\Gamma_-$, it vanishes in the limit
$\zeta''\to +\infty$ and approaches $2\pi i g(0,u)$
in the opposite limit $\zeta''\to -\infty$.
The difference from the limit value
is exponentially small for large $|\zeta''|$ as
$\pm\sqrt{2\pi}i\e^{-|\zeta''|^2/2}/|\zeta''|$ times $g(u,0)$.
Thus, in this scaling limit,
the index is a continuous function of $\zeta''$ that interpolates
the Higgs scaling result at
$\zeta'>0$ and the one at $\zeta'<0$.
According to a numerical evaluation of the integral,
the transition occurs essentially within the range $|\zeta''|< 5$.

This smooth transition appears to be different from
the expectation (\ref{CCindex}) that we have a sharp transition within
the range $|\zeta''|\ll 1$.
This discrepancy can be understood by noting a subtlety
in the definition of the index.\footnote{We thank Edward Witten for
a guide on this point.}
Recall that, when there is a non-trivial wall crossing,
the effective potential on the Coulomb branch approaches
$E_1=e^2|\zeta|^2/2=|\zeta''|^2/2\beta$
at infinity, which is finite
in the scaling limit under consideration.
In such a situation, the spectrum is continuous above 
$E_1=e^2|\zeta|^2/2$. As long as the gap $E_1$ is positive,
supersymmetric ground states must be normalizable,
and the index can be defined unambiguously as
\beq
\#(\mbox{bosonic zero energy states})
-\#(\mbox{fermionic zero energy states}).
\label{def0Index}
\eeq
(Here we suppress the flavor twist for simplicity.)
In particular, it is an integer and is deformation invariant.
However, there is an ambiguity if we try to define the index
as (\ref{Ih}), that is,
\beq
{\rm Tr}_{\mathcal H}^{}(-1)^F\e^{-\beta H}.
\label{def1Index}
\eeq
The expression (\ref{def1Index})
is not absolutely convergent due to the continuum above $E_1$, 
and hence depends on how it is defined. 
Our localization computation provides one definition to
(\ref{def1Index}), but it may not
agree with other possible definitions. In particular,
it does not have to agree with (\ref{def0Index}). 
One may estimate the ambiguity to be
of the order of $\e^{-\beta E_1}=\e^{-|\zeta''|^2/2}$. That is
in fact the order of the difference of
(\ref{CoulombSc}) from the Higgs scaling result, as
estimated above at large $|\zeta''|$.

In the Higgs scaling, on the other hand,
the asymptotic gap $E_1=|\zeta'|^2/(2e^2\beta^4)$
blows up to infinity as remarked earlier.
Hence, (\ref{def1Index}) is absolutely convergent
and must agree with (\ref{def0Index}).

It is the index as defined by (\ref{def0Index}) that is expected
to undergo a sharp transition by the runaway of Coulomb branch vacua.
Since it is deformation invariant as
long as the gap $E_1$ is bounded from below by a positive energy,
the total jump 
of our localization result must agree with the one from the picture of
runaway vacua.
In fact, we shall explicitly show the agreement in Section~\ref{sec:WC}
whenever the wall crossing on the Coulomb branch can be evaluated.

We now move on to
the computation of the index of the theory with a general gauge
group.
We shall take the ``Higgs scaling'':
$e^2\beta^3\to 0$ holding the distance of
$\zeta'=\beta^2e^2\zeta$ to the walls fixed.

\subsection{Higher Rank Theories}\label{subsec:higherrank}

We set $\beta=2\pi$.
We define $G(u,D):=N_{\ell}(-\half)^{\ell}g(u,D)
\e^{-{\pi\over e^2}D^2-2\pi i\zeta(D)}$
and consider it as a function of $(u,D)\in \cMMM\times \ttt_{\C}$.
Let us introduce a $\ttt^*_{\C}$-valued one form
$\nu$ on $\cMMM\times\ttt_{\C}$ by
\beq
\nu(\xi)\,\,:=\,\, -i\dd\bu_a h^{ab}\xi_b,\qquad\xi\in \ttt_{\C}.
\eeq
Both $G$ and $\nu$ are meromorphic in $D$.
In terms of these, the index formula (\ref{int}) reads
\beq
I={1\over |W|}
\lim\int_{(\cMMM\setminus\Deps)\times \Gamma}
\mu,\qquad
\mu:=G\,\dd^\ell u\wedge(\nu(\dd D))^{\wedge \ell}.
\label{Iexp}
\eeq
$\Gamma\subset \ttt_{\C}$ is a contour of $D$-integration.
The real locus $\Gamma=i\ttt$ was chosen in (\ref{int}), but
we shall deform it away so as not to hit the poles of the integrand,
as in the $U(1)$ case.
The identity (\ref{useful}) reads
\beq
\dbar G=\nu(D)G.
\label{idd1}
\eeq
One can easily show that $\partial_{\bu_c}h^{ab}$ is symmetric in $a,b,c$,
which means
\beq
\dbar\nu=0.
\label{idd2}
\eeq
For a set $\{Q_1,\ldots, Q_s\}\subset \ttt^*$, put
\beq
\mu_{Q_1,\ldots,Q_s}:=c_s\cdot G\,\dd^\ell u\wedge
(\nu(\dd D))^{\wedge (\ell-s)}
\wedge {\dd Q_1(D)\over Q_1(D)}\wedge\cdots
\wedge {\dd Q_s(D)\over Q_s(D)},
\eeq
with $c_s=(-1)^{s(\ell-1)+{s(s+1)\over 2}}$. Here $Q_{\alpha}$'s may or may not be
taken from the charges of the fields.
Note that it vanishes if $Q_1,\ldots, Q_s$ are linearly dependent.
For the empty set, we have $\mu_{\emptyset}=\mu$.
It follows from (\ref{idd1}) and (\ref{idd2})
that
\beq
\dd\mu_{Q_1,\ldots,Q_s}
=\sum_{\alpha=1}^s(-1)^{s-\alpha}
\mu_{Q_1,\ldots\widehat{Q_{\alpha}}\ldots,Q_s}\ ,
\label{Fid}
\eeq
where ``hat'' means omission.

We shall use this identity and Stokes theorem
successively to reduce the dimension of the $u$ integration.
To show the idea, let us pretend that $G$ and $\nu$ had
no-singularity and consider integration on
$\cMMM$ rather than $\cMMM\setminus \Deps$. (We ignore
the $D$ integration for now.)
%I.e. we consider a ``desert''.
We also pretend that $\cMMM$ is compact without boundary for now.
We choose a good cell decomposition of $\cMMM$. By ``good'' we mean
that a codimension $k$ cell is at the intersection of
$(k+1)$ codimension $(k-1)$ cells (e.g. codimension 1 cell
is at the intersection of 2 maximal dimensional cells).
Write $\{C_{\alpha}^{(n)}\}_{\alpha}$ for the set of all
$n$-dimensional cells. We choose, randomly, an element
$Q_{\alpha}\in i\ttt^*$ for each maximal dimensional cell $C_{\alpha}^{(2\ell)}$
and use $\mu=\dd\mu_{Q_{\alpha}}$ there.
Then,
\beqa
\int_{\cMMM}\mu&=&
\sum_{\alpha}\int_{C_{\alpha}^{(2\ell)}}\dd\mu_{Q_{\alpha}}
=\sum_{\alpha}\int_{\partial C^{(2\ell)}_{\alpha}}
\mu_{Q_{\alpha}}\nn\\
&=&\sum_{\beta}\int_{C^{(2\ell-1)}_{\beta}}
(\mu_{Q_{{\beta}^+}}-\mu_{Q_{{\beta}^-}})
=\sum_{\beta}\int_{C^{(2\ell-1)}_{\beta}}\dd\mu_{Q_{{\beta}^+},Q_{{\beta}^-}}
=\sum_{\beta}
\int_{\partial C^{(2\ell-1)}_{\beta}}\mu_{Q_{{\beta}^+},Q_{{\beta}^-}}
\nn\\
&=&\sum_{\gamma}\int_{C^{(2\ell-2)}_{\gamma}}
\left(\mu_{Q_{{\gamma}^1},Q_{{\gamma}^2}}
-\mu_{Q_{{\gamma}^1},Q_{{\gamma}^3}}+\mu_{Q_{{\gamma}^2},Q_{{\gamma}^3}}\right)
=\sum_{\gamma}\int_{\partial C^{(2\ell-2)}_{\gamma}}
\mu_{Q_{{\gamma}^1},Q_{{\gamma}^2},Q_{{\gamma}^3}}
\nn\\
&=&\cdots
\label{desert}
\eeqa
In this way, we can lower the dimension of $u$-integration, and
eventually we hit the middle dimension, i.e., $\ell$.
There we have complete cancellation
because of the identity (\ref{Fid}) and
$\mu_{Q_{{\gamma}_1},\ldots,Q_{{\gamma}_{\ell+1}}}=0$. Thus, %on the desert,
we get zero.

Now let us consider the real situation where $\cMMM$ is non-compact
and we delete $\Deps$ from $\cMMM$.
To avoid possible misunderstanding, we denote by $\Scr{I}_S$
the set labelling the singular hyperplanes.
$\Scr{I}_S$ is different from the set $\rmI$ labelling the
weight decomposition (\ref{wtChiral}) of $V_{\rm chiral}$
since different components may define the same singular hyperplane.
Thus, (\ref{Deps}) could also be written as
\beq
\Deps:=\bigcup_{i\in \Scr{I}_S}\Deps(H_i).
\eeq
We consider the non-degenerate case: all the intersections of hyperplanes
are transversal. In particular, there is no point where
distinct $(\ell+1)$ hyperplanes meet.
The boundary $\partial\Deps$ is separated into tubes with holes,
\beq
S_i:=\partial\Deps\cap\partial\Delta_{\varepsilon}(H_i).
\eeq
We give it the natural orientation.
We also introduce a cut-off at infinity,
$\cMMM_R:=i\ttt/{\rm Q}^{\vee}\times \ttt_R$, where
$\ttt_R$ consists of vectors of lengths $\leq R$.
In (\ref{Iexp}), we replace $\cMMM$ by $\cMMM_R$
and send $R$ to infinity at the end of the computation.
We put
\beq
S_{\infty}:=-\overline{(\partial \cMMM_R)\setminus \Deps}.
\eeq
Note that
\beq
\partial(\cMMM_R\setminus\Deps)=-\sum_{i\in\Scr{I}}S_i.
\eeq
Here and in what follows, unless otherwise stated,
the index $i$ runs over $\Scr{I}=\Scr{I}_S\cup \{\infty\}$. I.e.,
$i$ may be the label of a singular hyperplane or $i=\infty$.
We write $S_{i_1\ldots i_s}=S_{i_1}\cap\cdots\cap S_{i_s}$
and we also give it the natural orientation.
It is totally antisymmetric in $i_1,\ldots,i_s$ and
\beq
\partial S_{i_1\ldots i_s}=-\sum_{j\in\Scr{I}} S_{i_1\ldots i_s j}.
\label{bdryS}
\eeq

We would like to choose a cell decomposition of $\cMMM_R\setminus \Deps$.
Because it is a manifold with boundary and corners,
we cannot take a good one, but we try to take
it to be as good as possible. We require the following conditions (i)-(iii).
(i) Each open cell is either in the
interior or contained in the interior of exactly one $S_{i_1\ldots i_s}$.
(ii) It is good in the interior of $\cMMM_R\setminus \Deps$.
That is, the valence condition is satisfied
at the (closed) cell in the interior.
To describe the final condition, we note that a neighborhood of
$\cMMM_R\setminus\Deps$ of an interior point of a corner $S_{i_1\ldots i_s}$
is of the form $(\R_+)^s\times \R^{2\ell-s}$, which is the domain
in $\R^{2\ell}=\{(x_{i_1},\ldots,x_{i_s},y_{s+1},\ldots,y_{2\ell})\}$
defined by $x_{i_1}\geq 0,\ldots, x_{i_s}\geq 0$.
For $\{j_1,\ldots,j_p\}\subset\{i_1,\ldots, i_s\}$,
the corner $S_{j_1\ldots j_p}$ includes $S_{i_1\ldots i_s}$
and is identified as the region $x_{j_1}=\cdots=x_{j_p}=0$.
We introduce a cell decomposition of $(\R_+)^s\times \R^{2\ell-s}$ as follows.
For $\{j_1,\ldots,j_p\}\subset\{1,\ldots,s\}$,
we define the cell $C_{j_1\ldots j_p}$ by the condition.
\beq
0\leq x_{j_1}=\cdots=x_{j_p}\leq \mbox{\,\,all other $x_i$'s.}
\eeq
Note that the boundary of $C_{j_1\ldots j_p}$ consists of
$S_{j_1\ldots j_p}$ and $C_{j_1\ldots j_p i}$.
The orientation of $C_j$ is the one induced from $\cMMM$
and we can assign an orientation of $C_{j_1\ldots j_p}$ so that
\beq
\partial C_{j_1\ldots j_p}=- S_{j_1\ldots j_p}+
\sum_iC_{j_1\ldots j_p i}.
\label{bdryrel}
\eeq
We can now describe the condition (iii): a cell touching the interior of
$S_{i_1\ldots i_s}$ coincides with one of $C_{j_1\ldots j_p}$,
or its subdivision in the $\R^{2\ell-s}$ direction,
in a neighborhood of $S_{i_1\ldots i_s}$.

Under such a cell decomposition, we compute the integral
$\int_{\cMMM_R\setminus \Deps}\mu$ as follows.
To each maximal dimensional cell, we assign an element of $i\ttt^*$.
For a cell away from the boundary, the assignment is random.
If it touches the interior of $S_i$, with $i\in \Scr{I}_S$,
we assign the charge $Q_i$ of a field defining $H_i$.
If it touches the interior of $S_{\infty}$,
we assign an element $Q_{\infty}\in i\ttt^*$ which
may or may not be one of the charges.
More generally,
if it touches the interior of $S_{i_1\ldots i_s}$, it must be contained
in one of $C_j$ for $j\in\{i_1,\ldots, i_s\}$, then we assign $Q_j$.
The integral in the interior region can be processed as in (\ref{desert}), 
and we obtain zero.
From the cells touching the boundary, we have other contributions.
From the cells touching the interior $S^o_i$ of $S_i$, we have
\beq
\sum_{C^{(2\ell-1)}_{\beta}\subset S_i^o}\int_{C^{(2\ell-1)}_{\beta}}\mu_{Q_i}.
\eeq
To see what you obtain from the cells touching
the interior of higher codimension corners, let us look at
the region $(\R_+)^s\times \R^{2\ell-s}$ around $S_{i_1\ldots i_s}$.
\beqa
\int_{(\R_+)^s\times \R^{2\ell-s}}\mu&=&
\sum_i\int_{C_i}\dd\mu_{Q_i}=\sum_i\int_{\partial C_i}\mu_{Q_i}\nn\\
&=&\sum_i\int_{- S_i}\mu_{Q_i}
+\sum_i\sum_{j\ne i}\int_{C_{ij}}\mu_{Q_i}+[\cdots]\nn\\
&&~~~~~~~~~~~~~~~~~~~~~
\stackrel{\downarrow}{=}\sum_i\int_{- S_i}\mu_{Q_i}
+\sum_{i<j}\int_{\partial C_{ij}}\mu_{Q_i,Q_j}+[\cdots]\nn\\
&=&\sum_i\int_{- S_i}\mu_{Q_i}
+\sum_{i<j}\int_{-S_{ij}}\mu_{Q_i,Q_j}
+\sum_{i<j}\sum_{k\ne i,j}\int_{C_{ijk}}\mu_{Q_i,Q_j}+[\cdots]\nn\\
&=&\cdots\nn\\
&=&-\sum_i\int_{ S_i}\mu_{Q_i}
-\sum_{i<j}\int_{ S_{ij}}\mu_{Q_i,Q_j}
-\cdots
-\int_{ S_{i_1\ldots i_s}}\mu_{Q_{i_1},\ldots,Q_{i_s}}
+[\cdots] \ ,\nn\\
\eeqa
where we used (\ref{Fid}) as well as (\ref{bdryrel})
in the intermediate steps. For example, we used
the following in $\stackrel{\downarrow}{=}$,
\beq
\sum_i\sum_{j\ne i}\int_{C_{ij}}\mu_{Q_i}
=\sum_{i<j}\int_{C_{ij}}(\mu_{Q_i}-\mu_{Q_j})
=\sum_{i<j}\int_{C_{ij}}\dd\mu_{Q_i,Q_j}
=\sum_{i<j}\int_{\partial C_{ij}}\mu_{Q_i,Q_j}.
\eeq
The terms $+[\cdots]$ that appear after partial integration
consists of the boundary terms in the interior, which
will contribute to the complete cancellation as in (\ref{desert}).
Collecting all these, we have
\beq
\int_{(\tilde{\MMM}\setminus \Deps)\times\Gamma}\mu
\,\,=\,\,-\sum_{i}I_{i}
-\sum_{i<j}I_{ij}
-\cdots
-\sum_{i_1<\cdots <i_\ell}
I_{i_1\ldots i_\ell}
\,\,=\,\,-\sum_{p=1}^\ell\sum_{i_1<\cdots <i_p}I_{i_1\ldots i_p}\ ,
\label{eqeq}
\eeq
where
\beq
I_{i_1\ldots i_p}=\int_{S_{i_1\ldots i_p}\times\Gamma}
\mu_{Q_{i_1}\ldots Q_{i_p}}.
\eeq
We emphasize again that the index
$i_a$ runs over $\Scr{I}=\Scr{I}_S\cup\{\infty\}$.

\subsection{The $D$-contour}

So far, we have focused our attention to the integration over
$\cMMM_R\setminus \Deps$.
Let us now bring the $D$-integration back into our consideration.
An appropriate choice of the contour $\Gamma$ will allow us
 to process the integral further.
Our original choice is $\Gamma=i\ttt^*$ with a fixed orientation.
We may shift it to
\beq
\Gamma=i\delta+i\ttt,
\label{deltashift}
\eeq
for some $\delta\in i\ttt$ which is small enough so that the
integrand $\mu$ remains non-singular over
$\cMMM\setminus \Deps$ as the contour is shifted from
$i\ttt$ to $i\delta+i\ttt$.
Let us fix such a $\delta$ which is generic enough so that $Q_i(\delta)$
is non-zero for any $i\in\Scr{I}$.
Of course, the final result does not depend on the choice of $\delta$, but
each term $I_{i_1\ldots i_p}$ in the expansion (\ref{eqeq}) may depend on it.
So let us denote it by $I_{i_1\ldots i_p}(\delta)$.

Let us look at the term
$I_{i_1\ldots i_p}(\delta)$ where all the indices are from $\Scr{I}_S$.
If $Q_{i_a}(\delta)>0$ for some $i_a$, the integrand is regular
even if the neighborhood $\Deps(H_{i_a})$ is shrunk to the zero size.
Therefore, $I_{i_1\ldots i_p}(\delta)$ vanishes in the
$\varepsilon\to 0$ limit.
If $Q_{i_a}(\delta)<0$ for all of $i_1,\ldots, i_p$,
we deform $\Gamma$ upward in each $Q_{i_a}(D)$-plane so that
it decomposes into a union of
the infinite line in the upper-half plane
and the component that encircles the origin.
The former vanishes in the $\varepsilon\to 0$ limit, while
the latter picks up the simple pole at $Q_{i_a}(D)=0$.
We are therefore left with the $D$-integration along
the common kernel ${\rm Ker}\, Q_{i_1\ldots i_p}$
 of all $Q_{i_a}$'s.
Here, just as in (\ref{deltashift}),
there is a freedom to set the new contour at
\beq
i\delta_{i_1\ldots i_p}+{\rm Ker}\, Q_{i_1\ldots i_p},
\eeq
where $\delta_{i_1\ldots i_p}\in {\rm Ker}\, Q_{i_1\ldots i_p}$
is small enough. The result does not depend on
the choice of $\delta_{i_1\ldots i_p}$.
Taking one step backward, we may write the result as
\beq
I_{i_1\ldots i_p}(\delta)
=\prod_{a=1}^p\theta(-Q_{i_a}(\delta))
\int_{S_{i_1\ldots i_p}\times \Gamma_{i_1\ldots i_p}}
\mu_{Q_{i_1}\ldots Q_{i_p}},
\label{I...}
\eeq
where $\theta(x)$ is the step function ($1$ on $x> 0$ and $0$ on $x<0$)
and
\beq
\Gamma_{i_1\ldots i_p}:=C_{i_1\ldots i_p}\times
\left(i\delta_{i_1\ldots i_p}+{\rm Ker}\, Q_{i_1\ldots i_p}\right).
\label{defGamma...}
\eeq
Here $C_{i_1\ldots i_p}\subset \ttt_{\C}$ is any $p$-dimensional contour whose
$Q_{i_a}$-image is a circle around the origin.
We provide $\Gamma_{i_1\ldots i_p}$ the orientation that is induced from
$\Gamma$ in (\ref{deltashift}) by the above deformation.

The integrals in (\ref{I...}) can be processed further.
Let us take $\{Q_{j_1},\ldots, Q_{j_s}\}\subset i\ttt^*$
so that the $p+s$ elements $Q_{i_1},\ldots, Q_{i_p},Q_{j_1},\ldots, Q_{j_s}$
are linearly independent. We have from (\ref{Fid}),
\beq
\dd \mu_{Q_{i_1}\ldots Q_{i_p}Q_{j_1}\ldots Q_{j_s}}
=\sum_{a=1}^p(-1)^{p+s-a}\mu_{Q_{i_1}\ldots\widehat{Q_{i_a}}\ldots
 Q_{j_s}}
+\sum_{b=1}^s(-1)^{s-b}\mu_{Q_{i_1}\ldots
\widehat{Q_{j_b}}\ldots Q_{j_s}}.
\label{ddmumu}
\eeq
When integrated over $\Gamma_{i_1\ldots i_p}$,
the first $p$ terms of (\ref{ddmumu}) vanish
since the $a$-th term does not have a pole at $Q_{i_a}(D)=0$:
for any chain $C$ of $\cMMM_R\setminus \Deps$,
\beq
\int_{C\times\Gamma_{i_1\ldots i_p}}
\dd \mu_{Q_{i_1}\ldots Q_{i_p}Q_{j_1}\ldots Q_{j_s}}
=\sum_{b=1}^s(-1)^{s-b}\int_{C\times\Gamma_{i_1\ldots i_p}}
\mu_{Q_{i_1}\ldots Q_{i_p} Q_{j_1}\ldots
\widehat{Q_{j_b}}\ldots Q_{j_s}}.
\label{Fid2}
\eeq
In the same way as we obtained (\ref{eqeq}) using a cell decomposition
of $\cMMM_R\setminus \Deps$ and
the identity (\ref{Fid}), we find, via a cell decomposition of
$S_{i_1\ldots i_p}$ and the identity (\ref{Fid2}),
\beq
\int_{S_{i_1\ldots i_p}\times \Gamma_{i_1\ldots i_p}}
\mu_{Q_{i_1}\ldots Q_{i_p}}
=-\sum_{q=p+1}^\ell\sum_{i_{p+1}<\cdots <i_q}
I_{i_1\ldots i_p; i_{p+1}\ldots i_q}(\delta_{i_1\ldots i_p}),
\label{sum1}
\eeq
where
\beq
I_{i_1\ldots i_p; i_{p+1}\ldots i_q}(\delta_{i_1\ldots i_p})
=\int_{S_{i_1\ldots i_q}\times\Gamma_{i_1\ldots i_p}}
\mu_{Q_{i_1}\ldots Q_{i_q}}.
\label{indivi1}
\eeq
The total sum (\ref{sum1}) is independent of the choice of
$\delta_{i_1\ldots i_p}$ but the individual term
(\ref{indivi1}) may depend on the choice.
Let us assume that $Q_j(\delta_{i_1\ldots i_p})\ne 0$ for all charges
$Q_j$ which are linearly independent of
$Q_{i_1},\ldots, Q_{i_p}$.
Then, the term where all of $i_{p+1},\ldots i_q$ are from $\Scr{I}_S$
can be processed in the same way.

Repeating this procedure enough number of times,
the index can be written as the sum of
integrations of the form
\beq
\int_{S_{i_1\ldots i_\ell}\times\Gamma_{i_1\ldots i_\ell}}
\mu_{Q_{i_1}\ldots Q_{i_\ell}},\qquad
\mbox{for $\{i_1,\ldots, i_\ell\}\subset\Scr{I}_S$},
\label{fexpte}
\eeq
plus the sum of integrations at infinity of the form
\beq
\int_{S_{i_1\ldots i_q\infty}\times\Gamma_{i_1\ldots i_p}}
\mu_{Q_{i_1}\ldots Q_{i_q}Q_{\infty}},\qquad 0\leq p\leq q< \ell.
\label{inftu}
\eeq

The integration at infinity (\ref{inftu})
 depends on the value of $\zeta$ as well as
the $D$-contour.
It is not always possible to simplify the expression.
In each example which we studied, however, as long as we take the scaling limit
$e\to 0$, $\zeta\to \infty$, keeping $\zeta'=e^2\zeta$ fixed
at a generic value,
those unprocessible expressions either vanish or
cancel among each other,
and we find that the final result for the index
is a sum of the terms of the form (\ref{fexpte}).

\subsection{A systematic procedure}

In fact,
there is a systematic procedure to find a concise answer. It is to apply the
procedure employed in \cite{BEHT2} involving a choice of
$\eta\in i\ttt^*$ which is \underline{\it $\Scr{I}_S$-generic},
that is,
$\eta$ cannot be written as a positive span of $(\ell-1)$ or less
elements of $\{Q_i\}_{i\in \Scr{I}_S}$.
In the above procedure, we have successively chosen the shift parameter
$\delta_{\ldots}$:
\beq
\delta\to\delta_{i_1\ldots i_p}\to
\delta_{i_1\ldots i_p;i_{p+1}\ldots i_q}\to
\delta_{i_1\ldots i_p;i_{p+1}\ldots i_q;i_{q+1}\ldots i_r}\to
\cdots
\eeq
The choice was random, and in general route dependent:
$\delta_{i_1\ldots i_p;i_{p+1}\ldots i_q}$ does not have to be the same as
$\delta_{j_1\ldots j_s;j_{s+1}\ldots j_q}$ even if
$\{i_1,\ldots, i_q\}=\{j_1,\ldots, j_q\}$.
One feature of the systematic procedure is to remove the route dependence:
\beq
\delta_{i_1\ldots i_{p_1};i_{p_1+1}\ldots i_{p_2};\cdots;i_{p_l+1}\ldots i_q}
=\delta_{i_1\ldots i_q}.
\eeq
We recall that $\delta_{i_1\ldots i_p}$ is chosen
for each subset $\{i_1,\ldots,i_p\}$ of $\Scr{I}_S$ such that
$Q_{i_1},\ldots, Q_{i_p}$ are linearly independent: it vanishes
on $Q_{i_1},\ldots, Q_{i_p}$,
and takes a non-zero value on each charge
$Q_j$ which is linearly independent of
$Q_{i_1},\ldots , Q_{i_p}$. (The subset may be empty, for which
$\delta_{\emptyset}=\delta$.)
We also introduce a new notation:
\beq
I_{i_1\ldots i_p}[\delta_{i_1\ldots i_p}]
:=\int_{S_{i_1\ldots i_p}\times \Gamma_{i_1\ldots i_p}}
\mu_{Q_{i_1}\ldots Q_{i_p}} \ .
\label{defIbk}
\eeq
Here $\Gamma_{i_1\ldots i_p}$ is defined as a set by
 (\ref{defGamma...}), and is provided the orientation
which is induced from the orientation of
$\Gamma=i\delta'+i\ttt^*$ by deformation,
where $\delta'\in i\ttt^*$ has negative values on $Q_{i_1},\ldots, Q_{i_p}$.
Note that the orientation does not depend on the order of $i_1,\ldots, i_p$.
The notation is redundant --- the integral does not
really depend on the choice of $\delta_{i_1\ldots i_p}$. For example,
in the new notation, the index $I$ itself should be written as $I[\delta]$
---
$\delta$ enters into the integration contour $\Gamma$ via
(\ref{deltashift}) but the integral does not depend on it.
However, this notation is useful to keep track of what happens next.
As in (\ref{eqeq}) and (\ref{sum1}), we have
\beqa
I_{i_1\ldots i_p}[\delta_{i_1\ldots i_p}]
&=&-\sum_{q=p+1}^\ell\left(
\sum_{i_{p+1}<\cdots <i_q\atop {\rm in}\,\, \Scr{I}_S}
\prod_{a=p+1}^q\theta(-Q_{i_a}(\delta_{i_1\ldots i_p}))
\cdot I_{i_1\ldots i_q}[\delta_{i_1\ldots i_q}]\right.\nn\\
&&~~~~~~~~\left.+\sum_{i_{p+1}<\cdots <i_{q-1}\atop {\rm in}\,\, \Scr{I}_S}
\prod_{a=p+1}^{q-1}\theta(-Q_{i_a}(\delta_{i_1\ldots i_p}))
\cdot I_{i_1\ldots i_{q-1}\infty}[\delta_{i_1\ldots i_{q-1}}]\right),~~~
~~~
\label{recursion}
\eeqa
where
\beq
I_{i_1\ldots i_{q-1}\infty}[\delta_{i_1\ldots i_{q-1}}]
:=\int_{S_{i_1\ldots i_{q-1}\infty}\times\Gamma_{i_1\ldots i_{q-1}}}
\mu_{Q_{i_1}\ldots Q_{i_{q-1}}Q_\infty}.
\label{intinfty}
\eeq
The main ingredient of the systematic procedure of
\cite{BEHT2} is the choice of $\eta\in i\ttt^*$.
We require all $\delta_{i_1\ldots i_p}$ to have negative values on $\eta$:
\beq
\eta(\delta_{i_1\ldots i_p})<0,\qquad \forall \{i_1,\ldots, i_p\}.
\label{etacond}
\eeq

In the present case of 1d index, it is particularly useful to set
\beq
\eta=\zeta\quad \mbox{and}\quad Q_{\infty}=\zeta.
\label{thechoice}
\eeq
This is a valid choice if and only if $\zeta$ is $\Scr{I}_S$-generic.
For this choice, the integration at infinity (\ref{intinfty}) involves the
integration over $Q_{\infty}(D)=\zeta(D)$ of the following type:
\beq
\int_{\zeta(\Gamma_{i_1\ldots i_{q-1}})}
{\dd\zeta(D)\over \zeta(D)}f(\ldots,\zeta(D))\exp\left(
-{\pi\over e^2}D^2-2\pi i\zeta(D)\right).
\label{zDint}
\eeq
Note that (\ref{etacond}) and $\eta=\zeta$ means
$\zeta(\delta_{i_1\ldots i_{q-1}})<0$, which means that
the contour of integration (\ref{zDint}) goes below
the pole $\zeta(D)=0$. Hence, just as in the $U(1)$ case
(\ref{inftybdry}), the integral vanishes in the Higgs scaling:
$\,e\to 0$, $\zeta\to \infty$, keeping $\zeta'=e^2\zeta$ finite.
Therefore,
the integration at infinity (\ref{intinfty}) all vanish!

Thus, in the recursion
relation (\ref{recursion}) we are left with the bulk terms only.
As in \cite{BEHT2}, this recursion relation
leads to the following result:
\beq
I\,\,=\,\,{1\over |W|}\sum_{i_1<\cdots <i_\ell\atop {\rm in}\,\,\Scr{I}_S}
(-1)^\ell\prod_{j\in\{i_1,\ldots, i_\ell\}}
\theta(-Q_j(\delta_{i_1\ldots\widehat{j}
\ldots i_\ell}))
\int_{S_{i_1\ldots i_\ell}\times\Gamma_{i_1\ldots i_\ell}}
\mu_{Q_{i_1}\ldots Q_{i_\ell}}.
\eeq
The step function factor is non-zero if and only if
$\zeta=\eta$ belongs to the cone
${\mathcal C}_{Q_1\ldots Q_{i_{\ell}}}$ spanned by
$Q_{i_1},\ldots, Q_{i_{\ell}}$.
Therefore, the index can also be written as
\beq
I={1\over |W|}\sum_{\zeta\in {\mathcal C}_{Q_1\ldots Q_{i_{\ell}}}}
I_{\,i_1\ldots i_{\ell}},
\label{result1}
\eeq
\beq
I_{\,i_1\ldots i_{\ell}}:=
(-1)^{\ell}\int_{S_{i_1\ldots i_\ell}\times\Gamma_{i_1\ldots i_\ell}}
\mu_{Q_{i_1}\ldots Q_{i_\ell}}
=\epsilon_{i_1\ldots i_{\ell}}
\int_{S_{i_1\ldots i_\ell}}C_{\ell}\,g(u,0)\dd^{\ell}u,
\eeq
where $C_{\ell}:=N_{\ell}(-1)^{\ell(\ell+1)\over 2}(\pi i)^{\ell}$ and
\beq
\int_{\Gamma_{i_1\ldots i_{\ell}}}{\dd Q_{i_1}(D)\over Q_{i_1}(D)}\wedge
\cdots\wedge {\dd Q_{i_{\ell}}(D)\over Q_{i_{\ell}}(D)}
=(2\pi i)^{\ell}\epsilon_{i_1\ldots i_{\ell}} \ .
\label{defepsi}
\eeq
In view of the orientation of $\Gamma_{i_1\ldots i_{\ell}}$
as described below Eqn (\ref{defIbk}), $\epsilon_{i_1\ldots i_{\ell}}$ is
a sign that is antisymmetric in $i_1,\ldots, i_{\ell}$ and
odd under $Q_{i_1}\to -Q_{i_1}$.
Recall also that
$S_{i_1\ldots i_{\ell}}$ is antisymmetric in
$i_1,\ldots, i_{\ell}$ and even under $Q_{i_1}\to -Q_{i_1}$.
The product $\epsilon_{i_1\ldots i_{\ell}}S_{i_1\ldots i_{\ell}}$
is hence symmetric in $i_1,\ldots, i_{\ell}$  and
odd under $Q_{i_1}\to -Q_{i_1}$. This together with the
constraint that $\zeta$ must belong to the cone
${\mathcal C}_{Q_{i_1}\ldots Q_{i_{\ell}}}$
is the defining property of the JK residue \cite{SV}
(See \cite{BEHT2} for a review).
Therefore, the index is the sum of JK residue
of $C_{\ell} g(u,0)\dd^{\ell}u$
 with respect to $\zeta$ at the isolated intersections
of the singular hyperplanes. In the notation of
\cite{BEHT2}, it reads
\beq
I\,\,=\,\, {1\over |W|}\sum_{i_1<\cdots <i_\ell\atop {\rm in}\,\,\Scr{I}_S}
\sum_{p\in H_{i_1}\cap \cdots\cap H_{i_{\ell}}}
\mathop{\mbox{JK-Res}}_{p}
(\{Q_{i_1},\ldots,Q_{i_{\ell}}\},\zeta)\Bigl[
C_{\ell}
g(u,0)\dd^{\ell} u\Bigr].
\label{resultJK1}
\eeq

\subsection{The Result}

We now fix the normalization constant $N_{\ell}$.
The right choice turns out to be such that $C_{\ell}=\pm 1$,
where the choice of sign depends on the definition of $(-1)^F$.
As we will see in the next section,  this choice
yields the wall crossing formula that matches
with the Coulomb branch result.
To simplify the notation, we shall write $C_{\ell}g(u,0)$ as $g$,
or $g(u,z)$ if we want to make the parameter dependence explicit.
Also, we may write the result (\ref{resultJK1}) simply as
\beq
I\,\,=\,\,{1\over |W|}\,\mbox{JK-Res}_{\,\zeta}\, g\,\dd^{\ell} u.
\label{theresult}
\eeq
For $C_{\ell}=1$, we have
\beqa
g&=&\prod_{\alpha}2i\sin(-\pi\alpha(u))
{\displaystyle
\prod_{j\in\rmJ}2i\sin(-\pi(q_i(u)+q_i^F(z)))
\over  \displaystyle
\prod_{i\in\rmI}
2i\sin(\pi(Q_i(u)+Q_i^F(z)))}
\sum_{k\in\rmK}(-1)^{r_k}\e^{2\pi i(\rmq_k(u)+\rmq_k^F(z))}
\nn\\
&=&\prod_{\alpha}\left(x^{-{\alpha\over 2}}-x^{{\alpha\over 2}}\right)
{\displaystyle \prod_{j\in \rmJ}
\left(x^{-{q_j\over 2}}y^{-{q^F_j\over 2}}-x^{q_j\over 2}y^{q^F_j\over 2}
\right)
\over\displaystyle \prod_{i\in\rmI}
\left(x^{Q_i\over 2}y^{Q_i^F\over 2}-x^{-{Q_i\over 2}}y^{-{Q_i^F\over 2}}
\right)}
\sum_{k\in\rmK}(-1)^{r_k}x^{\rmq_k}y^{\rmq^F_k}
\label{2g}
\eeqa
with $x=\e^{2\pi i u}$ and $y=\e^{2\pi iz}$. The second line is easy to
remember, as it carries a clear meaning in the operator formalism,
and also will be useful for the actual computation.

\newcommand{\bfz}{{\bf z}}

An ${\mathcal N}=4$ theory is a special case, with a chiral multiplet
in the representation $\mathfrak{g}_{\C}\oplus V$,
a fermi multiplet in the representation $V$ and no Wilson line.
For an effectively compact ${\mathcal N}=4$ theory, we only have to consider
the canonical $\by^{2\bfsmR_-}$ twist,
with the charges as in (\ref{N4Rcharges}).
We shall often write $\by=\e^{\pi i\bfz}$ so that the twist is
by $\e^{2\pi i\bfz \bfsmR_-}$ (note the difference from
$y=\e^{2\pi i z})$.
If we choose $(-1)^F=(-1)^{2J_3}$ where $J_3$ is an $SU(2)$
R-symmetry generator, then, we need to set $C_{\ell}=(-1)^{\ell}$ which yields
\beqa
g&=&
\left({1\over 2i\sin(\pi\bfz)}\right)^{\ell}
\prod_{\alpha}{\sin(-\pi\alpha(u))\over
\sin(\pi(\alpha(u)-\bfz))}
\prod_{i\in\rmI}{\sin(-\pi(Q_i(u)+({R_i\over 2}-1)\bfz))\over
\sin(\pi(Q_i(u)+{R_i\over 2}\bfz))}
\nn\\[0.2cm]
&=&\left({1\over \by-\by^{-1}}\right)^{\ell}
\prod_{\alpha}{x^{-{\alpha\over 2}}-x^{\alpha\over 2}\over
x^{\alpha\over 2}\by^{-1}-x^{-{\alpha\over 2}}\by}
\prod_{i\in\rmI}{
x^{-{Q_i\over 2}}\by^{-\left({R_i\over 2}-1\right)}
-x^{{Q_i\over 2}}\by^{{R_i\over 2}-1}\over
x^{{Q_i\over 2}}\by^{R_i\over 2}
-x^{-{Q_i\over 2}}\by^{-{R_i\over 2}}} \ ,
\label{4g}
\eeqa
where $(Q_i,R_i)$ are the $T$-weights and $U(1)$ R-charges
of $V$. If the theory is not effectively compact, we need to include
an ${\mathcal N}=4$ flavor twist. The effect is to make the
replacement
\beq
x^{\pm{Q_i\over 2}}~\longrightarrow~
 x^{\pm{Q_i\over 2}}y^{\pm{Q_i^F\over 2}}
\label{4flavortwist}
\eeq
in the last factors of (\ref{4g}), where $y$ is the parameter of
flavor twist and $Q_i^F$ are the flavor weights of $V$.

\subsubsection*{Remarks}

{\bf 1}. The above result is valid under the condition that
$\zeta$ is $\Scr{I}_S$-generic. When the gauge group is Abelian,
this is always the case as long as $\zeta$ is in the interior of a phase.
However, this is not always the case in non-Abelian gauge theory.
A counter example is found in the triangle quiver with rank vector
$(k,1,1)$: The III-IV wall in Fig.~\ref{fig:qk11}
is in the interior of a phase when the condition (\ref{condrank}) is
satisfied, but the FI parameter there is not $\Scr{I}_S$-generic for any
$(a,b,c)$. If $\zeta$ is at such a point, we do not have a formula,
while the index is well-defined and is expected to be the same as the one
in the neighboring point.  In the next section, we show that
the result does not change if $\zeta$ goes across such
a point, when the non-genericity is of a simple type.

\noindent
{\bf 2}. The result (\ref{resultJK1}) can be written as
\beq
I\,\,=\,\,{1\over |W|}\sum_{p}\mathop{\mbox{JK-Res}}_{p}
(Q(p),\zeta)\Bigl[
g\,\dd^{\ell} u\Bigr],
\label{resultJK2}
\eeq
where $p$ runs over all isolated intersections of singular
hyperplanes and $Q(p)$ is the set of charges defining
the singular hyperplanes that meet at $p$.
This way of writing allows us to generalize the statement to the case
with degenerate intersections of hyperplanes, i.e., points
$p$ with $|Q(p)|>\ell$. 
This generalization is a proposal that remains to be justified.
Recall that the JK residue is defined \cite{BV,SV} by the property
\beq
\mathop{\mbox{JK-Res}}_{p}
(Q(p),\zeta)\left[
{\dd^{\ell} u \over Q_1(u)\cdots Q_{\ell}(u)}\right]
=\left\{\begin{array}{ll}
{(2\pi i)^{\ell}\over |\det (Q_1\cdots Q_{\ell})|}&\mbox{if $\zeta\in
{\mathcal C}_{Q_1\cdots Q_{\ell}}$}\\
0&\mbox{otherwise}
\end{array}\right.
\label{defJKres}
\eeq
for linearly independent elements $Q_1,\ldots, Q_{\ell}$ of $Q(p)$.
Note that dependence on $\zeta$ is only via the chamber (of $i\ttt^*$)
with respect to $Q(p)$ to which $\zeta$ belongs.
In fact, it is a function of a chamber, not of a particular element
of $i\ttt^*$ like $\zeta$,
in the original definition \cite{BV,SV}: the condition
``$\zeta\in{\mathcal C}_{Q_1\cdots Q_{\ell}}$'' in (\ref{defJKres})
is replaced by ``the chamber $\subset{\mathcal C}_{Q_1\cdots Q_{\ell}}$''.

%\newpage

\section{A Wall Crossing Formula}\label{sec:WC}

When the FI parameter $\zeta$ goes from one phase to another through a phase
boundary, the index may jump as some of the ground states may run away to
infinity of the Coulomb branch or of the Coulomb directions of
mixed branches. Our index computation
is consistent with this picture --- the change of the index
can be expressed as the integral along a cycle at infinity of the moduli
space $\MMM$. We have seen this explicitly in the analysis of $U(1)$
theories, see (\ref{inftybdry}) in Section~\ref{subsec:U1}.
For a general theory, this can be seen by keeping $\eta$ to be the
FI parameter before the wall crossing.
Then, the integrals at infinity (\ref{intinfty}), which are all zero before the
wall crossing, may become non-zero after the wall crossing, while
the residues in the bulk of $\MMM$ do not change as they
depend only on $\eta$, not on $\zeta$.

In this section, we evaluate this change of the index and compare the result
with the Coulomb branch analysis in Section~\ref{sec:Coulomb}, in the
simple wall crossing case: i.e. wall crossing across the phase boundary
which supports a mixed branch of rank 1.

\subsection{$U(1)$ Theories}

We consider a general $U(1)$ gauge theory
and look at the change of the index as the FI parameter $\zeta$
moves from positive to negative.
As we have seen in Section~\ref{subsec:U1},
the change of the index is given by the integration along the circle at
infinity of $\MMM=\C/\Z$.
In the single valued coordinate $x=\e^{2\pi i u}$,
the boundaries are the big circle at $x=\infty$ and the small circle at
$x=0$ and the wall crossing formula is given by
\beq
\Delta I={1\over 2\pi i}\left[\oint_{0}-\oint_{\infty}\right]{\dd x\over x}
g(x,y) \ ,
\eeq
where $g(x,y)$ is given in (\ref{2g}) (or (\ref{4g}) for ${\mathcal N}=4$
theories).

Let us first consider a general ${\mathcal N}=4$ theory.
The function $g$ given in (\ref{4g})
has limits at both ends,
\beq
g(x,\by)\longrightarrow {(-1)^{\neff^{(4)}}\over \by-\by^{-1}}\left\{
\begin{array}{ll}
\by^{-\neff^{(4)}}&x\to \infty\\
\by^{\neff^{(4)}}&x\to 0.
\end{array}\right.
\eeq
We recall that $\neff^{(4)}$ is the number of positively charged fields
minus the number of negatively charged fields.
Therefore, we find
\beqa
\Delta I&=&(-1)^{\neff^{(4)}}
{\,\by^{\neff^{(4)}}-\by^{-\neff^{(4)}}
\over \by-\by^{-1}}\nn\\
&=&(-1)^{\neff^{(4)}}{\rm sgn}(\neff^{(4)})\left(\,
\by^{|\neff^{(4)}|-1}+\cdots
+\by^{-(|\neff^{(4)}|-1)}\right).
\label{4wcU1gen}
\eeqa
The answer is the same even in the presence of an ${\mathcal N}=4$
flavor twist  (\ref{4flavortwist}).
This result agrees with the wall crossing formula based on
the Coulomb branch analysis (\ref{DeltaIC4}).

Let us next consider ${\mathcal N}=2$ theories. As a warm up, we
first look at the untwisted index of the $\CP^{N-1}$ model with Wilson line
$\rmq$, for which
\beq
g(x)={x^{\rmq}\over (x^{\half}-x^{-\half})^N}\ .
\eeq
Note that
\beqa
\rmq\geq {N\over 2}:&&g(x)\longrightarrow\left\{
\begin{array}{ll}
x^{\rmq-{N\over 2}}(1-x^{-1})^{-N}&x\to\infty,\\
0&x\to 0,
\end{array}\right.\\
|\rmq|<{N\over 2}:&&g(x)\longrightarrow 0\qquad x\to\infty\,\,\mbox{and}
\,\, 0,\\
\rmq\leq -{N\over 2}:&&g(x)\longrightarrow\left\{
\begin{array}{ll}
0&x\to \infty,\\
x^{\rmq+{N\over 2}}(x-1)^{-N}&x\to 0.
\end{array}\right.
\eeqa
Recall that $x$ is related to the zero mode of the scalar component
$\vs$ via (\ref{defu})
as
\beq
x=\e^{2\pi i u}=\exp\left(-i\beta v_{\tau}-\beta\vs\right),
\eeq
and hence $x\to \infty$ corresponds to $\vs\to -\infty$
while $x\to 0$ corresponds to $\vs\to +\infty$. In view of this,
the above behaviour of $g(x)$ is perfectly consistent with
the behaviour (\ref{wfCPN1})-(\ref{wfCPN2}) of the wavefunction for
the Coulomb branch vacua. The change of the index is
\beqa
\rmq\geq {N\over 2}:&&\Delta I=-\#\left\{
(\bar n_1,\ldots, \bar n_N)\,\,\Biggl|\,|\bar n|=\rmq-{N\over 2}
\right\},\label{wcCPN1}\\
|\rmq|<{N\over 2}:&&\Delta I=0,\label{wcCPN2}\\
\rmq\leq -{N\over 2}:&&\Delta I=(-1)^{N}\#\left\{
(n_1,\ldots, n_N)\,\,\Biggl|\, |n|=-\rmq-{N\over 2}
\right\}.\label{wcCPN3}
\eeqa
This matches with the Coulomb branch result,
$\Delta I=\Delta_CI$.

Let us now move on to a general ${\mathcal N}=2$ theory.
It is enough to consider one term of (\ref{2g}) with the Wilson line
charge $(\rmq,\rmq^F)$:
\beq
g=x^{\rmq}y^{\rmq^F}
{\displaystyle \prod_{j}
\left(x^{-{q_j\over 2}}y^{-{q^F_j\over 2}}-x^{q_j\over 2}y^{q^F_j\over 2}
\right)
\over\displaystyle \prod_{i}
\left(x^{Q_i\over 2}y^{Q_i^F\over 2}-x^{-{Q_i\over 2}}y^{-{Q_i^F\over 2}}
\right)}.
\eeq
Let us look at the behaviour at
$x\to \infty$ which corresponds to $\vs\to-\infty$.
We expand the denominator factors as
$\sum_{\bar n_i=0}^{\infty}(x^{-Q_i}y^{-Q_i^F})^{\bar n_i+\half}$
if $Q_i>0$
and $(-1)\sum_{n_i=0}^{\infty}(x^{Q_i}y^{Q_i^F})^{n_i+\half}$
if $Q_i<0$.
Also, we write the numerator factors as
$\sum_{m_j=0,1}(-1)^{m_j}(x^{q_j}y^{q^F_j})^{m_j-\half}$.
Then, the residue is picked for each $(\bar n_i,n_i,m_j)$
solving
\beq
\rmq+\sum_jq_j\left(m_j-\half\right)
-\sum_{Q_i>0}Q_i\left(\bar n_i+\half\right)+
\sum_{Q_i<0}Q_i\left(n_i+\half\right)=0,
\label{condres1}
\eeq
with the result
\beq
-(-1)^{\#\{j|m_j=1\}}(-1)^{\#\{i|Q_i<0\}}
y^{\rmq^F+\sum_jq_j^F(m_j-\half)-\sum_{Q_i>0}Q_i^F(\bar n_i+\half)
+\sum_{Q_i<0}Q_i^F(n_i+\half)}.
\label{res1}
\eeq
The equation (\ref{condres1}) is nothing but the condition
(\ref{gveqneg}) for the Coulomb branch vacuum on $\vs<0$, and the residue
(\ref{res1}) is equal to $-(-1)^{|m|}y^{Q^F}$
for $Q^F$ given in (\ref{gQneg}) and $(-1)^{|m|}=(-1)^{\sum_jm_j+\sum_im_i}$
in which $m_i=0$ for $Q_i>0$
and $m_i=1$ for $Q_i<0$. In particular, it is equal to
the $\vs<0$ contribution to the result (\ref{DeltaCI}).
The analysis for $x\to 0$ is similar and matches with the
$\vs>0$ contribution to the result (\ref{DeltaCI}).
Thus we conclude $\Delta I=\Delta_CI$.

Using this agreement in a general $U(1)$ gauge theory, we can check
that $\Delta I$ matches with the mixed branch result
in Section~\ref{subsec:simpleWC} for a simple
wall crossing in higher rank theories.
We separate the discussion to Abelian and non-Abelian theories.

\subsection{General Abelian Theory}

In a general Abelian theory,
any phase boundary ${\mathcal W}$ is simple, with a mixed branch
with unbroken gauge group $G_1\subset G$ isomorphic to $U(1)$.
It is a codimension one cone spanned by $(\ell-1)$ charges, or an open
cone therein.
We would like to compute the change of the index
when $\zeta$ goes through an interior of the wall ${\mathcal W}$.

There are two ways to find it: (i) see what happens to the derivation
of the formula (\ref{result1}), and
(ii) compare the results (\ref{result1}) for the two sides of the wall.
Both can be done, but we choose to do (ii) as it is easier to present.

The result (\ref{result1}) is the sum of
$I_{i_1\ldots i_\ell}$ over $\{i_1,\ldots, i_\ell\}$
such that $\zeta$ is inside the cone ${\mathcal C}_{Q_{i_1}\ldots Q_{i_\ell}}$
spanned by
$Q_{i_1},\ldots, Q_{i_\ell}$. Therefore, the contribution
$I_{i_1\ldots i_\ell}$ disappears or appears when
$\zeta$ goes out of or comes into the cone
${\mathcal C}_{Q_{i_1}\ldots Q_{i_\ell}}$,
respectively.
The boundary of ${\mathcal C}_{Q_{i_1}\ldots Q_{i_\ell}}$
consists of the cones spanned by
$(\ell-1)$ elements of $\{Q_{i_1},\ldots, Q_{i_\ell}\}$.
Therefore, there is a non-trivial contribution to the change $\Delta I$
of the index
when that cone includes the wall ${\mathcal W}$ under consideration.
This leads to the following formula.

Before writing down the formula, let us introduce some notation.
Let ${\mathcal H}_{\mathcal W}^+$ and ${\mathcal H}_{\mathcal W}^-$
be the two halves of $i\ttt^*$ separated by the hyperplane
$\R{\mathcal W}$ that includes the wall ${\mathcal W}$,
to which $\zeta$ belongs before and after the wall crossing.
Suppose ${\mathcal W}$ is included as an open subset of the cone
${\mathcal C}_{Q_1,\ldots,Q_{\ell-1}}$. Then $\zeta$ before ({\it resp}. after)
the wall crossing is included in the cone
${\mathcal C}_{Q_i,\ldots,Q_{\ell-1},Q_{i_{\ell}}}$ if and only if
$Q_{i_{\ell}}\in {\mathcal H}^+_{\mathcal W}$
({\it resp}. $Q_{i_{\ell}}\in {\mathcal H}^-_{\mathcal W}$).
Therefore, the change of the index is
\beq
\Delta I=
\sum_{{\mathcal W}\subset
{\mathcal C}_{Q_{i_1},\ldots, Q_{i_{\ell-1}}}}
\left(
-\sum_{Q_{i_\ell}\in {\mathcal H}_{\mathcal W}^+}
I_{i_1\ldots i_\ell}
+\sum_{Q_{i_\ell}\in {\mathcal H}_{\mathcal W}^-}
I_{i_1\ldots i_\ell}\right).
\label{wcfor1}
\eeq
This formula can be simplified.
Recall that
\beq
I_{i_1\ldots i_\ell}=\epsilon_{i_1\ldots i_{\ell}}
\int_{S_{i_1\ldots i_\ell}} \dd^{\ell}u\,g(u,z),
\eeq
where $\epsilon_{i_1\ldots i_{\ell}}$ is a sign defined by
(\ref{defepsi}).
The orientation of $\Gamma_{i_1\ldots i_\ell}$ is opposite between
the one with $Q_{i_\ell}\in {\mathcal H}_{\mathcal W}^+$ and
the one with $Q_{i_\ell}\in {\mathcal H}_{\mathcal W}^-$.
Note also that the integrand of (\ref{defepsi})
does not depend on $Q_{i_{\ell}}$ as long as it is
linearly independent of $Q_{i_1},\ldots ,Q_{i_{\ell-1}}$.
So, we may write
$\epsilon_{i_1\ldots i_{\ell}}
=\pm \epsilon^+_{i_1\ldots i_{\ell-1}\infty}$ for
$Q_{i_{\ell}}\in {\mathcal H}_{\mathcal W}^{\pm}$.
Then, inside the big parenthesis of (\ref{wcfor1})
can be written as
\beq
-\sum_{i_\ell\in \Scr{I}_S}\epsilon^+_{i_1\ldots i_{\ell-1}\infty}
\int_{S_{i_1\ldots i_{\ell-1}i_\ell}}
\dd^{\ell}u\,g(u,z).
\eeq
Using (\ref{bdryS}), this is equal to
\beq
\epsilon^+_{i_1\ldots i_{\ell-1}\infty}
\int_{S_{i_1\ldots i_{\ell-1}\infty}}
\dd^{\ell}u\,g(u,z)
=:I_{i_1\ldots i_{\ell-1}\infty}^+ \ .
\label{intf1}
\eeq
Therefore, the wall crossing formula (\ref{wcfor1}) can be written as
\beq
\Delta I=
\sum_{{\mathcal W}\subset {\mathcal C}_{Q_{i_1},\ldots, Q_{i_{\ell-1}}}}
I_{i_1\ldots i_{\ell-1}\infty}^+.
\eeq
This can be processed further.
Recall from Section~\ref{subsec:simpleWC} that
the effective theory on the mixed branch at the wall ${\mathcal W}$
has two sectors --- (C) with gauge group $G_1$ and
(H) with gauge group $H=C(G_1)/G_1$ which is $T/G_1$ when $G(=T)$ is Abelian.
Matters in (H) are those with charges in the hyperplane $\R{\mathcal W}$
and everything else belongs to (C). Note that $Q\in i\ttt^*$ belongs to
$\R{\mathcal W}$ if and only if it vanishes on $\mathfrak{g}_1$.
We denote the projection $\mathfrak{t}
\to\ttt/\mathfrak{g}_1=\mathfrak{h}$ by $u\mapsto u_{\mathfrak{h}}$.
We also choose an element $Q_{\infty}\in i\ttt^*\setminus \R{\mathcal W}$
such that $\e^{2\pi i\xi}\to \e^{2\pi i Q_{\infty}(\xi)}$
defines an isomorphism $G_1\subset T\to U(1)$.
Then, we have an isomorphism
\beqa
S_{i_1\ldots i_{\ell-1}\infty}&:=&\left\{
u\in \ttt_{\C}/\rmQv\,\Biggl|\,\,\begin{array}{ll}
|Q_{i_a}(u)-c_{i_a}|=\varepsilon,\,\,
a=1,\ldots,\ell-1\\
{\rm Im}Q_{\infty}(u)=\pm R
\end{array}
\right\}\nn\\
&\cong&\left\{u_{\mathfrak{h}}\in \mathfrak{h}_{\C}/{\rm Q}_H^{\vee}\,\,
\Bigl|\,\,|Q_{i_a}(u)-c_{i_a}|=\varepsilon,\,\,
a=1,\ldots,\ell-1\,\right\}\nn\\
&&~~~~~~~~
\times
\left\{\,Q_{\infty}(u)\in\C/\Z\,\,\Bigl|\,\,
{\rm Im}Q_{\infty}(u)=\pm R\,\right\}
\nn\\
&=&S^{({\rm H})}_{i_1\ldots i_{\ell-1}}\times S^{({\rm C})}_{\infty}.
\label{factoS}
\eeqa
This holds including the orientation provided
$\epsilon^+_{i_1\ldots i_{\ell-1}\infty}=\epsilon^{({\rm H})}_{i_1\ldots i_{\ell-1}}$.
The integral (\ref{intf1}) can then be written as
\beq
I_{i_1\ldots i_{\ell-1}\infty}^+
=\epsilon^{({\rm H})}_{i_1\ldots i_{\ell-1}}
\int_{S^{({\rm H})}_{i_1\ldots i_{\ell-1}}}\int_{S^{({\rm C})}_{\infty}}
\dd^{\ell} u \,\,g(u,z).
\eeq
We may write $g(u,z)=g^{({\rm H})}(u_{\mathfrak{h}},z)\cdot g^{({\rm C})}(u,z)$,
where $g^{({\rm H})}(u_{\mathfrak{h}},z)$ is the $g$-function for the theory (H)
and $g^{({\rm C})}(u,z)$ is
everything else --- we put all Wilson line factors into the latter even if
some of the charges may vanish on $\mathfrak{g}_1$.
The integration of $g^{({\rm C})}(u,z)$
along $S^{({\rm C})}_{\infty}$ yields the
change of the index in the theory (C),
\beq
\int_{S^{({\rm C})}_{\infty}}\dd^{\ell}u\,\, g^{({\rm C})}(u,z)
=\dd^{\ell-1}u_{\mathfrak{h}}\,\,\Delta I^{({\rm C})}(u_{\mathfrak{h}},z).
\eeq
Note that the result may depend on $u_{\mathfrak{h}}$ and provides
background Wilson lines for the theory (H).
Therefore, we have
\beq
\Delta I=\sum_{\zeta_0\in
{\mathcal C}^{({\rm H})}_{Q_{i_1}\ldots Q_{i_{\ell-1}}}}
\epsilon^{({\rm H})}_{i_1\ldots i_{\ell-1}}
\int_{S^{({\rm H})}_{i_1\ldots i_{\ell-1}}}
\dd^{\ell-1}u_{\mathfrak{h}}\,\,
g^{({\rm H})}(u_{\mathfrak{h}},z)\Delta I^{({\rm C})}(u_{\mathfrak{h}},z).
\eeq
Here we used the fact that
${\mathcal W}\subset {\mathcal C}_{Q_{i_1}\ldots
Q_{i_{\ell-1}}}$ is equivalent to
$\zeta_0\in {\mathcal C}^{({\rm H})}_{Q_{i_1}\ldots
Q_{i_{\ell-1}}}$
where $\zeta_0$ is a point on the wall ${\mathcal W}$ regarded
as an element of $i\mathfrak{h}^*$.
This yields the index for the theory (H)
with the FI parameter $\zeta_0$.
The result may be written as
\beq
\Delta I=(I^{({\rm H})}_{\zeta_0}\ast \Delta I^{({\rm C})})(z).
\eeq
Since we have shown $\Delta I=\Delta_C I$ for $U(1)$ theories,
we find that the above agrees with the mixed branch result (\ref{sWC1}).

\subsection{Simple Wall Crossing in a Non-Abelian theory}

Let us now discuss the simple wall crossing in a general non-Abelian
gauge theory. Before starting, we describe the set up
and introduce some notations and terminology.
The space of FI parameters $i(\ttt^*)^W$ can be regarded as a subspace
of $i\ttt^*$. %Its dimension, which we denote by $k$, is less than $\ell$.
A general phase boundary is a codimension one cone in
$i(\ttt^*)^W$, and it is the intersection of $i(\ttt^*)^W$
with cones in $i\ttt^*$ spanned by $(\ell-1)$ or less charges.
A simple phase boundary ${\mathcal W}$ has the following special property:
it is the intersection of $i(\ttt^*)^W$ with cones
spanned by exactly $(\ell-1)$ charges, and the hyperplanes spanned
by such charges are Weyl images of one another.
The unbroken subgroup $G_1$ may be defined as the subgroup of $T$ whose
Lie algebra $\mathfrak{g}_1$ is the orthogonal to one of such hyperplanes.
(Note that $G_1\subset G$ is defined only up to conjugation,
and other conjugation images in $T$ are the orthogonal to
the other hyperplanes.) The effective theory on the mixed branch
at the wall ${\mathcal W}$ has two sectors --- (C) with gauge group
$G_1$ and (H) with gauge group $H=C(G_1)/G_1$.
Let us collect some useful facts on $G_1$ and $H$:
\begin{itemize}
\item[(i)] $T/G_1=:T_H$ is a maximal torus of $H$.
\item[(ii)] The normalizer of $T_H$ in $H$ is $(N_T\cap C(G_1))/G_1$
and hence the Weyl group of $H$ is $W_H=[(N_T\cap C(G_1))/G_1]/[T/G_1]
\cong (N_T\cap C(G_1))/T$.
\item[(iii)] In particular, $W_H$ may be regarded as a subgroup of $W$
consisting of elements that does not move $\mathfrak{g}_1\subset \ttt$.
\item[(iv)] Therefore, $|W|/|W_H|=|W/W_H|$ is an integer,
 and that is the number of hyperplanes that meets $i(\ttt^*)^W$ at the
wall ${\mathcal W}$.
\end{itemize}
By (i), we may regard the hyperplane $\mathfrak{g}_1^{\perp}\subset \ttt^*$
as the dual of a Cartan subalgebra of $H$, $\mathfrak{g}_1^{\perp}\cong
\ttt_H^*$. Matters in the theory (H) consists of those whose charges
belong to this hyperplane $i\mathfrak{g}_1^{\perp}$, and everything
else belongs to (C).
We denote the projection $\ttt\to \ttt/\mathfrak{g}_1\cong \ttt_H$
by $u\mapsto u_{\mathfrak{h}}$.

Let us now evaluate the change of the index.
Let ${\mathcal H}_{\mathfrak{g}_1^{\perp}}^+$
and ${\mathcal H}_{\mathfrak{g}_1^{\perp}}^-$ be
the two halves of $i\ttt^*$ separated by the hyperplane
$i\mathfrak{g}_1^{\perp}$ to which $\zeta$ belongs before and
after the wall crossing.
Suppose the wall ${\mathcal W}$ is included as an open subset of the cone
${\mathcal C}_{Q_1,\ldots,Q_{\ell-1}}$ in $i\mathfrak{g}_1^{\perp}$.
Then $\zeta$ before ({\it resp}. after)
the wall crossing is included in the cone
${\mathcal C}_{Q_i,\ldots,Q_{\ell-1},Q_{i_{\ell}}}$ if and only if
$Q_{i_{\ell}}\in {\mathcal H}^+_{\mathfrak{g}_1^{\perp}}$
({\it resp}. $Q_{i_{\ell}}\in {\mathcal H}^-_{\mathfrak{g}_1^{\perp}}$).
Using this as well as the fact (iv) above, we find that
the change of the index is given by
\beqa
\Delta I&=&{1\over |W|}\left|{W\over W_H}\right|
\sum_{{\mathcal W}\subset {\mathcal C}_{Q_{i_1},\ldots, Q_{i_{\ell-1}}}\subset
i\mathfrak{g}_1^{\perp}}
\left(
-\sum_{Q_{i_\ell}\in {\mathcal H}_{}^{{}^+}
\!\!\!\!\!{}_{\mathfrak{g}_1^{\perp}}}
I_{i_1\ldots i_\ell}
+\sum_{Q_{i_\ell}\in {\mathcal H}_{}^{{}^-}
\!\!\!\!\!{}_{\mathfrak{g}_1^{\perp}}}
I_{i_1\ldots i_\ell}\right)\nn\\
&=&{1\over |W_H|}
\sum_{{\mathcal W}\subset
{\mathcal C}_{Q_{i_1},\ldots, Q_{i_{\ell-1}}}\subset i\mathfrak{g}_1^{\perp}}
I_{i_1\ldots i_{\ell-1}\infty}^+.
\label{wcfor22}
\eeqa
The definition of $I_{i_1\ldots i_{\ell-1}\infty}^+$ and the
 process of going to the second line is exactly the same as in the
Abelian case.
This can be processed further, again as in the Abelian theories,
based on the factorization
\beq
S_{i_1\ldots i_{\ell-1}\infty}\cong
S^{({\rm H})}_{i_1\ldots i_{\ell-1}}\times S^{({\rm C})}_{\infty},
\eeq
as in (\ref{factoS}), with
$\epsilon^+_{i_1\ldots i_{\ell-1}\infty}
=\epsilon^{({\rm H})}_{i_1\ldots i_{\ell-1}}$,
and the factorization
\beq
g(u,z)=g^{({\rm H})}(u_{\mathfrak{h}},z)g^{({\rm C})}(u,z).
\eeq
In the latter, the root factor of $g(u,z)$ in (\ref{2g}) and (\ref{4g})
should be included in $g^{({\rm H})}$ if $\alpha\in i\mathfrak{g}_1^{\perp}$
(i.e., if it is a root of $C(G_1)/G_1=H$)
and in $g^{({\rm C})}$ if $\alpha$ is non-zero on $\mathfrak{g}_1$
(i.e., if $\mathfrak{g}_{\alpha}$ do not commute with $\mathfrak{g}_1$).
This yields
\beqa
\Delta I&=&{1\over |W_H|}\sum_{\zeta_0\in
{\mathcal C}^{({\rm H})}_{Q_{i_1}\ldots Q_{i_{\ell-1}}}}
\epsilon^{({\rm H})}_{i_1\ldots i_{\ell-1}}
\int_{S^{({\rm H})}_{i_1\ldots i_{\ell-1}}}
\dd^{\ell-1}u_{\mathfrak{h}}\,\,
g^{({\rm H})}(u_{\mathfrak{h}},z)\Delta I^{({\rm C})}(u_{\mathfrak{h}},z)\nn\\
&=&(I^{({\rm H})}\ast \Delta I^{({\rm C})})(z).
\eeqa
Since we have shown $\Delta I=\Delta_C I$ for $U(1)$ theories,
we find that the above agrees with the mixed branch
result (\ref{sWC1}).

Note that the above computation applies to the case where ${\mathcal W}$
is not a real phase boundary. Recall that a non $\Scr{I}_S$-generic
point does not always admit a mixed branch solution, in which case
it must be in the interior of a phase.
For example, the III-IV wall in Fig.~\ref{fig:qk11}
of the triangle quiver in which the condition (\ref{condrank}) is satisfied.
When $k=2$, it is of simple type, i.e., it is the intersection
of $i(\ttt^*)^W$ with the cone spanned by exactly $\ell-1(=2)$ charges.
For such a case, since the mixed branch is empty, the index of theory (H)
must vanish, $I^{({\rm H})}_{\zeta_0}=0$, and hence the index
does not jump across ${\mathcal W}$, $\Delta I=0$.

\subsection*{${\mathcal N}=4$ Theories}

For ${\mathcal N}=4$ theories,
we have seen that $\Delta I$ is independent of the flavor twist
in the $U(1)$ theories. Therefore,
in either Abelian or non-Abelian theory,
$\Delta I^{({\rm C})}(u_{\mathfrak{h}},z)$ depends only on the $\bfR_{\!-}$
part of $z$ (denoted by ${\bf z}$),
\beq
\Delta I^{({\rm C})}(u_{\mathfrak{h}},z)
=\Delta I^{({\rm C})}({\bf z}).
\eeq
Therefore, the wall crossing formula has a factorized form,
\beq
\Delta I (z)=I^{({\rm H})}({\bf z},z_{\rm other})\times
\Delta I^{({\rm C})}({\bf z}),
\label{4factorization2}
\eeq
in agreement with the mixed branch analysis (\ref{4factorization}).

\section{Systems with ${\mathcal N}=4$ Supersymmetry}
\label{sec:4Ex}

In this section, we illustrate the index formula and wall crossing
in a few examples with ${\mathcal N}=4$ supersymmetry.

\subsection{Grassmannian}

The first example is the $U(k)$ gauge theory
with $N$ fundamental chiral multiplets.
We set $\zeta(D)=r\cdot {\rm tr}(D)$.
When $N\geq k$, $r>0$ is the geometric phase with
the Grassmannian $G(k,N)$ of $k$ planes in $\C^N$
as the target space. For $r<0$ or for $N<k$, supersymmetry is broken.

We shall compute the index in the geometric phase.
Since the model is compact, dependence of the flavor twist is trivial and
any assignment of R-charge yields the same result.
For simplicity, we set $R_i=0$ for all the chiral matters,
for which the $g$ function (\ref{4g}) reads
\beqa
g&=&\left({1\over \by-\by^{-1}}\right)^k
\prod_{a\ne b}{x_a^{-\half}x_b^{\half}-x_a^{\half}x_b^{-\half}\over
x_a^{\half}x_b^{-\half}\by^{-1}-x_a^{-\half}x_b^{\half}\by}
\left(\prod_{a=1}^k{
x_a^{-\half}\by-x_a^{\half}\by^{-1}\over
x_a^{\half}-x_a^{-\half}}\right)^N\nn\\
&=&(-1)^{k(N-k)}{\by^{-k(N-k)}\over (1-\by^2)^k}
{\prod_{a\ne b}(x_a-x_b)\over \prod_{a\ne b}(x_a-\by^2 x_b)}
{\prod_{a=1}^k(x_a-\by^2)^N
\over \prod_{a=1}^k(x_a-1)^N}.
\label{4gforgrass}
\eeqa
Singular hyperplanes are $H_a=(x_a=1)$ from the chiral multiplets
with the charge vector $e_a$ and
$H_{a,b}=(x_a=\by^2 x_b)$ from the root $e_a-e_b$ component of
the vector multiplet.
Here, $e_a$ are the standard
basis elements of $i\ttt^*\cong \R^k$.
Note that
\beq
\zeta=r(e_1+\cdots +e_k),\qquad r>0.
\eeq
The isolated intersection $H_1\cap\cdots \cap H_k$ should be taken
in the JK residue, as $\zeta$ is a positive span of the defining charges
$Q=\{e_1,\ldots, e_k\}$.
 What about those involving $H_{a,b}$'s? Consider for example,
$H_1\cap\cdots\cap H_{k-1}\cap H_{k,a}$ with
$Q=\{e_1,\ldots, e_{k-1},e_k-e_a\}$
and $H_1\cap\cdots\cap H_{k-1}\cap H_{a,k}$
with $Q=\{e_1,\ldots, e_{k-1},e_a-e_k\}$.
The former satisfy the condition that $\zeta$
is in the positive span of $Q$, while the
latter fails. On the other hand, for the former,
the residue vanishes because of the factor $(x_k-\by^2)^N$
in the numerator in (\ref{4gforgrass}). Similar reasoning applies to
intersections involving more than one vector singularities, because
in order to have co-dimension $k$ singularity at least on singularity
from the fundamental chiral must be involved.

Thus the only pole that contributes comes from $H_1\cap \cdots \cap H_k$,
\beq
I\,\,=\,\,{1\over k!}\cdot
{1\over (2\pi i)^k}\int\limits_{|x_1-1|=\cdots=|x_k-1|=\varepsilon}
{\dd x_1\cdots\dd x_k\over x_1\cdots x_k}\,g(x,\by).
\eeq
At this moment, we do not have a systematic way to carry out the
computation.
Instead, we can check in examples that it reproduces the known formula
\cite{Denef,Reineke},
\begin{equation}
I\,\,=\,\,(-1)^{k(N-k)}\cdot
\frac{\prod_{i=1}^N(\by^{-i}-\by^{i})}{\prod_{j=1}^k(\by^{-j}-\by^{j})
\prod_{l=1}^{N-k}(\by^{-l}-\by^{l})}\ .
\end{equation}

\subsection{Hypersurface in Projective Space}\label{subsec:4hypersurface}

Let us next consider
Example 2 in Section~\ref{subsec:LSMex}. $\zeta>0$ is the geometric phase
with the degree $d$ hypersurface $X_f\subset \CP^{N-1}$, while
$\zeta<0$ is the Landau-Ginzburg orbifold $W=f(X_1,\ldots,X_N)/\Z_d$.

The R-charge can be anything such that the superpotential $W=Pf(\Phi)$
has charge $2$. We take $R_P=2$ and $R_X=0$, for which
the $g$ function is
\beqa
g&=&{1\over \by-\by^{-1}}\left({x^{-\half}\by-x^{\half}\by^{-1}\over
x^{\half}-x^{-\half}}\right)^N{x^{d\over 2}-x^{-{d\over 2}}\over
x^{-{d\over 2}}\by-x^{d\over 2}\by^{-1}}\nn\\
&=&(-1)^{N-1}{\by^{1-N}\over \by-\by^{-1}}\left({x-\by^2\over x-1}\right)^N
{x^d-1\over x^d-\by^2}.
\eeqa
In the geometric phase $\zeta>0$, we select the $N$-th order pole at $x=1$:
\beqa
I&=&{1\over 2\pi i}\oint_1{\dd x\over x}\,g(x,\by)\nn\\
&=&(-1)^{N-1}{\by^{1-N}\over \by-\by^{-1}}\left.
{1\over (N-1)!}{\dd^{N-1}\over \dd x^{N-1}}\left[{(x-\by^2)^N(x^d-1)
\over x(x^d-\by^2)}\right]\right|_{x=1}.
\eeqa
For example,
\beqa
I_{N=3}&=&{d(d-3)\over 2}\left(\by^{-1}+\by\right),\label{Icurve}\\
I_{N=4}&=&{d(d^2-6d+11)\over 6}(\by^{-2}+\by^2)
+{d(2d^2-6d+7)\over 3},\label{Isurface}\\
I_{N=5}&=&{d(d^3-10d^2+35d-50)\over 24}(\by^{-3}+\by^3)
+{d(11d^3-50d^2+85d-70)\over 24}(\by^{-1}+\by),\nn\\
&&\label{Ithreefold}\\
&&\cdots\nn
\eeqa
These are the $\chi_y$ genus (\ref{chiy}) of the hypersurface $X_f$.
By the Lefschetz hyperplane theorem, the Hodge diamond must be of the form
\beq
\begin{array}{ccc}
&1& \\
\!\!h^{1,0}\!\!&&\!\!h^{0,1}\!\! \ \\
&1& \\
\end{array}\ \qquad
\begin{array}{ccccc}
&&1&& \\
&0&&0& \\
\!\!h^{2,0}\!\!&&\!\!h^{1,1}\!\!&&\!\!h^{0,2}\!\!\\
&0&&0& \\
&&1&& \\
\end{array} \  \qquad
\begin{array}{ccccccc}
&&&1&&& \\
&&0&&0&& \\
&0&&1&&0&\\
\!\!h^{3,0}\!\!&&\!\!h^{2,1}\!\!&&\!\!h^{1,2}\!\!&&\!\!h^{0,3}\!\!\\
&0&&1&&0&\\
&&0&&0&& \\
&&&1&&& \\
\end{array} \qquad\cdots
\label{HDgeom}
\eeq
Thus, we can read off the Hodge numbers from the index.
For example,  for $N=3$, $X_f$ is a curve of
genus $h^{1,0}={d(d-3)\over 2}+1$. For $N=4$ it is
a surface of $h^{2,0}={d(d^2-6d+11)\over 6}-1$
and $h^{1,1}={d(2d^2-6d+7)\over 3}$. Etc.
Check that $h^{N-2,0}=1$ in the Calabi-Yau case $d=N$.
Note also that $h^{1,1}=20$ for $d=N=4$ (K3 surface)
and $h^{2,1}=101$ for $d=N=5$ (quintic threefold).
To find the index in the Landau-Ginzburg phase $\zeta<0$, we use
the wall crossing formula (\ref{4wcU1gen}), in which we set $\neff^{(4)}=N-1$.
This yields
\beq
I_{\rm LG}=I_{\rm geometric}
+(-1)^{N-1}\left(\by^{-(N-2)}+\cdots +\by^{N-2}\right).
\eeq
This is also what we have seen in the Coulomb branch analysis.
There we have also seen that the wall crossing states form
the spin $j={|\neff^{(4)}|-1\over 2}={N-2\over 2}$
representation of the $SU(2)$ R-symmetry.
Since the vertical axis of the Hodge diamond shows the spin $J_3$ of
the same $SU(2)$, the $1$'s in the vertical middle of the Hodge diamond
(\ref{HDgeom}) drops. This leaves us with the following
Hodge diamond of the Landau-Ginzburg orbifold:
\beq
\begin{array}{ccc}
&0& \\
\!\!h^{1,0}\!\!&&\!\!h^{0,1}\!\! \ \\
&0& \\
\end{array}\ \qquad
\begin{array}{ccccc}
&&0&& \\
&0&&0& \\
\!\!h^{2,0}\!\!&&\!\!\!\!h^{1,1}\!-\!1\!\!\!\!&&\!\!h^{0,2}\!\!\\
&0&&0& \\
&&0&& \\
\end{array} \  \qquad
\begin{array}{ccccccc}
&&&0&&& \\
&&0&&0&& \\
&0&&0&&0&\\
\!\!h^{3,0}\!\!&&\!\!h^{2,1}\!\!&&\!\!h^{1,2}\!\!&&\!\!h^{0,3}\!\!\\
&0&&0&&0&\\
&&0&&0&& \\
&&&0&&& \\
\end{array} \qquad\cdots
\label{HDLG}
\eeq

Note that there is a non-trivial wall crossing even in the Calabi-Yau case
$d=N$ where $X_f$ is a Calabi-Yau manifold. In that case,
the elliptic genus of the two-dimensional version of the model
is well defined and does not exhibit wall crossing phenomenon, i.e.,
the Hodge diamond is the same as (\ref{HDgeom}) for both the
geometric phase and the Landau-Ginzburg orbifold phase.
On the other hand, in one-dimension, there is a wall crossing.
The difference comes from the twisted sector states, which are present in
2d orbifolds but absent in 1d orbifolds: In the $2d$ Landau-Ginzburg
orbifold, the ground states from the $d-1 (=N-1)$ twisted sectors are
nothing but the 1's in the diagonal middle. The horizontal middle
of (\ref{HDLG}) are the untwisted sector invariant states, which are
present both in 1d and 2d.

When $d\ne N$, the axial $U(1)$ R-symmetry
(which determines the vertical axis of Hodge diamond) is anomalous in 2d
and the elliptic genus is ill defined.\footnote{The vector $U(1)$ R-symmetry,
which determines the horizontal axis of the Hodge diamond,
is conserved for any $d$ versus $N$.}
However,
we can still consider the relation between the ground states
of the non-linear sigma model
and the Landau-Ginzburg orbifold. In 2d, the two theories are related by
the renormalization group.
For $d<N$, it is a flow from the sigma model to the Landau-Ginzburg
orbifold, and for $d>N$ the flow is in the opposite direction.
The spectrum of supersymmetric ground states is continuous along the flow,
provided we include the
$|N-d|$ isolated massive vacua on the Coulomb branch in the low energy side.
This is consistent with the above result on the 1d wall crossing
that $(N-1)$ states disappear as we go from the geometric phase
to the  Landau-Ginzburg orbifold phase.
In 2d, in the same direction,
$(N-d)$ states go to Coulomb branch vacua (counted with sign!)
and $(d-1)$ states becomes twisted sector states.

\subsection{Flavor Decoupling in Compact Models}

In Section~\ref{subsec:DefInd}, we proved that supersymmetric ground states
of an effectively compact ${\mathcal N}=4$ theory have charge zero under any
flavor symmetry. In particular, the index cannot depend on the flavor twist
parameter.
In fact, we have already seen this implicitly for the $\CP^{N-1}$ model.
That model has $SU(N)$ flavor symmetry and we may consider the index
twisted by that. Since the index vanishes for $\zeta<0$,
the index for $\zeta>0$ is determined by the wall
crossing part, $I=-\Delta I$, but we have seen that $\Delta I$
is independent of the flavor twist in any $U(1)$ theory
(\ref{4wcU1gen}).\footnote{In Section~\ref{subsec:2CPN},
we shall study ${\mathcal N}=2$ $\CP^{N-1}$ model,
where we will see that the $SU(N)$ flavor twist dependence of the index
is not only non-trivial
but carries important information about the ground states.}
In this subsection, we illustrate the twist independence in a more
complicated model. We shall see this in the form of
independence on the R-charge assignment.

Consider $U(1)^k$ theories with $k+1$ sets of chiral fields of charges,
$a_1$ number of $(1,0,0,\cdots,0)$, $a_2$ number of $(0,1,0,\cdots,0)$,
\dots, $a_k$ number of $(0,0,0,\cdots,1)$, and $a_{k+1}$ number of
$(1,1,1,\cdots,1)$. The superpotential is not possible.
The classical Higgs branch is present and is compact
when all the FI parameters are positive.
When some of the FI parameter is negative, supersymmetry is broken.
Since there is no superpotential, there is no constraint on
the R-charge assignment.
For example, we may assign $0$ to the first $k$ groups of matter fields,
but $q$ to the last group of $a_{k+1}$ fields.
Clearly, this is not a part of the gauge symmetry.
This gives
\begin{eqnarray}
g_{a_1a_2\cdots a_k;a_{k+1}}&=&\left(\frac{1}{2i\sin(\pi\bfz)}\right)^k
\times \prod_{i=1}^k
\left(\frac{\sin(-\pi(u_i-\bfz))}{\sin(\pi u_i)}\right)^{a_i}\cr\cr
&&\times
\left(\frac{\sin(-\pi(u_1+\cdots+u_k +(q-1)\bfz))}{\sin(\pi(u_1
+\cdots +u_k +q\bfz))}\right)^{a_{k+1}}.\nonumber
\end{eqnarray}

We of course look at the phase where
all the FI parameters $\zeta_i$ are positive.
Taking a simultaneous permutations of $a$'s and $\zeta$'s, we may further
assume that $\zeta_{i\neq k}>\zeta_k>0$.
There are exactly two relevant poles
$u_*^{(0)}\equiv \{u_1=u_2=\cdots=u_k=0\}$ and $u_*^{(k)}\equiv
\{u_1=u_2=\cdots=u_{k-1}=0=\sum_{i=1}^k u_i +q\bfz\}$,
and the index is
\begin{equation}
I_{a_1\cdots a_k;a_{k+1}}
={\rm res}_{u=u_*^{(0)}}\, g_{a_1a_2\cdots a_k;a_{k+1}}(u)
+{\rm res}_{u=u_*^{(k)}}\,g_{a_1a_2\cdots a_k;a_{k+1}}(u)\ .
\nonumber
\end{equation}
Both terms are explicitly dependent on $q$, yet the sum should
be independent of $q$.

This $q$-independence is easiest to see when $a_1=\cdots=a_{k-1}=1$.
Integrating  $u_1,\dots , u_{k-1}$ first, we find
\begin{eqnarray}
I_{1,1,\dots ,1,a_k;a_{k+1}}
=\left[\mathop{\rm res}_{u_{k}=0}+\mathop{\rm res}_{u_{k}=-q\bfz}\right]
f^{a_k a_{k+1}}_q(u_k,\bfz) \ ,
\end{eqnarray}
where
\begin{eqnarray}
f^{a_k a_{k+1}}_q(u_k,z)\equiv\frac{1}{2i\sin(\pi\bfz)}\cdot
\left(\frac{\sin(-\pi(u_k-\bfz))}{\sin(\pi u_k)}\right)^{a_k}\cdot
\left(\frac{\sin(-\pi(u_k +(q-1)\bfz))}{\sin(\pi(u_k +q\bfz))}\right)^{a_{k+1}} \ .
\end{eqnarray}
The integrand has two other poles, at $e^{2\pi i u_k}=0,\infty$, which are
both simple and with
which we may trade off the poles $e^{2\pi i u_k}=1,e^{-2\pi i q\bfz}$. Residues
at $e^{2\pi i u_k}=0,\infty$ are manifestly independent of $q$, proving
the assertion that the  index is independent of $q$.
This simple argument generalizes straightforwardly to general
$a_i$'s, by turning on most general flavor chemical potentials on
the first $k-1$ sets of chiral fields, which reduces all the
poles of $u_{i,\dots,k-1}$ planes to become simple.

We close with a few examples,
\begin{eqnarray}
I_{2,4;3}&=&
-\frac{1}{\by^{7}}-\frac{2}{\by^{5}}-\frac{2}{\by^{3}}
-\frac{2}{\by}-2\by-2\by^3-2\by^5-\by^7 \ ,  \cr \cr
I_{4,4;2}
&=& \frac{1}{\by^{8}}+\frac{2}{\by^{6}}
+\frac{3}{\by^{4}} +\frac{4}{\by^{2}}+4+4\by^2+3\by^4 +2\by^6+\by^{8} \ ,\\
\cr
I_{3,3,3;3}
&=& -\frac{1}{\by^{9}}-\frac{3}{\by^{7}}-\frac{6}{\by^{5}}
-\frac{8}{\by^{3}}-\frac{9}{\by}-9\by-8\by^3-6\by^5-3\by^7-\by^9
\ ,\nonumber
\end{eqnarray}
which indeed shows no $q$-dependence, showing that all ground
states are invariant under the flavor symmetry.

\subsection{Flavor Non-Decoupling in Non-Compact Models}

When the theory is not effectively compact,
the decoupling of flavor twist fails, or rather,
flavor twist is necessary to make the index well-defined.
The most illustrative example of this is already mentioned
in Section~\ref{subsec:oneloop}, embedded in the middle
of one-loop computation, though with
${\mathcal N=2}$ supersymmetry. Let us consider the
${\mathcal N}=4$ version, that is, the theory (with no gauge group)
of a single chiral multiplet on which the $U(1)$ flavor symmetry acts in the
standard way.
The Higgs branch is $\C$ and is non-compact.
The index with the twist parameter $y$ for the $U(1)$ is,
when we assign R-charge zero,
\beq
I\,\,=\,\,{y^{-\half}\by-y^{\half}\by^{-1}\over y^{\half}-y^{-\half}}.
\label{4Ifree}
\eeq
Indeed it depends on the twist parameter $y$ and is singular
when we turn it off, $y\to 1$.
As stated in Section~\ref{subsec:DefInd}, Hamiltonian way
to understand it is possible only when the real mass $m=m_3$ is non-zero, i.e.
when $|y|>1$ or $|y|<1$, which makes the theory compact by the Harmonic
oscillator potential. Then, (\ref{4Ifree}) can be expanded
in the geometric series in one way or the other,
and the result represents the count of states with
$0=\{{\bf Q}_+,\overline{\bf Q}_+\}=H_{m}-{\bf G}_{m}^F(m)$
for each pair of flavor and R-charges. It is interesting to note that
the index simplifies when the $\bfR_{\!-}$-twist is turned off,
\beq
I|_{\by=1}\,\,=\,\,-1.
\label{4Ifreered}
\eeq
How can this happen?

In fact, this holds in general and is another consequence of
${\mathcal N}=4$ supersymmetry,
which has another set of supercharges, ${\bf Q}_-$
and $\overline{\bf Q}_-$ with charge $-1$ and $+1$ under $\bfR_{\!-}$.
They obey the lower part of the algebra (\ref{defSAN41}), which yields
the following identity on states obeying $0=H_{m}-{\bf G}_{m}^F(m)$:
\beq
\{{\bf Q}_-,\overline{\bf Q}_-\}\,=\,
H_{m}+{\bf G}_{m}^F(m)\,\stackrel{\downarrow}{=}
2{\bf G}_{m}^F(m).
\eeq
Therefore, when $m\ne 0$,
 as far as the states that contribute to the index are concerned,
${\bf Q}_-+\overline{\bf Q}_-$ defines a one to one correspondence
between bosonic and fermionic states of non-zero flavor charge,
while it vanishes on states of zero flavor charge.
Therefore, $I|_{\by=1}$ receives contribution only from the states of
zero flavor charge, and hence is independent of the flavor twist parameter.

Let us illustrate these
in a system with gauge symmetry. We take the non-compact version of
 Example 2: $U(1)$ gauge theory with
charge 1 chiral multiplets $X_1,\ldots,X_N$ and a charge $-d$ chiral
multiplet $P$, and vanishing superpotential $W=0$.
This system has $U(N)$ flavor symmetry that acts
in the standard way on $X_1,\ldots, X_N$.
Positive $\zeta$ is the geometric phase with the total space of
${\mathcal O}(-d)$ over $\CP^{N-1}$ as the target.
Negative $\zeta$ is the orbifold phase where we have $\C^N/\Z_d$.
A generic element of the flavor group $g\in U(N)$ has
a compact set of fixed points --- the zero section of ${\mathcal O}(-d)$
in the geometric phase and the origin of $\C^N/\Z_d$ in the orbifold phase.
Special elements are those $g$ where $g^d$ has eigenvalue $1$, for which
the fixed point set is non-compact --- extending in the fibre direction of
${\mathcal O}(-d)$ or in the ray of $\C^N/\Z_d$.
Let us compute the index with the twist parameter $y=(y_1,\ldots, y_N)$
for the maximal torus of $U(N)$.
It must be well-defined when $y$ is generic, i.e.,
when none of $y_i^d$ is equal to $1$.
Assigning the trivial R-charge to all
fields, we have
\beqa
g&=&{1\over\by-\by^{-1}}\prod_{i=1}^N
\left({x^{-\half}y_i^{-\half}\by-x^{\half}y_i^{\half}\by^{-1}\over
x^{\half}y_i^{\half}-x^{-\half}y_i^{-\half}}\right)
\cdot{x^{d\over 2}\by-x^{-\half}\by^{-1}\over
x^{-{d\over 2}}-x^{d\over 2}}\nn\\
&=&(-1)^{N-1}{\by^{1-N}\over \by-\by^{-1}}
\prod_{i=1}^N\left({x-\vps_i\by^2
\over x-\vps_i}\right)\cdot {x^d-\by^{-2}\over x^d-1},
\eeqa
where we introduced $y_i^{-1}=\vps_i$ to simplify the expression.
The index in the geometric phase $\zeta>0$ is
\beq
I\,\,=\,\,
(-1)^{N}\by^{2-N}\sum_{i=1}^N\prod_{j\ne i}
\left({\vps_i-\vps_j\by^2\over \vps_i-\vps_j}\right)\cdot
{\vps_i^d-\by^{-2}\over \vps_i^d-1}.
\eeq
The diagonals $\vps_i=\vps_j$ for
$i\ne j$ are only apparent singularity. For example,
\beq
I_{N=2}={(\vps_1\vps_2s_{d-2}(\vps_1,\vps_2)+1)\by^{-1}
+(\vps_1^d\vps_2^d-1)
+(\vps_1\vps_2s_{d-2}(\vps_1,\vps_2)+\vps_1^d\vps_2^d)\by
\over (\vps_1^d-1)(\vps_2^d-1)},
\eeq
where $s_{d-2}(\vps_1,\vps_2)=(\vps_1^{d-1}-\vps_2^{d-1})/(\vps_1-\vps_2)$.
The index has singularity only at $\vps_i^d=1$ for some $i$, as expected.
The index in the orbifold phase $\zeta<0$ can be obtained from this
by the wall crossing formula (\ref{4wcU1gen}):
\beq
I_{\rm orbifold}\,\,=\,\,I_{\rm geometric}
+(-1)^{N-1}\left(\by^{-(N-2)}+\cdots +\by^{N-2}\right).
\eeq
Recall that the wall crossing part is independent of the flavor twist,
even when the theory is non-compact, since the non-compactness of
the Higgs branch does not affect the effective compactness of the
Coulomb branch. When we turn off the $\bfR_{\!-}$ twist,
we have
\beq
I_{\rm geometric}|_{\by=1}\,=\,(-1)^NN,\qquad
I_{\rm orbifold}|_{\by=1}\,=\,(-1)^N.
\eeq
Indeed they are independent of the flavor twist parameters.
Up to sign,
the result agrees with the Euler number of the fixed point set ---
the zero section ($\CP^{N-1}$) in the geometric phase and
the origin (one point) in the orbifold phase.
Why is that?
The real mass provides a quadratic
potential in the direction transverse to the fixed point set $X^F$,
and hence at energies  below the mass, the theory is well approximated
by the sigma model on $X^F$, whose index is
$(-1)^{{\rm dim}X^F}\chi(X^F)$.  The trivial action of $\Z_d$ on the origin
in the orbifold phase might
not cause a problem in 1d where there is no twisted sectors.
How about the sign?
In the limit of very large masses, the theory of the transverse modes
can be well approximated by the direct product of decoupled copies of
a free chiral multiplet with a large real mass.
Using the result (\ref{4Ifreered}) for the individual component,
we obtain $(-1)^{{\rm codim}X^F}$ as the index.
This explains the above result, $(-1)^1\cdot (-1)^{N-1}N$ for the geometric
phase and $(-1)^N\cdot 1$ in the orbifold phase.

From this exercise, we can extract the following general statement.
Suppose an ${\mathcal N}=4$ model has an
effective theory on the Higgs branch, which is a non-linear
sigma model on an orbifold $X$. Suppose there is a continuous flavor group
whose generic element has a compact set of fixed points $X^F$.
Then the index is well defined if it is twisted by the flavor symmetry,
and as we have learned above, the flavor twist dependence drops out when the
$\bfR_{\!-}$-twist is turned off. In particular, $I|_{\by=1}=I(y^{{\bf G}^F})$
must be an integer.
Let us assume that the local
orbifold group acts trivially on $X^F$.
Then, we have
\beq
I(y^{{\bf G}^F})\,\,=\,\,\sum_{i}(-1)^{\dim X}\chi(X_i^F),
\label{gLFT}
\eeq
where the sum is over the connected components of $X^F$.
When $X$ is a compact manifold, the index is independent of the flavor twist
and its value at $\by=1$ is $(-1)^{{\rm dim}X}\chi(X)$.
Thus, (\ref{gLFT}) is nothing but the Hopf theorem.
Therefore, (\ref{gLFT}) can be regarded as
a generalization of the Hopf theorem to ${\mathcal N}=4$
linear sigma models with non-compact Higgs branches.
From the derivation, the statement itself must hold
for any ${\mathcal N}=4$ non-linear sigma model on an orbifold $X$,
even without linear sigma model realization.

\subsection{``The Two Parameter Model''}

\begin{figure}[htb]
\psfrag{z1}{$\zeta^1$}
\psfrag{z2}{$\zeta^2$}
\psfrag{I}{I}
\psfrag{II}{II}
\psfrag{III}{III}
\psfrag{IV}{IV}
\centerline{\includegraphics{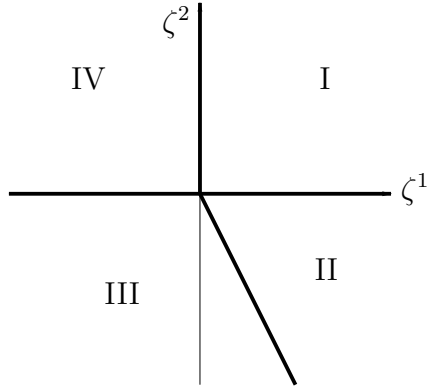}}
\caption{The phases of ``the two parameter model''}
\label{fig:tpmodel}
\end{figure}
The next example is
 the model with gauge group $U(1)\times U(1)$,
the following matter content
$$
\label{tab: WP model}
\begin{array}{c|cccc|c}
 & P & X_{1,2} & Y_{1,2,3} & Z & {\rm FI} \\
\hline
U(1)_1 & -4 & 0 & 1 & 1 & \zeta^1 \\
U(1)_2 & 0 & 1 & 0& -2 & \zeta^2
\end{array}
$$
and the superpotential $W=P\cdot f(X,Y,Z)$ where $f(X,Y,Z)$ is a polynomial
of charge $(4,0)$. This is the 1d version of the 2d (2,2) model which is
popular among physicists \cite{MP}.
There are four phases as shown in Fig.~\ref{fig:tpmodel}.
I is a geometric phase,
II is an orbifold phase,
III is a Landau-Ginzburg orbifold phase and IV is
a hybrid phase.
The geometry in Phase I is a smooth compact Calabi-Yau threefold
embedded as a hypersurface is a
toric fourfold of $h^{1,1}=2$.

We may set $R_P=2$ and $R_X=R_Y=R_Z=0$. Then the $g$
function is
\beqa
g&=&\left({1\over 2i\sin(\pi\bfz)}\right)^2
{\sin(-\pi(-4 u_1))\over\sin(\pi(-4u_1+\bfz))}
\left({\sin(-\pi(u_2-\bfz))\over\sin(\pi u_2)}\right)^2\nn\\
&&~~~~~~~~\left({\sin(-\pi(u_1-\bfz))\over\sin(\pi u_1)}\right)^3
{\sin(-\pi(u_1-2u_2-\bfz))\over\sin(\pi(u_1-2u_2))} \ .
\eeqa
Below we write the result of residue computation
in each phase.
\begin{itemize}
\item
Phase I:
\beq
I\,\,=\,\,84\by^{-1}+84\by.
\eeq
\item
Phase II:
\beq
I\,\,=\,\,82\by^{-1}+82\by.
\eeq
\item
Phase III:
\beq
I\,\,=\,\,\by^{-3}+83\by^{-1}+83\by+\by^3.
\eeq
\item
Phase IV:
\beq
I\,\,=\,\,\by^{-3}+86\by^{-1}+86\by+\by^3.
\eeq
\end{itemize}

We see that non-trivial wall crossing is happening at each phase boundary.
Since the gauge group is Abelian, all of them are simple walls.
Let us present the mixed branch analysis of the wall crossing.

\subsubsection*{\underline{\rm I $\to$ II}}

The unbroken gauge group at the mixed branch
along this phase boundary is $G_1=U(1)_2$ and the fields are decomposed as
\beqa
({\rm C})&&X_{1,2}\, (1),~ Z\, (-2),\nn\\
({\rm H})&&P\, (-4),~ Y_{1,2,3}\, (1).\nn
\eeqa
As in Section~\ref{subsec:simpleWC}, to the right of each field,
we put its charge under $G_1$ for (C)
and $G/G_1$ for (H).
The Coulomb branch of (C) has $\neff^{(4)}=2-1=1$ so that we have
$\Delta_CI^{({\rm C})}(\by)=-1$.
The Higgs branch of (H) is the degree four hypersurface in $\CP^2$:
 it is a curve of genus 3
and has $I^{({\rm H})}(\by)=2\by^{-1}+2\by^{1}$ (see
Section~\ref{subsec:4hypersurface}).
Thus,
\beq
\Delta I\,\,=\,\,(2\by^{-{1}}+2\by^{1})\cdot (-1)
\,\,=\,\,
-2\by^{-{1}}-2\by^{1}.
\eeq
Note also that the wall crossing states, which go out since
$\neff^{(4)}>0$, form the following Hodge diamond
$$
\begin{array}{ccc}
&1&\\
3&&3\\
&1&
\end{array}
$$

\subsubsection*{\underline{\rm II $\to$ III}}

The unbroken gauge group is $G_1=\{(g^2,g)\}$
and the fields are decomposed as
\beqa
({\rm C})&&P\, (-8),~ X_{1,2}\, (1),~ Y_{1,2,3}\, (2),\nn\\
({\rm H})&&Z\, (1).\nn
\eeqa
The Coulomb branch of (C) has $\neff^{(4)}=-1+2+3=4$ so that we have
$\Delta_CI^{({\rm C})}(\by)
=\by^{-{3}}+\by^{-{1}}+\by^{1}+\by^{3}$.
The Higgs branch of (H) is one point and hence has $I^{({\rm H})}(\by)=1$.
Thus,
\beq
\Delta I
\,\,=\,\,\by^{-{3}}+\by^{-{1}}+\by^{1}+\by^{3}.
\eeq
The wall crossing states, which go out since $\neff^{(4)}>0$,
form the Hodge diamond with $1,1,1,1$ in the vertical middle.

\subsubsection*{\underline{\rm III $\to$ IV}}

The unbroken gauge group is $G_1=U(1)_2$ and the fields are decomposed as
\beqa
({\rm C})&&X_{1,2}\, (1),~ Z\, (-2),\nn\\
({\rm H})&&P\, (-4),~ Y_{1,2,3}\, (1).\nn
\eeqa
The Coulomb branch of (C) has $\neff^{(4)}=-2+1=-1$ so that we have
$\Delta_CI^{({\rm C})}(\by)=1$.
The Higgs branch theory of (H) is the Landau-Ginzburg orbifold
$W=f_4(Y)/\Z_4$ where $f_4$ is a quartic polynomial.
To simplify the analysis let us take
 the Fermat polynomial, for which the relations are
$Y_1^3=Y_2^3=Y_3^3=0$.
Since $\Omega=\dd Y_1\wedge \dd Y_2\wedge \dd Y_3$ has charge $3$,
for $O(Y)\Omega$ to be $\Z_4$ invariant, $O(Y)$ must have charge 1 (mod $4$)
under $\Z_4$.
There are six states:
\beq
Y_1\Omega,~Y_2\Omega,~Y_3\Omega,~\,\,
Y_1Y_2^2Y_3^2\Omega,~Y_1^2Y_2Y_3^2\Omega,~
Y_1^2Y_2^2Y_3\Omega.
\eeq
The R-charge assignment which is conjugate symmetric is
$-{1\over 2}$ for the first three and ${1\over 2}$ for the last three.
Thus, $I^{({\rm H})}(\by)=3\by^{-1}+3\by^{1}$.
Therefore, the change of the index is
\beq
\Delta I\,\,=\,\,(3\by^{-{1}}+3\by^{1})\cdot 1
\,\,=\,\,
3\by^{-{1}}+3\by^{1}.
\eeq
The wall crossing states, which come in since $\neff^{(4)}<0$,
form the Hodge diamond with $0,3,3,0$ in the horizontal middle.

\subsubsection*{\underline{\rm IV $\to$ I}}

The unbroken gauge group is $G_1=U(1)_1$ and the fields are decomposed as
\beqa
({\rm C})&&P\, (-4),~ Y_{1,2,3}\, (1),~ Z\, (1),\nn\\
({\rm H})&&X_{1,2}\, (1).\nn
\eeqa
The Coulomb branch of (C) has $\neff^{(4)}=1-3-1=-3$ so that we have
$\Delta_CI^{({\rm C})}(\by)=\by^{-1}+1+\by$.
The Higgs branch of (H) is $\CP^1$ which has
$I^{({\rm H})}(\by)=-\by^{-{1}}-\by^{1}$.
Thus,
\beq
\Delta I\,\,=\,\,
(-\by^{-{1}}-\by^{1})\cdot (\by^{-1}+1+\by)
\,\,=\,\,-\by^{-{3}}-2\by^{-{1}}-2\by^{1}-\by^{3}.
\eeq
The wall crossing states, which come in since $\neff^{(4)}<0$,
form the Hodge diamond with $1,2,2,1$ in the vertical middle.

We see that $I_{\rm before}+\Delta I=I_{\rm after}$ indeed holds for each
wall crossing. We have also determined the Hodge diamond formed
by the wall crossing states.
Using this, we can determine the Hodge diamonds of all four phases starting
from the geometric phase, where the Hodge numbers are known
by the Lefschetz hyperplane theorem:
\begin{itemize}
\item
Phase I (input):
\beq
\begin{array}{ccccccc}
&&&1&&& \\
&&0&&0&& \\
&0&&2&&0&\\
1&&86&&86&&1\\
&0&&2&&0&\\
&&0&&0&& \\
&&&1&&& \\
\end{array}
\label{tpHDI}
\eeq
\item[$\longrightarrow$]
Phase II:
\beq
\begin{array}{ccccccc}
&&&1&&& \\
&&0&&0&& \\
&0&&1&&0&\\
1&&83&&83&&1\\
&0&&1&&0&\\
&&0&&0&& \\
&&&1&&& \\
\end{array}
\label{tpHDII}
\eeq
\item[$\longrightarrow$]
Phase III:
\beq
\begin{array}{ccccccc}
&&&0&&& \\
&&0&&0&& \\
&0&&0&&0&\\
1&&83&&83&&1\\
&0&&0&&0&\\
&&0&&0&& \\
&&&0&&& \\
\end{array}
\label{tpHDIII}
\eeq
\item[$\longrightarrow$]
Phase IV:
\beq
\begin{array}{ccccccc}
&&&0&&& \\
&&0&&0&& \\
&0&&0&&0&\\
1&&86&&86&&1\\
&0&&0&&0&\\
&&0&&0&& \\
&&&0&&& \\
\end{array}
\label{tpHDIV}
\eeq
\item[$\longrightarrow$]
back to Phase I.
\end{itemize}

The Hodge diamond (\ref{tpHDIII}) of Phase III
(Landau-Ginzburg orbifold phase) can be derived independently.
First of all, the 2d Hodge diamond is the same as (\ref{tpHDI})
at all four phases because the model is Calabi-Yau.
In the Landau-Ginzburg orbifold phase, it is known
that the $1,2,2,1$ in the diagonal middle and $3$ out of the $86$ in
each of the two horizontal middle entries
are twisted sector states (see e.g. Section 2.1.1 of \cite{BHHW}
based on \cite{IntVaf})
and should not be taken in 1d.
The others are untwisted sector states and should be included,
thus giving (\ref{tpHDIII}).

There is an alternative approach to extract information
on the ground states --- it is to look at the ``full Coulomb branch''
where we take $\bfx$ to be large generic and integrate out
all the matter multiplets.
This induces an effective potential
\beq
U_{\rm eff}={e^2_{\rm eff}\over 2}\Bigl(\zeta_{\rm eff}\Bigr)^2,
\eeq
with $\zeta_{\rm eff}=\zeta-\sum_iQ_i/2|Q_i(\bfx)|$,
plus accompanying vector potential and Yukawa terms.
The effective Lagrangian is reliable in the region where $|\bfx|$
is large enough $|\bfx|^3\gg e^2$. Since the bottom of the potential
is typically at $|\bfx|\sim 1/|\zeta|$, we need $e^2|\zeta|^3\ll 1$
to find a valid ground state on the full Coulomb branch.
That is, the FI parameter must be very close to the
origin $\zeta=0$. It is expected, however, that states that are found
there will not disappear if we scale up $\zeta$.
In the present example, the vacuum equation $\zeta_{\rm eff}=0$ reads
\beq
\zeta^1={1\over |\bfx_1|}+{1\over 2|\bfx_1-2\bfx_2|},
\qquad
\zeta^2={1\over |\bfx_2|}-{1\over |\bfx_1-2\bfx_2|}.
\label{VEtwopara}
\eeq
We see that there is no solution in Phases III and IV,
while in each of Phases I and II,
the solution space is well inside the safe region provided
$|\zeta|$ is small enough.
Applying the well-established routine of equivariant index counting,
we find that the contribution of the full
Coulomb branch vacua to the index is
\beq
I_{\rm Coulomb}=\left\{
\begin{array}{lr} -\by^{-3}-2\by^{-1}-2\by^1-\by^3 & \qquad \mbox{Phase I} \\
-\by^{-3}-\by^{-1}-\by^{1}-\by^{3} & \qquad \mbox{Phase II}\\
0 & \mbox{Phases III and IV}
\end{array}\right.
\label{ICtwopara}
\eeq
The Coulomb branch vacua form an $SU(2)$ multiplet and are expected to
have zero $U(1)$ R-charge.
We see that the result (\ref{ICtwopara})
correctly reproduces the vertical middle entries of
the Hodge diamond at each phase.

It should be emphasized that the mixed branch analysis in
Section~\ref{subsec:simpleWC} and the above full Coulomb branch analysis
are valid in totally different regimes and are good
for different purposes.
The former is valid deep inside the phase boundaries and can capture the
full detail of the wall crossing states there, while the latter is valid
only near the origin $\zeta=0$ and can capture a part of
the $SU(2)$ multiplets in each phase.
For example, if we go far out in the I-IV phase boundary,
i.e. $\zeta^1\to 0$ and $\zeta^2\to +\infty$,
the vacuum equation (\ref{VEtwopara}) forces $|\bfx_2|$ to vanish,
where the full Coulomb branch is totally invalid.

This is in sharp contrast to two-dimensional theories
where it is necessary and valid to take into account
the full Coulomb branch and all possible mixed branches at the same time,
for a common purpose.
For example, in the two-dimensional version of the present model \cite{MP},
the vacuum equation on the full Coulomb branch reads
\beq
\e^{t^1}=4^4{\sigma_1\over \sigma_1-2\sigma_2},\qquad
\e^{t^2}={(\sigma_1-2\sigma_2)^2\over\sigma_2^2},
\label{2dfC}
\eeq
where $t^a=\zeta^a-i\theta^a$ are the FI-theta parameter.
Also, there is just one additional flat direction --- from the mixed
branch where $U(1)_2$ is unbroken and $U(1)_1$ is Higgsed ---
which exists when
\beq
\e^{t^2}=2^2.
\label{2dMX}
\eeq
(\ref{2dfC}) and (\ref{2dMX}) together define the discriminant locus in the
FI-theta parameter space.
\begin{figure}[htb]
\psfrag{z1}{$\zeta^1$}
\psfrag{z2}{$\zeta^2$}
\psfrag{I}{I}
\psfrag{II}{II}
\psfrag{III}{III}
\psfrag{IV}{IV}
\centerline{\includegraphics{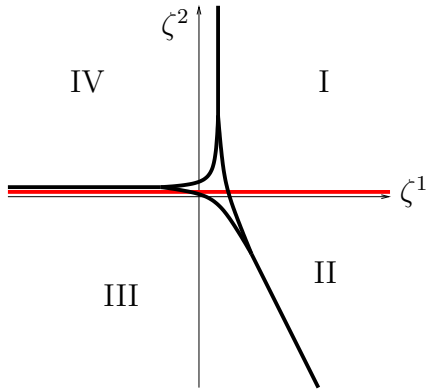}}
\caption{The $\zeta$-images of the discriminants (\ref{2dfC}) (black)
and (\ref{2dMX}) (red).}
\label{fig:2dtpmodel}
\end{figure}
Their $\zeta$-projection
asymptote to the phase boundaries.
% --- (\ref{2dfC}) yields I-IV, II-III and III-IV while (\ref{2dMX}) yields
%I-II and III-IV.
See Fig.~\ref{fig:2dtpmodel}.
 Note that the full Coulomb equation (\ref{2dfC})
allows valid solutions even in such asymptotic region.
For example, the asymptotic region of I-IV
phase boundary is $\e^{t^1}\to 4^4$ and $\e^{t^2}\to\infty$:
The vacuum equation does not force $\sigma_2\to 0$. We can take
$\sigma_1\to\infty$ holding $\sigma_2$ fixed and large.

In one dimension, to account for the wall crossing deep inside phase
 boundaries, we only need to consider the corresponding mixed branches.
This is supported by our index result which proves the
wall crossing formula obtained just from the mixed branch analysis.
There is no room for the full Coulomb branch to play any r\^ole there.
However, the full Coulomb branch is the key element of the recent
approach to index counting and wall crossing, in ${\mathcal N}=4$
quiver theories.
We turn to that subject next.

\section{${\cal N}=4$ Quivers and Quiver Invariants}
\label{sec:4Quiver}

Now we come to ${\cal N}=4$ quiver theories, which are
very important as they address 4d ${\mathcal N}=2$
wall-crossing phenomena. A given
quiver theory is a low energy dynamics of particular charge sector,
and our index computes the relevant protected
 spin character \cite{Gaiotto:2010be}
$${\rm Tr}\left[ (-1)^{2J_3}\by^{2(J_3+I_3)}\right]\ $$
with $I_3$ and $J_3$ being the helicity operators
of $SU(2)_R$ and
$SU(2)$ little group, respectively. The trace is supposed to
be taken after removing the universal half-hypermultiplet
factor in the supermultiplet, so that the value is +1 for a
half-hypermultiplet and $-\by-\by^{-1}$ for a vector
multiplet.
In particular the little group $SU(2)$ of $d=4$ descends
to $SU(2)_J$ $R$-symmetry while $U(1)_I$ is inherited
from $d=4$ $SU(2)_R$ $R$-symmetry \cite{Lee:2012sc}.

There has been extensive works for this class during last four years.
For quivers without loops, the index computation in the Coulomb
viewpoint have been carried out extensively, and lead to an
explicit and recursive and partition-based index formulae
\cite{Manschot:2010qz,Manschot:2011xc} based on a symplectic volume
hypothesis. This was later re-derived in an ab initio computation
from Seiberg-Witten theory \cite{Kim:2011sc}.
For loopless quivers,  it has been demonstrated \cite{Sen:2011aa}
that the
Coulomb branch analysis is sufficient in that the resulting
index agrees with an independent Higgs computation and also
with predictions coming from the Kontsevich-Soibelman
algebra \cite{KS}.

Quiver involving loops, and thus  superpotentials, pose more
challenges, as the Coulomb description often fails to
capture all
supersymmetric ground states. The additional states can be found
from the Higgs side \cite{Denef:2007vg,Lee:2012sc,Bena:2012hf}.
Some of these latter states, in particular, are impervious to
wall-crossing and have been dubbed quiver
 invariants \cite{Lee:2012sc,Lee:2012naa}.
These new kind of states serve as important building blocks in
general BPS state construction. They are also believed to represent
microstates of single-centered BPS black holes of $d=4$ ${\cal N}=2$ theories,
generalizing and fine-tuning Ashoke Sen's earlier proposal to
separate singe-center black holes from multi-center ones via angular
momentum content \cite{Sen:2009vz}.

With the new technology developed in this
paper, we are now at a position to address most general quivers
such as multi-loop quivers and non-Abelian quivers. In this section,
we address two particular classes; all cyclic Abelian quiver where
we reproduces results of Ref.~\cite{Lee:2012naa} and non-Abelian
triangle quivers where we will also test the generalized
Coulomb-like expansion of Refs.~\cite{Manschot:2013sya,Manschot:2014fua},
which fused the quiver invariants and the multi-center picture of
BPS states into a single practical machinery.

\subsection{Abelian $(k+1)$-Gon}\label{ngon}

Let us start exploring Abelian $(k+1)$-gon with arrows
$a=(a_1,a_2,\cdots, a_{k+1})$, which will serve as a prototype.
This class of theories were explored from both Higgs and Coulomb
branches \cite{Lee:2012sc,Bena:2012hf} and the complete equivariant
indices computed \cite{Lee:2012naa,Manschot:2012rx}. This also lead to
the notion of quiver invariants that we will come back to after
discussion of non-Abelian generalizations.

The model has gauge group $U(1)^{k+1}/U(1)_{\rm diag}$,
bifundamental matters $X_{1,\ldots,a_i}^{(i)}$ on the $i$-th edge and a generic
linear combination of monomials of the form
$X^{(k+1)}_{j_{k+1}}\cdots X^{(1)}_{j_1}$ as the superpotential.
The gauge group is isomorphic to $U(1)^k$, with $i\ttt^*\cong \R^k$,
under which the $i$-th group of matters have charge $Q_i=e_i$ for
$i=1,\ldots, k$, and the last group have charge $Q_{k+1}=-e_1-\cdots-e_k$.
There are $(k+1)$ phases ${\mathcal C}_1,\ldots, {\mathcal C}_{k+1}$
where ${\mathcal C}_i$ is the positive span of all the charges except
$Q_i$. We may assign R-charge 0 to the first $k$ groups of matters
and $2$ to the last group. Then, the $g$ function reads
\begin{eqnarray}\label{g}
&&g\,\,=\,\,\left(\frac{1}{2i\sin(\pi\bfz)}\right)^k
\times\\ \cr
&& \left(\frac{\sin(-\pi(u_1-\bfz))}{\sin(\pi u_1)}\right)^{a_1}
\left(\frac{\sin(-\pi(u_2-\bfz))}{\sin(\pi u_2)}\right)^{a_2}\cdots
\left(\frac{\sin(-\pi(-u_1-\cdots -u_k))}{\sin(\pi(-u_1-\cdots-u_k+\bfz))}
\right)^{a_{k+1}} \ .\nonumber
\end{eqnarray}
In the phase ${\mathcal C}_i$, we must take the residue at the
intersection of the singular hyperplanes of the fields except the $i$-group.
For ${\mathcal C}_{k+1}$, it is the standard residue at the origin,
\beq
I_{{\mathcal C}_{k+1}}\,\,=\,\,\int\limits_{|u_1|=\cdots=|u_k|=\vps}
\dd^k u\,\,g(u,\bfz).
\eeq
For ${\mathcal C}_i$ with $i=1,\ldots, k$, 
it is the residue at $u_j=0$ ($j\ne i$)
and $-u_1-\cdots -u_k+\bfz=0$.
Because of the sign in front of $u_i$ of the last equation,
we need to include an extra sign,
\beq
I_{{\mathcal C}_i}
\,\,=\,\,(-1)\int\limits_{|u_1|=\cdots\,\hat{i}\,\cdots=|u_k|=\vps\atop
|-u_1-\cdots -u_k+\bfz|=\vps}
\dd^k u\,\,g(u,\bfz).
\eeq
This exactly reproduces the results of Ref.~\cite{Lee:2012naa} which were
based on conventional index theorem for the Higgs branch,  also
demonstrating that the Higgs branch captures the entire index
for this class of theories in all chambers.

As a simple example, consider the case $k=2$ and $a=(6,5,4)$.
The is Example 3 in Section~\ref{subsec:LSMex} with rank vector
$(1,1,1)$ and the number vector $(a,b,c)=(6,5,4)$.
The three phases ${\mathcal C}_1$, ${\mathcal C}_2$, ${\mathcal C}_3$
are III, II, I respectively,
in the notation of Section~\ref{subsec:LSMex}
 (see Fig.~\ref{fig:q111}).
The index in these phases are
\beqa
I_{\rm I}&=&\int\limits_{|u_1|=|u_2|=\vps}\dd^2 u\,g(u,\bfz)
\,\,=\,\,-\frac{1}{\by^5}-\frac{2}{\by^3}+\frac{23}{\by}+23\by-2\by^3-\by^5,\\
I_{\rm II}&=&-\int\limits_{|u_1|=|-u_1-u_2+\bfz|=\vps}\dd^2 u\,g(u,\bfz)
\,\,=\,\,-\frac{1}{\by^3}+\frac{24}{\by}+24\by-\by^3,\\
I_{\rm III}&=&-\int\limits_{|u_2|=|-u_1-u_2+\bfz|=\vps}\dd^2 u\,g(u,\bfz)
\,\,=\,\,\frac{25}{\by}+25\by.
\eeqa
In Section~\ref{subsec:simpleWC},
we worked out the I $\to$ II wall crossing
of this example, as an illustration of the mixed branch analysis.
The formula obtained there, (\ref{wcAbTr}), reads
\beq
\Delta_{{\rm I}\,\to\,{\rm II}} I\,\,=
\,\,(-1)(\by^{-5}+\by^{-3}+\by^{-1}+\by+\by^3+\by^5)\cdot (-1).
\eeq
We see that $I_{\rm I}+\Delta_{{\rm I}\,\to\,{\rm II}} I=I_{\rm II}$
indeed holds. Of course, we can also look at the Hodge diamonds of
the phases and the wall crossing states, and find agreement.

\subsection{Non-Abelian Triangle Quivers of Ranks $(k,1,1)$}

Now we will generalize this to non-Abelian examples.
Here we will consider triangle quivers (Example 3,
Section~\ref{subsec:LSMex}). When a rank vector is $(k,1,1)$
and intersection numbers
are $(a,b,c)$, there exist four phases at most.
See Fig.~\ref{fig:qk11}.
The gauge group $[U(k)\times U(1)\times U(1)]/U(1)_{\rm diag}$
is isomorphic to $U(1)\times U(k)$ by
$[g,\omega^{-1},1]\longleftrightarrow (\omega,g)$.
With respect to a basis $\{e_0;e_1,\ldots,e_k\}$ of $i\ttt^*$,
where $e_0$ is for the $U(1)$ factor and
$e_{\alpha}-e_{\beta}$ are the roots of $U(k)$, the
matter charges and the FI parameter are written as
\beqa
&&Q_{X}=e_0, \qquad
Q_{Y^{\alpha}}=e_{\alpha},\qquad
Q_{Z_{\alpha}}=-e_0-e_{\alpha},\label{Qtriangle}\\
&&\zeta=\zeta^1e_0+{\zeta^2\over k}(e_1+\cdots+e_k).
\label{FItriangle2}
\eeqa
Here $\zeta^1$ and $\zeta^2$ are as in (\ref{FItriangle}).
We assign R-charge $2$ to $Z$'s and $0$ to others.
Then, the $g$ function reads
\beqa
g&=&\left({1\over 2i\sin(\pi\bfz)}\right)^{k+1}
\prod_{\alpha\ne\beta}{\sin(-\pi(u_{\alpha}-u_{\beta}))\over
\sin(\pi(u_{\alpha}-u_{\beta}-\bfz))}
\times\\
&&\left({\sin(-\pi(u_0-\bfz))\over\sin(\pi u_0)}\right)^{\!a}
\prod_{\alpha=1}^k
\left({\sin(-\pi(u_{\alpha}-\bfz))\over\sin(\pi u_{\alpha})}\right)^{\!b}
\prod_{\alpha=1}^k
\left({\sin(-\pi(-u_{\alpha}-u_0))\over\sin(\pi (-u_{\alpha}-u_0+\bfz)}
\right)^{\!c}.\nn
\eeqa
In chamber I, $\zeta$ is a positive span of
$Q_X$ and $Q_{Y^1},\ldots, Q_{Y^k}$, as is evident from (\ref{FItriangle2})
and (\ref{Qtriangle}). A non-zero contribution
comes only from the intersection of the corresponding hyperplanes,
$(u_0,u_1,\ldots,u_k)=(0,\ldots, 0)$,
\beq
I_{\rm I}\,\,=\,\,{1\over k!}\int\limits_{|u_0|=|u_1|=\cdots=|u_k|=\vps}
\dd^{k+1}u\,g(u,\bfz).
\eeq
In chamber II,
$\zeta=(\zeta^1-\zeta^2)Q_X+{\zeta^2\over k}(Q_{Z_1}+\cdots+Q_{Z_k})$
 is a positive span of $Q_X$ and $Q_{Z_{\alpha}}$'s.
A non-zero contribution
comes only from the intersection of the corresponding hyperplanes,
$(u_0,u_1,\ldots,u_k)=(0,\bfz,\ldots,\bfz)$,
\beq
I_{\rm II}\,\,=\,\,{(-1)^k\over k!}
\int\limits_{|u_0|=|-u_1-u_0+\bfz|=\cdots=|-u_k-u_0+\bfz|=\vps}
\dd^{k+1}u\,g(u,\bfz).
\eeq
In chamber IV,
 $\zeta=(-\zeta^1)(Q_{Y^{\alpha}}+Q_{Z_{\alpha}})
+{\zeta^2\over k}(Q_{Y^1}+\cdots +Q_{Y^k})$ is a positive span of
$Q_{Y^1},\ldots, Q_{Y^k}$ and $Q_{Y^{\alpha}}+Q_{Z_{\alpha}}$ for any $\alpha$,
while in chamber III,
$\zeta=(\zeta^2-\zeta^1)(Q_{Y^{\alpha}}+Q_{Z_{\alpha}})+{-\zeta^2\over k}
(Q_{Z_1}+\cdots+Q_{Z_k})$ is a positive span of
$Q_{Z_1},\ldots, Q_{Z_k}$ and $Q_{Y^{\alpha}}+Q_{Z_{\alpha}}$ for any $\alpha$.
In these chambers, a non-zero contribution can come only
from the point $p$ with $(u_0,u_1,\ldots,u_k)=(\bfz,0,\ldots,0)$,
which is a degenerate
intersection where $2k$ hyperplanes
$H_{Y^1}$, $\ldots,$ $H_{Y^k}$, $H_{Z_1}$, $\ldots,$ $H_{Z_k}$ meet.
At this point, we employ the constructive ``definition'' of the JK-residue
\cite{SV}. We first choose an element $\xi\in i\ttt^*$ which is
in the same chamber as
$\zeta$ with respect to the set
$Q(p)=\{Q_{Y^1},\ldots, Q_{Y^k},Q_{Z_1},\ldots, Q_{Z_k}\}$ 
and is generic with respect to all the partial sums of $Q(p)$.
Then, we set
\beq
\mathop{\mbox{JK-Res}}_{p}(Q(p),\zeta)
=\sum_{F\in \mathcal{FL}^+(Q(p),\xi)}\nu(F)\mathop{\rm Res}_F,
\label{sumflag}
\eeq
where the sum is over a certain set of flags of $i\ttt^*$ determined by
$Q(p)$ and $\xi$.
See \cite{SV} or its review given in \cite{BEHT2}.
Let us put
\beq
\xi(\epsilon)\,:=\,\zeta+\epsilon (ke_1+(k-1)e_2+\cdots +2e_{k-1}+e_k).
\eeq
For a small but non-zero $\epsilon$, this belongs to the same chamber as
$\zeta$ and is generic with respect to all the partial sums of $Q(p)$.
In chamber IV, let us choose $\xi=\xi(\epsilon)$ with a small positive
$\epsilon$. Then, the sum (\ref{sumflag}) consists of
a single term with the flag
\beq
\{0\}\subset \R Q_{Y^1}\subset \R Q_{Y^1}+\R Q_{Y^2}\subset\cdots\subset
\R Q_{Y^1}+\cdots +\R Q_{Y^k}\subset i\ttt^*,
\eeq
which yields
\beq
I_{\rm IV}\,\,=\,\,{(-1)\over k!}
\mathop{\mathrm{res}}_{u_0=\bfz}\cdot
\mathop{\mathrm{res}}_{u_k=0}\cdots
\mathop{\mathrm{res}}_{u_1=0} ~g(u,\bfz)\ .
\eeq
In chamber III, let us choose $\xi=\xi(\epsilon)$ with a small negative
$\epsilon$.
Then, the sum (\ref{sumflag}) consists of
a single term with the flag
\beq
\{0\}\subset \R Q_{Z_1}\subset \R Q_{Z_1}+\R Q_{Z_2}\subset\cdots\subset
\R Q_{Z_1}+\cdots +\R Q_{Z_k}\subset i\ttt^*,
\eeq
which yields
\beq
I_{\mathrm{III}}\,\,={(-1)^{k+1}\over k!}
~\mathop{\mathrm{res}}_{u_0=\bfz}\cdot
\mathop{\mathrm{res}}_{u_k=\bfz-u_0}\cdots
\mathop{\mathrm{res}}_{u_1=\bfz-u_0}
~g(u,\bfz)\ .
\eeq

Let us present the result of computation for low values of $k$ and $(a,b,c)$.
First, $k=2$. With $(a,b,c)=(4,1,4)$,
three chambers are empty and
\begin{eqnarray}\nonumber
I_{\mathrm{I}}&=&0\ ,\\\nonumber
I_{\mathrm{II}}&=&-\frac{1}{\by^5}-\frac{2}{\by^3}
-\frac{3}{\by}-3\by-2\by^3-\by^5\ ,\\\nonumber
I_{\mathrm{III}}&=&0\ ,\\\nonumber
I_{\mathrm{IV}}&=&0\ ,
\end{eqnarray}
which is consistent with the results derived previously from purely
geometrical method \cite{Lee:2013yka}. Let us look at the cases where
the indices are non-zero at all the four branches.
When the condition (\ref{condrank})
is obeyed, say $(a,b,c)=(7,3,4)$, the fourth phase boundary between
III and IV is actually moot, and no decay should be possible there.
In accordance we find, the indices at each chambers are
\begin{eqnarray}\nonumber
I_{\mathrm{I}}&=&10\ ,\\\nonumber
I_{\mathrm{II}}&=& \frac{1}{\by^4}+
\frac{2}{\by^2}+13+2\by^2+\by^4\ ,\\\nonumber
I_{\mathrm{III}}&=&\frac{1}{\by^2}+11+\by^2\ ,\\\nonumber
I_{\mathrm{IV}}&=&\frac{1}{\by^2}+11+\by^2\ .
\end{eqnarray}
If we take $(a,b,c)=(7,4,5)$, for which (\ref{condrank}) is violated,
the indices all four chambers are indeed different from one another as
\begin{eqnarray}
I_{\mathrm{I}}&=&50\ ,\label{745}\\\nonumber
I_{\mathrm{II}}&=& \frac{1}{\by^4}+
\frac{2}{\by^2}+87+2\by^2+\by^4\ ,\\\nonumber
I_{\mathrm{III}}&=&\frac{1}{\by^6}+
\frac{2}{\by^4}+\frac{4}{\by^2}+89
+4\by^2+2\by^4+\by^6\ ,\\\nonumber
I_{\mathrm{IV}}&=&\frac{1}{\by^6}+
\frac{2}{\by^4}+\frac{4}{\by^2}+54+4\by^2+2\by^4+\by^6\ .
\end{eqnarray}
Next, let us take $k=3$. For $(a,b,c)=(10,4,3)$, we find
\begin{eqnarray}
I_{\mathrm{I}}&=& -\frac{1}{\by^3}
-\frac{2}{\by}-2\by-\by^3\ ,\\\nonumber
I_{\mathrm{II}}&=&0\ ,\\\nonumber
I_{\mathrm{III}}&=&-\frac{1}{\by}-\by\ ,\\\nonumber
I_{\mathrm{IV}}&=&-\frac{1}{\by}-\by\ .
\end{eqnarray}
while for $k=3$ and $(a,b,c)=(10,5,3)$, we find
\begin{eqnarray}
I_{\mathrm{I}}&=& \frac{1}{\by^6}+
\frac{2}{\by^4}-\frac{2}{\by^2}
-7-2\by^2+2\by^4+\by^6\ ,\label{1053}\\\nonumber
I_{\mathrm{II}}&=&0\ ,\\\nonumber
I_{\mathrm{III}}&=&\frac{1}{\by^4}+
\frac{1}{\by^2}+1+\by^2+\by^4 \ ,\\\nonumber
I_{\mathrm{IV}}&=&\frac{1}{\by^4}-\frac{4}{\by^2}-9-4\by^2+\by^4\ .
\end{eqnarray}
The final example is for $k=3$ and $(a,b,c)=(10,4,4)$, which does have
its own quiver invariant. The indices are
\begin{eqnarray}
I_{\mathrm{I}}&=& 20\ ,\\\nonumber
I_{\mathrm{II}}&=&20\ ,\\\nonumber
I_{\mathrm{III}}&=&\frac{1}{\by^4}+
\frac{2}{\by^2}+22+2\by^2+\by^4 ,\\\nonumber
I_{\mathrm{IV}}&=& \frac{1}{\by^4}+
\frac{2}{\by^2}+22+2\by^2+\by^4  \ .
\end{eqnarray}

Note that I-IV and II-III walls are simple for any $k$,
and III-IV is also simple for $k=2$. There we can see how the simple
wall crossing formula works.

\subsubsection*{\underline{\rm I $\to$ IV}}

The data of the mixed branch is
\beqa
({\rm C})&U(1)\times \{{\bf 1}_k\}&
X_{1,\ldots,a}\, (1),~
Z^{1,\ldots, c}_{1,\ldots,k}\, (-1),\nn\\
({\rm H})&U(k)&Y_{1,\ldots,b}\, ({\bf k}).\nn
\eeqa
(C) has $\neff^{(4)}=a-kc$ and (H) has Higgs branch $G(k,b)$. This yields
\beq
\Delta I\,\,=\,\,I_{G(k,b)}\cdot
(-1)^{a-kc}{\rm sign}(a-kc)(\by^{-|a-kc|+1}+\cdots
+\by^{|a-kc|-1}).
\eeq

\subsubsection*{\underline{\rm II $\to$ III}}

The data of the mixed branch is
\beqa
({\rm C})&\{(\omega,\omega^{-1}{\bf 1}_k)\}&X_{1,\ldots,a}\, (1),~
Y_{1,\ldots, b}^{1,\ldots,k}\, (-1),\nn\\
({\rm H})&U(k)&Z^{1,\ldots,c}\, (\bar{\bf k}).\nn
\eeqa
(C) has $\neff^{(4)}=a-kb$ and (H) has Higgs branch $G(k,c)$. This yields
\beq
\Delta I\,\,=\,\,I_{G(k,c)}\cdot
(-1)^{a-kb}{\rm sign}(a-kb)(\by^{-|a-kb|+1}+\cdots
+\by^{|a-kb|-1}).
\eeq

\subsubsection*{\underline{\rm IV $\to$ III} {\rm ($k=2$ case)}}

The data of the mixed branch is
\beqa
({\rm C})&\{(1,{\omega~~\choose ~~1})\}&Y^1_{1,\ldots,b}\, (1),~
Z^{1,\ldots,c}_1\, (-1),\nn\\
({\rm H})&U(1)\times U(1)&X_{1,\ldots,a}\, (0,-1),~
Y^2_{1,\ldots, b}\,(1,0),~
Z_2^{1,\ldots,c}\,(-1,1);~\, W=\sum A^{ij}_hZ^h_2Y^2_jX_i.\nn
\eeqa
(C) has $\neff^{(4)}=b-c$ and (H) has complete intersection $M^a_{b,c}(A)$
of $a$ hypersurfaces in $\CP^{b-1}\times \CP^{c-1}$ as its Higgs branch.
Thus,
\beq
\Delta I\,\,=\,\,I_{M^a_{b,c}(A)}\cdot
(-1)^{b-c}{\rm sgn}(b-c)(\by^{-|b-c|+1}+\cdots+\by^{|b-c|-1}).
\eeq

These wall crossing formulae work
in all cases above. For example, for the I-IV wall in
the model with $k=3$, $(10,5,3)$,
we have $\Delta I=
-I_{G(3,5)}=-(\by^{-6}+\by^{-4}+2\by^{-2}+2+2\by^2+\by^4+\by^6)$,
in agreement with (\ref{1053}).
As another example, let us look at the IV-III wall in the model $k=2$,
$(7,4,5)$, where $\Delta I=I_{M^7_{4,5}(A)}$.
Note that $M^7_{4,5}(A)$ is the intersection of seven hyperplanes in
a seven dimensional manifold $\CP^3\times\CP^4$ and hence is a set of points,
as many as $(H_1+H_2)^7=35$.
This matches with (\ref{745}), as $54+35=89$.
The mixed branch analysis can also tell us on
the Hodge diamond of the wall crossing states.

\subsection{Quiver Invariants}

Now that we accumulated several examples, it is time to talk
about quiver invariant. It is instructive
to recall the explicit example given at the end of
subsection~\ref{ngon}.
The index and in fact the full cohomology of this example can be
worked out \cite{Lee:2012naa}. The Hodge diamond for the three
chambers are
\begin{equation}
\begin{array}{ccc}
&1& \\
26&&26 \ \\
&1& \\
\end{array}\ \quad
\begin{array}{ccccccc}
&&&1&&& \\
&&0&&0&& \\
&0&&2&&0&\\
0&&26&&26&&0\\
&0&&2&&0&\\
&&0&&0&& \\
&&&1&&& \\
\end{array} \  \quad
\begin{array}{ccccccccccc}
&&&&&1&&&&& \\
&&&&0&&0&&&& \\
&&&0&&2&&0&&&\\
&&0&&0&&0&&0&&\\
&0&&0&&3&&0&&0&\\
0&&0&&26&&26&&0&&0\\
&0&&0&&3&&0&&0&\\
&&0&&0&&0&&0&&\\
&&&0&&2&&0&&&\\
&&&&0&&0&&&& \\
&&&&&1&&&&& \\
\end{array}
\nonumber
\end{equation}
This set of Hodge diamond shows an interesting behavior.
The dimensions
of each chambers are different, and thus the size of the diamond. Yet,
notice how the nontrivial entries along the horizontal middle
remain the same throughout chambers. In fact, more detailed
analysis of states shows that the entries along the vertical middle
correspond to states that become destabilized
at the walls, while these
26+26 states remain stable even at the walls.

This kind of behavior has been observed
for all cyclic Abelian quivers,
as demonstrated precisely in Refs.~\cite{Lee:2012sc,Lee:2012naa}.
For this class, the Hodge diamond consists
of only two nontrivial lines,
vertical middle and horizontal middle, just
as in the above example; Again,
the vertical middle of the Hodge diamonds
correspond to states that are
detectable in the Coulomb description and
are invariably become
unstable at the walls, while states belonging to the horizontal
middle are visible only to the Higgs description and impervious to
wall-crossing instabilities. These wall-crossing-safe states have
been dubbed ``intrinsic Higgs states" \cite{Lee:2012sc}
or ``pure Higgs states" \cite{Bena:2012hf}.
The part of index counting these states only
shall be called
quiver  invariants, following
Refs.~\cite{Lee:2012sc,Lee:2012naa}, meaning that
for any given they represents part of spectrum invariant
across all chambers.

For these Abelian cyclic quiver, a state is either Coulomb-like,
meaning that it becomes a multi-center bound state at a marginal
stability wall, or ``intrinsic Higgs."   The split is clean,
so we can define  the quiver invariant as
\begin{equation}
I=\Omega_{\rm Inv}+I_{\rm Coulomb}\ .
\end{equation}
In the above Abelian triangle with linking numbers
$(6,5,4)$, for example,
the invariant part of the index is
\begin{equation}
\Omega^{1,1,1}_{6,5,4}\biggr\vert_{\rm Inv} = \frac{26}{\by}+26\by \ .
\end{equation}
There is a potential ambiguity at the center of the Hodge
diamond, i.e. $H^{(s/2,s/2)}$, when the vacuum manifold is of even
complex dimension $s$. For these cyclic Abelian quivers, the
ambiguity was neatly resolved by  either the Lefshetz hyperplane
theorem or by explicit counting/construction in the
Coulomb description \cite{Lee:2012sc}.

For general quiver theories,
such neat separation of states into two classes
cannot be true any more. As we saw earlier, for
theories with rank large than 1,
wall-crossing states are often  of hybrid type,
approximately a product of
Coulomb-like multi-center wavefunction and tightly
bound Higgs-like state.
This implies that wall-crossing states, i.e.
states that become unstable
at a wall can involve quiver invariants of
one or more subquivers as
building blocks. Generally, given a quiver,
one should  find purely
Coulomb-like states which are multi-center
states made from
elementary centers, mixed states which are
again multi-centered but with
one or more centers belong to the quiver
invariant of a subquiver,
and then finally a single-centered
wall-crossing-safe states, all
contributing to the full index.
The Hodge diamond will no longer be
as simple as above, yet it is clear
that there are states along the
horizontal middle that are impervious to wall-crossing.

This idea, much in the spirit of mixed phase
analysis we saw earlier,
is physically appealing but complex to
incorporate;  Ref.~\cite{Manschot:2013sya}
proposed a routine to do this, although
while keeping the quiver invariants
as unknown input data. The
idea is to break up a quiver to all
possible subquivers, labeled $A$,
and then consider the quiver invariant
$\Omega_{\rm Inv}^{A}$ for each
of them. Each of these will act as a
building block for wall-crossing
states of the full quiver. Such states
can no longer be considered invariant
under wall-crossing, yet cannot be
captured correctly if one only
count the ordinary Coulombic states.

 This tells us that
the full index, if we view it from the wall-crossing perspective,
will have to involve a sum over many partitions of the
quiver into subquivers, say, $\{A_i\}$. Schematically, we
have
\begin{eqnarray}
I=\sum_{\{A_i\}}
I^{\rm Coulomb}_{\{A_i\}}\times \prod_i \Omega^{A_i}_{\rm Inv}\ ,
\end{eqnarray}
where each summand treats a  subquiver $A_i$ as if it is a
single charge center with the intrinsic degeneracy
$\Omega^{A_i}_{\rm Inv}$,
and the index $I^{\rm Coulomb}_{\{A_i\}}$
counts the index of the resulting multi-center dynamics.
Much of machinery
developed in Refs.~\cite{Manschot:2010qz,Kim:2011sc,Manschot:2011xc}
can be used not only for Coulomb branches of  quiver theories but
also  for any multi-center bound state problems in $\mathbb{R}^3$
with four supersymmetries, so $I^{\rm Coulomb}_{\{A_i\}}$ makes
sense as a multi-center index.
Computational procedure for $I^{\rm Coulomb}$
that enter in this expansion is very involved but well-established.
Because the index must correctly reflect the necessary Weyl projections,
actual formulae are a bit more involved than this naive one;
Please see Ref.~\cite{Manschot:2014fua} for the complete prescription.

The sum includes the case where the subquiver is the original quiver
itself, which is precisely
where wall-crossing-safe states of the original quiver enters and
contributes a term $1\times \Omega_{\rm Inv}$. In the other extreme,
the sum also includes elementary subquivers, i.e., a single node with
rank 1, for each of which $\Omega_{\rm Inv}=1$. This shows that the
quiver invariant generalizes the notion of intrinsic degeneracy
of a hypermultiplet BPS particle to that of a single center
BPS state of large internal degeneracy.

Theory by theory,
isolating such invariant states is a well-defined, if tedious,
exercise. All one needs to do is to solve for ground states
and observe which states become unstable at marginal stability
walls and which state never does. In particular, one can send all
FI constant to ``zero" where all marginal stability walls collide,
leaving behind only  invariant states. Systematic counting is a
different matter since these theories are rather difficult to handle
at vanishing FI parameters.

In view of this situation, one could choose a practical viewpoint
and consider this equation as (recursive) definition of the quiver
invariants. Then one begins to compute left hand sides,
we can refer to the right hand side and extract quiver invariants.
The invariant of one quiver will appear repeatedly in the index of larger
quivers that can accommodate the former as a subquiver. As we
compute the index of quivers with increasing complexity and ranks,
we end up with a hugely over-determined problem therefore.
The fact that we always get consistent answers for $\Omega_{\rm Inv}$
in examples below nevertheless, goes a long way to confirm that
the concept of the quiver invariant is robust and also that the
above expansion scheme actually works and catalog faithfully
all possible Coulombic, mixed, and invariant
states.

Let us now scan through examples of the above triangle quivers.
For rank $(2,1,1)$ and intersection numbers $(4,1,4)$, three of
the four chambers are empty, implying vanishing quiver
invariant. In fact, the same is true of rank $(1,1,1)$ version
of the same quiver, and we find
\begin{eqnarray}
\Omega^{2,1,1}_{4,1,4}\biggr\vert_{\rm Inv}
=0=\Omega^{1,1,1}_{4,1,4}\biggr\vert_{\rm Inv}\ .
\end{eqnarray}
For intersections $(7,3,4)$, quiver invariant for
rank $(1,1,1)$ is known to be null,
and we have
\begin{eqnarray}
I_{\rm I,II,III,IV}=I_{\rm I,II,III,IV}^{\rm Coulomb}
+\Omega^{2,1,1}_{7,3,4}\biggr\vert_{\rm Inv}
\end{eqnarray}
from which we read off
\begin{equation}
\Omega^{2,1,1}_{7,3,4}\biggr\vert_{\rm Inv}=9 \ .
\end{equation}
Next, with intersections $(7,4,5)$, both $(1,1,1)$ and $(2,1,1)$ quivers
have nontrivial quiver invariants, and they show up for $(2,1,1)$ quiver index
as
\begin{eqnarray}
I_{\rm I,IV}=I_{\rm I,IV}^{\rm Coulomb}
+\Omega^{2,1,1}_{7,4,5}\biggr\vert_{\rm Inv}
\end{eqnarray}
and
\begin{eqnarray}
I_{\rm II,III}=I_{\rm II,III}^{\rm Coulomb}
+1\times\Omega^{1,1,1}_{7,4,5}
\biggr\vert_{\rm Inv}+\Omega^{2,1,1}_{7,4,5}
\biggr\vert_{\rm Inv} \ .
\end{eqnarray}
Thus, we find
\begin{eqnarray}
\Omega^{2,1,1}_{7,4,5}\biggr\vert_{\rm Inv}=49\ , \qquad
\Omega^{1,1,1}_{7,4,5}\biggr\vert_{\rm Inv}=34 \ ,
\end{eqnarray}
where the latter number is independently confirmed by studying  rank
$(1,1,1)$ quiver.

The least interesting examples among the rank $(3,1,1)$ quiver is
the one with intersection numbers $(10,4,3)$, because for each
ranks $(k\le 3,1,1)$ there is at least one empty chambers. This
is borne out in the index computation also, and we find
\begin{equation}
\Omega^{3,1,1}_{10,4,3}\biggr\vert_{\rm Inv}=
\Omega^{2,1,1}_{10,4,3}\biggr\vert_{\rm Inv}
=\Omega^{1,1,1}_{10,4,3}\biggr\vert_{\rm Inv} = 0  \ ,
\end{equation}
and $I_{\rm I,II,III,IV}$ are all purely Coulombic for $k=1,2,3.$
Intersection numbers $(10,5,3)$ is a little more interesting. Although,
rank $(3,1,1)$ and rank $(1,1,1)$ quivers both have an empty chamber
so that
\begin{equation}
\Omega^{3,1,1}_{10,5,3}\vert_{\rm Inv}=0
=\Omega^{1,1,1}_{10,5,3}\vert_{\rm Inv}\ .
\end{equation}
The quiver invariant of the $k=2$ subquiver is nontrivial and
enters the index of this $k=3$ quiver as
\begin{eqnarray}
I_{\rm I,IV}=I_{\rm I,IV}^{\rm Coulomb}
+\left(-\frac{1}{\by}-\by\right)\times
\Omega^{2,1,1}_{10,5,3}\biggr\vert_{\rm Inv}
\end{eqnarray}
and
\begin{eqnarray}
I_{\rm II,III}=I_{\rm II,III}^{\rm Coulomb}\ .
\end{eqnarray}
Again comparison with the Coulomb computation shows,
\begin{eqnarray}
\Omega^{2,1,1}_{10,5,3}\biggr\vert_{\rm Inv}=\frac{6}{\by}+6\by \ ,
\end{eqnarray}
which is again independently verified by directly
computing Coulomb and
Higgs indices for the $(2,1,1)$ quiver with $(a,b,c)=(10,5,3)$.
The final example is $(a,b,c)=(10,4,4)$, for which
\begin{equation}
\Omega^{3,1,1}_{10,4,4}\biggr\vert_{\rm Inv} =19 \ ,\quad
\Omega^{2,1,1}_{10,4,4}\biggr\vert_{\rm Inv}=
\Omega^{1,1,1}_{10,4,4}\biggr\vert_{\rm Inv}=0\ .
\end{equation}
All these examples exhibit the expected behavior of
how the quiver invariants  in a self-consistent manner.

The interesting dichotomy seen in the cyclic Abelian quivers,
between the Coulomb-like $SU(2)_R$ multiplets and the invariant $U(1)_R$
states suggest that we can compute not only the indices but even
the full Hodge diamonds, through  such recursive constructions of
 BPS states.\footnote{We thank Ashoke Sen for bringing this aspect
 to our attention.} Whether theses kind of ideas, including the notion of
 invariant part of spectrum and the recursive build-up of BPS states
 will work for more general gauged quantum mechanics remains to be seen.

\section{Systems with ${\mathcal N}=2$ Supersymmetry}
\label{sec:2Ex}

In this section, we illustrate the index formula and wall crossing
for ${\mathcal N}=2$ supersymmetric theories
which do not have ${\mathcal N}=4$ supersymmetry.
As explained in Section~\ref{subsec:DefInd}, dependence on the
flavor twist is the most interesting part.

\subsection{The $\mathbf{CP}^{N-1}$ Model}\label{subsec:2CPN}

The first example is the ${\cal N}=2$ linear sigma model for
$\CP^{N-1}$: the theory with gauge group $U(1)$,
$N$ chiral multiplet and Wilson line of charge $\rmq$,
such that $\frac{N}{2}-\rmq\in \Z$.
We have already computed the wall crossing formula $\Delta I$
in (\ref{wcCPN1})-(\ref{wcCPN3}). Since there is no negatively charged
chiral multiplets, we have $I=0$ for $\zeta<0$. Combining these,
we find that $I=-\Delta I$ for $\zeta>0$, which agrees with
the result in geometric phase (\ref{hCPN1})-(\ref{hCPN3}).

Here we would like to compute the index (for $\zeta>0$)
with the twist by the $SU(N)$ flavor symmetry under which the chiral
multiplets form the fundamental representation. The
twist parameter is written as $(y_1,\ldots, y_N)$ obeying
$y_1\cdots y_N=1$. Then, the $g$ function is
\beq
g\,=\,{x^{\rmq}\over
\displaystyle
\prod_{i=1}^N\left(x^{\half}y_i^{\half}-x^{-\half}y_i^{-\half}\right)}
={x^{\rmq+{N\over 2}}\over
\displaystyle
\prod_{i=1}^N\left(x-\vps_i\right)},
\eeq
where we write $y_i^{-1}=\varepsilon_i$ to simplify the expressions.
Therefore, the index is
\beqa
I&=&\sum_{i=1}^N{\vps_i^{\rmq+{N\over 2}-1}\over \displaystyle
\prod_{j\ne i}\left(\vps_i-\vps_j\right)}\nn\\
&=&{\displaystyle
\sum_{s\in \mathfrak{S}_N}(-1)^{\ell(s)}
\vps_{s(1)}^{\rmq+{N\over 2}-1}\vps_{s(2)}^{N-2}\cdots \vps_{s(N-1)}
\over \displaystyle \prod_{i<j}(\vps_i-\vps_j)} \ .
\label{ICPN}
\eeqa
For $\rmq\geq {N\over 2}$, the right hand side
is the Weyl character formula\footnote{The character of
the irreducible representation of highest weight $\lambda$
of a simple Lie group is
\beq
\chi_{\lambda}(\vps)={\sum_{w\in W}(-1)^{\ell(w)}\vps^{w(\lambda+\rho)}
\over \vps^{\rho}\prod_{\alpha>0}(1-\vps^{-\alpha})}
\eeq
where $\rho$ is half the sum of positive roots. For $SU(N)$, we have
$\vps^{\rho}=\vps_1^{N-1}\vps_2^{N-2}\cdots \vps_{N-1}$
and the denominator of this formula is $\prod_{i<j}(\vps_i-\vps_j)$.
}
for the irreducible representation of $SU(N)$ with highest weight
$\vps_1^{\rmq-{N\over 2}}$, i.e., the $\left(\rmq-{N\over 2}\right)$-th
symmetric power of the fundamental representation $\C^N$.
Since $\vps_i=y_i^{-1}$, the actual representation is its dual,
$S^{\rmq-{N\over 2}}(\C^N)^*$. In fact we had seen this already on the
Coulomb branch (\ref{wfCPN1}). Also, this is consistent with the fact
that holomorphic sections of ${\mathcal O}\left(\rmq-{N\over 2}\right)$
over $\CP^{N-1}$ are polynomials of degree $\left(\rmq-{N\over 2}\right)$
of the homogeneous coordinates, and hence transform as
$S^{\rmq-{N\over 2}}(\C^N)^*$ under $SU(N)$.
For $|\rmq|<{N\over 2}$, this vanishes since a pair of $\vps_i$'s have the
same power in each term of the numerator on the right hand side,
and hence there is a pairwise cancellation in the permutation sum.
The vanishing $I=0$ can also be seen easily by deforming the
$x$-integration contour to $0$ and $\infty$.
For $\rmq\leq-{N\over 2}$, it is better to write
the first line of (\ref{ICPN}) in terms of $y_i$'s as
\beq
I=(-1)^{N-1}\sum_{i=1}^N{y_i^{-\rmq+{N\over 2}-1}\over \displaystyle
\prod_{j\ne i}\left(y_i-y_j\right)}.
\eeq
Up to the sign,
this is the Weyl character of the representation $S^{-\rmq-{N\over 2}}\C^N$
of $SU(N)$. This is consistent with the expressions of the wavefunctions on
the Coulomb branch (\ref{wfCPN2}).

The above result is consistent with a particular case
of the Borel-Weil-Bott theorem \cite{BW,BWB}: As a representation
of $SL(N,\C)$,
\beq
H^i(\CP^{N-1},{\mathcal O}(j))\,\,\cong\,\,\left\{
\begin{array}{ll}
S^j(\C^N)^*&i=0,\,\,j\geq 0\\
S^{-j-N}\C^N&i=N-1,\,\,j\leq -N,\\
0&\mbox{else}.
\end{array}\right.
\eeq

\subsection{Grassmannian}

We next consider the model with gauge group $U(k)$,
 $N$ chiral multiplets in the fundamental representation,
and the Wilson line in the $\det^{\rmq}$ representation.
The anomaly free condition is
$\rmq+{N\over 2}\in\Z$. When $N\geq k$,
the chamber $\zeta>0$ is the geometric phase with
 the Grassmannian $G(k,N)$ supporting the line bundle
$\mathfrak{F}={\mathcal O}(\rmq-{N\over 2})$.
For $N<k$ or for $\zeta<0$, supersymmetry is broken.

Let us compute the index with the twist by the $SU(N)$ flavor symmetry.
The $g$ function is
\beqa
g&=&\prod_{a\ne b}\left(x_a^{\half}x_b^{-\half}-x_a^{-\half}x_b^{\half}\right)
{\displaystyle (x_1\cdots x_k)^{\rmq}\over
\displaystyle
\prod_{i=1}^N\prod_{a=1}^k
\left(x_a^{\half}y_i^{\half}-x_a^{-\half}y_i^{-\half}\right)}\nn\\
&=&\prod_{a\ne b}(x_a-x_b)
{\displaystyle (x_1\cdots x_k)^{\rmq+{N\over 2}-k+1}\over \displaystyle
\prod_{i,a}(x_a-\vps_i)},
\eeqa
where we set $y_i^{-1}=\vps_i$.
The singular hyperplanes are at $x_a=\vps_i$ and the isolated intersections
are where, for each $a$, $x_a=y_{i_a}$ for some $i_a$.
Note that the residue vanishes when $i_a=i_b$ for $a\ne b$. In particular,
the index vanishes when $N<k$ where this is unavoidable.
Since we have $1/|W|=1/|\mathfrak{S}_k|$ in front,
we just have to sum over a subset $\{i_1,\ldots, i_k\}$
of $\{1,\ldots, N\}$ consisting of $k$ distinct (unordered) elements:
\beqa
I&=&\sum_{\{i_1,\ldots,i_k\}\subset \{1,\ldots, N\}}
\prod_{a\ne b}(\vps_{i_a}-\vps_{i_b})
{\displaystyle(\vps_{i_1}\cdots \vps_{i_k})^{\rmq+{N\over 2}-k}
\over \displaystyle
\prod_{a=1}^k\prod_{j\ne i_a}(\vps_{i_a}-\vps_j)}
\nn\\
&=&\sum_{\rmI\subset \{1,\ldots, N\}\atop
|\rmI|=k}{\displaystyle
\Bigl(\prod_{i\in \rmI}\vps_i\Bigr)^{\rmq+{N\over 2}-k}\over\displaystyle
\prod_{i\in\rmI}\prod_{j\not\in\rmI}(\vps_i-\vps_j)}
\,\,=\,\,{\displaystyle
\sum_{s\in \mathfrak{S}_N}(-1)^{\ell(s)}
s\left[(\vps_1\cdots\vps_k)^{\rmq-{N\over 2}}\vps^{\rho}\right]
\over\displaystyle\prod_{i<j}(\vps_i-\vps_j)}.
\label{IGkN}
\eeqa
For $\rmq\geq {N\over 2}$, the right hand side
is the Weyl character formula
for the irreducible representation of $SU(N)$ with highest weight
$(\vps_1\cdots\vps_k)^{\rmq-{N\over 2}}$.
Since $\vps_i=y_i^{-1}$, it is the character of the dual of that
representation.
By the Weyl dimension formula, the untwisted index is
\beq
I|_{y_1=\cdots y_N=1}=\prod_{i=1}^k{\prod_{j=k+1}^N(\rmq-{N\over 2}+j-i)
\over \prod_{j=k+1}^N(j-i)}
=\prod_{i=1}^k{(\rmq+{N\over 2}-i)!(k-i)!\over
(\rmq-{N\over 2}-i)!(N-i)!}.
\eeq
For $|\rmq|<{N\over 2}$, the index vanishes, $I=0$,
since a pair of $\vps_i$'s have the same power in each term of the numerator
of (\ref{IGkN}),
yielding pairwise cancellation of the permutation sum.
For $\rmq\leq -{N\over 2}$, it is more convenient to use $y_i$'s,
in terms of which
the middle expression of (\ref{IGkN}) can be written as
\beq
I\,\,=\,\,(-1)^{k(N-k)}
\sum_{\rmI\subset \{1,\ldots, N\}\atop
|\rmI|=k}{\displaystyle
\Bigl(\prod_{i\in \rmI}y_i\Bigr)^{-\rmq+{N\over 2}-k}\over\displaystyle
\prod_{i\in\rmI}\prod_{j\not\in\rmI}(y_i-y_j)}.
\eeq
Up to the sign $(-1)^{k(N-k)}$,
this is the character of the irreducible representation of $SU(N)$
with the highest weight $(y_1\cdots y_k)^{-\rmq-{N\over 2}}$.

The above result is consistent with a particular case
of the Borel-Weil-Bott theorem \cite{BW,BWB}: As a representation
of $SL(N,\C)$,
\beq
H^i(G(k,N),{\mathcal O}(j))\,\,\cong\,\,\left\{
\begin{array}{ll}
V_{j(e_1+\cdots+e_k)}^*&i=0,\,\,j\geq 0\\
V_{(-j-N)(e_1+\cdots+e_k)}&i=k(N-k),\,\,j\leq -N,\\
0&\mbox{else}.
\end{array}\right.
\eeq

\subsection{Distler-Kachru Model}

We next consider the Distler-Kachru model (Example 2,
Section~\ref{subsec:LSMex}). We recall the data of the model in the
table below:
\beq
\begin{array}{c|cccc}
&X_{1,\ldots,N}&P&\xi&\eta^{1,\ldots, M}\\
\hline
U(1)&1&-\ell~~&-d~~&q_{1,\ldots, M}\\
U(1)_F&0&1&0&-1~~
\end{array}\label{dataDK}
\eeq
The $g$ function is
\beq
g=
{(x^{d\over 2}-x^{-{d\over 2}})
\prod_{\alpha}(x^{-{q_{\alpha}\over 2}}y^{1\over 2}-
x^{q_{\alpha}\over 2}y^{-{1\over 2}})\over
(x^{1\over 2}-x^{-{1\over 2}})^N
(x^{-{\ell\over 2}}y^{1\over 2}-x^{\ell\over 2}y^{-{1\over 2}})}
=(-1)^{M+1}y^{1-M\over 2}x^{-k}{(x^d-1)\prod_{\alpha}(x^{q_{\alpha}}-y)
\over (x-1)^N(x^{\ell}-1)}
\eeq
where
\beq
k:={1\over 2}\left(d+\sum_{\alpha}q_{\alpha}-N-\ell\right).
\label{DKkdef}
\eeq
Note that the anomaly free condition is $k\in \Z$.
Index computation is strightforward.
Let us consider the case $M=N-1$, $q_{\alpha}=1$, $\ell=2$, in which case
$k={d-3\over 2}$  ($d$ must be odd).
Then, the index $I_{N,d}$ is, in the geometric phase,
\beqa
I^+_{2,d}&=&d,\label{DK:Iisd}\\
I^+_{3,d}&=&-d y^{-\half}+dy^{\half},\\
I^+_{4,d}&=&{d^3+23d\over 24}y^{-1}-{d^3-d\over 12}+{d^3-d\over 24}y,\\
I^+_{5,d}&=&-{d^3+11 d\over 12}y^{-{3\over 2}}
+{d^3-d\over 6}y^{-\half}-{d^3-13d\over 12}y^{\half}+0\cdot y^{3\over 2},\\
&&\cdots\nn
\eeqa
and in the Landau-Ginzburg phase,
\beqa
I^-_{N,d}&=&\half y^{-{d-1\over 4}}\left(\,
{y^{d\over 2}-1\over y^\half -1}\,
+\,(-1)^{{d-1\over 2}+N}\,{y^{d\over 2}+1\over y^\half +1}\,\right)
\label{IDKLG}\\
&=&
\left\{\begin{array}{ll}
\displaystyle y^{-[{d-1\over 4}]}+\cdots
+y^{-1}+1+y+\cdots +y^{[{d-1\over 4}]}&\mbox{$N$ even},\\
\displaystyle y^{-[{d+1\over 4}]+\half}+\cdots
+y^{-\half}+y^{\half}+\cdots +y^{[{d+1\over 4}]-\half}&\mbox{$N$ odd}.
\end{array}\right.
\nn
\eeqa

The simplicity in the Landau-Ginzburg phase is remarkable.
Let us try to understand it. Recall that the model has
gauge group $\Z_{\ell}=\Z_2$ and has chiral multiplets
$X_1,\ldots,X_N$ of $\Z_2\times U(1)_F$ charge $(1,\half)$,
a fermi multiplet $\xi$ of charge $(d,-{d\over 2})$
and fermi multiplets
$\eta^1,\ldots, \eta^{N-1}$ of
charge $(1,-{1\over 2})$. The $U(1)_F$ charges are different from
(\ref{dataDK}) but that is because $P$ has a non-zero value and
its $U(1)_F$ charge has to be cancelled by dressing it with the original
$U(1)$ gauge symmetry. The model also has the superpotential
$\mathfrak{W}=f(X)\xi+\sum_{\alpha}g_{\alpha}(X)\eta^{\alpha}$.
In the present model, $g_{\alpha}(X)$ are linear in $X$
and $f(X)$ is of degree $d$ in $X$.
From the genericity requirement stated in Section~\ref{subsec:LSMex},
we can find a coordinate such that $g_1(X)=X_2,\ldots, g_{N-1}(X)=X_N$,
and that the coefficient of $X_1^d$ in $f(X)$ is non-zero.
The space of supersymmetric ground states is the $\mathfrak{W}$-cohomology
space before the orbifold projection,
and it is easy to see that they are generated by elements of the form
$X_1^j|0\rangle$ ($j=0,1,\ldots, d-1$),
where $|0\rangle$ is the Clifford vacuum annihilated by $\xi$
and $\eta^{\alpha}$'s.
Orbifold projection is done by selecting $\Z_2$ invariant elements,
but the problem is to find how $\Z_2$ acts on $|0\rangle$.
To see this we look at the charge of the Clifford vacuum under the original
$U(1)$ gauge group. The vacuum $|0\rangle$ is annihilated also by
$\psi_i$'s and $\psi_P$. Then, the conjugate invariant charge assignment
fixes the charge of $|0\rangle$ to be $N-{d-1\over 2}$.
In this way, we find that the generator of $\Z_2$ acts as
\beq
|0\rangle\,\,\longmapsto\,\,(-1)^{{d-1\over 2}+N}|0\rangle.
\eeq
This determines which of $X_1^j|0\rangle$ should be selected.
It is evident that the resulting index, with $(-1)^F=1$
on $|0\rangle$, is equal to (\ref{IDKLG}).

Let us see how the wall crossing happens, using our analysis on
the Coulomb branch in Section~\ref{subsec:CgenU1}.
We consider the case $N=2$ in the above series.

The index in the geometric phase is $d$ as in (\ref{DK:Iisd}).
This can be directly checked as follows.
For $\zeta\gg 0$, the low energy theory is the non-linear sigma model
on the hypersurface $f(X)=0$ of $\CP^1$ with the fermi multiplet with values in
the kernel of $g_1(X)$. The hypersurface consists of $d$ distinct points
for a generic choice of $f(X)$, and $g_1(X)$ is non-zero at each of them.
Thus, the target space is the set of $d$ points and there is no fermi
multiplet. We see that there are $d$ supersymmetric ground states
which are all bosonic and have $U(1)_F$ charge zero. 

Let us now work out the Coulomb branch. The vacuum equation
(\ref{gveqneg})-(\ref{gveqpos}) reads
\beqa
\vs<0:&&n_{X}=0,\quad \bar n_P=0,\quad m_{X}=0,\quad m_P=1,\nn\\
&&|\bar n_{X}|+2 |n_P|+d m_{\xi}
+(1-m_{\eta})=k.\label{DKveq2}\\
\vs>0:&&\bar n_{X}=0,\quad n_P=0,\quad |m_{X}|=2,\quad m_P=0,\nn\\
&&|n_{X}|+2 |\bar n_P|+d(1-m_{\xi})
+m_{\eta}=k,\label{DKveq1}
\eeqa
Let us discuss each $d$ one by one.

When $d=1$, $k$ is negative, and there is no solution.
That is, there should be no wall crossing.  This is indeed consistent with
$I^+_{2,1}=I^-_{2,1}=1$. Also, we have seen that
there is a unique ground state of $U(1)_F$ charge zero,
for both $\zeta\gg 0$ and $\zeta\ll 0$. It is natural to expect that
it is the unique ground state of the system all the way from $\zeta\gg 0$
to $\zeta\ll 0$.

When $d=3$, $k$ is zero, and the above equation has a unique
solution with all $n$'s zero.
This means $\neff=1>0$. Therefore, for $\zeta>0$,
there is a unique Coulomb branch vacuum
for each of $\vs<0$ and $\vs>0$, both bosonic and with $Q_F=0$.
For $\zeta<0$ there is none.
That is, there are two wall crossing states that go out as
$\zeta$ goes from positive to negative.
This is consistent with $I^+_{2,3}=3$ and $I^-_{2,3}=1$, and also to
the fact that there are three supersymmetric
ground states for $\zeta\gg 0$ and one for $\zeta\ll 0$,
all with vanishing $U(1)_F$ charge.
As $\zeta$ goes from $+\infty$ to $-\infty$,
two of the three ground states become wall crossing states and go out,
while one of them stays till the end.
We may write the process as
\beq
%d=5:
\begin{array}{c||c}
Q_F&0\\
\hline\hline
\zeta\gg 0&\mbox{$\circ$ $\circ$ $\circ$}\\
\zeta\sim +0&
\mbox{$\circ$ $\underline{\circ}$ $\underline{\circ}$}\\
\hline
\zeta\sim -0&\circ\\
\zeta\ll&\circ
\end{array}
\eeq
Here, a white circle means a bosonic supersymmetric 
state and an underlined circle means
a wall crossing state that goes out as $\zeta\searrow 0$.

When $d=5$, then $k=1$, and there are six solutions in total,
all having $\neff>0$.
Thus, they correspond to wall crossing states at small positive $\zeta$
that go out as $\zeta\searrow 0$. Their profiles are as follows:
\beq
\begin{array}{c|cccc}
{\rm support}&{\rm state}&(-1)^F&Q_F&{\rm number}\\
\hline
\vs<0&|\vs|\e^{\zeta\vs}a_{\bar X}^{\dag}\,\bareta\,\bpsi_P
|0\rangle_{{}_{\rm OSC}}
&+1&0&2\\
&|\vs|^{{1\over 2}}\e^{\zeta\vs}\bpsi_P
|0\rangle_{{}_{\rm OSC}}
&-1&1&1\\
\vs>0&
|\vs|\e^{-\zeta\vs}a_{X}^{\dag}\,\bxi\,\bpsi_1\bpsi_2\blambda
|0\rangle_{{}_{\rm OSC}}
&+1&0&2\\
&|\vs|^{1\over 2}\e^{-\zeta\vs}\,\bxi\,\bareta\,\bpsi_1\bpsi_2\blambda
|0\rangle_{{}_{\rm OSC}}&-1&-1&1
\end{array}
\eeq
They induce the following change of the index,
\beq
\Delta_CI(y^{Q_F})=y^{-1}-4+y.
\eeq
This is consistent with $I^+_{2,5}=5$
and $I^-_{2,5}=y^{-1}+1+y$. However, the total number of states (six)
is bigger than the number (five) of the ground states at the starting
point, and some of them are even fermionic.
This may be understood as follows:
As $\zeta$ is decreased from large positive to small positive,
two boson-fermion pairs of states, one pair with charge $1$ and another pair
with charge $-1$, descend to have zero energy.
As $\zeta\searrow 0$, four of the bosonic states of charge $0$ and
the fermionic states of charges $1$ and $-1$ run away to
infinity.
For $\zeta<0$, there remain three bosonic states of charge $-1,0,1$
supported at the origin $\vs=0$.
This process may be written as
\beq
%d=5:
\begin{array}{c||c|c|c}
Q_F&-1&0&1\\
\hline\hline
\zeta\gg 0&&\mbox{$\circ$ $\circ$ $\circ$ $\circ$ $\circ$}&\\
\zeta\sim +0&\mbox{$\circ$ $\underline{\bullet}$}
&\mbox{$\circ$ $\underline{\circ}$ $\underline{\circ}$
$\underline{\circ}$ $\underline{\circ}$}&
\mbox{$\circ$ $\underline{\bullet}$}\\
\hline
\zeta\sim -0&\circ&\circ&\circ\\
\zeta\ll 0&\circ&\circ&\circ
\end{array}
\eeq
Here, a black circle stands for a fermionic ground state.

We can continue in the same way. Here are the next few:
\beq
d=7:
\begin{array}{c||c|c|c}
Q_F&-1&0&1\\
\hline\hline
\zeta\gg 0&&\mbox{$\circ$ $\circ$ $\circ$ $\circ$ $\circ$
$\circ$ $\circ$}&\\
\zeta\sim +0&\mbox{$\circ$ $\underline{\circ}$ $\underline{\bullet}$
$\underline{\bullet}$}
&\mbox{$\circ$ $\underline{\circ}$ $\underline{\circ}$
$\underline{\circ}$ $\underline{\circ}$
$\underline{\circ}$ $\underline{\circ}$}&
\mbox{$\circ$ $\underline{\circ}$ $\underline{\bullet}$
$\underline{\bullet}$}\\
\hline
\zeta\sim -0&\circ&\circ&\circ\\
\zeta\ll 0&\circ&\circ&\circ
\end{array}
\eeq
\beq
d=9:
\begin{array}{c||c|c|c|c|c}
Q_F&-2&-1&0&1&2\\
\hline\hline
\zeta\gg 0&&&\mbox{9$\,\circ$'s}&&\\
\zeta\sim +0&&\mbox{$\circ$ $\underline{\circ}$ $\underline{\circ}$
$\underline{\bullet}$ $\underline{\bullet}$ $\underline{\bullet}$}&
\mbox{$\circ$ $\!+8\,\underline{\circ}$'s}
&\mbox{$\circ$ $\underline{\circ}$ $\underline{\circ}$
$\underline{\bullet}$ $\underline{\bullet}$ $\underline{\bullet}$}&
\\
\hline
\zeta\sim -0&\overline{\circ}&\circ&\circ&\circ&\overline{\circ}\\
\zeta\ll 0&~~\circ~~&\circ&\circ&\circ&~~\circ~~
\end{array}
\eeq
\beq
d=11:
\begin{array}{c||c|c|c|c|c}
Q_F&-2&-1&0&1&2\\
\hline\hline
\zeta\gg 0&&&\mbox{11$\,\circ$'s}&&\\
\zeta\sim +0&\mbox{$\circ$ $\circ$ $\underline{\bullet}$
$\underline{\bullet}$}
&\mbox{$\circ$ $\!+3\,\underline{\circ}$'s $\!+4\,\underline{\bullet}$'s}&
\mbox{$\circ$ $\!+10\,\underline{\circ}$'s}
&\mbox{$\circ$ $\!+3\,\underline{\circ}$'s $\!+4\,\underline{\bullet}$'s}&
\mbox{$\circ$ $\circ$ $\underline{\bullet}$ $\underline{\bullet}$}
\\
\hline
\zeta\sim -0&\mbox{$\circ$ $\circ$ $\overline{\bullet}$}
&\circ&\circ&\circ&\mbox{$\circ$ $\circ$ $\overline{\bullet}$}\\
\zeta\ll 0&\circ&\circ&\circ&\circ&\circ
\end{array}
\eeq
Overlined circles are wall crossing states that come in as
$\zeta$ enters the negative region. (Equivalently, states
that go out as $\zeta\nearrow 0$.)
To be precise, there is an ambiguity in $d=7$ and $d=9$.
The $d=7, Q_F=\pm 1$ and
$d=9, Q_F=\pm 2$ entries involve $\neff=0$ states, whose fate is
not clear just from the Coulomb branch analysis. The above is a minimal
possibility. The actual process could have been
another minimal possibility:
\beq
{d=7\atop
Q_F=\pm 1}:
\begin{array}{|c|}
\hline
\mbox{$\circ$ $\underline{\circ}$ $\underline{\bullet}$
$\underline{\bullet}$}\\
\hline
\circ\\
\hline
\end{array}~\longrightarrow~
\begin{array}{|c|}
\hline
\mbox{$\circ$ $\circ$ $\underline{\bullet}$
$\underline{\bullet}$}\\
\hline
\mbox{$\circ$ $\circ$ $\overline{\bullet}$}\\
\hline
\end{array}\qquad\qquad
{d=9,\atop Q_F=\pm 2}:
\begin{array}{|c|}
\hline
\\
\hline
~~\overline{\circ}~~\\
\hline
\end{array}
~\longrightarrow~
\begin{array}{|c|}
\hline
\mbox{$\circ$ $\underline{\bullet}$}\\
\hline
\mbox{$\circ$}\\
\hline
\end{array}
\eeq

\subsection{Triangle Quiver}

We consider the ${\mathcal N}=2$ triangle quiver, with the
gauge group $G=U(1)_1\times U(1)_2$, three groups of chiral and three
groups of fermi multiplets,
\beq
\begin{array}{c|cccccc}
&X_{1,\ldots,a}&Y_{1,\ldots,b}&Z_{1,\ldots,c}&
\eta^{1,\ldots,M_3}_{3}
&\eta^{1,\ldots,M_1}_{1}
&\eta^{1,\ldots,M_2}_{2}\\
\hline
U(1)_1&1&0&-1&-1&1&0\\
U(1)_2&0&1&-1&-1&0&1\\
\hline
U(1)_{F_1}&1&0&0&-1&0&-1\\
U(1)_{F_2}&0&1&0&-1&-1&0
\end{array}
\label{Neq2tri}
\eeq
and the superpotential
\beq
\mathfrak{W}=\sum_{\gamma=1}^{M_3}f_{3\gamma}(X,Y)\eta^{\gamma}_3
+\sum_{\alpha=1}^{M_1}f_{1\alpha}(Y,Z)\eta^{\alpha}_1
+\sum_{\beta=1}^{M_2}f_{2\beta}(Z,X)\eta^{\beta}_2,
\eeq
where $f_{3\gamma}(X,Y)$, $f_{1\alpha}(Y,Z)$ and $f_{2\beta}(Z,X)$
are generic bilinear polynomials of the two entries.
The system has $U(1)_{F_1}\times U(1)_{F_2}$ flavor symmetry
whose charges had been listed above in advance.
The anomaly free condition is
\beq
a+c+M_3+M_1\in 2\Z,\quad
b+c+M_3+M_2\in 2\Z.
\eeq
The D-term equations read
\beq
||X||^2-||Z||^2=\zeta^1,\qquad
||Y||^2-||Z||^2=\zeta^2,
\eeq
and the F-term equations read
\beq
f_{3\gamma}(X,Y)=0~\forall\gamma,\quad
f_{1\alpha}(Y,Z)=0~\forall\alpha,\quad
f_{2\beta}(Z,X)=0~\forall\beta.
\eeq
The theory has three phases, just as in
Fig.~\ref{fig:q111}, and their nature
depends very much on the numbers $(a,b,c)$ and $(M_3,M_1,M_2)$.
The $g$ function reads
\beqa
g&=&{(2i\sin(\pi(u_1+u_2+z_1+z_2)))^{M_3}(2i\sin(\pi(-u_1+z_2)))^{M_1}
(2i\sin(\pi(-u_2+z_1)))^{M_2}\over
(2i\sin(\pi(u_1+z_1)))^a(2i\sin(\pi(u_2+z_2)))^b(2i\sin(\pi(-u_1-u_2)))^c}
\nn\\
&=&(-1)^{c+M_2+M_3}y_1^{-{a+M_2+M_3\over 2}}y_2^{-{b+M_1+M_3\over 2}}\times
\nn\\
&&~~~~~~~x_1^{a+c-M_1-M_3\over 2}x_2^{b+c-M_2-M_3\over 2}
{(x_1x_2y_1y_2-1)^{M_3}(x_1-y_2)^{M_1}(x_2-y_1)^{M_2}\over
(x_1-y_1^{-1})^a(x_2-y_2^{-1})^b(x_1x_2-1)^c}
\label{gN2trQ}
\eeqa

We consider the case $(a,b,c)=(1,3,3)$, $(M_3,M_1,M_2)=(2,2,2)$
in detail.
In Phase I, the vacuum equations require
$Z=0$ and that $(X,Y)$ is at a point of $\CP^2$
determined by $f_{31}(X,Y)=f_{32}(X,Y)=0$.
Of the six fermi multiplets,
five mixes with the fermions of the chiral multiplets and
only one of the $\eta_1^{1,2}$ remains massless.
With respect to the two flavor symmetries,
which are modified by the gauge symmetry so that $X$ and $Y$ are
neutralized, $\eta_1^{1,2}$ has charge $(-1,-1)$.
Therefore, the index is
\beq
I_{\rm I}=y_1^{1\over 2}y_2^{1\over 2}-y_1^{-{1\over 2}}y_2^{-{1\over 2}}.
\label{ItrN2I}
\eeq
Phase II has a similar structure, and the index is the same as in Phase I,
\beq
I_{\rm II}=y_1^{1\over 2}y_2^{1\over 2}-y_1^{-{1\over 2}}y_2^{-{1\over 2}}.
\label{ItrN2II}
\eeq
In Phase III, the vacuum equations require $X=0$ and that
$(Y,Z)$ is in the complete intersection ${\mathcal X}$ of the hypersurfaces
$f_{11}(Y,Z)=f_{12}(Y,Z)=0$ in $\CP^2\times \CP^2$.
(${\mathcal X}$ is the del Pezzo surface of degree $6$ ---the
blow up at three general points of $\CP^2$.)
The fermions $\eta_1^{1,2}$ mixes with the superpartner of the two
transverse directions of ${\mathcal X}$ in $\CP^2\times\CP^2$.
Of the four remaining fermions, $\eta_3^{1,2}$ and $\eta_2^{1,2}$,
one mixes with the superpartner of $X$
and we are left with the three spanning the vector bundle
${\mathcal E}$ defined by
\beq
0\,\to\, {\mathcal E}\,\longrightarrow\,
{\mathcal O}(0,1)^{\oplus 2}\oplus {\mathcal O}(1,0)^{\oplus 2}\,
\stackrel{(f_3(Y),f_2(Z))}{\longlongrightarrow}\,
{\mathcal O}(1,1)\,\to\,0.
\label{exseq1}
\eeq
Ground states correspond to
Dolbeault cohomology classes with values in\footnote{In
the final equality, we use $\wedge^3{\mathcal E}={\mathcal O}(1,1)$
that follows from (\ref{exseq1}).}
\beqa
\mathfrak{F}&=&\sqrt{K}_{\mathcal X}\otimes {\det}^{-{1\over 2}}{\mathcal E}
\otimes\wedge {\mathcal E}
\,\,=\,\,(\wedge{\mathcal E})(-1,-1)\nn\\
&=&{\mathcal O}(-1,-1)\,\oplus\,
{\mathcal E}(-1,-1)\,\oplus\, (\wedge^2{\mathcal E})(-1,-1)
\oplus {\mathcal O}.
\eeqa
Using Riemann-Roch formula, we find that the Dolbeault indices of the four
terms of the last expression
are $1$, $-1$, $-1$, $1$ respectively.
Since $\eta_3$'s and $\eta_2$'s both have the
$U(1)_{F_1}\times U(1)_{F_2}$ flavor charge $(-1,-1)$,
again after neutralizing $y$ and $z$, we find that
the index in this phase is
\beq
I_{\rm III}
=y_1^{3\over 2}y_2^{3\over 2}
+y_1^{1\over 2}y_2^{1\over 2}
-y_1^{-{1\over 2}}y_2^{-{1\over 2}}
-y_1^{-{3\over 2}}y_2^{-{3\over 2}}.
\label{ItrN2III}
\eeq
It is straightforward to check that the above indices in the three phases
agrees with the ones obtained from (\ref{gN2trQ}) by the residue computation.

Let us now discuss the wall crossing. We first consider the
I $\to$ III crossing. There we have a mixed branch with
the unbroken gauge group $G_1=U(1)_1$ and broken gauge group
$G/G_1\cong U(1)_2$.
The Coulomb and Higgs parts of the fields are
\beqa
({\rm C})&&X~(1,0|1,0),~\,Z_{1,2,3}~(-1,-1|0,1),
~\,\eta_3^{1,2}~(-1,-1|-1,0),~\,\eta_1^{1,2}~(1,0|0,-1),\nn\\
({\rm H})&&Y_{1,2,3}~(1|0,0),~\,\eta_2^{1,2}~(1|-1,-1).\nn
\eeqa
Inside the parenthesis are the gauge and the flavor charges, copied
from (\ref{Neq2tri}) except that the
$U(1)_{F_2}$ charges are modified so that the Higgs fields $Y_{1,2,3}$
are neutral. For the fields in (H), we omit the $G_1=U(1)_1$ charges
which are trivial by definition.
 The charges under $U(1)_2$
are regarded as those under $G/G_1\cong U(1)_2$.
Let us first analyze the ${\vs}_1<0$ mixed branch.
The equation (\ref{gveqneg}) has a unique solution which yields
$\neff=-2$. Thus, there is no supersymmetric ground state for $\zeta^1>0$
and a unique supersymmetric ground state of the form
(\ref{profilesnZns}) for $\zeta^1<0$:
\beq
|\mbox{C-vac}\rangle
=|{\vs}_1|\e^{-\zeta^1{\vs}_1}\blambda\,\bpsi_{z_1}\bpsi_{z_2}\bpsi_{z_3}
\bareta_1^1\bareta_1^2|0\rangle_{\!{}_{\rm OSC}}
\qquad \left(-{1\over 2}\Bigr|{1\over 2},{1\over 2}\right).
\label{gs17}
\eeq
We listed the charge of the state
under $G/G_1\times U(1)_{F_1}\times U(1)_{F_2}$, from (\ref{gQneg}),
which provides the background charge in
the effective theory on the Higgs branch.
The Higgs branch theory is thus the $\CP^2$ model with Wilson line
$\rmq=-{1\over 2}$ and the fermi-multiplet with values in
${\mathcal E}={\mathcal O}(1)^{\oplus 2}$.
Ground states correspond to Dolbeault cohomology classes with values in
\beqa
\mathfrak{F}&=&
\sqrt{K}_{\CP^2}\otimes {\det}^{-{1\over 2}}{\mathcal E}
\otimes\wedge {\mathcal E}\otimes {\mathcal O}\left(-{1\over 2}\right)
\,=\,\left(\wedge{\mathcal E}\right)(-3)\nn\\
&=&
{\mathcal O}(-3)\oplus {\mathcal O}(-2)^{\oplus 2}\oplus {\mathcal O}(-1).
\label{F18}
\eeqa
There is a unique Higgs vacuum corresponding to
$H^{0,2}(\CP^2,{\mathcal O}(-3))\cong \C$.
Since $\eta_2^{1,2}$ have $U(1)_{F_1}\times U(1)_{F_2}$ charge $(-1,-1)$,
their oscillator vacuum, which corresponds to the term ${\mathcal O}(-3)$
in (\ref{F18}), has charge $(1,1)$. Taking the background
charge (\ref{gs17}) into account, we find that the ground state has charge
$({3\over 2},{3\over 2})$. Note also that it has $(-1)^F=1$ if
$|0\rangle_{\!{}_{\rm OSC}}$ is even.
Therefore, it contributes to the change in the index by
$+y_1^{3\over 2}y_2^{3\over 2}$.
Let us next analyze the ${\vs}_1>0$ mixed branch.
The equation (\ref{gveqpos}) has a unique solution with
$\neff=-2$. Thus, there is a supersymmetric ground state only when
$\zeta^1<0$ which is of the form (\ref{profilesnZps}):
\beq
|\mbox{C-vac}\rangle
=|{\vs}_1|\e^{\zeta^1{\vs}_1}\bpsi_{x}
\bareta_3^1\bareta_3^2|0\rangle_{\!{}_{\rm OSC}}
\qquad \left({1\over 2}\Bigr|-{1\over 2},-{1\over 2}\right).
\label{gs19}
\eeq
The Higgs branch theory is the $\CP^2$ model with
with Wilson line
$\rmq={1\over 2}$ and the fermi-multiplet with values in
${\mathcal E}={\mathcal O}(1)^{\oplus 2}$.
Ground states correspond to Dolbeault cohomology classes with values in
\beqa
\mathfrak{F}&=&
\sqrt{K}_{\CP^2}\otimes {\det}^{-{1\over 2}}{\mathcal E}
\otimes\wedge {\mathcal E}\otimes {\mathcal O}\left({1\over 2}\right)
\,=\,\left(\wedge{\mathcal E}\right)(-2)\nn\\
&=&
{\mathcal O}(-2)\oplus {\mathcal O}(-1)^{\oplus 2}\oplus {\mathcal O}.
\label{F20}
\eeqa
There is a unique Higgs vacuum corresponding to
$H^{0,0}(\CP^2,{\mathcal O})\cong \C$.
The state $\bareta_2^1\bareta_2^2|0\rangle$, which corresponds to
the term ${\mathcal O}$
in (\ref{F20}), has charge $(-1,-1)$ under $U(1)_{F_1}\times U(1)_{F_2}$.
Taking the background charge (\ref{gs19}) into account,
we find that the ground state has charge
$(-{3\over 2},-{3\over 2})$. Note also that it has $(-1)^F=-1$ if
$|0\rangle_{\!{}_{\rm OSC}}$ is even.
Therefore, it contributes to the change in the index by
$-y_1^{-{3\over 2}}y_2^{-{3\over 2}}$.
To summarize, the change in the index for the I $\to$ III
wall crossing is
\beq
\Delta_{{\rm I}\to {\rm III}} I\,\,=\,\,y_1^{3\over 2}y_2^{3\over 2}
-y_1^{-{3\over 2}}y_2^{-{3\over 2}}.
\label{WCtrN2ItoIII}
\eeq
The II $\to$ III crossing is similar and we find the same expression for
$\Delta I$ as above,
\beq
\Delta_{{\rm II}\to {\rm III}} I\,\,=\,\,y_1^{3\over 2}y_2^{3\over 2}
-y_1^{-{3\over 2}}y_2^{-{3\over 2}}.
\label{WCtrN2IItoIII}
\eeq
Finally, let us consider the I $\to$ II crossing.
There we have a mixed branch with
the unbroken gauge group $G_1=U(1)_2$ and broken gauge group
$G/G_1\cong U(1)_1$.
The Coulomb and the Higgs parts of the fields are
\beqa
({\rm C})&&Y_{1,2,3}~(0,1|0,1),~\,Z_{1,2,3}~(-1,-1|1,0),
~\,\eta_3^{1,2}~(-1,-1|0,-1),~\,\eta_2^{1,2}~(0,1|-1,0),\nn\\
({\rm H})&&X~(1|0,0),~\,\eta_1^{1,2}~(1|-1,-1).\nn
\eeqa
The gauge and the flavor charges are copied
from (\ref{Neq2tri}) except that the
$U(1)_{F_1}$ charges are modified so that the Higgs field $X$
is neutral. For the fields in (H), we omit the $G_1=U(1)_2$ charges
which are trivial by definition.
 The charges under $U(1)_1$
are regarded as those under $G/G_1\cong U(1)_1$.
The supersymmetry equation for the matter sector,
(\ref{gveqneg}) for ${\vs}_2<0$ and
(\ref{gveqpos}) for ${\vs}_2>0$, has no solution. Thus, there is no
supersymmetric ground states in the mixed branch theory for both
$\zeta^2>0$ and $\zeta^2<0$. In particular, the index does not change,
\beq
\Delta_{{\rm I}\to{\rm II}} I\,\,=\,\,0.
\label{WCtrN2ItoII}
\eeq
The results for the index, (\ref{ItrN2I}), (\ref{ItrN2II})
and (\ref{ItrN2III}), are consistent with the wall crossing formulae,
(\ref{WCtrN2ItoIII}), (\ref{WCtrN2IItoIII}) and (\ref{WCtrN2ItoII}).

\section*{Acknowledgement}

We would like to thank
Francesco Benini,  Jyotirmoy Bhattacharya, Richard Eager,
Seung-Joo Lee, Yu Nakayama, Tomoki Ohtsuki,
 Mauricio Romo, Ashoke Sen, Yuji Tachikawa, Zhao-Long Wang,
Chris Woodward
for discussions, conversations, instructions, collaboration on
related works, and encouragement. HK and PY thank Kavli IPMU
for the hospitality during their visit. KH and PY are also
grateful to organizers of RIKKYO MathPhys 2014, by which time
the main results of this work have emerged, for providing a
stimulating environment for the collaboration. This work
is supported in part by JSPS Grant-in-Aid
for Scientific Research No. 21340109 and WPI Initiative, MEXT,
Japan at Kavli IPMU, the University of Tokyo.
We would like to thank the referee for
spotting many mistakes in the manuscript.
The paper and our understanding have improved significantly 
by the revision. We would also like to thank Edward Witten for great
help during the revision process.

\appendix{1d ${\mathcal N}=2$ supersymmetry}\label{app:SUSY}

\subsection{1d ${\mathcal N}=2$ superspace}

The ${\mathcal N}=2$ superspace has time coordinate $t$
and fermionic coordinates $\theta$ and $\btheta$. Supersymmetry is
\beq
\delta=-\epsilon{\rm Q}+\bepsilon\overline{\rm Q},
\label{SUSYtrSF}
\eeq
where ${\rm Q}$ and $\overline{\rm Q}$ are differential operators
\beq
{\rm Q}\,=\,{\partial\over \partial\theta}
+{i\over 2}\btheta {\partial\over \partial t},\qquad
\overline{\rm Q}\,=\,-{\partial\over \partial\btheta}
-{i\over 2}\theta {\partial\over\partial t},
\eeq
which satisfy
${\rm Q}^2=\overline{\rm Q}^2=0$, $\{{\rm Q},\overline{\rm Q}\}
=-i\partial_t$. They commutes with another set of operators
\beq
{\rm D}\,=\,{\partial\over \partial\theta}
-{i\over 2}\btheta {\partial\over\partial t},\qquad
\overline{\rm D}\,=\,-{\partial\over \partial\btheta}
+{i\over 2}\theta{\partial\over\partial t},
\eeq
which satisfy ${\rm D}^2=\overline{\rm D}^2=0$ and
$\{{\rm D},\overline{\rm D}\}=i\partial_t$.

A chiral superfield $\Phi$ is a superfield satisfying
$\overline{\rm D}\Phi=0$. It can be expanded as
\beq
\Phi=\phi+\theta\psi-{i\over 2}\theta\btheta\dot{\phi}.
\eeq
Supersymmetry variation of the components fields
follows from (\ref{SUSYtrSF})
\beqa
&&\delta\phi=-\epsilon\,\psi,\nn\\
&&\delta\psi=i\bepsilon\,\dot{\phi}.
\label{susychiral}
\eeqa
A fermi superfield $\mathfrak{Y}$ is a fermionic superfield satisfying
$\overline{\rm D} \,\mathfrak{Y}=E(\Phi)$
with $E(\Phi)$ a holomorphic function of a chiral superfield $\Phi$.
It is expanded as
\beq
\mathfrak{Y}=\eta-\theta F-{i\over 2}\theta\btheta\dot{\eta}
-\btheta E(\Phi).
\eeq
Supersymmetry variation of the components is
\beqa
&&\delta\eta=\epsilon F+\bepsilon E(\phi),\nn\\
&&\delta F=\bepsilon\left(-i\dot{\eta}+\psi^i\partial_iE(\phi)\right).
\label{susyfermi}
\eeqa
For the chiral and fermi superfields as above, we have
($\dot{=}$ stands for equality up to total derivatives)
\beqa
\int\dd\theta\dd\btheta\,\,i\bPhi\dot{\Phi}
&\dot{=}&
\dot{\bphi}\dot{\phi}\,\,
+\,\,i\bpsi\dot{\psi},\\
\int\dd\theta\dd\btheta\,\,\overline{\mathfrak{Y}}\mathfrak{Y}&\dot{=}&
i\bareta\dot{\eta}+\overline{F} F
-|E(\phi)|^2-\bareta\partial_iE(\phi)\psi^i-\bpsi^{\bi}
\partial_{\bi}\overline{E(\phi)}\eta.
\eeqa
For a set of fermi and chiral superfields
$\mathfrak{Y}^\alpha,J_\alpha(\Phi)$
with $\overline{\rm D} \mathfrak{Y}^\alpha=E^\alpha(\Phi)$ and
$E^\alpha(\phi)J_\alpha(\phi)=0$,
the combination $\mathfrak{Y}^\alpha J_\alpha(\Phi)$
is a fermionic chiral superfield,
for which we can take
\beq
\int\dd\theta\Bigl(\,\mathfrak{Y}^\alpha J_\alpha(\Phi)\,\Bigr)_{\btheta=0}
=-\eta^\alpha\partial_i J_{\alpha}(\phi)\psi^i-F^\alpha J_{\alpha}(\phi).
\eeq
This is the F-term associated to the superpotential
$\mathfrak{W}=\mathfrak{Y}^\alpha J_\alpha(\Phi)$.

\subsection{Gauge theory}

The vector multiplet consists of
a $\frakg_{\C}$ and an $i\frakg$ valued superfields $\Omega$ and $V^{(-)}$
which transform as
\beq
\e^{\Omega}\,\longrightarrow\, h^{-1} \e^{\Omega}\, k,\qquad
iV^{(-)}\,\longrightarrow\, k^{-1}iV^{(-)} k+k^{-1}\dot{k},
\eeq
where $k$ is a $G$-valued superfield and $h$ is a $G_{\C}$-valued
chiral superfield,
\beq
k^{\dag}=k^{-1},\qquad \overline{\rm D}h=0.
\eeq
One may define the covariant derivatives
\beq
\overline{\mathcal{D}}:=\e^{-\Omega}\overline{\rm D}\e^{\Omega},\quad
\mathcal{D}:=\e^{\Omega^{\dag}} {\rm D} \e^{-\Omega^{\dag}},\quad
\mathcal{D}_t^{(-)}:=\partial_t+iV^{(-)},
\eeq
which are invariant under the chiral gauge transformation by $h$
but transform covariantly under the unitary gauge transformation by $k$:
\beq
\overline{\mathcal{D}}\to k^{-1}\overline{\mathcal{D}} k,\quad
\mathcal{D}\to k^{-1}\mathcal{D} k,\quad
\mathcal{D}_t^{(-)}\to k^{-1}\mathcal{D}_t^{(-)}k.
\eeq
The first two of them satisfy the relations
\beq
\overline{\mathcal{D}}^2=\mathcal{D}^2=0,\qquad
\{\mathcal{D},\overline{\mathcal{D}}\}=i\mathcal{D}_t^{(+)}
:=i(\partial_t+iV^{(+)}),
\eeq
for some superfield $V^{(+)}$.
We define the field strength superfield by
\beq
\Upsilon:=[\overline{\mathcal D},{\mathcal D}_t^{(-)}].
\eeq
It is unitary covariant, $\Upsilon\to k^{-1}\Upsilon k$,
and (covariantly) chiral, $\overline{\mathcal D}\Upsilon=0$.
The gauge kinetic term is
\beq
L_{\rm guage}\,=\,\int\dd\theta\dd\btheta\,{1\over 2e^2}{\rm Tr}\,
\overline{\Upsilon}\Upsilon.
\eeq
For a $G$-invariant linear form $\zeta:i\frakg\to \R$,
we have the FI term
\beq
L_{\rm FI}\,=\,
\int\dd\theta\dd\btheta\,\zeta(V^{(-)}) \ .
\label{app:FI}
\eeq
For unitary covariant chiral and fermi superfields $\Phi$
and $\mathfrak{Y}$, the kinetic terms are
\beq
L_{\rm chiral}\,=\,\int\dd\theta\dd\btheta\,\,i
\bPhi\,{\mathcal D}^{(-)}_t\Phi,\qquad
L_{\rm fermi}\,=\,\int\dd\theta\dd\btheta\,
\overline{\mathfrak{Y}}\mathfrak{Y}.
\eeq

\subsubsection*{Component Expressions}

Let us define the Wess-Zumino gauge as
\beq
\Omega=-{1\over 2}\theta\btheta (v_t+\vs).
\eeq
The residual gauge symmetry is
\beq
k=g,\qquad h=g-{i\over 2}\theta\btheta\dot{g},
\label{02residual}
\eeq
for a $G$-valued function $g$ of $t$.
For this choice, we have
\beq
\overline{\mathcal D}=-\partial_{\btheta}
+{i\over 2}\theta D_t^{(+)},\qquad
{\mathcal D}=\partial_{\theta}
-{i\over 2}\btheta D_t^{(+)}.
\eeq
where
\beq
D_t^{(\pm)}:=D_t\pm i\vs=\partial_t+i(v_t\pm\vs).
\eeq
We may write
\beqa
V^{(-)}&=&v_t-\vs
-i\theta\blambda-i\btheta\lambda+\theta\btheta D,\\
\Phi&=&\phi+\theta\psi-{i\over 2}\theta\btheta D_t^{(+)}\phi,\\
\mathfrak{Y}&=&
\eta-\theta F-{i\over 2}\theta\btheta D_t^{(+)}\psi_-
-\btheta E(\Phi).
\eeqa
They transform covariantly under the residual gauge transformations
(\ref{02residual}) except $iv_t\to g^{-1}iv_tg+g^{-1}\dot{g}$.
The field strength superfields is then written as
\beq
\Upsilon\,\,=\,\,-\lambda+\theta(D_t\vs+iD)
+{i\over 2}\theta\btheta D_t^{(+)}\lambda_-.
\eeq
Under the supersymmetry
$\delta=-\epsilon{\rm Q}+\bepsilon\overline{\rm Q}$,
the Wess-Zumino gauge is not preserved. To bring things back to the Wess-Zumino
gauge, we need to perform the following unitary and chiral gauge transforms
(to the first order):
\beq
k=1+\theta\bepsilon {v_t+\vs\over 2}
+\btheta\epsilon{v_t+\vs\over 2},\qquad
h=1+\theta\bepsilon (v_t+\vs).
\eeq
This way, we find
the supersymmetry transform of the component fields:
\beqa
\delta v_t&=&-\delta\vs~=~{i\over 2}\epsilon\blambda
+{i\over 2}\bepsilon\lambda,\nn\\
\delta \lambda&=&\epsilon(D_t\vs+iD),\nn\\
\delta D&=&{1\over 2}\epsilon D_t^{(+)}\blambda
-{1\over 2}\bepsilon D_t^{(+)}\lambda,
\label{stVector}\\[0.2cm]
\delta\phi&=&-\epsilon\psi,\nn\\
\delta\psi&=&i\bepsilon D_t^{(+)}\phi,
\label{stChiral}\\[0.2cm]
\delta\eta&=&\epsilon F+\bepsilon E(\phi),\nn\\
\delta F&=&\bepsilon \left(-iD_t^{(+)}\eta+\psi^i\partial_iE(\phi)\right).
\label{stFermi}
\eeqa
The supersymmetric Lagrangians have the following expressions:
\beqa
L_{\rm gauge}&\dot{=}&
{1\over 2e^2}{\rm Tr}\Bigl[\, (D_t\vs)^2+i\blambda\,D_t^{(+)}\lambda+D^2
\,\Bigr],\\[0.2cm]
L_{\rm FI}&=&-\zeta(D),\\[0.2cm]
L_{\rm chiral}&\dot{=}&
D_t\bphi D_t\phi
+i\bpsi\,D_t^{(-)}\psi
+\bphi \left\{D-\vs^2\right\}\phi
-i\bphi\lambda\psi+i\bpsi\blambda\phi,\\[0.2cm]
L_{\rm fermi}&\dot{=}&
i\bareta\,D_t^{(+)}\eta+\overline{F} F
-\overline{E(\phi)}E(\phi)
-\bareta\partial_iE(\phi)\psi^i
-\bpsi^{\bi}\partial_{\bi}\overline{E(\phi)}\eta.
\eeqa

\subsection{Non-linear sigma model}
\label{app:NLSM}

We consider a K\"ahler manifold $(X,g)$ and a holomorphic vector bundle
with a hermitian metric $({\mathcal E},h)$ on $X$.
Locally, the K\"ahler metric is written
using a K\"ahler potential as $g_{i\bj}=\partial_i\partial_{\bj}K$,
and the fiber metric $h$ is expressed as a hermitian matrix
$h_{\alpha\bar\beta}=h(e_{\beta},e_{\alpha})$ for a choice of local
holomorphic frame $\{e_{\alpha}\}$.
The variables are $X$-valued chiral superfield $\Phi$
and ${\mathcal E}$-valued chiral fermi superfield $\mathfrak{Y}$.
Their kinetic terms are
\beqa
\int\dd\theta\dd\btheta\,i\partial_iK(\Phi,\bPhi)
\dot{\Phi}^i&\dot{=}&
g_{i\bj}\dot{\bphi}^{\bj}\dot{\phi}^i
+ig_{i\bj}\bpsi^{\bj}D_t\psi^i,\\[0.3cm]
\int\dd\theta\dd\btheta\,h_{\alpha\bar\beta}(\Phi,\bPhi)
\bXi_-^{\bar\beta}\mathfrak{Y}_-^{\alpha}&\dot{=}&
ih_{\alpha\bar\beta}\bareta^{\bar\beta}D_t\eta^{\alpha}
+h_{\alpha\bar\beta}\bareta^{\bar\beta}{F^{}_{i\bj}}^{\!\alpha}_{}
{}_{\!\gamma}
\psi^i\bpsi^{\bj}
\eta^{\gamma}
+h_{\alpha\bar\beta}
\overline{\widetilde{F}}^{\bar\beta}\widetilde{F}^{\alpha},
\eeqa
where
\beq
D_t\psi^i
=\dot{\psi}^i+\dot{\phi}^j\Gamma^i_{jk}\psi^k,\qquad
D_t\eta^\alpha=\dot{\eta}^{\alpha}+\dot{\phi}^j
A^{\alpha}_{j\beta}\eta^{\beta}
\eeq
with $\Gamma^i_{jk}:=g^{i\bl}\partial_jg_{k\bl}$
and $A^{\alpha}_{j\beta}:=h^{\alpha\bar\gamma}\partial_jh_{\beta\bar\gamma}$
which are the connection forms of $T_X$ and ${\mathcal E}$
in the holomorphic frames.
Also, ${F^{}_{i\bj}}^{\!\alpha}_{}{}_{\!\beta}$ is the curvature of
${\mathcal E}$,
\beq
{F^{}_{i\bj}}^{\!\alpha}_{}{}_{\!\beta}:=
\left(\partial_iA_{\bj}-\partial_{\bj}A_i+[A_i,A_{\bj}]
\right)^{\alpha}_{\,\beta}
=-h^{\alpha\bar\gamma}\left(\partial_i\partial_{\bj}h_{\beta\bar\gamma}
-\partial_{\bj}h_{\delta\bar\gamma}h^{\delta\bar\alpha}
\partial_ih_{\beta\bar\alpha}\right)
\eeq
and $\widetilde{F}^{\alpha}$ are the covariant auxiliary fields,
\beq
\widetilde{F}^{\alpha}
:=F^{\alpha}-\psi^jA^{\alpha}_{j\beta}\eta^{\beta}.
\eeq
$\widetilde{F}^{\alpha}$'s
are covariant because $F^\alpha$'s transform inhomogeneously
under the holomorphic frame change
$e_{\beta'}=e_{\alpha}g^{\alpha}_{\,\,\beta'}$
as $F^\alpha
=g^\alpha_{\,\,\beta'}F^{\beta'}-\psi^i
\partial_ig^\alpha_{\,\,\beta'}\eta^{\beta'}$, which follows from
$\mathfrak{Y}^\alpha=g^\alpha_{\,\,\beta'}(\Phi)\mathfrak{Y}^{\beta'}$.

\subsection{${\mathcal N}=4$ theories}

An ${\mathcal N}=4$ vector multiplet consists of an ${\mathcal N}=2$
vector multiplet $(\Omega,V^{(-)})$ and an ${\mathcal N}=2$ chiral multiplet
$\Sigma$ in the adjoint representation.
An ${\mathcal N}=4$ chiral multiplet consists of ${\mathcal N}=2$
chiral and fermi multiplets in a representation, $\Phi$ and $\mathfrak{Y}$,
 which obey $\overline{\mathcal D}\mathfrak{Y}=\Sigma\Phi$.
An ${\mathcal N}=4$ supersymmetric Lagrangians are
\beqa
\bfL_{\rm gauge}
&=&\int\dd\theta\dd\btheta\,{1\over 2e^2}{\rm Tr}\left[\,
i\,\bSigma\,{\mathcal D}_t^{(-)}\Sigma\,+\,\overline{\Upsilon}\Upsilon\,
\right],
\label{Neq4gaugekin}
\\
\bfL_{\rm chiral}
&=&\int\dd\theta\dd\btheta\,\left[\,
i\bPhi\,{\mathcal D}_t^{(-)}\Phi\,+\,
\overline{\mathfrak{Y}}\mathfrak{Y}\,\right],
\label{Neq4chikin}\\
\bfL_{W}
&=&-{\rm Re}\int \dd\theta\,\mathfrak{Y}^i\partial_iW(\Phi)\,\Bigr|_{\btheta=0},
\eeqa
where $W(\Phi)$ is a gauge invariant and holomorphic function of $\Phi$.
They are the dimensional reductions of the
gauge kinetic term, matter kinetic term,
and superpotential term, respectively, in 4d ${\mathcal N}=1$ or
2d $(2,2)$ gauge theories.
The FI term (\ref{FI}) is also ${\mathcal N}=4$ supersymmetric.


\begin{thebibliography}{99}

\small
\parskip=0pt plus 2pt



\bibitem{Windex}
E.~Witten,
  ``Constraints on Supersymmetry Breaking,''
  Nucl.\ Phys.\ B {\bf 202} (1982) 253.


\bibitem{Wphases}
E.~Witten,
  ``Phases of N=2 theories in two-dimensions,''
  Nucl.\ Phys.\ B {\bf 403} (1993) 159
  [hep-th/9301042].


\bibitem{IntSei}
K.~Intriligator and N.~Seiberg,
  ``Aspects of 3d N=2 Chern-Simons-Matter Theories,''
  JHEP {\bf 1307} (2013) 079
  [arXiv:1305.1633 [hep-th]].


%\cite{Witten:1995im}
\bibitem{Witten:1995im}
  E.~Witten,
  ``Bound states of strings and p-branes,''
  Nucl.\ Phys.\ B {\bf 460} (1996) 335
  [hep-th/9510135].
  %%CITATION = HEP-TH/9510135;%%

\bibitem{DouglasMoore}
M.~R.~Douglas and G.~W.~Moore,
  ``D-branes, quivers, and ALE instantons,''
  hep-th/9603167.

\bibitem{DouglasADHM}
M.~R.~Douglas,
  ``Gauge fields and D-branes,''
  J.\ Geom.\ Phys.\  {\bf 28} (1998) 255
  [hep-th/9604198].

\bibitem{DGM}
M.~R.~Douglas, B.~R.~Greene and D.~R.~Morrison,
  ``Orbifold resolution by D-branes,''
  Nucl.\ Phys.\ B {\bf 506} (1997) 84
  [hep-th/9704151].


\bibitem{Yi:1997eg}
  P.~Yi,
  ``Witten index and threshold bound states of D-branes,''
  Nucl.\ Phys.\ B {\bf 505} (1997) 307
  [hep-th/9704098].
  %%CITATION = HEP-TH/9704098;%%

%\cite{Sethi:1997pa}
\bibitem{Sethi:1997pa}
  S.~Sethi and M.~Stern,
  ``D-brane bound states redux,''
  Commun.\ Math.\ Phys.\  {\bf 194} (1998) 675
  [hep-th/9705046].
  %%CITATION = HEP-TH/9705046;%%


%\cite{Cecotti:1992qh}
\bibitem{Cecotti:1992qh}
  S.~Cecotti, P.~Fendley, K.~A.~Intriligator and C.~Vafa,
  ``A New supersymmetric index,''
  Nucl.\ Phys.\ B {\bf 386} (1992) 405
  [hep-th/9204102].
  %%CITATION = HEP-TH/9204102;%%

%\cite{Seiberg:1994rs}
\bibitem{Seiberg:1994rs}
  N.~Seiberg and E.~Witten,
  ``Electric - magnetic duality, monopole condensation, and confinement in N=2 supersymmetric Yang-Mills theory,''
  Nucl.\ Phys.\ B {\bf 426} (1994) 19
   [Erratum-ibid.\ B {\bf 430} (1994) 485]
  [hep-th/9407087].
  %%CITATION = HEP-TH/9407087;%%

%\cite{Seiberg:1994aj}
\bibitem{Seiberg:1994aj}
  N.~Seiberg and E.~Witten,
  ``Monopoles, duality and chiral symmetry breaking in N=2 supersymmetric QCD,''
  Nucl.\ Phys.\ B {\bf 431} (1994) 484
  [hep-th/9408099].
  %%CITATION = HEP-TH/9408099;%%

%\cite{Ferrari:1996sv}
\bibitem{Ferrari:1996sv}
  F.~Ferrari and A.~Bilal,
  ``The Strong coupling spectrum of the Seiberg-Witten theory,''
  Nucl.\ Phys.\ B {\bf 469} (1996) 387
  [hep-th/9602082].
  %%CITATION = HEP-TH/9602082;%%
  %\cite{Denef:2000nb}


%\cite{Bergman:1997yw}
\bibitem{Bergman:1997yw}
  O.~Bergman,
  ``Three pronged strings and 1/4 BPS states in N=4 superYang-Mills theory,''
  Nucl.\ Phys.\ B {\bf 525} (1998) 104
  [hep-th/9712211].
  %%CITATION = HEP-TH/9712211;%%

%\cite{Lee:1998nv}
\bibitem{Lee:1998nv}
  K.~-M.~Lee and P.~Yi,
  ``Dyons in N=4 supersymmetric theories and three pronged strings,''
  Phys.\ Rev.\ D {\bf 58}, 066005 (1998)
  [hep-th/9804174].
  %%CITATION = HEP-TH/9804174;%%

%\cite{Bak:1999da}
\bibitem{Bak:1999da}
  D.~Bak, C.~-k.~Lee, K.~-M.~Lee and P.~Yi,
  ``Low-energy dynamics for 1/4 BPS dyons,''
  Phys.\ Rev.\ D {\bf 61} (2000) 025001
  [hep-th/9906119].
  %%CITATION = HEP-TH/9906119;%%

%\cite{Gauntlett:1999vc}
\bibitem{Gauntlett:1999vc}
  J.~P.~Gauntlett, N.~Kim, J.~Park and P.~Yi,
  ``Monopole dynamics and BPS dyons N=2 superYang-Mills theories,''
  Phys.\ Rev.\ D {\bf 61}, 125012 (2000)
  [hep-th/9912082].
  %%CITATION = HEP-TH/9912082;%%



  %\cite{Stern:2000ie}
\bibitem{Stern:2000ie}
  M.~Stern and P.~Yi,
  ``Counting Yang-Mills dyons with index theorems,''
  Phys.\ Rev.\ D {\bf 62} (2000) 125006
  [hep-th/0005275].
  %%CITATION = HEP-TH/0005275;%%



\bibitem{Denef:2000nb}
  F.~Denef,
  ``Supergravity flows and D-brane stability,''
  JHEP {\bf 0008} (2000) 050
  [hep-th/0005049].
  %%CITATION = HEP-TH/0005049;%%



\bibitem{DouglasStab}
M.~R.~Douglas,
  ``D-branes, categories and N=1 supersymmetry,''
  J.\ Math.\ Phys.\  {\bf 42} (2001) 2818
  [hep-th/0011017].


\bibitem{KachruMcgreevy}
S.~Kachru and J.~McGreevy,
  ``Supersymmetric three cycles and supersymmetry breaking,''
  Phys.\ Rev.\ D {\bf 61} (2000) 026001
  [hep-th/9908135].



\bibitem{Denef}
F.~Denef,
  ``Quantum quivers and Hall / hole halos,''
  JHEP {\bf 0210} (2002) 023
  [hep-th/0206072].


      %\cite{Denef:2007vg}
\bibitem{Denef:2007vg}
  F.~Denef and G.~W.~Moore,
  ``Split states, entropy enigmas, holes and halos,''
  JHEP {\bf 1111} (2011) 129
  [hep-th/0702146 [HEP-TH]].


\bibitem{Reineke}
M.~Reineke,
``The Harder-Narasimhan system in quantum groups and cohomology
of quiver moduli,''
Invent. Math. {\bf 152} (2003) 349-368; arXiv:math/0204059 [math.QA].






   %\cite{deBoer:2008zn}
\bibitem{deBoer:2008zn}
  J.~de Boer, S.~El-Showk, I.~Messamah and D.~Van den Bleeken,
  ``{Quantizing N=2 multicenter solutions},''
  JHEP {\bf 0905} (2009) 002
  [arXiv:0807.4556 [hep-th]].
  %%CITATION = JHEPA,0905,002;%%


 %\cite{Manschot:2010qz}
\bibitem{Manschot:2010qz}
  J.~Manschot, B.~Pioline and A.~Sen,
  ``Wall crossing from Boltzmann black hole halos,''
  JHEP {\bf 1107}, 059 (2011)
  [arXiv:1011.1258 [hep-th]].
  %%CITATION = ARXIV:1011.1258;%%
  %46 citations counted in INSPIRE as of 27 Sep 2013

\bibitem{Lee:2011ph}
  S.~Lee and P.~Yi,
  ``Framed BPS states, moduli dynamics, and wall-crossing,''
  JHEP {\bf 1104} (2011) 098
  [arXiv:1102.1729 [hep-th]].


%\cite{Kim:2011sc}
\bibitem{Kim:2011sc}
  H.~Kim, J.~Park, Z.~Wang and P.~Yi,
  ``Ab initio wall-crossing,''
  JHEP {\bf 1109}, 079 (2011)
  [arXiv:1107.0723 [hep-th]].
  %%CITATION = ARXIV:1107.0723;%%



  %\cite{Manschot:2011xc}
\bibitem{Manschot:2011xc}
  J.~Manschot, B.~Pioline and A.~Sen,
  ``A Fixed point formula for the index of multi-centered N=2 black holes,''
  JHEP {\bf 1105}, 057 (2011)
  [arXiv:1103.1887 [hep-th]].
  %%CITATION = ARXIV:1103.1887;%%
  %20 citations counted in INSPIRE as of 28 Sep 2013

  %\cite{Lee:2012sc}
\bibitem{Lee:2012sc}
  S.~-J.~Lee, Z.~-L.~Wang and P.~Yi,
  ``Quiver invariants from intrinsic Higgs states,''
  JHEP {\bf 1207}, 169 (2012)
  [arXiv:1205.6511 [hep-th]].
  %%CITATION = ARXIV:1205.6511;%%

\bibitem{Lee:2012naa}
  S.~-J.~Lee, Z.~-L.~Wang and P.~Yi,
  ``BPS states, refined indices, and quiver invariants,''
  JHEP {\bf 1210} (2012) 094
  [arXiv:1207.0821 [hep-th]].

  %\cite{Manschot:2012rx}
\bibitem{Manschot:2012rx}
  J.~Manschot, B.~Pioline and A.~Sen,
  ``From black holes to quivers,''
  JHEP {\bf 1211} (2012) 023
  [arXiv:1207.2230 [hep-th]].



\bibitem{Lee:2013yka}
  S.~-J.~Lee, Z.~-L.~Wang and P.~Yi,
  ``Abelianization of BPS quivers and the refined Higgs index,''
  arXiv:1310.1265 [hep-th].

%\cite{Sen:2011aa}
\bibitem{Sen:2011aa}
  A.~Sen,
  ``Equivalence of three wall-crossing formulae,''
  Commun.\ Num.\ Theor.\ Phys.\  {\bf 6}, 601 (2012)
  [arXiv:1112.2515 [hep-th]].
  %%CITATION = ARXIV:1112.2515;%%
  %12 citations counted in INSPIRE as of 27 Sep 2013


  \bibitem{KS}
  M. Kontsevich and Y. Soibelman,
``{Stability structures, motivic Donaldson-Thomas invariants
and cluster transformations},'' [arXiv:0811.2435]


%\cite{Gaiotto:2008cd}
\bibitem{GMN1}
  D.~Gaiotto, G.~W.~Moore and A.~Neitzke,
``Four-dimensional wall-crossing via three-dimensional field theory,''
  Commun.\ Math.\ Phys.\  {\bf 299} (2010) 163
  [arXiv:0807.4723 [hep-th]].
  %%CITATION = ARXIV:0807.4723;%%

%\cite{Gaiotto:2009hg}
\bibitem{GMN2}
D.~Gaiotto, G.~W.~Moore and A.~Neitzke,
``Wall-crossing, Hitchin Systems, and the WKB approximation,''
arXiv:0907.3987 [hep-th].
%%CITATION = ARXIV:0907.3987;%%


  %\cite{Bena:2012hf}
\bibitem{Bena:2012hf}
  I.~Bena, M.~Berkooz, J.~de Boer, S.~El-Showk and D.~Van den Bleeken,
  ``Scaling BPS solutions and pure-Higgs states,''
  JHEP {\bf 1211}, 171 (2012)
  [arXiv:1205.5023 [hep-th]].
  %%CITATION = ARXIV:1205.5023;%%






\bibitem{BEHT1}
F.~Benini, R.~Eager, K.~Hori and Y.~Tachikawa,
  ``Elliptic genera of two-dimensional N=2 gauge
theories with rank-one gauge groups,''
  Lett.Math.Phys. (2013)
  [arXiv:1305.0533 [hep-th]].

\bibitem{BEHT2}
F.~Benini, R.~Eager, K.~Hori and Y.~Tachikawa,
  ``Elliptic genera of 2d N=2 gauge theories,''
  arXiv:1308.4896 [hep-th].


\bibitem{HoTo}
K.~Hori and D.~Tong,
 ``Aspects of non-Abelian gauge dynamics in two-dimensional N=(2,2) theories,''
  JHEP {\bf 0705}, 079 (2007)
  [hep-th/0609032].

\bibitem{Hduality}
K.~Hori,
 ``Duality in two-dimensional (2,2) supersymmetric non-Abelian gauge theories,''
  JHEP {\bf 1310} (2013) 121
  [arXiv:1104.2853 [hep-th]].

\bibitem{JK}
L. C. Jeffrey and F. C. Kirwan, ``Localization for nonabelian group actions,''
Topology {\bf 34} (1995) 291-327, arXiv:alg-geom/9307001.

\bibitem{BV}
M. Brion and M. Vergne, ``Arrangement og hyperplanes I: Rational
functions and Jeffrey-Kirwan residue,'' 
Ann. Sci. ENS {\bf 32} (1999) 715-741,
arXiv:math/9903178 [math.DG].

\bibitem{SV}
  A. Szenes and M. Vergne, ``Toric reduction and a conjecture of batyrev
and materov,''
Invent. Math. {\bf 158} (2004) 453-495, arXiv:math/0306311 [math.AT].

\bibitem{Talks}
K. Hori, ``1d index and wall-crossing", StringMath 2014, Alberta, June 2014;
P. Yi ``Index theorems for d=1 GLSM and BPS states," Recent
Development in Theoretical Physics, Seoul, June 2014.

%\cite{Cordova:2014oxa}
\bibitem{Cordova}
  C.~Cordova and S.~-H.~Shao,
  ``An index formula for supersymmetric quantum mechanics,''
  arXiv:1406.7853 [hep-th].
  %%CITATION = ARXIV:1406.7853;%%

  %\cite{Hwang:2014uwa}
\bibitem{Hwang}
  C.~Hwang, J.~Kim, S.~Kim and J.~Park,
  ``General instanton counting and 5d SCFT,''
  arXiv:1406.6793 [hep-th].
  %%CITATION = ARXIV:1406.6793;%%
  %1 citations counted in INSPIRE as of 08 Jul 2014




\bibitem{HHP}
M.~Herbst, K.~Hori and D.~Page,
  ``Phases Of N=2 theories In 1+1 dimensions with boundary,''
  arXiv:0803.2045 [hep-th].





\bibitem{WessBagger}
J.~Wess and J.~Bagger,
  {\it Supersymmetry and supergravity},
  Princeton, USA: Univ. Pr. (1992) 259 p


\bibitem{HoKn1}
K.~Hori and J.~Knapp,
  ``Linear sigma models with strongly coupled phases - one parameter models,''
  JHEP {\bf 1311} (2013) 070
  [arXiv:1308.6265 [hep-th]].


\bibitem{Alv}
L.~Alvarez-Gaume,
  ``Supersymmetry and the Atiyah-Singer index theorem,''
  Commun.\ Math.\ Phys.\  {\bf 90} (1983) 161.

\bibitem{FriWin}
D.~Friedan and P.~Windey,
  ``Supersymmetric derivation of the Atiyah-Singer index
and the chiral anomaly,''
  Nucl.\ Phys.\ B {\bf 235} (1984) 395.


  \bibitem{IntVaf}
K.~A.~Intriligator and C.~Vafa,
  ``Landau-ginzburg Orbifolds,''
  Nucl.\ Phys.\ B {\bf 339} (1990) 95.


 \bibitem{JKLMR}
 H.~Jockers, V.~Kumar, J.~M.~Lapan, D.~R.~Morrison and M.~Romo,
  ``Nonabelian 2D gauge theories for determinantal Calabi-Yau varieties,''
  JHEP {\bf 1211} (2012) 166
  [arXiv:1205.3192 [hep-th]].




\bibitem{MP}
D.~R.~Morrison and M.~R.~Plesser,
  ``Summing the instantons: Quantum cohomology
and mirror symmetry in toric varieties,''
  Nucl.\ Phys.\ B {\bf 440} (1995) 279
  [hep-th/9412236].






\bibitem{BHHW}
I.~Brunner, K.~Hori, K.~Hosomichi and J.~Walcher,
  ``Orientifolds of Gepner models,''
  JHEP {\bf 0702} (2007) 001
  [hep-th/0401137].




%\cite{Gaiotto:2010be}
\bibitem{Gaiotto:2010be}
  D.~Gaiotto, G.~W.~Moore and A.~Neitzke,
  ``Framed BPS states,''
  arXiv:1006.0146 [hep-th].
  %%CITATION = ARXIV:1006.0146;%%
  %92 citations counted in INSPIRE as of 27 Sep 2013


%\cite{Manschot:2013sya}
\bibitem{Manschot:2013sya}
  J.~Manschot, B.~Pioline and A.~Sen,
  ``On the Coulomb and Higgs branch formulae for multi-centered black holes and quiver invariants,''
  JHEP {\bf 1305}, 166 (2013)
  [arXiv:1302.5498 [hep-th]].

%\cite{Manschot:2014fua}
\bibitem{Manschot:2014fua}
  J.~Manschot, B.~Pioline and A.~Sen,
  ``The Coulomb branch formula for quiver moduli spaces,''
  arXiv:1404.7154 [hep-th].
  %%CITATION = ARXIV:1404.7154;%%



\bibitem{Sen:2009vz}
  A.~Sen,
  ``Arithmetic of quantum entropy function,''
  JHEP {\bf 0908} (2009) 068
  [arXiv:0903.1477 [hep-th]].
  %%CITATION = ARXIV:0903.1477;%%
  %47 citations counted in INSPIRE as of 08 Jul 2014





\bibitem{KacWit}
S.~Kachru and E.~Witten,
  ``Computing the complete massless spectrum of a Landau-Ginzburg orbifold,''
  Nucl.\ Phys.\ B {\bf 407} (1993) 637
  [hep-th/9307038].

\bibitem{DK}
J.~Distler and S.~Kachru,
  ``(0,2) Landau-Ginzburg theory,''
  Nucl.\ Phys.\ B {\bf 413} (1994) 213
  [hep-th/9309110].







\bibitem{BW}
A.~Borel and A.~Weil (presented by J-P.~Serre),
``Repr\'esentations lin\'eaires et espaces homog\`enes K\"ahl\'eriens
des groupes de Lie compacts,'' S\'eminaire Bourbaki (1954).

\bibitem{BWB}
R.~Bott, ``Homogeneous vector bundles,''
Ann. Math. {\bf 66} (1957) 203-248.



\end{thebibliography}
\end{document}